\documentclass[a4paper]{article}
\pdfoutput=1
\usepackage{jcappub}
\usepackage{amsmath}
\usepackage{bm}
\usepackage{graphicx}
\usepackage{enumitem}
\usepackage{geometry}
\usepackage{xspace}
\usepackage{graphicx}

\usepackage[compat=1.1.0]{tikz-feynman}

\usepackage{mathtools}

\usepackage{relsize}

\allowdisplaybreaks 

\makeatletter
\gdef\@fpheader{}
\g@addto@macro\bfseries{\boldmath}
\makeatother

\newcommand{\ie}{{i.e.~}}
\newcommand{\cf}{\textsl{cf.~}}

\let\oldsqrt\sqrt
\def\sqrt{\mathpalette\DHLhksqrt}
\def\DHLhksqrt#1#2{%
\setbox0=\hbox{$#1\oldsqrt{#2\,}$}\dimen0=\ht0
\advance\dimen0-0.2\ht0
\setbox2=\hbox{\vrule height\ht0 depth -\dimen0}%
{\box0\lower0.4pt\box2}}

\newcommand{\dd}{\mathrm{d}}

\newcommand{\uin}{\mathrm{in}}

\newcommand{\bmk}{\boldmathsymbol{k}}

\newcommand{\Rea}{\Re \mathrm{e}\,}

\newcommand{\GeV}{\mathrm{GeV}}

\newcommand{\Mp}{M_\usssPl}

\newcommand{\efolds}{$e$-folds}

\newcommand{\beq}{\begin{equation}}
\newcommand{\eeq}{\end{equation}}
\newcommand{\bea}{\begin{equation}\begin{aligned}}
\newcommand{\eea}{\end{aligned}\end{equation}}

\newlength{\wsingfig}
\newlength{\wdblefig}
\newlength{\wquadfig}
\newlength{\wtriplefig}
\newcommand{\Eq}[1]{eq.~(\ref{#1})}
\newcommand{\Eqs}[1]{eqs.~(\ref{#1})}
\newcommand{\Fig}[1]{fig.~{\ref{#1}}}

\newcommand{\RRef}[1]{ref.~{\cite{#1}}}
\newcommand{\Refs}[1]{refs.~{\cite{#1}}}
\newcommand{\Sec}[1]{sec.~\ref{#1}}

\newcommand{\App}[1]{appendix~\ref{#1}}

\newcommand{\deflen}[2]{%
    \expandafter\newlength\csname #1\endcsname
    \expandafter\setlength\csname #1\endcsname{#2}%
}

\usepackage{caption}
\usepackage{subcaption}
\usepackage{mathrsfs}
\usepackage{graphicx}

\newcommand{\exd}{{ \mathrm{d}} }
\def\bea{\begin{eqnarray}}
\def\eea{\end{eqnarray}}

\def\cO{{\mathcal{O}}}

\def\cG{{\mathcal{G}}}
\def\cJ{{\mathcal{J}}}
\def\cP{{\mathcal{P}}}
\def\cQ{{\mathcal{Q}}}
\def\cR{{\mathcal{R}}}
\def\ssA{{\scriptscriptstyle A}}
\def\ssB{{\scriptscriptstyle B}}
\def\ssP{{\scriptscriptstyle P}}

\def\ssS{{\scriptscriptstyle \mathrm{S}}}
\def\ssT{{\scriptscriptstyle \mathrm{T}}}

\def\mfF{{\mathfrak{F}}}
\def\kUV{{k_{\UV}}}
\def\zin{{z_{\mathrm{in}}}}

\def\Trenv{\underset{B}{\mathrm{Tr}}}
\def\smath#1{\text{\scalebox{.85}{$#1$}}}
\def\sfrac#1#2{\smath{\frac{#1}{#2}}}

\def\pref#1{(\ref{#1})}
\def\Mp{M_\mathrm{p}}
\def\ellp{\ell_\mathrm{p}}
\def\slrl{\varepsilon_1}
\def\mfp{{\mathfrak p}}

\def\bmk{{\bm{k}}}
\def\cH{{\cal H}}
\def\cI{{\cal I}}

\def\cV{{\cal V}}
\def\IR{{\scriptscriptstyle \mathrm{IR}}}
\def\UV{{\scriptscriptstyle \mathrm{UV}}}
\def\nn{\nonumber}
\newcommand{\roughly}[1]{\mathrel{\raise.3ex\hbox{$#1$\kern-0.55em
\lower1ex\hbox{$\sim$}}}}
\newcommand{\lsim}{\roughly<}
\newcommand{\gsim}{\roughly>}
\def\w{\tilde v}
\def\p{\tilde p}

\PassOptionsToPackage{cmyk}{xcolor}

\subheader{}

\title{Minimal decoherence from inflation}

\author[a,b,c,d]{C.P.~Burgess,}
\author[e]{R.~Holman,}
\author[a,b,f]{Greg Kaplanek,}
\author[g]{\\J\'er\^ome Martin}
\author[h,i]{and Vincent Vennin}

\affiliation[a]{Department of Physics \& Astronomy, 
McMaster University, Hamilton, ON, Canada, L8S 4M1}

\affiliation[b]{Perimeter Institute for Theoretical Physics, 
Waterloo, ON, Canada, N2L 2Y5}

\affiliation[c]{Department of Theoretical Physics, CERN, Gen\`eve 23,
  Switzerland}

\affiliation[d]{School of Theoretical Physics, Dublin Institute for
  Advanced Studies,\\ \phantom{Bigskip} 10 Burlington Rd., Dublin,
  Co. Dublin, Ireland}

\affiliation[e]{Minerva University, 14 Mint Plaza, San Francisco,
  CA 94103, USA}

\affiliation[f]{Theoretical Physics, Blackett Laboratory, Imperial
  College, London, SW7 2AZ, UK}

\affiliation[g]{Institut d'Astrophysique de Paris, UMR 7095-CNRS,
  Universit\'e Pierre et Marie Curie,\\ \phantom{Bigskip} 98bis
  boulevard Arago, 75014 Paris, France}

\affiliation[h]{Laboratoire de Physique de l'\'Ecole Normale
  Sup\'erieure, ENS, CNRS,\\ \phantom{Bigskip} Universit\'e PSL,
  Sorbonne Universit\'e, Universit\'e Paris Cit\'e, F-75005 Paris,
  France}

\affiliation[i]{Laboratoire Astroparticule et Cosmologie, CNRS,
  Universit\'e de Paris,\\ \phantom{Bigskip} 10 rue Alice Domon et
  L\'eonie Duquet, 75013 Paris, France}

\emailAdd{cburgess@perimeterinstitute.ca}
\emailAdd{rholman@minerva.edu}
\emailAdd{g.kaplanek@imperial.ac.uk}
\emailAdd{jmartin@iap.fr}
\emailAdd{vincent.vennin@ens.fr}

\date{today}

\begin{document}

\sloppy

\abstract{We compute the rate with which super-Hubble cosmological
  fluctuations are decohered during inflation, by their gravitational
  interactions with unobserved shorter-wavelength scalar and tensor
  modes.  We do so using Open Effective Field Theory methods, that
  remain under control at the late times of observational interest,
  contrary to perturbative calculations.  Our result is minimal in the
  sense that it only incorporates the self-interactions predicted by
  General Relativity in single-clock models (additional interaction
  channels should only speed up decoherence). We find that decoherence
  is both suppressed by the first slow-roll parameter and by the
  energy density during inflation in Planckian units, but that it is
  enhanced by the volume comprised within the scale of interest, in
  Hubble units. This implies that, for the scales probed in the Cosmic
  Microwave Background, decoherence is effective as soon as inflation
  proceeds above $\sim 5\times 10^{9}\, \GeV$. Alternatively, if
  inflation proceeds at GUT scale decoherence is incomplete only for
  the scales crossing out the Hubble radius in the last $\sim 13$
  \efolds \, of inflation. We also compute how short-wavelength scalar modes decohere primordial tensor perturbations, finding a faster rate unsuppressed by slow-roll parameters.
Identifying the parametric dependence of decoherence, and the rate at
which it proceeds, helps suggest ways to look for quantum
effects. 
}

\subheader{CERN-TH-2022-174}
 
\maketitle

\section{Introduction}
\label{sec:intro}

One of the most remarkable consequences of cosmology's recent
conversion into a precision observational science has been the
detection of large-scale correlations in the way that matter and
radiation are distributed across the observable universe
\cite{eBOSS:2020yzd, Planck:2018vyg}. These correlations point to a
pattern of nearly scale-invariant primordial fluctuations inherited
from the much-earlier universe about which otherwise little is known.

Even more remarkably these primordial fluctuations share the spectral
properties expected of quantum fluctuations, if these were stretched
across the sky in the remote past by some sort of accelerated
universal expansion
\cite{Mukhanov:1981xt,Guth:1982ec,Hawking:1982cz,Starobinsky:1982ee,Bardeen:1983qw,Mukhanov:1988jd}. A
natural question in any such a picture is how initial quantum
fluctuations become the classical fluctuations that are known to
describe the later observations so well
\cite{Brandenberger:1990bx,Kiefer:2008ku,Polarski:1995jg,Kiefer:1998qe,Lombardo:2005iz,Burgess:2006jn,Martineau:2006ki,Sharman:2007gi,Burgess:2014eoa}.
Part of this question concerns precisely what is meant by `classical'
in this context ({\it e.g.}~tree-level vs loop-level; WKB-squeezed vs
generic quantum states; decoherence and so on).

In this paper our interest is the evolution of a quantum field's
reduced density matrix, $\langle \varphi_1 \, | \, \varrho \, | \,
\varphi_2 \rangle$, in the field basis, where $\varrho =
\hbox{Tr}_{\rm env}\, \rho$ is obtained by tracing out certain
`unobserved' degrees of freedom (in the language of open systems what
is called the `environment'). Classicalization will mean decoherence,
in the sense that $\varrho$ evolves from a pure to a mixed state; in
particular one whose off-diagonal elements rapidly fall to zero in the
field basis. Once $\varrho$ becomes diagonal in this way it is
indistinguishable from a non-quantum statistical ensemble of classical
field configurations with probability distribution $P[\varphi] =
\langle \varphi \, | \, \varrho \, | \, \varphi \rangle$.
 
Although it has long been recognized that existing cosmological
measurements are largely insensitive to any off-diagonal components
$\langle \varphi_1 \, | \, \varrho \, | \, \varphi_2 \rangle$
\cite{Polarski:1995jg,Kiefer:1998qe}, proposals are now being made to
circumvent this to seek observational evidence for a quantum origin
for primordial fluctuations
\cite{Martin:2015qta,Martin:2017zxs,Campo:2005sv,Maldacena:2015bha}. The
efficiency of any primordial decoherence is likely relevant to such
searches, and in particular rapid decoherence can make it unlikely
that quantum effects survive to the present day to be
seen~\cite{Martin:2021znx}.

An obstacle must be surmounted and a choice must be made in order to
describe such primordial quantum-to-classical transitions. Decoherence
requires an environment: not all degrees of freedom should be
measured. So the choice is to identify the environmental modes whose
removal decoheres the fluctuations we can see. Since super-Hubble
modes play a special role by freezing in early-universe effects,
smaller wavelength sub-Hubble modes are natural candidates for a
decohering environment. We therefore study how rapidly super-Hubble
metric fluctuations are decohered by shorter-wavelength metric
perturbations, using only the mutual gravitational self-interactions
predicted by General Relativity.

The obstacle to be surmounted is more technical: the time interval
between fluctuation generation and detection can be extremely large,
and effects that were initially small can have enough time to become
large during the long wait in between. This can cause a breakdown of
perturbative methods; what are often called `secular growth' effects
in
cosmology~\cite{Tsamis:2005hd,Burgess:2009bs,Giddings:2011zd}. Similar
late-time breakdown of perturbative methods are ubiquitous elsewhere
in physics because no matter how small a perturbing interaction
$\cH_{\rm int} \ll \cH_0$ is, it is always true that
$e^{-i(\cH_0+\cH_{\rm int})t}$ is not well-approximated by $e^{-i\cH_0
  t}(1 - i \cH_{\rm int}t + \cdots)$ at sufficiently late times. The
good news is: because these problems are ubiquitous, tools for
circumventing them and making reliable late-time predictions are also
very well-developed \cite{Burgess:2020tbq}. All that is required is to
adapt these tools to cosmology~\cite{Colas:2022hlq}.

In this paper we follow up on earlier work adapting to gravity
well-developed tools from the quantum theory of open
systems,\footnote{\Refs{Kaplanek:2019dqu, Kaplanek:2019vzj, Kaplanek:2020iay, Kaplanek:2021fnl} explore the use of Open EFT
  techniques for the much simpler case where late-time predictions are
  only sought for an Unruh-DeWitt qubit detector~\cite{Unruh:1976db,
    DeWitt:1980hx}, rather than for the entire $\phi$ field, for which
  very explicit calculations can be performed.} whose use to explore
late-time evolution we call Open Effective Field Theory (Open
EFT)~\cite{Burgess:2014eoa, Agon:2014uxa, Burgess:2015ajz, Braaten:2016sja, Martin:2018zbe, Martin:2018lin, Burgess:2021luo, Colas:2022hlq, brahma2022universal, Hammou:2022sol}. These tools show how the evolution equation for the reduced density matrix in the
interaction picture can be brought into an approximate (schematic)
form
\begin{equation}
\label{LindbladCartoon}
\partial_t \varrho = -i \left[ \overline\cH_{\rm int} \,, \varrho \right]
+  \mathscr{L}_{2}\left(\varrho \right)+ \cO\left(\cH_{\rm int}^3\right)
\, \hbox{terms}\,,
\end{equation}
where $\cH_{\rm int}$ denotes the terms in the interaction Hamiltonian
that couple the environment to the measured degrees of freedom and
$\overline\cH_{\rm int}$ denotes its average over the
environment. $\mathscr{L}_2$ contains all terms second order in
$\cH_{\rm int}$ and in many circumstances has a Lindblad
form~\cite{Lindblad:1975ef, Gorini:1976cm}, which is derived below in
some detail. Lindblad equations can have solutions that re-sum
late-time behaviour even if the evolution equation itself is only
computed perturbatively. Any terms not written explicitly are at least
cubic in $\cH_{\rm int}$.

We here apply these tools to compute the decoherence of super-Hubble
scalar fluctuations of the metric in the simplest single-clock near-de
Sitter (inflationary) cosmologies that are the best-explored
explanations of primordial fluctuations \cite{Baumann:2009ds}. An
important observation is that the linear term in $\cH_{\rm int}$ never
contributes to decoherence because it simply represents Liouville
evolution (which can never take pure states to mixed states). All
decoherence effects necessarily first arise at second order in
$\cH_{\rm int}$, and this plays an important role when identifying the
dominant interactions.

We work within the standard joint slow-roll and semi-classical
expansions that track powers of small slow-roll parameters,
$\varepsilon_i$, and powers of the small loop-counting parameter $GH^2
= H^2/(8\pi \Mp^2)$ where $G$ is Newton's constant, $H$ is the
inflationary Hubble scale and $\Mp$ is the reduced Planck
mass.\footnote{We follow the power-counting estimates of
  \Refs{Adshead:2017srh, Babic:2019ify} for these two expansion
  parameters and use fundamental units throughout (for which $\hbar =
  c = 1$).}  For super-Hubble modes of co-moving momentum $\bm{k}$
these are supplemented by an additional expansion in powers of $k/(a
H)$ where $k = |\bm{k}|$.

General relativity predicts that in such an expansion fluctuations of
the metric self-interact.
Of these only the cubic interactions -- whose detailed form is worked
out for near-de Sitter geometries in \RRef{Maldacena:2002vr}
(summarized for convenience in Appendix \ref{App:operators}) --
contribute at leading order when considering how interactions with
shorter wavelength modes decohere the long-wavelength super-Hubble
modes relevant for primordial fluctuations.

There is a simple reason why only cubic interactions dominate. As
argued above, decoherence first arises at second order in $\cH_{\rm
  int}$ and consequently does so at order $1/\Mp^2$. Although quartic
interactions can also contribute to fluctuation evolution at order
$1/\Mp^2$ they do so in \pref{LindbladCartoon} only through terms
linear in $\cH_{\rm int}$, and so cannot cause decoherence. Quick
inspection of the interactions listed in \RRef{Maldacena:2002vr} (and
Appendix \ref{App:operators}) shows that all but two of these are
additionally suppressed for super-Hubble modes, either by additional
factors of slow-roll parameters or by additional powers of $k/(aH)$.

For super-Hubble scalar metric fluctuations the two relevant
interactions involve either $v\, \partial^i v \, \partial_i v$ or $v
\, \partial^k v^{ij} \, \partial_k v_{ij}$ (where $v$ denotes the
Mukhanov-Sasaki scalar perturbation and $v_{ij}$ is the tensor
perturbation) and both contribute with equal strength. These
respectively describe decoherence generated by short-wavelength scalar
and tensor fluctuations. In passing we also compute how the
interaction $v^{ij} \, \partial_i v \, \partial_j v$ allows
short-wavelength scalar modes to decohere super-Hubble tensor modes,
finding them to be less suppressed by slow-roll parameters. We do not
compute the similar-sized contribution of short-wavelength tensor
modes towards the decoherence of long-wavelength tensors.

To determine the effect of the dominant cubic interactions we compute
the evolution equation for the reduced density matrix describing the
quantum state of the subset of modes visible to late-time observers
like ourselves. We show why this evolution is very quickly
well-approximated by a Lindblad equation describing Markovian
evolution for super-Hubble modes during inflation. We then integrate
this equation to identify the late-time evolution of $\varrho$ where
perturbation theory naively breaks down. We use this to compute a
mode's decoherence over time and show that it is already very rapid
despite the feeble gravitational strength of the
interaction. Inclusion of other interactions is likely only to speed
up the decoherence process.

As has been remarked elsewhere \cite{Kiefer:2006je, Burgess:2014eoa}
the squeezing of modes during inflation \cite{Albrecht:1992kf}
explains in a simple way why the density matrix diagonalizes in a
basis of field eigenstates; making these the system's natural
`pointer' basis. It is the surviving diagonal elements $P[\varphi] =
\langle \varphi \, | \,\varrho\, | \,\varphi \rangle$ at which
stochastic
\cite{Starobinsky:1986fx,Starobinsky:1994bd,Mijic:1994vv,Seery:2010kh,Prokopec:2008gw}
and de Sitter EFT \cite{Cohen:2020php, Cohen:2021fzf,
  Baumgart:2019clc} methods ultimately aim.

Our result for the amplitude of decoherence is given by
\Eq{eq:purity:exactr} and arises suppressed by the gravitational
loop-counting parameter $(H/\Mp)^2$ and by the first slow-roll
parameter~\cite{Liddle:1994dx, Schwarz:2001vv, Leach:2002ar}, $\slrl =
-\dot H/H^2$, leading to an amplitude controlled by\footnote{The same
  arguments imply tensor modes decohere with an amplitude $H^2/\Mp^2$
  ({\it i.e.}~unsuppressed by $\slrl $).}
\begin{equation}
\frac{\slrl  H^2}{8\pi \Mp^2} \sim \slrl ^2
\, \mathcal{P}_\zeta\lesssim 10^{-4} \times 10^{-10}
\end{equation}
where $\mathcal{P}_\zeta(k)\simeq H^2/(8\pi^2 \slrl \Mp^2) \sim
10^{-10}$ is the observed size of scalar perturbations and $\slrl
\lsim 10^{-2}$ is bounded above by the non-observance of primordial
tensor perturbations \cite{Planck:2018jri}. But this small amplitude
is abundantly compensated by an exponential growth since
\pref{eq:purity:exactr} grows during inflation proportional to
$(aH/k)^3 \propto e^{3Ht}$.

For $\rho_\mathrm{inf}^{1/4}\gsim 5\times 10^{9}\mbox{GeV}$, or
equivalently a tensor to scalar ratio $r\gsim 6.5\times 10^{-28}$,
this predicts classicalization of CMB scales is long completed before
inflation ends.  Alternatively, if $r \sim 10^{-3}$ (and so the
discovery of primordial tensor fluctuations is within reach) then
decoherence becomes important for modes that spend more than around
$\sim 13$ $e$-folds outside the Hubble scale during inflation. All of
these estimates assume no additional decoherence occurs (or
disappears) after inflation ends, or occurs during inflation due to
other interactions with short-wavelength modes, or due to interactions
with other environmental degrees of freedom.

Our presentation is structured as follows. \Sec{sec:OpenEFT} starts by
setting up the Open EFT relevant to the scalar fluctuations of the
metric using only the standard building blocks of single-clock
inflation. We focus on the implications of the dominant cubic
self-interactions, taken from amongst the cubic interaction terms
outlined in \RRef{Maldacena:2002vr}. We focus initially on
interactions involving only scalar modes (returning to include tensors
in \Sec{sec:Lindblad}) setting up the system and environmental degrees
of freedom in terms of the super-Hubble and sub-Hubble modes of the
Mukhanov-Sasaki field $v$.

\Sec{sec:Lindblad} derives the relevant late-time evolution equation
for super-Hubble modes, showing that it has the form of a Lindblad
equation for each mode $\bm{k}$ of the field. This is done by first
passing through the intermediate step of deriving a Nakajima-Zwanzig
master equation and then carefully identifying the regime in which it
becomes approximately Markovian. The environmental correlation
functions appearing in this Lindblad evolution are evaluated
explicitly and it is shown how the ultraviolet (UV) divergences
encountered when doing so can be re-normalized. We also provide here a
preliminary discussion of the issues of the gauge-dependence of our
formalism.

\Sec{sec:Observables} then applies these results to compute some
implications for observable modes from the removal of their shorter
wavelength counterparts. Two observables computed are ($i$) very small
corrections that are predicted for the power spectrum and ($ii$) the
late-time decoherence that is implied for super-Hubble modes by the
tracing out of these unobserved short-wavelength degrees of freedom.

We conclude in \Sec{sec:Conclusions} with a brief discussion of the
open ends that our calculation does not resolve and possible next
steps. 
Included in this
discussion is a calculation of how short-wavelength scalar modes
decohere super-Hubble tensor modes during inflation. This suffices to
confirm the dependence on small parameters predicted by power-counting
arguments but leaves open the contribution of short-wavelength tensor
modes to the decoherence of primordial tensor fluctuations.

\section{Open system of super-Hubble metric modes}
\label{sec:OpenEFT}

This section sets up the open-system framework for describing the
self-interactions of metric fluctuations in a near-de Sitter
geometry. The system of interest is as found in many of the simplest
single-clock inflationary models, with the metric $g_{\mu\nu}$ coupled
to a real scalar field $\varphi$ through
\begin{equation}
\label{actionstart}
S = \int \dd^4 x\; \sqrt{ - g} \bigg[ \frac{\Mp^2}{2} \, R
    - \frac{1}{2} g^{\mu\nu} \, \partial_{\mu} \varphi \,
    \partial_{\nu} \varphi  - V(\varphi) \bigg]
\end{equation}
where $\Mp^{-2} = 8 \pi G$, $R$ is the Ricci scalar and $V(\varphi)$
is the potential energy of the inflaton $\varphi$.  Our focus is on
the late-time evolution of fluctuations about a homogeneous near de
Sitter geometry given by $\varphi = \phi(t)$ and the metric
\begin{equation}
\label{metric}
\dd s^2 = - \dd t^2 + a^2(t) \, \dd \bm{x}^2 = a^2(\eta)
  \left( - \dd \eta^2 + \dd\bm{x}^2 \right)
\end{equation}
with scale factor
\begin{equation}
\label{eq:defscalefactor}
a \simeq e^{Ht} \simeq-\frac{1}{H\eta} \,.
\end{equation}
Here $H$ is the usual Hubble scale, $H = \dot a/a$,
computed\footnote{We denote derivatives with respect to $t$ with
  overdots and derivatives with respect to $\eta$ using primes.} using
cosmic time $t$ (to which conformal time $\eta$ is related by $a \,\dd
\eta = \dd t$).

\subsection{Curvature perturbation and self-interactions}

To this end we expand the scalar field about its homogeneous
background, $\varphi(t,\bm{x}) = \phi(t) + \delta \varphi(t,\bm{x})$,
and employ the ADM decomposition to describe small metric fluctuations
about metric by foliating the space-time into a family of space-like
hyper-surfaces,
\begin{eqnarray} \label{ADMmetric}
  \dd s^2 = - N^2 \dd t^2 + h_{ij} \big( \dd x^i + N^i \dd t \big)
  \big( \dd x^j + N^j \dd t \big) \ .
\end{eqnarray}
After picking a gauge to fix time and spatial reparametrizations,
standard arguments reveal that the scalar fluctuations described by
the action (\ref{actionstart}) end up being described by a single
physical scalar degree of freedom plus tensor fluctuations.  More
specifically, we follow \RRef{Maldacena:2002vr} and write the metric
fluctuation to second order as
\begin{equation} \label{metricsplit}
     h_{ij} = a^2 e^{2\zeta} \hat h_{ij} \quad\hbox{with}\quad
     \hat h_{ij} = \delta_{ij} + \gamma_{ij} + \frac12 \,
     \delta^{kl} \gamma_{ik} \gamma_{lj} + \cdots \,,
\end{equation}
where $\det \hat h_{ij} = 1$ and $\delta^{ij} \partial_i \gamma_{jk} =
\delta^{ij} \gamma_{ij} = 0$. Two convenient gauge choices are then
obtained by either setting $\delta \varphi = 0$ (co-moving gauge) or
setting $\zeta = 0$ (spatially-flat gauge). Dependence on the
slow-roll parameters is easier to follow when $\delta \varphi$ is the
scalar variable (since its expected amplitude does not involve the
slow-roll parameters) but super-Hubble time-evolution is clearer using
the variable $\zeta$ (because $\zeta$ as defined in \pref{metricsplit}
becomes time-independent to all orders in fluctuations in the limit $k
\to 0$). This is why hereafter we work in the co-moving gauge. We here
temporarily drop the tensor fluctuation,\footnote{{The variable
    $v_{ij}$ used above is equivalent to $\gamma_{ij}$, just
    normalized differently -- see \pref{canonicaltensor}.}}
$\gamma_{ij}$, and focus exclusively on scalar perturbations, since
these are the ones that have arguably been observed through
cosmological measurements, but circle back to reconsider tensor
fluctuations in \Sec{ssec:Tensor}.

The leading (quadratic) part of the action governing fluctuations
comes from expanding \pref{actionstart} in powers of $\zeta$ and has
the form (see for example \Refs{Kodama:1985bj, Mukhanov:1990me,
  Maldacena:2002vr} -- with some details also given in Appendix
\ref{App:operators})
\begin{equation}
\label{freescalaraction}
  ^{(2)}S = \int \dd t \; \dd^3 \bm{x}\;
  \frac{\dot{\phi}^2}{2H^2}\bigg[ a^3 \dot{\zeta}^2
    - a (\partial \zeta )^2 \bigg] \ ,
\end{equation}
where we recall that $\phi(t)$ denotes the background value of the
inflaton
and $(\partial \zeta)^2 = \delta^{ij} \partial_i \zeta \, \partial_j
\zeta$. The kinetic term can be made canonical by re-expressing in
terms of the Mukhanov-Sasaki
variable~\cite{Mukhanov:1981xt,Kodama:1985bj}, given by
\begin{equation}
\label{eq:defzeta}
v(\eta, {\bm x})=a\Mp \sqrt{2\slrl } \, \zeta(\eta,{\bm x})\,,
\end{equation}
where 
\begin{equation}
  \slrl =-\frac{\dot{H}}{H^2} 
\end{equation}
is the first slow-roll parameter -- related to the field velocity
through $\dot{\phi}^2=2H^2\Mp^2\slrl $ -- which is small if the
background geometry is near de Sitter. In terms of $v$ the quadratic
part of the action given in \Eq{freescalaraction} takes the canonical
form \cite{Mukhanov:1990me}
\begin{equation}
\label{eq:action}
{}^{\left(2\right)} S=\frac{1}{2}
\int{\dd \eta\; \dd^3 \bm{x} \;
\bigg[\left(v^\prime\right)^2
-\delta^{ij}\partial_i v \, \partial_jv
+\frac{\left(a\sqrt{\slrl }\right)^{\prime\prime}}{a\sqrt{\slrl }}\;
v^2 \bigg]}\, .
\end{equation}

Interactions arise 
at cubic and higher orders in the fluctuations, with $S_{\rm int} =
{}^{(3)}S + {}^{(4)}S + \cdots$ where ${}^{(n)}S$ involves $n$ powers
of the fluctuation fields. Amongst the self-interactions involving
just $\zeta$ obtained in this way is
\begin{equation}
\label{cubiczeta}
{}^{(3)}S \supset \int\ \dd t \; \dd^3 \bm{x} \;
\frac{\dot{\phi}^4 a}{4H^4\Mp^2} \left(\partial\zeta\right)^2\zeta
\end{equation}
where the symbol ``$\supset$'' emphasizes that there are other cubic
interactions in ${}^{(3)}S$ that are not explicitly written (for a
full list of the cubic scalar interactions see Eq.~(3.9)
of~\RRef{Maldacena:2002vr}, or~\Eq{allscalarcubics} in
\App{App:operators}).

As argued in the introduction, quartic and higher interactions beyond
those cubic in $v$ need not be considered when computing decoherence
of super-Hubble modes because they can contribute only sub-dominantly
in $1/\Mp$. The variable $v$ is convenient when counting factors of
$1/\Mp$ in this way because its lowest-order correlations functions
are independent of $\Mp$.  In terms of $v$ the cubic interaction
(\ref{cubiczeta}) becomes

\begin{equation}
\label{cubic:int}
{}^{\left(3\right)}S \supset
\int\dd \eta\; \dd^3 \bm{x}\; \frac{\sqrt{\slrl }}{2\sqrt{2}\Mp\, a}
\left(\delta^{ij}\partial_iv\partial_j v\right) v\,,
\end{equation}
revealing it also to be order $\sqrt{\slrl }$ in slow-roll. What is
important for our later purposes is that all of the other cubic
interactions listed in \RRef{Maldacena:2002vr} are either higher-order
in slow-roll parameters or trade two spatial derivatives for two time
derivatives.\footnote{The neglect of time derivatives relative to
  spatial derivatives acting on environmental modes is only justified
  when the environmental modes are also super-Hubble and need not be
  a good approximation for modes with $k/a \sim H$. Our calculations shed
  some light on the validity of this approximation, though the issue
  remains a partially open problem. See
  \Sec{sec:Observables} for further discussion.} The freezing of $\zeta$ on super-Hubble scales implies
$\dot\zeta/\zeta \propto k^2/(aH)^2$ for $k \ll aH$ and so implies
time derivatives contribute only sub-dominantly in powers of $k/(aH)$
for super-Hubble modes.

The momentum conjugate to $v$ is $p= \delta S / \delta v' =
v'$, and the Hamiltonian is
\begin{equation}
\label{eq:Hamiltonian:cubic}
\cH(\eta) = \cH_0(\eta) + \cH_{\mathrm{int}}(\eta)
\end{equation}
with free part
\begin{equation}
\label{freeHclassicalx}
\cH_0 := \frac{1}{2} \int\dd^3 \bm{x} \; \left[ p^2+
\delta^{ij}\partial_iv \, \partial_jv  - \frac{\big(a\,
\sqrt{\slrl }\big)^{\prime\prime}}{a\sqrt{\slrl }} \; v^2 \right]\,.
\end{equation}
The interaction corresponding to \pref{cubic:int} is 
\begin{equation}
\label{Hclassicalint}
\cH_{\mathrm{int}} \supset - \frac{\sqrt{\slrl }}{2\sqrt{2} \Mp a}
\int\dd^3 \bm{x}\; \delta^{ij} v\, \partial_iv \,\partial_j v + \cdots \,.
\end{equation}
This form of the interaction Hamiltonian is obtained in the co-moving
gauge. Although gauge-independent observables are difficult to
construct at cubic and higher order in cosmological perturbation
theory, in \RRef{Maldacena:2002vr} it was checked that a calculation
performed in the spatially-flat gauge gives the same result for the
bispectrum. This supports the idea that physical results obtained from our
calculations using this interaction Hamiltonian will be gauge
independent.

Our goal is to describe dynamics perturbatively in $\cH_{\rm int}$ and
so it is useful first to diagonalize $\cH_0$. This is achieved in
momentum space,
\begin{equation}
\label{eq:v:Fourier}
v(\eta,\bm{x}) = \int \frac{\dd^3 \bm{k}}{(2\pi)^{3/2}} \; 
v_{\bm{k}}(\eta) \, e^{i \bm{k}\cdot\bm{x}}\,,
\end{equation}
and writing $ v_{\bm{k}}(\eta)$ in terms of mode functions $u_{\bm{k}}(\eta)$
\begin{equation} \label{vhatk}
  { v}_{\bm{k}}(\eta) = u_{\bm{k}}(\eta) {c}_{\bm{k}}
  + u^{\ast}_{-\bm{k}}(\eta) {c}^{\dagger}_{-\bm{k}} \,.
\end{equation}
shows that hermiticity in real space $ v(\eta,\bm{x}) =
v^\dagger(\eta,\bm{x})$ implies $ v_{-\bm{k}}(\eta) =
v_{\bm{k}}^\dagger(\eta)$ in momentum space, and these are both
equivalent to having particles and antiparticles not being independent
of one another. A similar expression holds for the conjugate momentum
field $ p_{\bm{k}}$. Equal-time commutation relations for the
operators ${ v}_{\bm{k}}(\eta)$ and $ p_{\bm{k}}(\eta)$
\begin{equation}
\label{equaltimeCCR}
{\big[} { v}_{\bm{k}}(\eta),  p_{\bm{q}}(\eta) \big] = i \delta(\bm{k} + \bm{q}),
\end{equation}
are equivalent to $[{c}_{\bm{k}}, {c}^{\dagger}_{\bm{q}}] = \delta(\bm{k} -
\bm{q})$, provided the $u_{\bm{k}}(\eta)$ are normalized by $u_{\bm{k}}
u_{-\bm{k}}^{*\prime}-u_{\bm{k}}^* u_{-\bm{k}}'=i$.

Plugging this decomposition into \pref{freeHclassicalx} gives, at the
classical level,
\begin{equation}
\label{freeHclassical}
\cH(\eta)  :=  \frac{1}{2}\int  \dd^3 \bm{k} \;
\big[  p_{\bm{k}}(\eta)  p_{\bm{k}}^*(\eta)
+ \omega^2(\bm{k},\eta)  v_{\bm{k}}(\eta)  v_{\bm{k}} ^*(\eta) \big] \,.
\end{equation}
On quantization noncommuting operators are replaced by their
symmetrized product -- such as $ p_{\bm{k}} p^*_{\bm{k}} \to \frac12
\{ p_{\bm{k}} \,, p^\dagger_{\bm{k}} \}$ -- so that hermiticity is
preserved. The time-dependent frequency is
\begin{equation}
\label{omegadef}
\omega^2(\bm{k},\eta) := k^2 - \frac{(a\sqrt{\slrl })''}{a\sqrt{\slrl }} \ .
\end{equation}
In the limit $\slrl \to 0$ this frequency function
$\omega(\bm{k},\eta)$ takes the well-known de Sitter form
\begin{equation}
\label{eq:defomega}
\omega^2(\bm{k},\eta) \simeq k^2 - \frac{2}{\eta^2} \,,
\end{equation}
and describes adiabatic evolution in the regime $k^2 \eta^2 \gg 1$.

In what follows we quantize using a field basis rather than the
particle Fock space built using the creation and annihilation
operators $c_{\bm{k}}$ and $c^\dagger_{\bf{k}}$.  For the free system
this is the analog of treating harmonic oscillators using states
$\langle x | \Psi \rangle$ and density matrices $\langle x | \rho | y
\rangle$ described in the position basis (rather than using
occupation-number representations $\langle n | \Psi \rangle$ and
$\langle n | \rho | m \rangle$ built from the raising and lowering
operators $c$ and $c^\dagger$).  We briefly pause here to clarify an
issue that arises due to the reality condition $v_{-\bm{k}} =
v^\dagger_{\bm{k}}$.

Normally the position eigenvalue $x$ for an oscillator is real and so
having a complex field $v_{\bm{k}}$ for each $\bm{k}$ sounds like it
contains too many coordinates to describe a single oscillator for each
$\bm{k}$. For a complex field (for which particles and antiparticles
are not identified) this is correct: complex coordinates correspond to
two sets of real position coordinates and these correspond to the two
types of oscillator -- one each for particle and antiparticle -- that
exist for each $\bm{k}$.

For real fields the condition $v_{-\bm{k}} = v^\dagger_{\bm{k}}$ cuts
the number of oscillators in half and so leaves a single oscillator
for each $\bm{k}$. There are two equivalent ways to frame the field
representation in this case. We can either restrict ourselves to only
half of the total available momentum labels and keep the complex
variables $v_{\bm{k}}$ arbitrary, or we can keep all momentum labels
and use the reality condition to have effectively only a single real
field for each $\bm{k}$.

To see how these are related in detail we follow \RRef{Martin:2018zbe}
and write the real and imaginary parts of $v_{\bm{k}}$
\begin{equation} \label{alphaRI}
v_{\bm{k}}(\eta) =: \frac{v^{(\mathrm{R})}_{\bm{k}}(\eta)
+i \, v^{(\mathrm{I})}_{\bm{k}}(\eta)}{\sqrt{2}},
\end{equation}
for which $v_{-\bm{k}} = v^\dagger_{\bm{k}}$ implies $v^{\rm
  R}_{\bm{k}} = v^{\rm R}_{-\bm{k}}$ while $v^{\rm I}_{\bm{k}} =
-v^{\rm I}_{-\bm{k}}$. $v^{\rm R}_{\bm{k}}$ and $v^{\rm I}_{\bm{k}}$
evolve separately under linear evolution and this evolution is
identical provided the Hamiltonian is invariant under reflections in
$\bm{k}$, which is true in particular if the physics involved is
parity invariant or if it is invariant under arbitrary rotations. We
may therefore treat the system as if it involves a single real field,
$\w_{\bm{k}} = \w^\dagger_{\bm{k}}$ for {\it all} $\bm{k}$ and then
identify $\sqrt2 \; v^{\rm R}_{\bm{k}} = \w_{\bm{k}}+\w_{-\bm{k}}$ and
$\sqrt2 i \; v^{\rm I}_{\bm{k}} = \w_{\bm{k}}-\w_{-\bm{k}}$ respectively
as its even and odd parts under reflection. The evolution equation for
$\w$, $v^{\rm R}$ and $v^{\rm I}$ are all identical in the
applications below. In what follows our interest is in the matrix
elements of the density matrix $\rho_{\bm{k}}$ for the single
oscillator that arises for each $\bm{k}$, and for fixed $\bm{k}$ we
compute their evolution in an eigenbasis of the real field
$\w_{\bm{k}}$, since this simplifies the notation by allowing us to
drop the superscripts `R' and `I' on the fields. A similar story
applies also to the momentum which we denote $\p_{\bm{k}}$.

\subsection{The system and the environment}

\begin{figure}
\centering
  \includegraphics[width=0.65\linewidth]{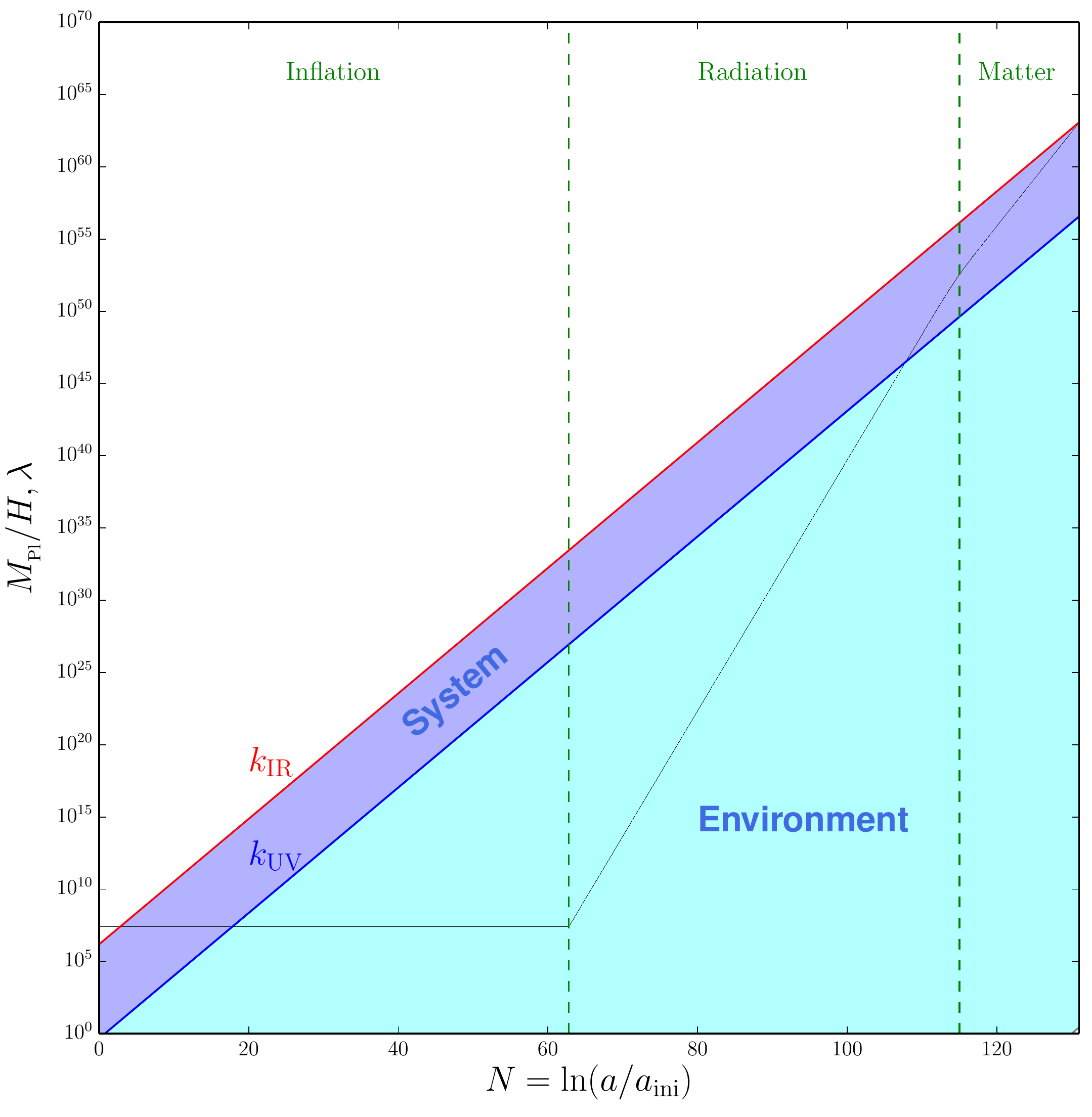}
  \caption{We sketch out the domain of the system and environment
    modes. The black line denotes the Hubble radius and the coloured
    lines stand for the mode wavelengths.  The system is comprised of
    co-moving scales between $k_{\IR}$ and $k_{\UV}$, both of which are
    outside the Hubble radius at the end of inflation. The environment
    is made of all scales such that $k>k_\UV$.
   }
\label{fig:defsystem}
\end{figure}

We next divide the Hilbert space of states for this system into the
`system' ({\it i.e.}~degrees of freedom we choose to follow because
they appear in observations made at late times) and an `environment'
(consisting of all modes that are not observed), see also
Fig.~\ref{fig:defsystem}. Having made this split we trace out over the
environment modes and follow only state evolution within the observed
sector.

Present-day measurements only sample primordial fluctuations whose
co-moving momenta have magnitudes $k = |\bm{k}|$ that lie within a
finite range 
\begin{equation}
\label{eq:defsystemintermsofk}
k_{\IR} < k < k_{\UV} 
\end{equation} 
where $k_{\IR}/a_0 \sim 0.05 \, a_0 \mathrm{Mpc}^{-1}$ and
$k_{\UV}/a_0$ (with $k_\UV \sim 2500 \, k_\IR$) are the smallest and
largest currently observable physical momenta (such as
through CMB or large-scale structure observations) and $a_{0}$ is the
present-day scale factor. \Fig{fig:defsystem} shows this
schematically (not to scale).

In what follows we define the observed system (system $A$) to be those
modes whose co-moving momenta satisfy\footnote{In practice we define our
system to include both observable modes and those with $k < k_\IR$ whose
  wavelengths are too long to have been observed. Such modes are
  expected to be absorbable into the definition of the background
  geometry, and their inclusion should not affect our later discussion
  since they do not contribute to the observables ({\it
    e.g.}~decoherence) on which we ultimately focus. (See
  \RRef{Shandera:2017qkg} for an open-system treatment of
IR modes.)} $k <
k_{\UV}$ while the environment (system $B$) satisfies $k > k_\UV$, and
so write the position-space field as
\begin{equation}
\label{vdecomp}
      {v}(\eta,\bm{x}) = {v}_{\ssA}(\eta,\bm{x}) \otimes {\mathcal{I}}_{\ssB} \
      +  \mathcal{I}_{\ssA} \otimes {v}_{\ssB}(\eta,\bm{x})
\end{equation}
with $v_\ssA$ denoting the system and $v_\ssB$ representing the environment:
\begin{align}
\label{vclassicalsplit}
v_{\ssA}(\eta,\bm{x}) & : = \int \frac{\dd^3 \bm{k}}{(2\pi)^{3/2}} \;
\Theta\big( k_{\UV} - k \big) v_{\bm{k}}(\eta) e^{i \bm{k}\cdot\bm{x}}
=  \int_{k < k_{\UV} } \frac{\dd^3 \bm{k}}{(2\pi)^{3/2}} \;
v_{\bm{k}}(\eta) e^{i \bm{k}\cdot\bm{x}} \, ,\\
v_{\ssB}(\eta,\bm{x}) & : = \int \frac{\dd^3 \bm{k}}{(2\pi)^{3/2}} \;
\Theta\big( k - k_{\UV} \big) v_{\bm{k}}(\eta) e^{i \bm{k}\cdot\bm{x}}
=  \int_{k > k_{\UV}} \frac{\dd^3 \bm{k}}{(2\pi)^{3/2}} \;
v_{\bm{k}}(\eta) e^{i \bm{k}\cdot\bm{x}} 
\end{align}
where $\Theta(x)$ is the Heaviside step function and $\cI_\ssA$ and
$\cI_\ssB$ representing the appropriate unit operators. The free
Hamiltonian similarly becomes
\begin{equation}
  {\cH}_0(\eta) = {\cH}_{\ssA}(\eta) \otimes {\mathcal{I}}_{\ssB}
  + {\mathcal{I}}_{\ssA} \otimes {\cH}_{\ssB}(\eta)
\end{equation}
where $\cH_\ssA$ and $\cH_\ssB$ are both given by
\pref{freeHclassical} but with the momentum range respectively
restricted to the intervals $k < k_\UV$ and $k > k_\UV$.

Inserting the decomposition (\ref{vdecomp}) into the cubic interaction
Hamiltonian $\cH_{\mathrm{int}}$ defined in \Eq{Hclassicalint} gives
several contributions, of the schematic form $v_{\ssA}^3$, $v_{\ssA}^2
v_{\ssB}$, $v_{\ssA} v_{\ssB}^2$ and $v_{\ssB}^3$. Of these, only the
cross terms ($v_{\ssA}^2 v_{\ssB}$ and $v_{\ssA} v_{\ssB}^2$) couple
the system to the environment (and so contribute, say, to
decoherence), and of these momentum conservation suppresses the
$v_{\ssA}^2 v_{\ssB}$ interactions because it is impossible to sum two
small momenta to get a large one. The largest interactions also come
when derivatives act only on large-momentum (environment) fields. For
these reasons we focus primarily on the contribution that has the form
$v_{\ssA} \delta^{ij} (\partial_i v_{\ssB} ) (\partial_j v_{\ssB})$ and
so take
\begin{align}
\label{Hintmom}
\cH_{\mathrm{int}}(\eta) & \supset - \frac{\sqrt{\slrl }}{2\sqrt{2} \Mp
  a(\eta)} \int\dd^3 \bm{x}\; v_{\ssA} (\eta, \bm{x})
\otimes \delta^{ij}\partial_i v_{\ssB} (\eta, \bm{x})
\partial_j v_{\ssB} (\eta, \bm{x})  \ .
\end{align}
We therefore seek an open-system description of evolution using the
Hamiltonian
\begin{equation}
\label{fullquantumH}
      {\cH}(\eta) = {\cH}_{\ssA}(\eta) \otimes {\mathcal{I}}_{\ssB}
      +{\mathcal{I}}_{\ssA} \otimes {\cH}_{\ssB}(\eta) + {\cH}_{\mathrm{int}}(\eta) 
\end{equation}
with Hamiltonians given explicitly by \Eqs{freeHclassical} and
\eqref{Hintmom}. We circle back to include mixed tensor-scalar cubic
interactions in~\Sec{ssec:Tensor}.

\subsection{State evolution}
\label{sec:conversion}

We seek to predict the evolution of the system's state ${\rho}(\eta)$
given the above choice of Hamiltonian, at first working
perturbatively. We do so within both Schr\"odinger picture and
interaction picture, since each can be more convenient for some kinds
of questions.

The Schr\"odinger-picture density matrix for the full
system-plus-environment, $\rho_{\ssS}(\eta)$, evolves through the
standard Liouville equation,
\begin{equation} \label{SpicVN}
\frac{\partial {\rho}_{\ssS}}{\partial \eta} = - i \Bigl[ {\cH}_{\ssS}(\eta)   , {\rho}_{\ssS}(\eta) \Bigr] \ .
\end{equation}
The interaction picture density matrix is defined relative to this by
\begin{equation}
  {\rho}(\eta) =  {U}_0^{\dagger}(\eta,\eta_{\mathrm{in}}) {\rho}_{\ssS}(\eta) \,
  {U}_0(\eta,\eta_{\mathrm{in}}) \ ,
\end{equation}
where
\begin{equation}
  {U}_{0}(\eta_1, \eta_2) : = \mathcal{T}
  \exp\left( - i \int_{\eta_2}^{\eta_1} \dd\eta\; {\cH}_{0\ssS}(\eta) \right)  \,,
\end{equation}
and so satisfies
\begin{equation} \label{INTpicVN}
  \frac{\partial {\rho}}{\partial \eta}
  = - i \Bigl[ {\cH}_{\mathrm{int}}(\eta) , {\rho}(\eta) \Bigr] \,.
\end{equation}
In particular, both pictures agree at the initial time $\rho(\eta_{\rm
  in}) = \rho_\ssS(\eta_{\rm in})$.

Field operators $v_\ssS(\bm{x})$ are time-independent in Schr\"odinger
picture, but interaction-picture fields
\begin{equation}
  {v}(\eta,\bm{x}) \ := \ {U}_0^{\dagger}(\eta,\eta_{\mathrm{in}})
  {v}_{\ssS}(\bm{x}) \, {U}_0(\eta,\eta_{\mathrm{in}}) 
\end{equation}
evolve only via the free part of the Hamiltonian with initial
condition ${v}(\eta_{\mathrm{in}},\bm{x}) = v_\ssS(\bm{x})$. Because
the free evolution factorizes between system and environment
\begin{equation}
  {U}_0(\eta_1,\eta_2) : =  {U}_{0\ssA}(\eta_1, \eta_2)
  \otimes {U}_{0\ssB}(\eta_1, \eta_2)
\end{equation}
where
\begin{equation}
{U}_{0\ssA}(\eta_1, \eta_2) : = \mathcal{T}
\exp\left( - i \int_{\eta_2}^{\eta_1} \dd\eta\;
{\cH}_{\ssA\ssS}(\eta) \right)  \quad \mathrm{and}
\quad  {U}_{0\ssB}(\eta_1, \eta_2)
:= \mathcal{T}\exp\left( - i \int_{\eta_2}^{\eta_1}
\dd\eta\; {\cH}_{\ssB\ssS}(\eta) \right) \ ,
\end{equation}
the system and environment parts of the fields
\begin{equation}
{v}_{\ssS}(\bm{x}) = {v}_{\ssA\ssS}(\bm{x}) \otimes
{\mathcal{I}}_{\ssB} + {\mathcal{I}}_{\ssA} \otimes {v}_{\ssB\ssS}(\bm{x})
\end{equation}
evolve independently under free evolution:
\begin{equation}
{v}_{\ssA}(\eta,\bm{x}) = {U}_{0\ssA}^{\dagger}(\eta,\eta_{\mathrm{in}})
{v}_{\ssA\ssS}(\bm{x})  {U}_{0\ssA}(\eta,\eta_{\mathrm{in}})
\quad \mathrm{and} \quad {v}_{\ssB}(\eta,\bm{x})
= {U}_{0\ssB}^{\dagger}(\eta,\eta_{\mathrm{in}})
{v}_{\ssB\ssS}(\bm{x})  {U}_{0\ssB}(\eta,\eta_{\mathrm{in}}) \,.
\end{equation}

We evolve the state assuming that the system and environment are
uncorrelated at $\eta=\eta_{\rm in}$ and all modes are prepared in the
Bunch-Davies vacuum~\cite{Bunch:1978yq}:
\begin{equation}
\label{ICfull}
{\rho}(\eta_{\mathrm{in}}) = {\rho}_{\mathrm{\ssS}}(\eta_{\mathrm{in}})
= | 0 \rangle \langle 0 | = | 0_{\ssA} \rangle \langle 0_{\ssA} |
\otimes  | 0_{\ssB} \rangle \langle 0_{\ssB} | \,.
\end{equation}
Here
\begin{equation}
\label{vacuumsplitting}
| 0_{\ssA} \rangle := \bigotimes_{k<k_{\UV} } | 0_{\bm{k}} \rangle
\quad  \mathrm{and} \quad | 0_{\ssB} \rangle :=
\bigotimes_{ k>k_{\UV} } | 0_{\bm{k}} \rangle \quad \hbox{with} \quad
{c}_{\bm{k}}(\eta_{\mathrm{in}}) | 0_{\bm{k}} \rangle = 0
\quad \mathrm{for\ all\ }\bm{k} \,,
\end{equation}
with the mode functions $u_{\bm{k}}(\eta)$ appearing in \Eq{vhatk}
given (for massless states) by
\begin{equation}
\label{eq:deSitter:BD:vk}
u_{\bm{k}}(\eta) = \frac{e^{-i k\eta}}{\sqrt{2 k}}
\left(1-\frac{i}{k \eta}\right)\, .
\end{equation}

The time evolution for observations restricted to be only in sector
$A$ is completely determined by the evolution of the reduced density
matrix, $\varrho$, obtained by tracing out all the environment degrees
of freedom of sector $B$ from the full density matrix.
\begin{equation} \label{reducedSpic}
{\varrho}(\eta) := \Trenv \Bigl[ {\rho}(\eta) \Bigr] \,.
\end{equation}
In the absence of interactions the free evolution of the density
matrix factorizes in momentum space, with separate momenta remaining
uncorrelated
\begin{equation} \label{rhofactor}
{\varrho}(\eta) = \bigotimes_{k < k_{\UV} }
{\varrho}_{\bm{k}}(\eta) \ ,
\end{equation}
at all times. The time-dependence of each factor describes the
squeezing of super-Hubble modes due to their non-adiabatic evolution
in the presence of the time-dependent Hamiltonian. It is the
deviations from this that are of most interest in what follows.

\section{Evolution equations}
\label{sec:Lindblad}

This section contains the core derivation on which our results
ultimately depend: we derive here how the interaction~\eqref{Hintmom}
alters the late-time evolution of the reduced density matrix for the
observed modes. Because this interaction is linear in the
long-wavelength field our evolution equation remains quadratic in this
field even at second order, and so evolution still proceeds separately
for each super-Hubble mode $\bm{k}$, greatly simplifying the analysis.

We do so because this proves to be the dominant interaction through
which short-wavelength scalar modes decohere long-wavelength scalar
fluctuations. Simplified evolution emerges for super-Hubble modes at
late times and its domain of validity is studied in some detail
because within it predictions can be made at unusually late
times. Finally, these arguments are repeated for gravitational
interactions coupling scalar and tensor metric modes to determine ones
through which these modes dominantly decohere one another.

\subsection{Nakajima-Zwanzig equation}

As an intermediate step we first derive the Nakajima-Zwanzig equation
for the reduced density matrix $\varrho$.  This equation explicitly
eliminates the unseen environmental degrees of freedom to rewrite the
Liouville equation purely in terms of the observed degrees of freedom.

Although very general, this equation computes $\partial_t \varrho$ as
a convolution of the earlier values of $\varrho$ throughout its past
history, and so is not so in itself. We show how this equation
simplifies when specializing to super-Hubble modes at very late times
(compared with the Hubble time), because it then becomes
Markovian. The resulting Lindblad-type evolution equation for the
reduced density matrix $\varrho$ lends itself to making reliable
late-time predictions that would otherwise lie beyond the reach of
perturbative methods.

\subsubsection{General derivation}

To simplify later formulae it is useful to write the
interaction-picture interaction Hamiltonian (\ref{Hintmom}) as
\begin{equation}
\label{HintRdef}
      {\cH}_{\mathrm{int}}(\eta) = G(\eta) \int\dd^3 \bm{x}\;
      {v}_{\ssA} (\eta, \bm{x}) \otimes {B}(\eta,\bm{x}) \,,
\end{equation}
where $B$ denotes the relevant environmental field combination
\begin{equation}
\label{eq:B:def}
      {B}(\eta,\bm{x})  : =  \delta^{ij} \partial_{i} {v}_{\ssB}(\eta,\bm{x})
      \partial_{j} {v}_{\ssB}(\eta,\bm{x}),
\end{equation}
and the coupling strength is 
\begin{equation}
\label{couplingdef}
G(\eta) :=  - \frac{\sqrt{\slrl }}{2\sqrt{2}\;  \Mp \, a(\eta) } \,.
\end{equation}

To derive the Nakajima-Zwanzig equation (see \RRef{Kaplanek:2019dqu}
for a similar derivation in a simpler setting) we define the
projection super-operator $\mathcal{P}$ to act on an arbitrary
operator $\cO$ in the Hilbert space by
\begin{equation}
  \cP\{ \cO \} = \Trenv\big[ \cO \big] \otimes
  \left \vert 0_{\ssB} \rangle \langle 0_{\ssB} \right \vert \ , 
\end{equation}
where $ | 0_{\ssB} \rangle \langle 0_{\ssB} |$ is the Bunch Davies
vacuum for the environment sector as given in
\Eq{vacuumsplitting}. This satisfies $\cP^2 = \cP$ as does its
complement $\cQ = 1 - \cP$, which is also a projection
super-operator. In particular, $\cP$ maps the full density matrix
$\rho(\eta)$ onto the reduced density matrix $\varrho(\eta)$ as
follows:
\begin{equation}
\label{Pproj}
\cP\{ {\rho}(\eta) \}  = {\varrho}(\eta) \otimes
\left\vert 0_{\ssB} \rangle \langle 0_{\ssB} \right \vert \,,
\end{equation}
where $\varrho$ is the reduced density matrix for the system defined
by \pref{reducedSpic}.

The Nakajima-Zwanzig equation for $\varrho(\eta)$ is derived by
applying $\cP$ to the interaction picture Liouville equation, written
in terms of a Liouville super-operator:
\begin{align}
  \partial_{\eta} {\rho}(\eta) =
  \mathscr{L}_{\eta}\{ {\rho}(\eta) \} \quad \mathrm{with}
  \quad \mathscr{L}_{\eta}\{ {\rho}(\eta) \}
  := - i \Bigl[ {\cH}_{\mathrm{int}}(\eta) , {\rho}(\eta) \Bigr] \,,
\end{align}
with the goal of expressing it purely in terms of $\varrho$. Using
$\cP + \cQ = 1$ this leads to
\begin{align}
\label{Peq}
\cP\left\{ \partial_{\eta} {\rho}(\eta) \right\} & = 
\cP \mathscr{L}_{\eta} \cP \{ {\rho}(\eta) \} +
\cP \mathscr{L}_{\eta} \cQ\{   {\rho}(\eta) \}  \\
\label{Qeq}
\cQ\{ \partial_{\eta}  {\rho}(\eta) \}
& =  \cQ \mathscr{L}_{\eta} \cP \{  {\rho}(\eta) \}
+ \cQ \mathscr{L}_{\eta} \cQ\{  {\rho}(\eta) \} \ .
\end{align}
These can be regarded as evolution equations for $\cP\{\rho\}$ and
$\cQ\{\rho\}$ provided $\cP\{\partial_\eta \rho\} =
\partial_{\eta}\cP\{\rho\}$, which is true because the Bunch-Davies
vacuum used in the definition \pref{Pproj} is time-independent in the
interaction picture.

\pref{Qeq} can then be employed to eliminate $\cQ\{\rho(\eta)\}$,
using the formal solution
\begin{equation}
  \cQ\{  {\rho}(\eta) \} =  \cG(\eta,\eta_{\mathrm{in}})
  \cQ\{  {\rho}(\eta_{\mathrm{in}}) \} + \int_{\eta_{\mathrm{in}}}^{\eta}
  \dd \tau\ \mathcal{G}(\eta,\tau) \cQ \mathscr{L}_{\eta}
  \cP\{  {\rho} (\tau) \} 
\end{equation}
where
\begin{equation}
  \mathcal{G}(\eta,\tau) := 1 + \sum_{n=1}^{\infty}
  \int_{\tau}^{\eta} \dd \tau_1 \ \cdots \int_{\tau}^{\tau_{n-1}}
  \dd \tau_{n}\ \cQ \mathscr{L}_{\tau_1} \cdots \cQ \mathscr{L}_{\tau_{n}}  \ .
\end{equation}
Inserting this into \Eq{Peq} yields  
\begin{equation}
\label{NZfull} 
  \cP\{ \partial _{\eta}  {\rho}(\eta) \} =  \cP \mathscr{L}_{\eta}
  \cP \{  {\rho} (\eta) \}  + \cP \mathscr{L}_{\eta} \cG(\eta,0)
  \cQ\{  {\rho}(\eta_{\mathrm{in}}) \} + \int_{\eta_{\mathrm{in}}}^{\eta}
  \dd s\ \mathcal{K}(\eta,s)\{  {\rho} (s) \} \quad 
\end{equation}
with kernel
\begin{equation}
  \mathcal{K}(\eta,s) = \cP \mathscr{L}_{\eta} \cG(\eta,s)
  \cQ \mathscr{L}_{s} \cP \,.
\end{equation}
For the uncorrelated initial state used here the second term in
\Eq{NZfull} vanishes, since
\begin{equation}
  \cQ\big\{ | 0_{\ssA} \rangle \langle 0_{\ssA} |
  \otimes | 0_{\ssB} \rangle \langle 0_{\ssB} |  \big\} = 0\, .
\end{equation}

This gives an evolution equation that involves only $\cP\{\rho\}$,
which can be expanded to any desired order in $\cH_{\rm
  int}$. Stopping at second order -- for which we may use $\cG(\eta,s)
\simeq 1$ -- we find
\begin{equation}
  \partial_{\eta} \cP \{ \rho(\eta) \} = \cP \{ \partial_{\eta}
  \rho(\eta) \}  \simeq   \cP \mathscr{L}_{\eta} \cP\{ \rho(\eta) \}
  + \int_{\eta_{\mathrm{in}}}^{\eta} \dd s\ \cP \mathscr{L}_{\eta}
  \cQ \mathscr{L}_{s} \cP \{ \rho(s) \} \,,
\end{equation}
which, using the definitions of $\cP$, $\cQ$ and $\mathscr{L}_{\eta}$,
becomes the following equation for $\varrho$:
\begin{align}
  \frac{\partial  {\varrho}}{\partial \eta} &\simeq - i \;
  \Trenv\Bigl\{  \Bigl[  {\cH}_{\rm int}(\eta) ,
    {\varrho}(\eta) \otimes | 0_{\ssB} \rangle \langle 0_{\ssB} | \Bigr] \Bigr\} 
  -\int_{\eta_{\mathrm{in}}}^{\eta} \dd \eta'\
  \Trenv\bigg\{ \bigg[  {\cH}_{\rm int}(\eta) \,,\,
    \Bigl[  {\cH}_{\rm int}(\eta') \,,\,
      {\varrho}(\eta') \otimes | 0_{\ssB} \rangle \langle 0_{\ssB} | \Bigr]
    \nonumber \\
    & - \Trenv\Bigl\{ \Bigl[  {\cH}_{\rm int}(\eta'),
      {\varrho}(\eta') \otimes |0_{\ssB} \rangle \langle 0_{\ssB}
      | \Bigr] \Bigr\} \otimes | 0_{\ssB} \rangle \langle 0_{\ssB}
    | \bigg] \bigg\} \,. 
\end{align}
Specializing to the interaction \eqref{HintRdef} the evolution equation for $\varrho$ finally
simplifies to
\begin{align}
\label{NZ}
\frac{\partial {\varrho}}{\partial \eta} &\simeq
- i G(\eta) \mathscr{B}(\eta) \int \dd^3\bm{x} \;
\left[  {v}_{\ssA}(\eta,\bm{x}),  {\varrho}(\eta) \right]
-  \int \dd^3\bm{x} \int \dd^3\bm{x}' \int_{\eta_{\mathrm{in} } }^{\eta}
\dd \eta' \; G(\eta) G(\eta')
\nonumber \\ & \times
\bigg\lbrace \left[  {v}_{\ssA}(\eta,\bm{x}) ,
  {v}_{\ssA}(\eta',\bm{x}')  {\varrho}(\eta') \right] C_\ssB(\eta, \eta';
\bm{x} - \bm{x}') 
+ \left[  {\varrho}(\eta') {v}_{\ssA}(\eta',\bm{x}')  ,
  {v}_{\ssA}(\eta,\bm{x}) \right]
C^{\ast}_{\ssB}(\eta, \eta';\bm{x} - \bm{x}') \bigg\rbrace 
\end{align}
whose right-hand side neglects $\cO(G^3)$ terms. The required
expectation values of the environment operator ${B}$ are
\begin{equation}
\label{R_1pt_def}
\mathscr{B}(\eta) := \langle 0_{\ssB} | {B}(\eta,\bm{x}) | 0_{\ssB} \rangle
= \int_{k>k_{\UV}} \frac{\dd^3 \bm{k}}{(2\pi)^3} |\bm{k}|^2 |u_{\bm{k}}(\eta)|^2
\end{equation}
and
\begin{equation}
\label{2pt_text}
C_\ssB(\eta,\eta' ; \bm{x} - \bm{x}') =  \langle 0_{\ssB} |
\left[ {B}(\eta,\bm{x}) - \mathscr{B}(\eta) \right]
\left[ {B}(\eta',\bm{x}') - \mathscr{B}(\eta')  \right] | 0_{\ssB} \rangle .
\end{equation}
The second equality in \pref{R_1pt_def} uses the
translation-invariance of the Bunch-Davies state as well as the
specific for the operator given in \pref{eq:B:def}, and a similar
expression for $C_\ssB$ is given in
\Eq{2ptfunctionfluctuatingvarying}. The integral in \pref{R_1pt_def}
diverges and part of the later discussion shows how such divergences
are handled.

\subsubsection{Nakajima-Zwanzig equation for each mode}

The second-order Nakajima-Zwanzig equation (\ref{NZ}) simplifies
considerably when the interaction is linear in $v_\ssA$ -- as it is in
\eqref{HintRdef} -- because evolution does not mix modes, similar to
free evolution. To make this explicit define the momentum-space
correlation function $\mathscr{C}_{\bm{k}}(\eta,\eta')$ using
\begin{equation}
\label{CR_FT}
C_\ssB(\eta, \eta' ; \bm{y}) = \int \frac{\dd^{3} \bm{k}}{(2\pi)^{3/2}}
\mathscr{C}_{\bm{k}}(\eta,\eta') e^{i \bm{k} \cdot \bm{y} } \,.
\end{equation}
$\mathscr{C}_{\bm{k}}$ is computed explicitly in Appendix
\ref{App:Cketaetap} for $k < k_{\UV}$, which shows in particular that
$\mathscr{C}_{\bm{k}}(\eta,\eta')$ only depends on the modulus $k =
|\bm{k}|$ -- and so in particular $\mathscr{C}_{-\bm{k}}(\eta,\eta') =
\mathscr{C}_{\bm{k}}(\eta,\eta')$.

Factorizing the density matrix as in \Eq{rhofactor} shows that
$\varrho_{\bm{k}}$ for each mode $\bm{k}$ evolves independently.
\Eq{NZ} implies that (for $\bm{k} \neq 0$) each factor separately
satisfies
\begin{align}
\label{NZmodes}
\frac{\mathcal{V}}{(2\pi)^3}
\frac{\partial {\varrho}_{\bm{k}}}{\partial \eta}
 &= - (2\pi)^{3/2} \int_{ \eta_{\mathrm{in}} }^{\eta} \dd \eta' \;
G(\eta) \, G(\eta') \bigg\lbrace \left[ {\w}_{\bm{k}}(\eta) ,
  \w_{\bm{k}}(\eta')  {\varrho}_{\bm{k}}(\eta') \right]
\mathscr{C}_{\bm{k}}(\eta,\eta')
\nonumber \\
& + \left[  {\varrho}_{\bm{k}}(\eta')
  {\w}_{\bm{k}}(\eta') ,
  {\w}_{\bm{k}}(\eta) \right] \mathscr{C}^{\ast}_{\bm{k}}(\eta,\eta')
\bigg\rbrace  ,
\end{align}
where $\w_{\bm{k}}$ is the proxy for $v^{\rm R}_{\bm{k}}$ and $v^{\rm
  I}_{\bm{k}}$ defined below \pref{alphaRI} and the factor $\cV$
denotes the volume of space and arises when keeping track of the
normalization of momentum modes in the continuum limit (see
\App{App:Dimensions}). Its presence ensures the final expressions
remains finite as $\cV \to \infty$ when $\bm{k}$ is taken to be
continuum normalized. A similar expression also holds for $\bm{k} = 0$
but also includes a contribution from $\mathscr{B} = \langle B
\rangle_{\rm env}$.

\subsubsection{Environmental correlations}
\label{ssec:EnvCorr}

\begin{figure}
\begin{center}
\includegraphics[width=0.70\textwidth]{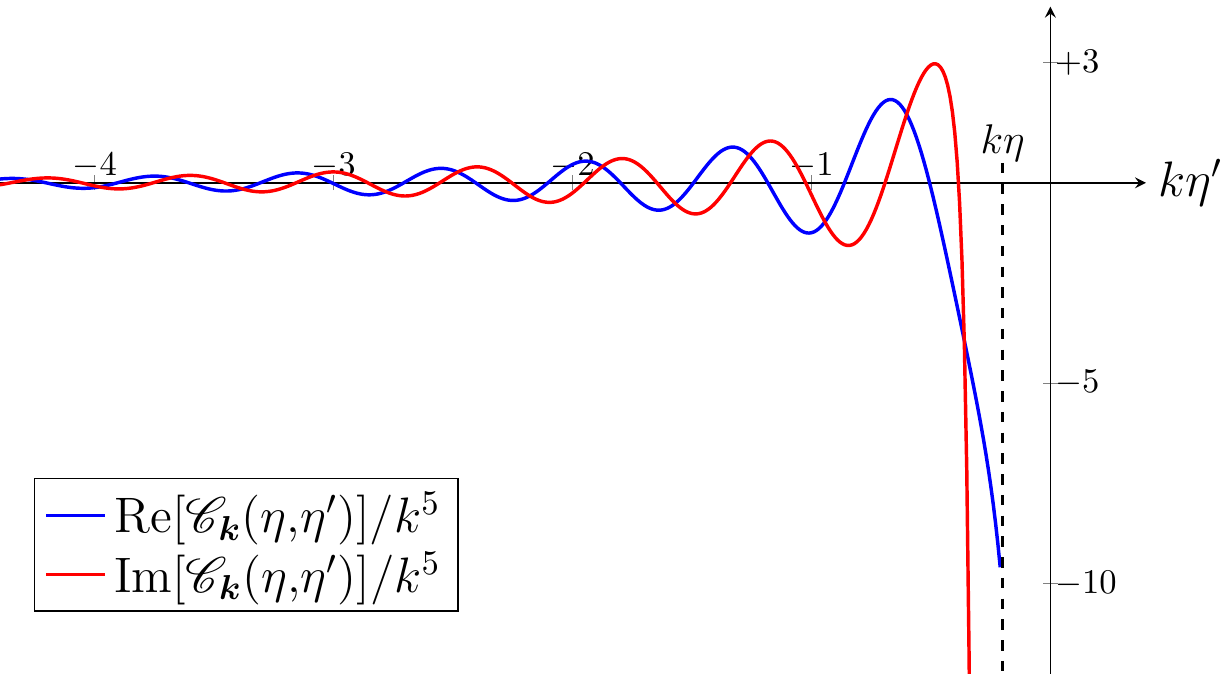}
\caption{$\mathrm{Re}[\mathscr{C}_{\bm{k}}(\eta,\eta') ]$ and
  $\mathrm{Im}[\mathscr{C}_{\bm{k}}(\eta,\eta') ]$ as a function of $k
  \eta'$ for $k \eta = -0.2$ and $\kUV/k =5$. Note the
  singularity at $\eta' \simeq \eta$.}
\label{fig:etaPlot}
\end{center}
\end{figure}

Later sections explore implications of the Nakajima-Zwanzig equation
(\ref{NZmodes}) and in particular what it predicts for very late
times. The result depends in detail on the environmental correlator
$\mathscr{C}_{\bm{k}}(\eta,\eta')$, which is computed explicitly in
\App{App:envcorr_eta}, see \Eq{Ck_answer_Appendix}. It is plotted as a
function of $k \eta'$ in \Fig{fig:etaPlot}. For later use this
section summarizes several useful limits of the result found there.

In the coincidence limit $\eta' \to \eta$, the correlator is singular
and can be expanded as
\begin{align}
\label{Ck_coincident_etap}
\mathscr{C}_{\bm{k}}(\eta,\eta') & \simeq
\frac{k^5}{32(2\pi)^{7/2}} \bigg[ \frac{1}{( - k \eta )^4 }
\left(-  32 \kappa - 8 - \frac{64}{3\kappa}
+ \frac{4}{\kappa^2}\right)
+ \frac{1}{( - k \eta )^2}\biggl(- \frac{64}{3}\kappa^{3}
- 16 \kappa^{2} + 32\kappa
\nonumber \\ &
  + \frac{16}{3} + \frac{64}{15\kappa}\biggr)
  - \frac{32}{ 5}\kappa^{5} - 8\kappa^{4}
+ \frac{32}{ 9 }\kappa^{3} + 4 \kappa^{2}
- \frac{64  }{ 15  }\kappa - \frac{44}{45}  \bigg] +
\frac{\pi}{32(2\pi)^{7/2}} \bigg[ \delta^{\prime\prime\prime\prime}(\eta - \eta')
\nonumber \\ &
  + \frac{4(\eta - \eta')}{\eta\eta'} \delta^{\prime\prime\prime}(\eta - \eta')
+ \frac{4\big[\eta^2 + (\eta')^2 + \big( \frac{5}{6} k^2 \eta \eta'
      - 4 \big) \eta \eta' \big] }{\eta^2 (\eta')^2}
  \delta^{\prime\prime}(\eta - \eta') 
\nonumber \\ &
  +\frac{4(\eta - \eta') \big( 3 k^2 \eta \eta' - 4 \big)}{\eta^2 (\eta')^2}
  \delta^{\prime}(\eta - \eta')
  + \frac{\frac{43}{15} k^4 \eta^2 (\eta')^2
    + \frac{4}{3} k^2 \big[9 \eta^2 - 32 \eta \eta'
      + 9 (\eta')^2 \big] + 16}{\eta^2 (\eta')^2}
\nonumber \\ & 
  \times \delta(\eta - \eta') \bigg] 
+\frac{i }{32(2\pi)^{7/2}}   \bigg[ \frac{24}{(\eta'-\eta)^5}
+ \frac{20(k^2 \eta^2 - 6)}{3 \eta^2 (\eta' - \eta)^3}
+ \frac{40}{\eta^3 (\eta' - \eta)^2}
\nonumber \\ &
+\frac{k^2(43 k^2 \eta^2 - 460)}{15 \eta^2 (\eta' - \eta)}
- \frac{40}{\eta^5} + \frac{92 k^2}{3 \eta^3} \bigg].
\end{align}
where we define 
\begin{equation}
\kappa := \frac{\kUV}{k}  > 1\, .
\end{equation} 
The above expression assumes $\eta > \eta'$ and the result for $\eta <
\eta'$ is found using $\mathscr{C}_{\bm{k}}(\eta,\eta') =
\mathscr{C}^{\ast}_{\bm{k}}(\eta',\eta)$. Notice that the imaginary
part is completely contained in the last lines of this expression, and
is totally independent of the parameter $\kUV$.

Alternatively, for $\eta' \to -\infty$ (for fixed $\eta$ and
parametrically making $| k \eta' | \gg 1$) the correlator instead has
the form
\begin{align}
\label{Ck_early_etap}
\mathscr{C}_{\bm{k}}(\eta,\eta') & \simeq
\frac{k^5}{32(2\pi)^{7/2}} \bigg\{ - \frac{4
 \left[ i + \kappa (-k\eta) \right]^2 \left(2 \kappa^2 -1 \right)^2 }
{\kappa^2(-k\eta)^2 (- k \eta')^2 } \, e^{- 2 i \kUV (\eta - \eta')} \\
& + \frac{16 \kappa \left( \kappa + 1  \right) \left[ i + \kappa
    (-k\eta) \right]
  \left[ i +  (-k\eta) ( \kappa+1 ) \right] }{ (-k\eta)^2 (- k \eta')^2 }
\, e^{- i (1 + 2 \kappa) k (\eta - \eta')} \bigg\}.
\end{align} 
which falls off for large $\eta'$ proportional to $(\eta')^{-2}$.

\subsection{Markovian approximation}

Now comes the main approximation. The main observation is that the
Nakajima-Zwanzig equation (\ref{NZmodes}) for each mode simplifies if
the contribution to the kernel $G(\eta) G(\eta')
\mathscr{C}_{\bm{k}}(\eta,\eta')$ is so sharply peaked about $\eta'
\simeq \eta$ that $\varrho_{\bm{k}}(\eta)$ does not vary
significantly in the integration region where the kernel has
appreciable support. When this is true the evolution becomes Markovian
in the sense that $\partial_\eta \varrho_{\bm{k}}$ depends
only on $\varrho_{\bm{k}}$ evaluated at the same time
(rather than on its entire earlier history).

\subsubsection{A false start}
\label{sssec:falsestart}

Before proceeding we first pause to describe a common method often
used in the literature, which in this case does not reveal the correct
Markovian limit. In this method one expands the density matrix in a
Taylor series,
\begin{equation} \label{TaylorUnphys}
\varrho_{\bm{k}}(\eta' )\simeq \varrho_{\bm{k}}(\eta)
+ (\eta' - \eta) \partial_\eta \varrho_{\bm{k}}(\eta )
+ \mathcal{O}\big[ (\eta'-\eta)^2 \big]
\end{equation}
with the result truncated at leading order to obtain the Markovian
regime. Such an expansion seems very likely to be a good approximation
because \Eq{INTpicVN} shows that in the interaction picture all
contributions to $\partial_\eta \varrho_\bmk$ are suppressed by the
perturbative coupling (in our case $H^2/\Mp^2$).

Interestingly, we find that this derivation leads in the present
instance to an unphysical Markovian limit, whose evolution equation
can violate the fundamental positivity conditions that density
matrices must satisfy. This signals a failure of the approximations
used. Since this procedure is frequently used in the literature, we
here describe in more detail the way in which it fails in the current
setting.

Inserting the leading term of \Eq{TaylorUnphys} into
\Eq{NZmodes} leads to the evolution equation\footnote{In order to
arrive at this equation using these steps, one must expand
$v_{\bm{k}}(\eta')$ in terms of ladder operators which
are then inverted with a Bogoliubov transformation to give rise
to the operators in \Eq{Ok_operators}.}
\begin{align}
\label{Unphys_Lindblad}
\frac{\mathcal{V}}{(2\pi)^3}
\frac{\partial {\varrho}_{\bm{k}}}{\partial \eta}
= - i \Bigl[ {\mathcal{H}}_{\mathrm{eff}\bm{k}}(\eta) ,
{\varrho}_{\bm{k}} \Bigr]
+\sum_{n,m = 1}^{2} h_{\bm{k}, nm} \left[ {O}_{\bm{k}, n}
{\varrho}_{\bm{k}}(\eta) {O}^{\dagger}_{\bm{k}, m}
- \frac{1}{2} \left\{ {O}^{\dagger}_{\bm{k}, m}
{O}_{\bm{k}, n}, {\varrho}_{\bm{k}}(\eta) \right\} \right],
\end{align}
where we define
\begin{equation}
\label{Ok_operators}
{O}_{\bm{k}, 1} := k^2 \, {\w}_{\bm{k}}\, , \qquad
{O}_{\bm{k}, 2} := k \, {\p}_{\bm{k}} \, ,
\end{equation}
and 
\begin{equation}
{\mathcal{H}}_{\mathrm{eff}\bm{k}}(\eta) =
\mathrm{Im}\left[ \mathfrak{A}_{\bm{k}}(\eta,\eta_{\mathrm{in}}) \right]\,
\left( {\w}_{\bm{k}}\right)^2
+\mathrm{Im}\left[ \mathfrak{B}_{\bm{k}}(\eta,\eta_{\mathrm{in}}) \right]
{\w}_{\bm{k}} \p_{\bm{k}} \,.
\end{equation}
The matrix of coefficients is
\begin{equation}
h_{\bm{k}, nm} = \frac{1}{k^4}
\begin{pmatrix}
\displaystyle
 2\mathrm{Re}\left[ \mathfrak{A}_{\bm{k}}(\eta,\eta_{\mathrm{in}})\right]  
&&
\displaystyle
 k\, \mathfrak{B}^*_{\bm{k}}(\eta,\eta_{\mathrm{in}})  
\\
\\
\displaystyle
k\, \mathfrak{B}_{\bm{k}}(\eta,\eta_{\mathrm{in}}) 
&& 0
\end{pmatrix} 
\,, 
\end{equation}
where 
\begin{align}
\mathfrak{A}_{\bm{k}}(\eta,\eta_\mathrm{in}) &:=  i (2\pi)^{3/2}
\int_{ \eta_{\mathrm{in}} }^{\eta} \dd \eta' \, G(\eta) G(\eta')
\mathscr{C}_{\bm{k}}(\eta,\eta') \Bigl[- u_{\bm{k}}^{\ast\prime}(\eta)
u_{\bm{k}}(\eta') +  u_{\bm{k}}^{\prime}(\eta) u_{\bm{k}}^{\ast}(\eta') \Bigr]\, ,
\\
\mathfrak{B}_{\bm{k}}(\eta,\eta_\mathrm{in}) &:= i (2\pi)^{3/2}
\int_{ \eta_{\mathrm{in}} }^{\eta} \dd \eta' \, G(\eta) G(\eta')
\mathscr{C}_{\bm{k}}(\eta,\eta') \Bigl[u_{\bm{k}}^{\ast}(\eta) u_{\bm{k}}(\eta')
- u_{\bm{k}}(\eta) u_{\bm{k}}^{\ast}(\eta') \Bigr] \,.
\end{align}

\begin{figure}
\centering
  \includegraphics[width=0.65\linewidth]{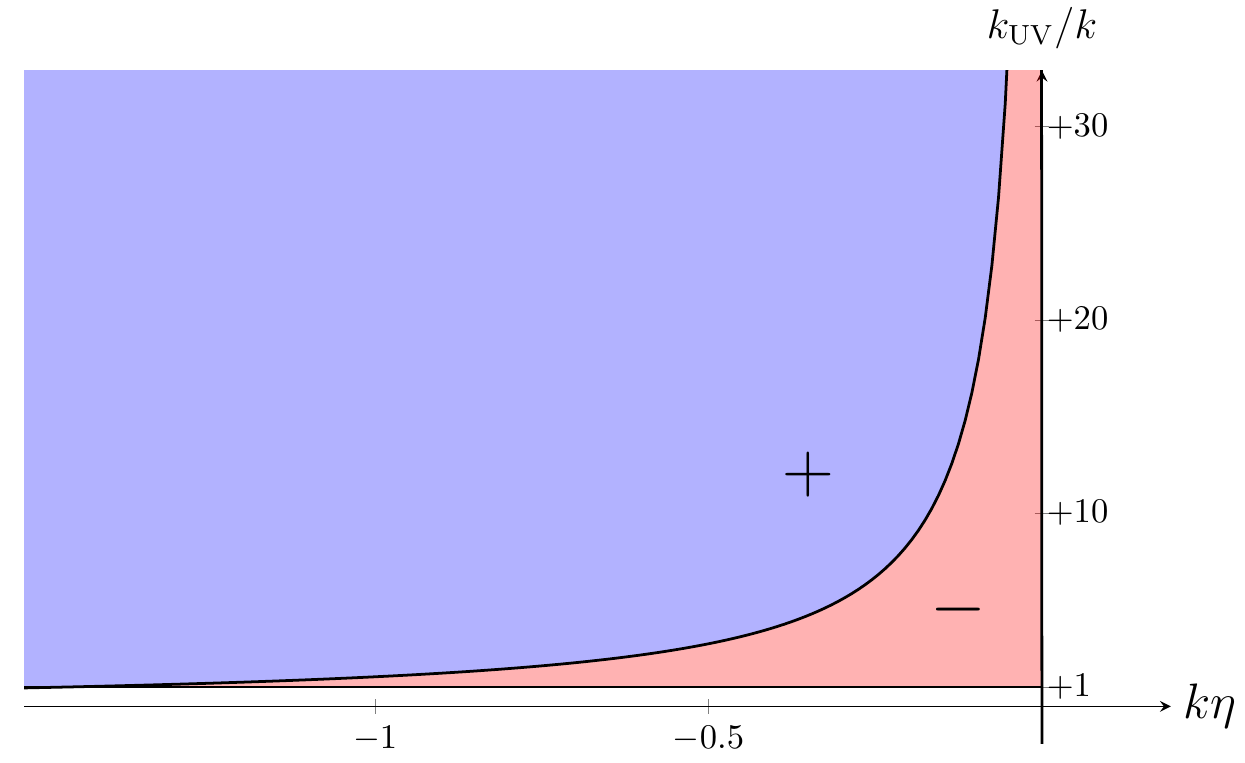}
  \caption{A plot of the sign of Re $\mathfrak{A}_\bmk$ as a function
    of $k\eta$ and $\kappa = \kUV/k$ (for the case $k \eta_{\mathrm{in}} \to -\infty$), with blue representing positive
    and red being negative. The boundary between the two signs
    follows roughly the curve $\kUV \eta \simeq \sqrt3$.}
\label{Fig:sign}
\end{figure}

Now comes the main point: evolution using \Eq{Unphys_Lindblad} only
keeps the eigenvalues of $\varrho$ real and between 0 and 1 (as
required for probabilities) if the eigenvalues
\begin{equation}
\lambda_{\bm{k}}^{\pm} = k^{-4}
\mathrm{Re}\left[ \mathfrak{A}_{\bm{k}}(\eta) \right]
\pm k^{-4} \sqrt{ \mathrm{Re}\left[ \mathfrak{A}_{\bm{k}}(\eta) \right]^2
+  k^2 \, \mathrm{Re}\left[ \mathfrak{B}_{\bm{k}}(\eta) \right]^2
+ k^2 \, \mathrm{Im}\left[ \mathfrak{B}_{\bm{k}}(\eta) \right]^2 }
\end{equation}
of $h_{\bmk,nm}$ are strictly non-negative. But in the super-Hubble
limit $- k \eta \ll 1$ with $-k\eta_{\mathrm{in}}$ fixed and $\kappa=\kUV/k\gg 1$ one finds
\begin{equation}
\mathrm{Re}[ \mathfrak{A}_{\bm{k}} ] \simeq
- \frac{3 \epsilon_1 H^2}{256 \pi^2 \Mp^2 }
\frac{k^3}{\kUV (-k \eta)^{3} } + \cdots 
\end{equation}
is negative and this implies $\lambda_{\bm{k}}^{-}$ is also
negative. The sign of $\mathrm{Re}(\mathfrak{A}_\bmk)$ is evaluated
numerically and shown in \Fig{Fig:sign} as a function of
$\kUV/k$ and $k \eta$ (for the case of $k\eta_{\mathrm{in}} \to-\infty$); showing that the leading small $-k\eta$ limit
is negative whenever $-\kUV \eta \lsim \sqrt3$ (and so $\kUV$ is
super-Hubble) but is positive otherwise.

\subsubsection{A Markovian regime}
\label{sec:Markov}

Another strategy is to jointly expand all terms that multiply the sharply peaked kernel $G(\eta)
G(\eta') \mathscr{C}_{\bm{k}}(\eta,\eta')$ in the integrand of \Eq{NZmodes}, 
\begin{align}
\label{Taylor_1}
\Bigl[ \w_\bmk(\eta), {\w}_{\bm{k}}(\eta')  {\varrho}_{\bm{k}}(\eta') \Bigr]
&\simeq \Bigl[ \w_\bmk(\eta) ,  {\w}_{\bm{k}}(\eta)
  {\varrho}_{\bm{k}}(\eta) \Bigr]
+ (\eta' - \eta)\Bigl[ \w_\bmk(\eta) ,
  \left[   \w_{\bm{k}}(\eta) \partial_\eta
{\varrho}_{\bm{k}}(\eta)
+   {\p}_{\bm{k}}(\eta)  {\varrho}_{\bm{k}}(\eta)  \right] \Bigr]
+ \cdots , 
\end{align}
and
\begin{align}
\label{Taylor_1}
\Bigl[ {\varrho}_{\bm{k}}(\eta')  {\w}_{\bm{k}}(\eta'), \w_\bmk(\eta) \Bigr]
&\simeq \Bigl[ {\varrho}_{\bm{k}}(\eta)  {\w}_{\bm{k}}(\eta), \w_\bmk(\eta) \Bigr]
+ (\eta' - \eta) \Bigl[ \left[\partial_\eta  {\varrho}_{\bm{k}}(\eta)
\w_{\bm{k}}(\eta)  +  {\varrho}_{\bm{k}}(\eta)
{\p}_{\bm{k}}(\eta) \right], \w_\bmk(\eta) \Bigr]+ \cdots  ,
\end{align}
and seek the regime where the first term dominates the integral. 

When this is a good approximation \Eq{NZmodes} becomes the following
interaction-picture Lindblad equation,
\begin{equation}
\label{Lindblad_1}
\frac{\mathcal{V}}{(2\pi)^3}
\frac{\partial  {\varrho}_{\bm{k}}}{\partial \eta}
\simeq   - \, \mathrm{Re}[\mathfrak{F}_{\bm{k}}(\eta,\eta_{\mathrm{in}})]
\, \left[  {\w}_{\bm{k}}(\eta) , \big[  {\w}_{\bm{k}}(\eta),
{\varrho}_{\bm{k}}(\eta) \big] \right]
- i \, \mathrm{Im}[\mathfrak{F}_{\bm{k}}(\eta,\eta_{\mathrm{in}})]
\, \left[ \left[{\w}_{\bm{k}}(\eta) \right]^2 ,
{\varrho}_{\bm{k}}(\eta) \right]  \ ,
\end{equation}
where we define the integrated environmental coefficient
\begin{equation}
\label{Ffrakdef}
\mathfrak{F}_{\bm{k}}(\eta,\eta_{\mathrm{in}})  : =
(2\pi)^{3/2} \int_{ \eta_{\rm in} }^{\eta} \dd \eta' \,
G(\eta) G(\eta') \mathscr{C}_{\bm{k}}(\eta,\eta')
\qquad \hbox{(scalar environment)}\,.
\end{equation} 
\Eq{Lindblad_1} describes Markovian evolution and, for the same
reasons as described above for the false start, the evolution
\pref{Lindblad_1} is only consistent with a probability interpretation
for $\varrho_{\bm{k}}$ if $\mathrm{Re}[\mathfrak{F}_{\bm{k}}] > 0$, so
this must be true in any valid approximation. We verify that it {\it
  is} true for several explicit limits below.

Experience with open systems suggests that the solutions to
\Eq{Ffrakdef} can sometimes be trusted well into the future in a way
that those of \Eq{NZmodes} cannot. A necessary condition for
this extended domain of validity is that the right-hand side not make
explicit reference to the initial time $\eta_{\rm in}$, since it is
only then that one can expect the evolution equation to have a broader
domain of validity than its perturbative derivation (for the reasons
outlined in detail in \RRef{Burgess:2020tbq}). For this to be useful
in the present instance would require the function
$\mathfrak{F}_\bmk(\eta,\eta_{\rm in})$ to be approximately
independent of $\eta_{\rm in}$.

We therefore evaluate \Eq{Ffrakdef} in some detail in \App{App:Lindblad_F}, encountering along the way ultraviolet
divergences that we renormalize after first regularizing using a
variant of dimensional regularization. The general expression for
$\mathfrak{F}_\bmk$ we obtain is somewhat unwieldy for general values
of its arguments $\eta$, $\eta_{\mathrm{in}}$ and $\kUV > k$ and so we
quote here only several useful limiting forms. In particular we find
that
$\mathrm{Re}\left[\mathfrak{F}_{\bm{k}}(\eta,\eta_\mathrm{in})\right]$
is UV finite and becomes universal in the late-time super-Hubble limit
$ - k \eta \ll 1$, with
\begin{equation}
\label{ReFSHresult}
\mathrm{Re}\left[ \mathfrak{F}_{\bm{k}}(\eta, \eta_{\mathrm{in}}) \right]
\simeq  \frac{\slrl  H^2 k^2}{1024 \pi^2 \Mp^2 }
\left\lbrace \frac{20 \pi}{( - k \eta)^2}
+\frac{g\left(\kappa, k \eta_{\mathrm{in}} \right)}
{( - k \eta)} + \mathcal{O}\left[ (- k \eta)^0 \right]\right\rbrace \,.
\end{equation}
This is universal in the sense that all dependence on the parameters
$k_\UV$ and $\eta_{\rm in}$ first appear at subdominant order in
$k\eta$; within the known function $g(\kappa,
k\eta_{\mathrm{in}})$. We show below that
$\mathrm{Re}\left[\mathfrak{F}_{\bm{k}}(\eta,\eta_\mathrm{in})\right]$
is the quantity relevant to decoherence and its universal form for
late times opens up the possibility of also trusting its solutions at
very late times.

In the slightly more restrictive regime $ - k \eta \ll - k
\eta_{\mathrm{in}} \ll 1$ we similarly have -- see formula
(\ref{F_superHubble_App}) -- $\mathfrak{F}_\bmk = \mathfrak{F}^{\rm
  (div)}_\bmk + \mathfrak{F}^{(\mathrm{fin})}_\bmk$ where the UV-divergent part is
\begin{equation}
\label{F_superHubble_bodydiv}
\mathfrak{F}^{\rm (div)}_{\bm{k}}(\eta, \eta_{\mathrm{in}})
= \frac{i\slrl  H^2 k^2}{1024 \pi^2 \Mp^2 }
 \left[ \frac{40}{(- k\eta)^2} - \frac{92}{3} + \frac{43}{15 } (-k\eta)^2   \right]
\left\lbrace \frac{1}{\epsilon}
+ \log\left[ \frac{\kUV}{\mu}\left(2+\frac{1}{\kappa}\right)
\right] \right\rbrace 
\end{equation}
and so only contributes to Im $\mathfrak{F}_{\bm{k}}$. The remaining
finite part is
\begin{align}
\label{F_superHubble_body}
\mathfrak{F}_{\bm{k}}^{(\mathrm{fin})}(\eta, \eta_{\mathrm{in}}) & \simeq
\frac{\slrl  H^2 k^2}{1024 \pi^2 \Mp^2 }
\Biggl(
\frac{20\pi}{(-k\eta)^2}+\frac{4i}{(-k\eta)^2}\left\{ 7-10 
\log\left[  e^{\gamma} (2\kappa+1)(-k\eta) \right]\right\}
\nonumber \\ &
+\frac{1}{(-k\eta)}\biggl[\frac{4}{3}\left(24\kappa+6+\frac{16}{\kappa}
-\frac{3}{\kappa^2}\right)\log \left(\frac{\eta}{\eta_\mathrm{in}}\right)
+\frac{40i}{\zin}\biggr]  
-\frac{46\pi}{3}-\frac{128i}{3}
\nonumber \\ &
+\frac{92i}{3}\log\left[  e^{\gamma} (2\kappa+1)(-k \eta) \right]
+{\cal O}\left(-k\eta,-k\eta_\mathrm{in}\right)\Biggr),
\end{align}
where $\gamma$ is Euler-Mascheroni constant. The divergence is visible in the limit that the regularization parameter $0 < |\epsilon| \ll 1$
tends to zero, and $\mu > 0$ is the usual associated arbitrary mass
scale. Although the formula (\ref{F_superHubble_body}) above neglects
terms $\mathcal{O}(-k\eta)$, we explicitly write out the divergence
proportional to $(-k\eta)^2$ in formula (\ref{F_superHubble_bodydiv})
for completeness, and for later use when asking whether and how the
divergences appearing in \Eq{F_superHubble_bodydiv} can be
renormalized by appropriate choice of counter-term.

The goal is to solve \Eq{Lindblad_1} and extract the physical
observables from it, such as the decoherence rate and the (very small)
corrections to the power spectrum. Before pursuing this we must tie up
two loose ends: understand how to renormalize the divergences
appearing in \Eq{F_superHubble_body} (so that we can understand why
they do not affect the physical predictions) and verify the validity
of the underlying Markovian approximation.

\subsubsection{Domain of validity of the Markovian approximation}
\label{sec:validity:Markov}

As the example of \Sec{sssec:falsestart} shows, truncating a Taylor
expansion inside the integral need not always be a good
approximation. It should be a good approximation however if the
time-scale $T$ over which the Taylor expanded quantity varies is much
longer than the width $\tau$ of the correlation function's peak, since
then subsequent terms should be suppressed by powers of $\tau/T$. We
here show that the leading corrections to \Eq{Lindblad_1} are
suppressed by powers of $k\eta$ in the late-time super-Hubble limit
(for which $k \eta \to 0$).

To see why, insert the subdominant term of \Eq{Taylor_1} into
\Eq{NZmodes}, yielding
\begin{align}
& \frac{\mathcal{V}}{(2\pi)^3}
\frac{\partial  {\varrho}_{\bm{k}}}{\partial \eta} =
-\,\mathrm{Re}\left[\mathfrak{F}_{\bm{k}}(\eta,\eta_{\mathrm{in}})\right]
\, \left[  \w_{\bm{k}}(\eta) , \left[  \w_{\bm{k}}(\eta),
{\varrho}_{\bm{k}}(\eta) \right] \right] - i \,
\mathrm{Im}\left[\mathfrak{F}_{\bm{k}}(\eta,\eta_{\mathrm{in}})\right]
\, \left[\big( \w_{\bm{k}}(\eta) \big)^2  ,\,
{\varrho}_{\bm{k}}(\eta) \right]
\nonumber \\
& -\mathrm{Re}\left[ \mathfrak{M}_{\bm{k}}(\eta,\eta_{\mathrm{in}})\right]
\, \left[\w_{\bm{k}}(\eta) , \left[{\p}_{\bm{k}}(\eta),
{\varrho}_{\bm{k}}(\eta) \right] \right]
- i \, \mathrm{Im}\left[ \mathfrak{M}_{\bm{k}}(\eta,
\eta_{\mathrm{in}}) \right]  \, \left[{\w}_{\bm{k}}(\eta) ,
\left\{  {\p}_{\bm{k}}(\eta),   {\varrho}_{\bm{k}}(\eta)
\right\} \right] \,,
\end{align}
where the first line is the same as \Eq{Lindblad_1} and
$\mathfrak{F}_{\bm{k}}$ is as defined in \Eq{Ffrakdef} while
\begin{equation}
\label{Mfrak_def}
\mathfrak{M}_{\bm{k}}(\eta,\eta_{\mathrm{in}}) \ : =
(2\pi)^{3/2} \int_{ \eta_{\mathrm{in}} }^{\eta} \dd \eta' \,
G(\eta) G(\eta') \mathscr{C}_{\bm{k}}(\eta,\eta') \, (\eta' - \eta) \ .
\end{equation}
We seek the regime where the terms involving $\mathfrak{F}_{\bm{k}}$ dominate
those proportional to $\mathfrak{M}_{\bm{k}}$.

The function $\mathfrak{M}_{\bm{k}}$ is computed in Appendix
\ref{App:ValidityM}, where we find --- {\it c.f.}~formula
(\ref{M_superHubble_App}) --- that in the super-Hubble limit $0 < - k
\eta \ll - k \eta_{\mathrm{in}} \ll 1$ 
\begin{align}
\mathfrak{M}_{\bm{k}}(\eta,\eta_\mathrm{in}) &\simeq
\frac{\slrl  H^2 k}{1024 \pi ^2 \Mp^2} \Biggl[
- \frac{40 i \log \left({\eta}/{\eta_{\mathrm{in}}}\right)
+ \mathcal{O}(- k \eta_{\mathrm{in}}) }{-k \eta}
+  \frac{40 i}{(- k \eta_{\mathrm{in}})}
\nonumber \\ &
+ 4 \left(8\kappa+2+\frac{16}{3\kappa} -\frac{1}{\kappa^2} \right)
\log \left(\frac{\eta}{e  \eta_{\mathrm{in}}}\right)
+ \mathcal{O}(-k\eta) \Biggr]
+\mathfrak{M}^{\rm (div)}_{\bm{k}}(\eta,\eta_\mathrm{in})\,,
\end{align}
where we again encounter a $1/\epsilon$ divergence in the imaginary
part of the form
\begin{equation}
\mathfrak{M}^{\rm (div)}_{\bm{k}}(\eta,\eta_\mathrm{in})
\simeq  -\frac{5i\slrl  H^2 k}{768 \pi ^2 \Mp^2}  (-k\eta) \, 
\left\lbrace \frac{1}{\epsilon}
+\log \left[ \frac{2\kUV+ k}{\mu}
\right] \right\rbrace  \,. 
\end{equation}
This can be absorbed by a counter-term in the same way as can the
divergences in $\mathfrak{F}_{\bmk}$ (see below). By contrast, the
real part $\mathrm{Re}[\mathfrak{M}_{\bm{k}}(\eta,\eta_\mathrm{in})]$
is finite and $\mathcal{O}\left[( - k \eta)^0 \right]$ in the
super-Hubble limit. What is important is that this is subdominant to
$\mathrm{Re}[\mathfrak{F}_{\bm{k}}(\eta,\eta_\mathrm{in})] \propto
(k\eta)^{-2}$ in this regime; putting late times and super-Hubble
scales squarely within the domain of validity of the Markovian
methods.

\subsubsection{Renormalization of the Lindblad equation}
\label{sec:renormalization}

Next consider the issue of renormalization. How renormalization works
is easier to see if we convert \Eq{Lindblad_1} to Schr\"odinger
picture, since this reintroduces the free Hamiltonian (whose
parameters are presumably the ones that get renormalized) into the
evolution.

Repeating the above steps leads to the Schr\"odinger picture Lindblad
equation,
\begin{align}
\label{Lindblad_2}
\frac{\mathcal{V}}{(2\pi)^3}
\frac{\partial  {\varrho}_{\mathrm{\ssS}\bm{k}}}{\partial \eta}  & \simeq
- \mathrm{Re}[\mathfrak{F}_{\bm{k}}(\eta,\eta_{\mathrm{in}})] \,
\left[  {\w}_{\mathrm{\ssS}\bm{k}} , \big[  {\w}_{\mathrm{\ssS}\bm{k}},
    {\varrho}_{\mathrm{\ssS}\bm{k}}(\eta) \big] \right]
\\ &
- i \left[ \;  {\mathcal{H}}_{\mathrm{\ssS}\bm{k}}(\eta)
+ \mathrm{Im}[\mathfrak{F}_{\bm{k}}(\eta,\eta_{\mathrm{in}})]
{\w}^2_{\mathrm{\ssS}\bm{k}}  , \,
{\varrho}_{\mathrm{\ssS}\bm{k}}(\eta) \, \right],  
\end{align}
where the Hamiltonian in momentum space is
${\mathcal{H}}_{\mathrm{\ssS}\bm{k}}(\eta) =
{\mathcal{H}}^{(0)}_{\mathrm{\ssS}\bm{k}}(\eta) + \delta
{\mathcal{H}}_{\mathrm{\ssS}\bm{k}}(\eta)$ where
\begin{equation}
\label{laterHA}
{\mathcal{H}}^{(0)}_{\mathrm{\ssS}\bm{k}}(\eta) = \frac{1}{2} \left[
  {\p}^2_{\mathrm{\ssS}\bm{k}}  
 + \left( k^2 - \frac{2}{\eta^2} \right)
   {\w}^2_{\mathrm{\ssS}\bm{k}}  \right] \,,
\end{equation}
is the free part coming from the action of \Eq{actionstart} and
$\delta{\mathcal{H}}_{\mathrm{\ssS}\bm{k}}(\eta)$ contains any order
$1/\Mp^2$ terms coming from local corrections to this action. The
counterterms that cancel the divergence in
$\mathrm{Im}\left[\mathfrak{F}_\bmk(\eta,\eta_\mathrm{in})\right]$
must come from the quadratic term $\w_\bmk$ in the expansion of
$\delta{\mathcal{H}}_{\mathrm{\ssS}\bm{k}}(\eta)$. What is important
in \Eq{Lindblad_2} is that this is possible because these
counter-terms only appear together with the imaginary part of
$\mathcal{F}_{\bm{k}}$ that contains the $1/\epsilon$ divergences in
\Eq{F_superHubble_body}.

But what are the counter-terms into which divergences might go? This
ultimately is determined by the parameters appearing in the
Lagrangian, and for a renormalizable theory (like QED) this would
simply be our starting Lagrangian \pref{actionstart}. However General
Relativity is famously {\it not} renormalizable in the same way, and
so for it divergences must be handled within an effective field theory
treatment in which \Eq{actionstart} is regarded as the leading part
of a low-energy derivative expansion. In this case standard
power-counting arguments~\cite{Burgess:2003jk} determine what kinds of
terms must be added to it at any perturbative order to capture
divergences. For the gravitational systems of interest here this means
that counter-terms arise either as renormalizations of Newton's
constant (or $\Mp$) or as parameters appearing in four-derivative
interactions, like curvature-squared or mixed
scalar-derivative/curvature terms.

Although we cannot here definitely prove that divergences all get
renormalized into these parameters, we can provide several consistency
checks. We cannot be definitive because it was only when we specialize
to decoherence that we are free to ignore all other interactions
beyond our specific cubic interaction, even at order $1/\Mp^2$. But
these other interactions can contribute UV divergences and it is only
the complete set of divergences at a fixed order in the small
couplings -- powers of $H^2/\Mp^2$ and $\slrl $ in the present
instance~\cite{Adshead:2017srh, Babic:2019ify} -- that are guaranteed
to cancel. What we can do is check that the divergences we encounter
have the dependence on $k$ and $\eta$ that is required if they are to
be absorbable into the expected Einstein-Hilbert or curvature-squared
counter-terms. Along the way we show that the decoherence calculation
is UV finite, and so in particular is independent of the values of
these renormalized parameters.

To see how this works\footnote{Notice that these types of consistency
  checks are easiest to do when using dimensional regularization since
  this preserves the underlying gauge symmetries of the gravitational
  action.} notice that formula (\ref{F_superHubble_bodydiv}) shows
that the divergences encountered in \Eq{Lindblad_2} all have the
form
\begin{equation}
\mathrm{Im}\left[\mathfrak{F}_{\bm{k}}(\eta,\eta_\mathrm{in})\right]\;  
{\w}^2_{\mathrm{\ssS}\bm{k}}  \propto  \frac{\slrl  H^2}{\Mp^2 }
\left(\frac{c_{1}}{\eta^2}  + c_2 \, k^2
+ c_3 \, k^4 \eta^2 \right)  
{\w}^2_{\mathrm{\ssS}\bm{k}}  
\end{equation}
for some real constants $c_{1,2,3}$ that all contain a $1/\epsilon$
divergence in dimensional regularization. Keeping in mind that
$a(\eta) = -(H \eta)^{-1}$, these have the $k$- and $\eta$-dependence
appropriate to the renormalization of terms in the Lagrangian of the
form
\begin{equation}
\label{renorm_operators}
\frac{c_1 }{\eta^2} {\w}^2_{\mathrm{\ssS}\bm{k}} 
\subset \left[ \frac{\mathrm{d}}{\mathrm{d} t} ( a v ) \right]^2
\,, \quad
c_2 \, k^2{\w}^2_{\mathrm{\ssS}\bm{k}} 
\subset c_2 (\partial v)^2 \quad \hbox{and} \quad
c_3 \, k^4 \eta^2{\w}^2_{\mathrm{\ssS}\bm{k}}  \subset c_3
\, (\partial^2 v)^2 \,.
\end{equation}
These are among the kinds of operators that arise in
the fluctuation expansion of the Einstein-Hilbert Lagrangian $\sqrt{-g
} R$, or of a curvature-squared term like $\sqrt{ - g }
R^2$. (For more detail on this see Appendix \ref{App:operators}).

The same EFT reasoning also explain why terms proportional to $\eta^2$
-- such as in \Eq{F_superHubble_bodydiv} -- are not problems even at
early times where $\eta$ can be big (a limit that would also require
choosing $\eta_{\mathrm{in}}$ to be big as well). These naively seem
to be in danger of interfering with the physical arguments that select
the adiabatic Bunch-Davies initial conditions at early times. To see
why they are not a worry, consider for example a term of the form
\begin{equation}
\frac{\slrl H^2}{\Mp^2}(k \eta )^2 \sim \frac{\slrl
  \ellp^2}{\lambda_{\mathrm{phys}}^2}
\end{equation}
where $\ellp = 1/\Mp$ is the Planck length and
$\lambda_{\mathrm{phys}} \sim a/k = (-H k \eta)^{-1}$. Such terms can
only be important for physical wavelengths $\lambda_{\mathrm{phys}}
\lesssim \sqrt{\slrl } \ellp$, and so lie well outside the domain of
validity of any EFT of gravity.

\subsection{Gaussian transport}
\label{ssec:Gausian hot}

Because the right-hand side of \Eq{Lindblad_2} is quadratic in the
fields it follows that an initially gaussian state -- such as the
Bunch-Davies vacuum -- remains gaussian under evolution. When this is
true \Eq{Lindblad_2} can be converted into a direct late-time
evolution equation for field correlations (rather than for the density
matrix), as we now show. This alternative formulation is possible
because for gaussian systems ${\varrho}_{\mathrm{\ssS}\bm{k}}$ is
completely characterized by the one- and two-point functions of fields
and canonical momenta:
\begin{align}
\left\langle  \w_{\mathrm{\ssS}\bm{k}} \w_{\mathrm{\ssS}\bm{k}'}  \right\rangle
&= P_{vv}(k) \, \delta(\bm{k}-\bm{k}') \,, \quad
 \left\langle  \p_{\mathrm{\ssS}\bm{k}}  \p_{\mathrm{\ssS}\bm{k}'}   \right\rangle
 = P_{pp}(k) \, \delta(\bm{k}-\bm{k}'),
\\ 
 \left\langle \w_{\mathrm{\ssS}\bm{k}} {\p}_{\mathrm{\ssS}\bm{k}'}  \right\rangle
 &=  \left[P_{v p}(k)+\frac{i}{2}\right] \, 
\delta(\bm{k}-\bm{k}')  \,.
\end{align}

Directly differentiating the definitions -- {\it e.g.}~$\langle
\w_{\bm{k}} \w_{\bm{k}} \rangle = \mathrm{Tr} [
  \w^2_{\mathrm{\ssS}\bm{k}} \, {\varrho}_{\mathrm{\ssS}\bm{k}} ]$ --
and evaluating the time derivative of $\varrho_{\ssS\bmk}$ using
\Eq{Lindblad_2}, together with commutation relations like
\Eq{equaltimeCCR}, leads to the following transport equations for
the power spectra
\begin{eqnarray}
\label{eq:Pvv'} 
P_{vv}'(k,\eta) &=& 2\, P_{vp}(k,\eta)\,, \\
\label{eq:Pvp'}
P_{vp}'(k,\eta) &=& P_{pp}(k,\eta) - \left\lbrace \omega^2(k,\eta) + 
\mathrm{Im}[\mathfrak{F}_{\bm{k}}(\eta,\eta_{\mathrm{in}})]
 \right\rbrace P_{vv}(k,\eta)\,, \\
 \label{eq:Ppp'}
P_{pp}'(k,\eta) &=& - 2\left\lbrace \omega^2(k,\eta) + \mathrm{Im}
[\mathfrak{F}_{\bm{k}}(\eta,\eta_{\mathrm{in}})]
 \right\rbrace P_{vp}(k,\eta)
+2\, \mathrm{Re}[\mathfrak{F}_{\bm{k}}(\eta,\eta_{\mathrm{in}})] \,. 
\end{eqnarray}
Like the Lindblad equation these equations can be integrated to late
times when the function $\mathfrak{F}_\bmk$ is independent of the
initial time $\eta_{\rm in}$.

The effect of the cubic interaction in this evolution is
twofold. First, it ensures that the frequency $\omega^2(k)$ only
appears in the combination
\begin{align}
  \widetilde{\omega}^2(k,\eta) :=\omega^2(k,\eta)
  +\mathrm{Im}[\mathfrak{F}_{\bm{k}}(\eta,\eta_{\mathrm{in}})] \,,
\end{align}
and so effectively alters the $k$-dependence of the dispersion
relation by shifting $\omega^2(k) \to \widetilde \omega^2(k)$. This
shift would also be expected from the form of \Eq{Lindblad_2} since
there $\mathrm{Im}(\mathfrak{F}_{\bm{k}})$ appears as an additive
contribution proportional to ${\w}^2_{\mathrm{\ssS}\bm{k}}$ in the
Hamiltonian.\footnote{In an unfortunate nomenclature corrections like
  these to the lowest-order dispersion relation have come to be
  referred to in the literature as `Lamb shift' terms.} As described
above, it is because UV divergences only appear in
$\mathrm{Im}(\mfF_\bmk)$ that they can be renormlized into parameters
appearing in $\omega^2(k)$. Such terms cannot drive decoherence
because they contribute to evolution as would a correction to the
Hamiltonian appearing in the Liouville equation, and so cannot evolve
pure states into mixed states.

The cubic interaction's other effect is to add a source term
$\mathrm{Re}(\mathfrak{F}_{\bm{k}})$ into the evolution equation for
$P_{pp}$. \Sec{sec:Observables} below shows that this term is the
one responsible for decoherence and because
$\mathrm{Re}(\mathfrak{F}_\bmk)$ is UV finite so must be the
decoherence rate. Showing this involves solving these equations
explicitly and this is facilitated by eliminating two of the variables
to obtain a single third-order differential equation for $P_{vv}$:
\begin{align}
\label{eq:Pvv:ode}
P_{vv}'''(k,\eta)+4\, \widetilde{\omega}^2(k,\eta) \, P_{vv}'(k,\eta)
+4\, \widetilde{\omega}(k,\eta)\, \widetilde{\omega}'(k,\eta)\,P_{vv}
=4\, \mathrm{Re}[\mathfrak{F}_{\bm{k}}(\eta,\eta_{\mathrm{in}})]\,.
\end{align}
Once this is solved the remaining correlators $P_{vp}$ and $P_{pp}$
are found from \Eqs{eq:Pvv'} and (\ref{eq:Pvp'}).

\subsection{Contribution from the tensor environment}
\label{ssec:Tensor}

Before finding solutions we conclude this section by computing the
rate with which an environment of short-wavelength tensor modes
decoheres the long-wavelength scalars, showing that it gives twice the
rate found above from a scalar environment. The full rate for
decohering visible scalar fluctuations is the sum of the contributions
from smaller scalar and tensor modes.

We first show that a tensor environment contributes to decoherence
with the same leading parametric dependence on $\varepsilon_1$,
$H/\Mp$ and $-k\eta = k/(aH)$ as does a scalar environment. The same
arguments as given above again imply that decoherence first arises at
order $(H/\Mp)^2$ and does so only through cubic
interactions. Furthermore, short-wavelength tensor modes can only
decohere super-Hubble scalar fluctuations through the
tensor-tensor-scalar interactions listed in \RRef{Maldacena:2002vr}
[see also \Eq{othergravitonscalarcubics} and \Eq{action:tss}],
and of these only the interaction
\begin{equation}
\label{leadingTTScubic}
{} ^{(3)}S  \supset  \frac{\Mp^2}{8}
\int \dd t\, \dd^3 \bm{x}\, a \, \slrl \, \zeta \,
\partial_l\gamma_{ij} \partial_l \gamma_{ij} \,,
\end{equation}
contributes at leading order in slow-roll parameters and in powers of
$k/(aH)$.

Appendix~\ref{sec:decotensor} shows in detail how scalar evolution is
modified by the presence of the tensor environment, again leading to
an evolution equation of the Nakajima-Zwanzig form~\eqref{NZmodes},
but with the correlator $\mathscr{C}_{\bm{k}}$ replaced by
$\mathscr{C}_{\bm{k}} + \mathscr{T}_{\bm{k}}$, with
$\mathscr{T}_{\bm{k}}$ defined by
\begin{align}
\label{CR_FTz}
C_{\ssT}(\eta, \eta' ; \bm{y}) := \langle 0_{\ssB} | \left[ {B_\ssT}(\eta,\bm{x})
- \mathscr{B}_\ssT(\eta) \right]  \left[ {B_\ssT}(\eta',\bm{x}')
- \mathscr{B}_\ssT(\eta')  \right] | 0_{\ssB} \rangle 
=\int \frac{\dd^{3} \bm{k}}{(2\pi)^{3/2}}
\mathscr{T}_{\bm{k}}(\eta,\eta') e^{i \bm{k} \cdot \bm{y} } \,,
\end{align}
where $\mathscr{B}_\ssT(\eta) := \langle 0_{\ssB} |
{B_\ssT}(\eta,\bm{x}) | 0_{\ssB} \rangle$ and
\begin{equation}
\label{2pt_textz}
 {B_\ssT}(\bm{x}) : = \delta^{ij} \delta^{kl} \delta^{rs} \partial_{i}
    {v}_{kr}(\eta,\bm{x}) \partial_{j} {v}_{ls}(\eta,\bm{x})
\end{equation}
is the new (tensor) environmental operator implied by
\Eq{leadingTTScubic}.

Appendix~\ref{sec:decotensor} also evaluates the leading behaviour of
$\mathscr{T}_{\bm{k}}$ for small $(-k\eta)$, which turns out to be
twice the contribution from $\mathscr{C}_{\bm{k}}$ alone. The combined
tensor and scalar contributions are therefore three times larger than
the scalar result alone, with the combination of scalar and tensor
environments leading to a late-time limit $ - k \eta \ll 1$ limit that
is three times larger than in \Eq{ReFSHresult} [see
  \Eq{eq:FappendixE}]:
\begin{equation}
\label{ReFSHresultz}
\mathrm{Re}[ \mathfrak{F}_{\bm{k}}(\eta,\eta_{\mathrm{in}}) ]
\simeq \frac{\slrl  H^2 k^2}{1024 \pi^2 \Mp^2 } \left[\frac{60 \pi}{(-k\eta)^2} +
  \mathcal{O}\left(\frac{1}{-k\eta} \right)\right]
\qquad \hbox{(scalar+tensor environment)} \,.
\end{equation}
Again all dependence on $\eta_{\mathrm{in}}$ and $\kUV$ appear only at
subdominant order in $(-k\eta)$. We comment here that the $\kUV$-independence of the result follows because the most important scales for decoherence are the ones that are closest to the scale $k$ being decohered (both much larger than the Hubble length). Because the important scales are not at the cutoff, the decoherence is largely insensitive to the value used for $\kUV$. This also helps to further underline why the $v_{\ssA} \otimes (\partial v_{\ssB})^2$ interaction is the dominant contribution in going from \pref{Hclassicalint} to \pref{Hintmom} --- all other neglected interactions there contribute most near the value of $\kUV$ and so are unimportant for decoherence (for the same reason the value of $\kUV$ itself is not).

\section{Late-time solutions}
\label{sec:Observables}

We now solve the Lindblad equation~\eqref{Lindblad_2} to extract some
of its physical implications.
 
\subsection{Solution to the Lindblad equation}

Because the right-hand side of the Lindblad equation is quadratic in
the system field $v(\eta,\bm{x})$, the solutions for the reduced
density matrix in the field eigenstate basis remain Gaussian. In
Schr\"odinger picture this means
\begin{align}
\label{eq:rho:Gaussian}
\left\langle \w_{\mathrm{\ssS}\bm{k},1} \right\vert \varrho_{\mathrm{\ssS}\bm{k}}
\left\vert \w_{\mathrm{\ssS}\bm{k},2} \right\rangle =
\sqrt{\frac{ \mathrm{Re} (a_k) -  c_k }{\pi}} \; \exp\left(-\frac{a_k}{2}
  \, \w^2_{\mathrm{\ssS}\bm{k},1}  -\frac{a^*_k}{2}
 \, \w^2_{\mathrm{\ssS}\bm{k},2} 
+c_k \, \w_{\mathrm{\ssS}\bm{k},1} \w_{\mathrm{\ssS}\bm{k},2} \right)\,,
\end{align}
for time-dependent functions $a_k(\eta)$ and $c_k(\eta)$. This state
is properly normalised inasmuch as
$\mathrm{Tr}(\varrho_{\mathrm{\ssS}\bm{k}}) = 1$ and the requirement
$\varrho_{\mathrm{\ssS}\bm{k}}^\dagger =
\varrho_{\mathrm{\ssS}\bm{k}}$ implies $c_k$ is real.

The Lindblad equation determines the functions $a_k(\eta)$ and
$c_k(\eta)$, and because the state is Gaussian these are completely
determined by the two-point functions $P_{vv}(k)$, $P_{vp}(k)$ and
$P_{pp}(k)$, through the formulae
\begin{align}
P_{vv}(k)=\frac{1}{2 \left[\mathrm{Re} (a_k)-c_k\right]}\, ,\quad
P_{vp}(k)=-\frac{\mathrm{Im} (a_k)}
{2 \left[\mathrm{Re} (a_k)-c_k\right]}\, ,\quad
P_{pp}(k)=\frac{\left\vert a_k\right\vert^2-c_k^2}
{2 \left[\mathrm{Re} (a_k)-c_k\right]}\, ,
\end{align}
which invert to give
\begin{align}
\label{eq:ab:from:Pspectra:1}
\mathrm{Re} (a_k)&= \frac{1}{ P_{vv}(k)}\left[P_{vv}(k) P_{pp}(k)-P_{vp}^2(k)
+\frac{1}{4}\right] \, ,\quad 
\mathrm{Im} (a_k) = - \frac{P_{vp}(k)}{P_{vv}(k)}  \\
c_k & = \frac{1}{ P_{vv}(k)} \left[P_{vv}(k) P_{pp}(k)-P_{vp}^2(k)
  -\frac{1}{4}\right] \, .
\label{eq:ab:from:Pspectra:2}
\end{align}

A measure of the state's decoherence is given by its `purity', defined
by
\begin{equation}
\label{purity}
\mfp_{\bm{k}}(\eta) :=
\mathrm{Tr}\left[ {\varrho}^2_{\mathrm{\ssS}\bm{k}} (\eta)\right]  \,.
\end{equation} 
This quantity satisfies $0 \leq \mfp_\bmk \leq 1$ and $\mfp_\bmk = 1$
if and only if $\varrho_\bmk$ is a pure state. Decoherence is said to
be effective when $\mfp_{\bm{k}} \ll 1$. For a Gaussian state the
purity (\ref{eq:rho:Gaussian}) evaluates to~\cite{Serafini:2003ke,
  Grain:2019vnq, Colas:2021llj,Martin:2021znx}
\begin{align}
\label{eq:gamma:Cov}
\mfp_{\bm{k}} = \sqrt{\frac{\mathrm{Re} (a_k)-c_k}{\mathrm{Re} (a_k)+c_k} }
=\frac{1}{2\sqrt{ P_{vv}(k) P_{pp}(k)-P_{vp}^2(k) }}\, ,
\end{align}
where the second equality uses
\Eqs{eq:ab:from:Pspectra:1}-\eqref{eq:ab:from:Pspectra:2}.

The state is pure if and only if $c_k=0$, or equivalently $P_{vv}(k)
P_{pp}(k) - P_{vp}^2(k) = 1/4$. By contrast, the state is strongly
decohered when $\mfp_\bmk \ll 1$ and so $P_{vv}(k) P_{pp}(k) -
P_{vp}^2 \gg 1/4$. This corresponds to the case $c_k \simeq
\mathrm{Re}(a_k)$ and so \Eq{eq:rho:Gaussian} becomes
\begin{align}
\label{eq:rho:Gaussiandeco}
 \Bigl| \left\langle \w_{\mathrm{\ssS}\bm{k},1} \right\vert
 \varrho_{\mathrm{\ssS}\bm{k}} \left\vert \w_{\mathrm{\ssS}\bm{k},2}
 \right\rangle \Bigr| \propto \exp\left[-\frac{\mathrm{Re}(a_k)}{2} (
   \w_{\mathrm{\ssS}\bm{k},1} -\w_{\mathrm{\ssS}\bm{k},2} )^2
   \right]\,,
\end{align}
showing in particular that decoherence occurs in the field basis --
{\it i.e.}~$|\langle \w_1 | \varrho \,| \w_2 \rangle | \to \delta(\w_1
- \w_2)$ -- if $\Rea(a_k)$ also grows in this limit.

For later use we notice that \Eqs{eq:Pvv'}-\eqref{eq:Ppp'} imply the
following evolution equation for the combination of correlators that
controls the purity:
\begin{align}
\label{eq:det:Cov:eom}
\frac{\partial}{\partial\eta}
\left[P_{vv}(k,\eta) P_{pp}(k,\eta)-P_{vp}^2(k,\eta) \right]
=2\, \mathrm{Re}[\mathfrak{F}_{\bm{k}}(\eta,\eta_{\mathrm{in}})]\, P_{vv}(k)\, .
\end{align}
Among other things this confirms that decoherence is driven purely by
the UV-finite quantity $\mathrm{Re}(\mathfrak{F}_{\bm{k}})$, as
foreshadowed in earlier sections. Its late-time behaviour is reliable
in the regime $-k\eta \ll 1$ because in this regime
\Eq{ReFSHresultz} shows $\mathrm{Re}( \mathfrak{F}_\bmk)$ is
approximately independent of $\eta_{\rm in}$ and $k_\UV$.
    
\subsection{Solution to the transport equations}

It remains to solve the Lindblad equation to determine the functions
$a_k(\eta)$ and $c_k(\eta)$. We exploit the Gaussianity to do so
directly using the equivalent transport equations (\ref{eq:Pvv'})
through (\ref{eq:Ppp'}) or their equivalent (\ref{eq:Pvv:ode}).

\Eq{eq:Pvv:ode} can be integrated when there is no source term on
its right-hand side, with solution given by $P_{vv}=\vert
\tilde{u}_{\bm{k}}\vert^2$, where $\tilde{u}_{\bm{k}}$ solves the
Mukhanov-Sasaki equation $\tilde{u}_{\bm{k}}''+\widetilde{\omega}^2(k)
\tilde{u}_{\bm{k}}=0$ obtained using the modified dispersion relation
$\omega(k) \to \widetilde \omega(k)$. This solution builds in the
initial condition that it approaches the Bunch-Davies vacuum $\tilde
u_\bmk \to u_\bmk$ in the limit of vanishing cubic coupling.

Nonzero source term can then be included using the Green's function
formalism, leading to
\begin{align}
\label{eq:Pvv:exact}
P_{vv}(k,\eta)=\left\vert \tilde{u}_{\bm{k}}(\eta)\right\vert^2
+ 8 \int_{\eta_{\uin}}^\eta\dd\eta'\, 
\mathrm{Re} [\mathfrak{F}_{\bm{k}}(\eta',\eta_{\mathrm{in}})]
\, \mathrm{Im}^2 \left[\tilde{u}_{\bm{k}}(\eta')
\tilde{u}_{\bm{k}}^*(\eta)\right]  \, .
\end{align}
\Eqs{eq:Pvp'} and~\eqref{eq:Ppp'} then give the two other power
spectra,
\begin{align}
\label{eq:Ppp:exact}
P_{vp}(k,\eta) &= \mathrm{Re} \left[ \tilde{u}_{\bm{k}}'(\eta)
  \tilde{u}_{\bm{k}}^*(\eta)\right]  
+ 8 \int_{\eta_{\uin}}^\eta\dd\eta'\, 
\mathrm{Re} [\mathfrak{F}_{\bm{k}}(\eta',\eta_{\mathrm{in}})]
\,\mathrm{Im} \left[\tilde{u}_{\bm{k}}(\eta')
\tilde{u}_{\bm{k}}^*(\eta)\right]
\, \mathrm{Im} \left[\tilde{u}_{\bm{k}}(\eta')
  \tilde{u}_{\bm{k}}^{*\prime}(\eta)\right]
\\
P_{pp}(k,\eta) &= \left\vert \tilde{u}'_{\bm{k}}(\eta)\right\vert^2
+ 8 \int_{\eta_{\uin}}^\eta\dd\eta'\, 
\mathrm{Re} [\mathfrak{F}_{\bm{k}}(\eta',\eta_{\mathrm{in}})]
\, \mathrm{Im} ^2\left[\tilde{u}_{\bm{k}}(\eta')
\tilde{u}_{\bm{k}}^{*\prime}(\eta)\right] \,.
\end{align}
These in principle solve the Lindblad equation entirely once
\Eqs{eq:Pvv:exact}-\eqref{eq:Ppp:exact} are used in
\Eqs{eq:ab:from:Pspectra:1}-\eqref{eq:ab:from:Pspectra:2} to evaluate
the density matrix (\ref{eq:rho:Gaussian}). We do not write the
corresponding expression here explicitly because our purposes are
already well served by \Eqs{eq:Pvv:exact}-\eqref{eq:Ppp:exact}.

There are two distinct regimes in which these solutions can be
used. In straight-up perturbation theory they can be used directly
provided one works only to lowest order in the semiclassical
expansion. This in turn requires replacing $\tilde u_\bmk$ with
$u_\bmk$ inside the integrals given that $\mathfrak{F}_\bmk$ is
already order $(H/\Mp)^2$. An extended domain of validity could apply
in circumstances where $\mathrm{Re}(\mathfrak{F}_\bmk)$ and
$\mathrm{Im}(\mathfrak{F}_\bmk)$ are independent of $\eta_{\rm in}$,
but this must be checked on a case-by-case basis.

\subsection{Quantifying decoherence}

We may now compute the time-dependence of the state's purity by
directly integrating \Eq{eq:det:Cov:eom}. Assuming that the state is
pure at $\eta = \eta_{\rm in}$ we have
\begin{equation}\label{purityXi}
   \mfp_\bmk(\eta) = \frac{1}{\sqrt{1 + \Xi_\bmk(\eta)}}
\end{equation}
with
\begin{align} 
\label{eq:purity:exact}
\Xi_\bmk(\eta) = 8 \int_{\eta_\uin}^\eta \dd\eta'\; 
\mathrm{Re}[\mathfrak{F}_{\bm{k}}(\eta',\eta_{\mathrm{in}})]P_{vv}(k,\eta') \,,
\end{align}
where $P_{vv}$ is given in \Eq{eq:Pvv:exact}. We next evaluate this
expression and assess its domain of validity.

To that end we first remark that $P_{vv}$ never deviates much from its
counterpart in the free theory, namely $P_{vv}^{\mathrm{free}}=\vert
u_{\bm{k}}\vert^2$ where $u_{\bm{k}}$ is the Bunch-Davies mode
function~\eqref{eq:deSitter:BD:vk}. Although \Eq{F_superHubble_body}
shows that $\mathrm{Im}(\mathfrak{F}_{\bm{k}}) \propto (H/\Mp)^2
\eta^{-2} \propto a^2H^4/\Mp^2$ (up to logarithms) grows strongly on
super-Hubble scales, it does not grow faster than $\omega^2\propto
\eta^{-2} \propto a^2 H^2$. Up to logarithms both quantities grow at
the same rate and so $\mathrm{Im}(\mathfrak{F}_{\bm{k}})$ gives a
correction to the effective mass of super-Hubble fluctuations (and so
to the tilt of the power spectrum) of relative size
$\mathrm{Im}(\mathfrak{F}_{\bm{k}})/\omega^2\propto (H/\Mp)^2$ (and
therefore remains negligible even at late times).

The integral in \Eq{eq:Pvv:exact} can thus be evaluated by
letting $\tilde{u}_{\bm{k}}\simeq u_{\bm{k}}$. Using the approximate
form~\eqref{ReFSHresultz} (which includes the contribution of
small-scale tensors) for Re $\mathfrak{F}_{\bf{k}}$ on
super-Hubble scales, together with the super-Hubble limit
\begin{equation}
\mathrm{Im}\left[ u_{\bm{k}}(\eta') u_{\bm{k}}^{\ast}(\eta) \right]
\simeq  \frac{(\eta^3 - \eta^{\prime 3})}{6 \eta \eta^{\prime }}
+ \mathcal{O}(k^2\eta^3),
\end{equation}
which follows from \Eq{eq:deSitter:BD:vk}, leads to the power-spectrum
correction
\begin{eqnarray}
\Delta P_{vv}&\simeq& 
\frac{5 \slrl  H^2}{192 \pi \Mp^2} \left[\frac{1}{6}
\left( \frac{\eta^4}{\eta_{\uin}^3}
-\frac{\eta_{\uin}^3}{\eta^2} \right)
-\eta \log\left(\frac{\eta}{\eta_{\mathrm{in}}}\right) \right].
\end{eqnarray}
Here, $\eta_\uin$ corresponds to the time at which the integration
starts, which in this expression is assumed to be already
super-Hubble, so that the Markovian approximation holds (see
\Sec{sec:validity:Markov}). In practice, suppose $k=\sigma a_\uin
H_\uin$, where $\sigma<1$ denotes the ratio between the Hubble radius
and the mode wavelength at the initial time. The above thus implies a
very small, time-independent and scale-invariant fractional correction
to the power spectrum
\begin{equation}
\label{powerchange}
\frac{\Delta P_{vv}}{P_{vv}}   \simeq
\frac{5 \slrl  H^2 \sigma^3}{576 \pi \Mp^2} .
\end{equation}
In summary, as expected gravitational mode coupling leaves only a
tenuous imprint on the power spectrum: an extremely small correction
to the its tilt that is suppressed by $(H/\Mp)^2$, and a correction to
its amplitude that is suppressed by $\slrl (H/\Mp)^2$ (see also \cite{brahma2022universal}).

Returning now to the decoherence, we therefore approximate $P_{vv}$ in
\Eq{eq:purity:exact} by its Bunch-Davies counterpart, $P_{vv} =
|u_\bmk|^2$. Recall that \Eq{eq:purity:exact} is derived by
integrating the Markovian evolution equations and so strictly speaking
its validity requires $-k\eta' \ll 1$ throughout the entire
integration region. But within this region the small $k\eta'$ form of
$\mathfrak{F}_\bmk(\eta',\eta_{\rm in})$ shows that the integrand
strongly peaks\footnote{\label{techfoot} As a technical aside, one
  might worry that because Re $[\mathfrak{F}_\bmk(\eta,\eta_{\rm
      in})]$ is singular in the coincident limit, integrating through
  the singular region away from small $k\eta$ might give a competing
  contribution. Appendix C.3 shows that these contributions are in
  fact subdominant in $k \eta$. } in the regime $-1 \ll k\eta' \leq
k\eta$, with the dominant contribution coming at late times driven by
the universal leading behaviour shown in \Eq{ReFSHresultz}. Using
this and the super-Hubble form $P_{vv} \simeq a^2 H^2/(2k^3) =
1/(2k^3\eta^2)$ in \Eq{eq:purity:exact} leads to 
\begin{equation} 
\label{eq:purity:exactr}
\Xi_\bmk(\eta) \simeq    \frac{5 \slrl  }{64 \pi}
\left( \frac{ H}{ \Mp} \right)^2 \frac{1}{(-k\eta)^3}
=   \frac{5 \slrl  }{64 \pi} \left( \frac{ H}{ \Mp } \right)^2
\left( \frac{aH}{k} \right)^3  \,,
\end{equation}
showing how $\Xi_\bmk(\eta)$ grows strongly at late times. This type
of growth proportional to $a^3$ is as expected fairly generally when
decoherence is driven by a short-distance environment
\cite{Burgess:2006jn, Burgess:2014eoa}, and is also seen in the
strictly perturbative calculation of \cite{Nelson:2016kjm}.

Eq.~\ref{eq:purity:exactr} leads to several interpretational
questions. First, the strong $a^3$ growth implies that
$\Xi_\bmk(\eta)$ can easily become order unity over 60 $e$-foldings of
inflation despite the presence of the extremely small factors $\slrl
H^2/\Mp^2$. Can \Eqs{purityXi} and (\ref{eq:purity:exactr}) still be
trusted once $\Xi_\bmk$ is no longer small? We argue that they can
because in the late-time regime where this growth dominates the
evolution is controlled by \Eq{Lindblad_1} -- or \Eq{Lindblad_2}
-- with a coefficient function $\mathfrak{F}_\bmk$ that is independent
of $\eta_{\rm in}$. This is the regime for which the arguments of
\cite{Burgess:2015ajz, Kaplanek:2019dqu, Kaplanek:2019vzj} (see also
\cite{Burgess:2020tbq}) allow the domain of validity of the Lindblad
evolution equation to be broader than the naive perturbative domain on
which the Lindblad evolution itself is derived.\footnote{This is much
  the same way that the evolution equation $\exd n/\exd t = - \Gamma
  n$ reliably predicts the exponential decay law when $\Gamma t$ is
  greater than unity, despite the decay rate $\Gamma$ itself usually
  being computed only perturbatively.} This is why we use the full
form \pref{purityXi} rather than expanding $\mfp_\bmk$ out to linear
order in $\Xi_\bmk$.

Another conceptual question concerns the restriction $-k\eta_{\rm in}
\ll 1$ that is required to keep the entire integration regime of
\Eq{eq:purity:exact} within the Markovian regime. Ideally we'd
instead like to take the limit $\eta_{\rm in} \to - \infty$, pushing
the initial epoch when environment and system are uncorrelated to the
distant past where the uncorrelated Bunch-Davies modes for the
environment are deeply sub-Hubble and effectively behave as if they
are in flat space. 

Physically one expects that no decoherence arises for early times because 
at these times both the system modes of interest and their shorter-wavelength
brethren in the environment are all sub-Hubble and so behave largely as they would
in flat space. The vacuum then doesn't decohere for the same reason that short-wavelength vacuum modes don't spontaneously decohere long-wavelength quantum systems around us
all the time in flat space. In principle this can be demonstrated explicitly in perturbation theory without the need for any late-time resummation (such as by going back
to the full Nakajima-Zwanzig evolution equation \pref{NZmodes} to
compute how things evolve once $k\eta' \ll -1$ in the presence of the
full correlation function $\mathscr{C}_\bmk +
\mathscr{T}_\bmk$).\footnote{It is important when doing so to keep in mind that
all times are evaluated along a time contour that has a small negative imaginary 
part (as is also done here), since this is what projects the initial state onto the 
vacuum along the lines described in \cite{Maldacena:2002vr}.} 

One therefore expects decoherence to begin only once modes become of order Hubble size, and because decoherence rates are usually suppressed relative to the enivronment's underlying correlation scale (the Hubble scale in the de Sitter example) by factors of the system-environment coupling (in our case gravitational in size: $H^2/\Mp^2$) it is plausible that most of the decoherence happens only once modes are deep in the super-Hubble regime. In this paper we compute for technical reasons the late-time decoherence (with $k\eta_{\rm in}$ also chosen to be super-Hubble) but the fact that our leading results are universal (independent of $k_\UV$ and $\eta_{\rm in}$) strongly suggests that the decoherence rate we find also captures the dominant super-Hubble evolution experienced by modes prepared in the Bunch Davies vacuum at $\eta_{\rm in} \to - \infty$.

A final dangling question involves the relative importance of cubic interactions involving environmental fields differentiated with respect to time rather than space. As argued earlier, these are subdominant when the environmental modes are super-Hubble, so the validity of neglecting them depends on whether Hubble-sized or sub-Hubble modes in the environment contribute significantly to the decoherence of super-Hubble system states. At face value our calculation does not rule this out because it shows that the decoherence of super-Hubble sytem modes at time $\eta$ is mostly driven by environmental correlations between times $\eta$ and $\eta' \to \eta$. Since this receives contributions from very short wavelength fluctuations it is conceivable that sub-Hubble modes are significant, and so we are continuing to examine whether additional environmental operators can contribute significanly to the decoherence rate.

\subsubsection{Numerical estimates}

Let us re-express the small factor $\slrl H^2/\Mp^2$ appearing in
the decoherence parameter~(\ref{eq:purity:exact}) in
terms of common observables, in order to better understand its size. 

In the single-field slow-roll inflationary models of interest here,
the amplitude of the primordial scalar power spectrum is given by
\begin{equation}
 \mathcal{P}_\zeta \simeq \frac{H^2}{8\pi^2\slrl \Mp^2} \simeq 2.2\times
10^{-9}
\end{equation}
where the numerical size is what is required for this to explain the
observed primordial power spectrum~\cite{Planck:2018jri}. Furthermore,
in these models the first slow-roll parameter is directly related to
the tensor-to-scalar ratio $r$ by $r=16\slrl$. It follows that
\begin{align} \label{eq:purityfinal0}
\Xi _{\bm k}(N)\simeq 1.7\times 10^{-17}
\left(\frac{\mathcal{P}_\zeta}{2.2\times 10^{-9}}\right)
\left(\frac{r}{10^{-3}}\right)^2 e^{3(N_\mathrm{end}-N_*)-3(N_\mathrm{end}-N)}
\,,
\end{align}
where we trade time for the number of inflationary $e$-folds $N$, with
$N_\mathrm{end}$ denoting the number of $e$-folds at the end of
inflation while $N_*=N_*(k)$ is the number of $e$-folds evaluated at
the horizon exit during inflation for mode $k$. The same expression
can alternatively be written in terms of the energy scale at which
inflation proceeds, $\rho_\mathrm{inf}^{1/4}$, leading to
\begin{align}
\label{eq:purityfinal}
  \Xi _{\bm k}(N)=1.6 \times 10^5
  \left(\frac{\mathcal{P}_\zeta}{2.2\times 10^{-9}}\right)^{-1}
  \left(\frac{\rho_\mathrm{inf}^{1/4}}{\Mp}\right)^8
  e^{3(N_\mathrm{end}-N_*)-3(N_\mathrm{end}-N)}
 \,.
\end{align}

For scales probed in CMB experiments one typically has
$N_\mathrm{end}-N_*\simeq 50$ and if so \Eq{eq:purityfinal0} or
\pref{eq:purityfinal} show the purity for CMB scales at the end of
inflation is given by
\begin{align}
  \mfp_\mathrm{CMB}(N_\mathrm{end})\simeq 
  6.5 \times 10^{-25} \left(
  \frac{10^{-3}}{r}\right)  
  \simeq 2.1
  \times 10^{-35} \left(\frac{\Mp}{\rho_\mathrm{inf}^{1/4}} \right)^4 .
  \end{align}
Gravitational decoherence is efficient if inflation proceeds at energy
scales larger than $\rho_\mathrm{inf}^{1/4}>5.2 \times 10^{9}$ GeV or
(equivalently) $r>6.5\times 10^{-28}$. Alternatively, a future
detection of the tensor-to-scalar ratio $r$ at a level above $r\sim
10^{-3}$ as targeted by future CMB polarization
experiments~\cite{Hazumi:2019lys} (which puts the energy scale of
inflation above $10^{15}\GeV$) implies decoherence proceeds quickly
after Hubble crossing during inflation.

\begin{figure}
\centering
  \includegraphics[width=0.65\linewidth]{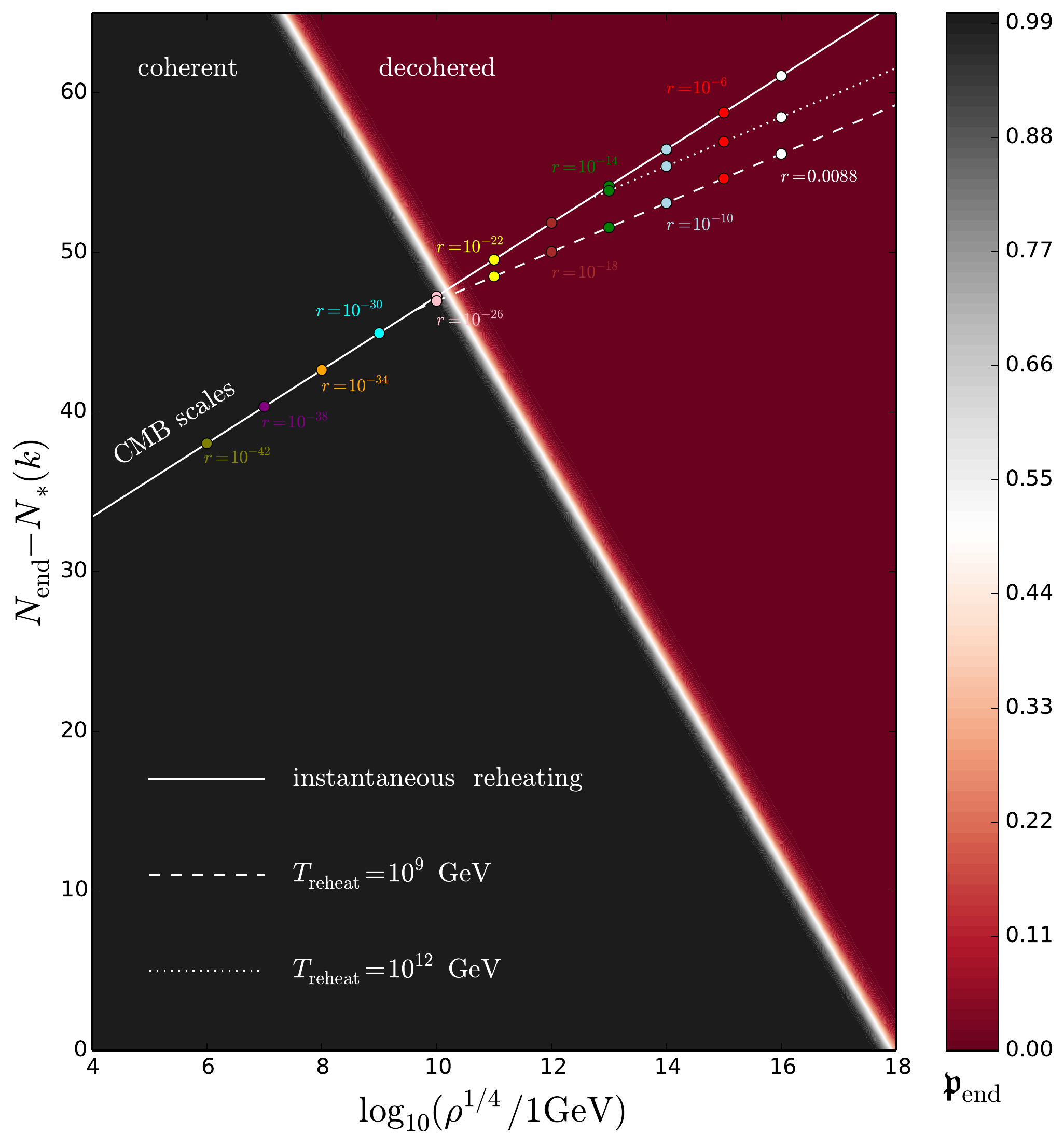}
  \caption{Purity as a function of the scale (namely, number of
    $e$-folds before the end of inflation) and the energy scale of
    inflation. Also plotted are predictions for how these two
    quantities are related for the value of $k$ that is just
    re-entering the Hubble scale today, as a function of assumptions
    made about the post-inflationary reheat epoch, with the solid,
    dashed and dotted lines respectively representing instantaneous
    reheating at the end of inflation,
    $T_\mathrm{reh}=10^{12}\mathrm{GeV}$ and
    $T_\mathrm{reh}=10^{9}\mathrm{GeV}$ (with $g_*\simeq 1000$ and
    $w=0$ during reheating for the latter two).
    }
\label{fig:purity:map}
\end{figure}

More generally, the purity at the end of inflation, $\mfp_\mathrm{end}
:= \mfp_{\bm{k}}(N_\mathrm{end})$, depends on two parameters: the mode
$k$ of interest -- characterized by $N_\mathrm{end}-N_*(k)$ -- and the
energy scale of inflation, $\rho_\mathrm{inf}^{1/4}$. This dependence
is shown as a colour scale in Fig.~\ref{fig:purity:map}, which reveals
two regions -- one for which $\mfp_\mathrm{end} \ll 1$ (red
`decohered' region) and one for which $\mfp_\mathrm{end}\simeq 1$
(black `coherent' region), separated by the relatively abrupt
transition represented by the thin white region. This transition
corresponds to where $\Xi_\bmk$ in \Eq{eq:purityfinal} is order
unity; {\it i.e.}~by the straight line $N_\mathrm{end}-N_*\simeq
110-6.14 \ln \left[\rho_\mathrm{inf}^{1/4}/(1 \mathrm{GeV})\right]$.

For comparison, for any fixed $k$ a relationship is also predicted
between $N_\mathrm{end}-N_*(k)$ and $\rho_\mathrm{inf}^{1/4}$ by
following the mode's post-inflationary evolution through the reheating
epoch up to the present day. For $k$ corresponding to a physical
wavelength that today equals the Hubble radius,
$k/a_\mathrm{now}=H_\mathrm{now} = h\, (100 \hbox{ km/sec/Mpc})$ the
prediction is
\begin{align}
\label{Nvsrho0}
N_*-N_\mathrm{end} &=
\ln \left[\left(\Omega_{\gamma 0}\right)^{-1/4}
\frac{\rho_\mathrm{cri}^{1/4}}{1\mathrm{GeV}}
\left(\frac{\pi^2 g_*}{30}\right)^{\frac{-1+3 w}{12(1+w)}}
\right]-\frac{2}{3} \left( \frac{1+3w}{1+w} \right)
\ln \left(\frac{\rho_\mathrm{inf}^{1/4}}{1\mathrm{GeV}}\right)
\nonumber \\ &
-\frac{1-3w}{3(1+w)}\ln \left(\frac{T_\mathrm{reheat}}{1 \mathrm{GeV}}\right)\, ,
\end{align}
where $\Omega_{\gamma 0} \simeq 2.471 \times 10^{-5}/h^2$ is the
fraction of radiation today, $\rho_\mathrm{cri} \simeq 8.099 \, h^{2}
\times 10^{-47}\mathrm{GeV}^{-4}$ is the critical energy density
today, $g_*$ the number of degrees of freedom after the end of
inflation, $T_\mathrm{reheat}$ the reheat temperature and $w = p_{\rm
  rh}/\rho_{\rm rh}$ the equation of state parameter during
reheating. The curves expressing \Eq{Nvsrho0} are also shown in
Fig.~\ref{fig:purity:map} for several choices of reheating properties.

If CMB polarization experiments detect a cosmic gravitational wave
background in the foreseeable future then $r$ cannot be too much
smaller than $r \simeq 10^{-3}$ and the above expressions show that
only modes that leave the Hubble radius less than $\simeq 12.9$
$e$-folds before the end of inflation do not have time to
decohere. Whether the very small scales associated with such modes can
be probed observationally is of course an open question.

\section{Discussion}
\label{sec:Conclusions}

The main conceptual result of this paper is to apply open-system
techniques to the evolution equation for the quantum state of observed
metric modes during single-field inflation, including effects
generated by their gravitational interaction with shorter wavelength
modes. We identify those gravitational interactions that dominate the
decoherence of long-wavelength modes, and show that their contribution
to the evolution of longer-wavelength modes is given by equations like
\pref{NZmodes}.

\subsection{Decoherence of scalar modes}

We compute the relevant environmental correlation functions -- {\it
  i.e.}~the function $\mathscr{C}_\bmk(\eta)$ appearing in
\Eq{NZmodes} and its tensor counterpart $\mathscr{T}_\bmk(\eta)$
once short-wavelength tensors are included in the environment -- and
show that their peaked form implies the evolution becomes Markovian
for super-Hubble modes during inflation, leading to the approximate
evolution equations \pref{Lindblad_1} or \pref{Lindblad_2} (in
interaction and Schr\"odinger pictures, respectively) whose
coefficient function $\mathfrak{F}_\bmk(\eta,\eta_{\rm in})$ we also
compute explicitly. In particular $\mathfrak{F}_\bmk(\eta,\eta_{\rm
  in})$ turns out to be universal (depend only on $k\eta$ to leading
approximation in the super-Hubble, late-time regime and grows strongly
with dominant support at late times when $k\eta$ is small). All
dependence on other parameters (like $\eta_{\rm in}$ and $k_\UV$)
arising only at subdominant order in $k\eta$. We also derive an
equivalent set of evolution equations -- \Eqs{eq:Pvv'} through
\pref{eq:Ppp'} -- for correlation functions in this Markovian regime.

Because $\mathfrak{F}_\bmk$ peaks at small $k\eta$ it is super-Hubble
modes in the environment that dominate the decoherence process. This
in turn simplifies the selection of the dominant interactions because
it implies that interactions can be neglected if they involve time
derivatives (as opposed to spatial derivatives) acting on
environmental fields.

Following \Refs{Burgess:2015ajz, Kaplanek:2019dqu, Kaplanek:2019vzj, Burgess:2020tbq, Colas:2022hlq}
we argue that the universality of
$\mathfrak{F}_\bmk(\eta,\eta_{\rm in})$ allows the solutions to
\Eqs{Lindblad_1} and \pref{Lindblad_2} to be trusted to later times
than usual for perturbative reasoning, and so allows the late-time
resummation of the secular growth found within the perturbative
predictions for decoherence. We compute the late-time behaviour
implied by the leading Markovian evolution and use it to evaluate the
evolution of the mode purity.

The main practical results that emerge from this reasoning are given
in \Eqs{eq:purity:exactr} and \eqref{powerchange} that respectively
describe the rapid decoherence of observable primordial scalar metric
fluctuations and the (very small) change to their power spectrum
arising from their interactions with the environment of
shorter-wavelength metric modes. Both scalar and tensor metric
fluctuations in the environment act to decohere super-Hubble scalar
modes and do so with the same dependence on inflationary parameters
and the same efficiency per mode. The two spin states then ensure that
tensor modes give twice the decoherence of scalars.

Because super-Hubble modes during inflation decohere at a rate
proportional to $(aH/k)^3 \propto e^{3Ht}$, they grow exponentially
quickly in cosmic time. This growth is initially small because it is
suppressed by a small coupling prefactor $\propto \slrl G H^2$, and so
starts off at most $10^{-14}$ in single-field models. But it can
become order unity during the 40-60 $e$-foldings of inflation that
follow horizon exit and so can easily allow decoherence to be complete
well before inflation ends. We believe our predictions continue to be
valid even when decoherence is not small because the Lindblad
evolution resums this late time growth.  The associated modification
to the power spectrum for scalar and tensor fluctuations remains very
small (being loop-suppressed) at all times.

\subsection{Decoherence of tensor modes by a scalar environment}

We note in passing that the slow-roll suppression found above is not
generic for all fluctuations. In particular super-Hubble tensor modes
decohere {\it more quickly} than scalar modes because their dominant
interactions are unsuppressed by slow-roll parameters. Since the
factors of $H/\Mp$ and $(aH/k)$ arise on very general grounds (leading
order in the semiclassical expansion and the environment having
shorter wavelength than the decohering modes) one is led to expect the
decoherence of tensor modes to be of order $(H/\Mp)^2 (aH/k)^3$, and
so be larger than our scalar result by a factor of $1/\slrl $.

This expectation can be tested using a relatively minor extension of
the calculations described above that give the decoherence rate of
tensor modes due to an environment of shorter-wavelength scalar metric
fluctuations. We here provide a partial calculation of this; computing
only the decoherence caused by interactions with environmental
short-wavelength scalar metric fluctuations. A full calculation of the
total rate with which tensor modes are decohered by their couplings to
the short-wavelength tensor environment is in
preparation.

\subsubsection{Lindblad evolution}

Repeating the same arguments as above shows that the leading cubic
tensor-scalar-scalar interactions from the list in
\RRef{Maldacena:2002vr} -- {\it c.f.}~\Eq{allgravitonscalarcubics}
-- is
\begin{equation}
\label{allgravitonscalarcubicsy}
^{(3)}S  \supset  \int \dd t\, \dd^3 \bm{x}\,
\Mp^2  \slrl  a \gamma^{ij} (\partial_i \zeta)( \partial_j \zeta)  \,,
\end{equation}
which contributes to the interaction hamiltonian the slow-roll
unsuppressed interaction
\begin{equation}
\label{HintRdefy_body}
{\cH}_{\mathrm{int}}(\eta) = - \frac{1}{\Mp a} \int\dd^3 \bm{x}\,
{v}_{ij} (\eta, \bm{x}) \otimes {B^{ij}}(\eta,\bm{x}) \,.
\end{equation}
Here $v_{ij}$ is the canonical field related to $\gamma_{ij}$ by
\Eq{canonicaltensor} and the environmental operator
\begin{equation}
\label{couplingdefy}
{B}^{ij}(\bm{x})  : =  \delta^{ik} \delta^{jl}
\partial_{k} {v}(\eta,\bm{x}) \partial_{l} {v}(\eta,\bm{x}) \,,
\end{equation}
is related to the operator $B$ of \Eq{eq:B:def} by $\delta_{ij}
B^{ij} = B$. Crucially, the absence of slow-roll suppression in
\Eq{HintRdefy_body} gives it an effective coupling
\begin{equation}\label{tildeGvsG}
\widetilde G(\eta) = - \frac{1}{\Mp a(\eta)}
 = 2 \sqrt{\frac{2}{\slrl }} \; G(\eta) \,,
\end{equation}
where $G(\eta)$ is the effective coupling defined in \Eq{couplingdef}. 

Rotation invariance implies $\mathscr{B}^{ij} = \langle 0_\ssB |
B^{ij} | \,0_\ssB \rangle \propto \delta^{ij}$ and so because
$\delta^{ij} v_{ij} = 0$ the mean of $B^{ij}$ can be ignored in what
follows. Similarly, the required correlation function is
\begin{align}
\label{2pt_texty}
C^{iajb}(\eta,\eta' ; \bm{x} - \bm{x}') &:=
\langle 0_{\ssB} | \left[ {B}^{ia}(\eta,\bm{x})
  - \mathscr{B}^{ia}(\eta) \right]  \left[ {B}^{jb}(\eta',\bm{x}')
  - \mathscr{B}^{jb}(\eta')  \right] | 0_{\ssB} \rangle  \\
& =  \int  \frac{\dd^3\bm{k}}{(2\pi)^{3/2}}\; \mathscr{C}^{iajb}_{\bm{k}}
\;  e^{ i \bm{k} \cdot ( \bm{x} - \bm{x}' )}  \,,   
\end{align}
with
\begin{equation}
\mathscr{C}^{iajb}_{\bm{k}} =\frac{2}{(2\pi)^{9/2}}
\int_{q,p>k_{\UV}} \dd^3\bm{q}\; \dd^{3}\bm{p} \; p^i q^a p^j q^b \,
u_{{q}}(\eta) u_{{p}}(\eta) u^{\ast}_{{q}}(\eta') u^{\ast}_{{p}}(\eta')
\, \delta(\bm{p} + \bm{q} - \bm{k})   \,,
\end{equation}
and comparing this with \Eq{Ck1} shows that $\mathscr{C}_{\bm{k}} =
\delta_{ia} \delta_{jb}\mathscr{C}^{iajb}_{\bm{k}}$ is the quantity
computed in \Sec{ssec:EnvCorr} above. Appendix
\ref{App:Tensordecohere} shows how rotation invariance can also be
used to reduce this to a set of scalar integrals, related to those
appearing in $\mathscr{C}_{\bm{k}}(\eta,\eta')$. Once contracted with
the graviton polarization tensors the required correlator at leading
order in $k\eta$ is again universal -- {\it i.e.}~independent of
$\eta_{\rm in}$ and $k_\UV$ -- and related to our earlier result by
\begin{equation}
\mathscr{C}_{\bm{k}}^{iajb}  \epsilon^{\ssP}_{ia} \epsilon^{\ssP'}_{jb}
\simeq \frac{2}{15} \; \mathscr{C}_{\bm{k}} \delta^{\ssP\ssP'}
\end{equation}
up to subdominant order in $k\eta$. 

Keeping in mind the coupling~\eqref{tildeGvsG}, repeating the same
steps as in earlier sections imply the analog of the Nakajima-Zwanzig
evolution equation \eqref{NZmodes} involves
$\mathscr{C}^{ijkl}_{\bm{k}}$, and in the super-Hubble limit again
takes the Lindblad form \eqref{Lindblad_1} but with \Eq{Ffrakdef}
now given by the leading universal form
\begin{equation}
\label{Ffrakdefy}
\mathrm{Re}\left[ \mathfrak{T}_{\bm{k}}(\eta, \eta_{\mathrm{in}}) \right]
\simeq  \frac{\slrl  H^2 k^2}{1024 \pi^2 \Mp^2 }
\left\lbrace \frac{20 \pi}{( - k \eta)^2} +
\mathcal{O}\left[ (- k \eta)^{-1} \right]\right\rbrace
\left( \frac{16}{15 \slrl } \right)
\simeq \frac{ H^2 }{48 \pi \Mp^2 \eta^2} \, ,
\end{equation}
with $\mathfrak{T}_{\bm{k}}$ the Lindblad coefficient analogous to $\mathfrak{F}_{\bm{k}}$ for the scalar (defined in eq.~\eqref{LindbladT_def}).

Comparing this with the result \pref{ReFSHresult} for how scalars
modes decohere other scalar modes reveals the expected lack of
slow-roll suppression.

\subsubsection{Decoherence}

The contribution of scalars already shows that tensor modes decohere
more quickly than do scalar modes because their self-interactions are
less suppressed by slow-roll parameters. The purity of tensor modes is
given by an expression like \Eq{purityXi} with $\Xi_\bmk$ replaced by
\begin{align} 
\label{eq:purity:exactT}
\Xi^\ssT_\bmk(\eta) = 8 \int_{\eta_\uin}^\eta \dd\eta'\; 
\mathrm{Re}[\mathfrak{T}_{\bm{k}}(\eta',\eta_{\mathrm{in}})]
|u_\bmk(\eta)|^2 \,,
\end{align}
and the universal late-time contribution of the short-wavelength
scalar environment to $\mathfrak{F}_\bmk$ given by
\Eq{Ffrakdefy}. This implies a leading late-time contribution to
$\Xi^\ssT_\bmk$ of size
\begin{equation}
\label{ReFSHresulty}
\Xi^\ssT_\bmk(\eta) \simeq    \frac{1}{36\pi}
\left( \frac{ H}{ \Mp} \right)^2 \frac{1}{(-k\eta)^3}
=   \frac{1 }{36\pi} \left( \frac{ H}{ \Mp} \right)^2
\left( \frac{aH}{k} \right)^3  \,,
\end{equation}
which is to be compared with equation \pref{eq:purity:exactr}. In
particular it is unsuppressed by slow-roll parameters, as claimed (see also \cite{gong2019quantum,ye2018quantum}). The
full expression for tensor decoherence requires adding to this the
contribution of the tensor environment.

\subsection{Open questions}

The calculation described herein is clearly only the first step in a
potentially long journey. Many things remain to be pinned down, and
there is much value in doing so particularly if primordial
fluctuations turn out to have a quantum origin (as they so far seem to
do). We next list some of the open issues for scalar-mode decoherence,
and then turn to what our result might imply for proposed late-time
searches for observational quantum
signatures~\cite{Campo:2005sv,Maldacena:2015bha,Martin:2015qta,Martin:2017zxs}.

One open issue concerns post-inflationary evolution of the reduced
density matrix. We have shown how super-Hubble evolution during
inflation acts to diagonalize the density matrix in a field basis, but
we also know that the large factor $(aH/k)^3$ shrinks after inflation
until it is again order unity at horizon re-entry. When saying that
decoherence is efficient in the later universe we take for granted
that the reduced density matrix remains effectively diagonal during
the subsequent post-inflationary evolution before horizon
re-entry. This seems intuitively reasonable: entanglement can be
fragile and one rarely expects initially classical mixed systems to
spontaneously self-purify. But this should be possible to prove, and
we have not yet done so.

A related question concerns the effects of other degrees of freedom
with other (possibly stronger) interactions or other variations like possible changes to the speed of sound (as in the EFT of inflation \cite{Cheung:2007st}). We have no general statements as to how more complicated interactions might change our result \pref{eq:purity:exactr} apart from the general observation that most interactions are stronger than gravity and so likely produce a larger effect, making the decoherence progress faster --- it is in this sense that our calculation is minimal. One
suspects that many of these issues can benefit from further exchange
of ideas between cosmology and the physics of open quantum systems.

\subsection{Loopholes}

At face value the efficiency of gravitational decoherence we find
makes observing quantum coherence in measurements of primordial
fluctuations much harder, assuming that these are generated by quantum
fluctuations in single-field inflationary models and that the scale of
inflation is not extremely low. So if evidence for coherent
fluctuations should appear tomorrow, what might this mean?

Calculations like ours are useful for this question because they
identify how decoherence depends on a theory's parameters. Even within
the domain of validity of our result the inflationary scale might turn
out to be very small. Or the modes in question might be short enough
that they do not spend as much time outside the Hubble scale (even for
$\slrl \sim 10^{-2}$ modes can spend about 10 $e$-foldings outside the
Hubble scale during inflation before decoherence stops being
negligible).

There are also a variety of assumptions on which our calculations
rely, at least one of which would have to break down. Perhaps the
model involves more than the single inflaton; if so then $\zeta$ need
not be conserved and this conservation played an important role in
suppressing interactions involving time derivatives. Perhaps the
important environmental scales are not at short wavelengths compared
with observable modes; if so then their interactions need not be local
and so the general argument of \RRef{Burgess:2014eoa} need no longer
imply the same dependence on $a^3$ as found here. Perhaps additional
interactions can re-cohere initially decohered states. Inquiring minds
need to know.

Let us also note that the precise amount of decoherence it takes to
erase a particular quantum feature can vary. For instance, in
\RRef{Martin:2021znx}, it was shown that decohered states in a de
Sitter universe still carry a large quantum discord if decoherence is
slow enough, \ie if $\mathfrak{p}_{\bm{k}}\propto a^{-p}$ with
$p<4$. Given that we have found $p=3/2$ in the case of gravitational
decoherence, this implies that, although decoherence is very
effective, the erasure of quantum discord is not, which might still
leave open the possibility to detect quantum signatures.  There is
also the possibility to look for quantum effects at small scales, that
spend too few \efolds~ outside the Hubble radius during inflation to
efficiently decohere. If inflation proceeds at GUT scale, such
wavelengths cannot be probed in the CMB, but they might be accessible
in smaller-scale structures.

In many ways finding observational evidence for quantum coherence
amongst primordial fluctuations is the most attractive option; by our
present lights it would be the most surprising and so likely teach us
the most.

\section*{Acknowledgements}
We thank Tim Cohen, Thomas Colas, Julien Grain, Amaury Micheli,
Mehrdad Mirbabayi, Enrico Pajer, Eva Silverstein and Junsei Tokuda for
helpful conversations, and the Corfu Summer Institute for providing
such pleasant environs in which some of them took place. CB's research
was partially supported by funds from the Natural Sciences and
Engineering Research Council (NSERC) of Canada. Research at the
Perimeter Institute is supported in part by the Government of Canada
through NSERC and by the Province of Ontario through MRI. G.K.~is supported by the Simons Foundation award ID 555326 under the Simons Foundation Origins of the Universe initiative, Cosmology Beyond Einstein's Theory as well as by the European Union Horizon 2020 Research Council grant 724659 MassiveCosmo ERC2016COG.

\appendix

\section{Higher-derivative operators in the action}
\label{App:operators}

This appendix fleshes out the details of the derivation of the
standard EFT of scalar and tensor fluctuations for single-field
inflation (defined in \RRef{Maldacena:2002vr}), as well as the
expansion of higher-curvature squared operators when this is
understood as an EFT of gravity. The purpose of this appendix is
two-fold: first, it is to review the set of cubic interactions
considered in this work and in particular to further identify which
interactions are dominant in driving decoherence. Secondly, it is to
provide evidence that the counter-term operators renormalizing
divergences in \Sec{sec:renormalization} are indeed amongst those in
the EFT of gravity in single-field inflation.

\subsection{Single-field inflation}

We start with a brief review of the standard fluctuation formalism
that is to be used. As described in \Sec{sec:OpenEFT}, single-field
inflation consists of gravity and a single scalar field in the form of
the action (\ref{actionstart}), repeated here for convenience,
\begin{equation}
\label{actionAppstart}
S[g_{\mu \nu},\varphi] = \int \dd^4 x\; \sqrt{ - g}
\biggl[ \frac{\Mp^2}{2} R - \frac{1}{2}
g^{\mu\nu} \, \partial_{\mu} \varphi \, \partial_{\nu} \varphi
- V(\varphi) \biggr] \ .
\end{equation}
Homogeneous classical solutions for the inflaton $\phi(t)$ and Hubble
parameter $H = \dot{a}/a$ (recall that a dot means derivative with
respect to cosmic time) therefore obey
\begin{equation}
\label{background_App}
3 \Mp^2 H^2  =  \tfrac{1}{2} \dot{\phi}^2 + V(\phi) \,, \quad
\Mp^2 \dot{H}  =  - \tfrac{1}{2} \dot{\phi}^2  \quad \hbox{and} \quad
\ddot{\phi} + 3 H \dot{\phi} + V^{\prime}(\phi) = 0 \,.
\end{equation}
We perturb about a near-de Sitter spacetime, $\dd s^2 = - \dd t^2 +
a^2(t) \dd \bm{x}^2$, working in the Arnowitt-Deser-Misner (ADM)
formalism as in \RRef{Maldacena:2002vr} using the perturbed metric
\begin{equation}
  \dd s^2  = - N^2 \dd t ^2+ h_{ij} \left( N^i \dd t + \dd x^i \right)
  \left( N^j \dd t + \dd x^j \right) \ , 
\end{equation}
with $N$ the lapse function and $N^{i}$ the shift vector and the
inverse of spatial metric $h^{ij}$ defined by $h^{ij}h_{jk} =
\delta^{i}_{\; k}$. In terms of these variables the action
(\ref{actionAppstart}) becomes
\begin{align}
\label{actionApp}
S & = \int \dd^4 x\; \sqrt{h} N \left[\frac{\Mp^2}{2}
  \left( \cR  + \frac{E_{ij}E^{ij} - E^2}{N^2} \right) - V(\varphi)
  + \frac{\big( \dot{\varphi} - N^{i} \partial_{i} \varphi\big)^2}{2 N^2}
  - \frac{1}{2} h^{ij} \partial_{i} \varphi \, \partial_{j} \varphi\right]\, 
\end{align}
with $\cR$ the 3D Ricci scalar built from the spatial metric $h_{ij}$
and $K_{ij} = E_{ij} / N$ is the extrinsic curvature of these spatial
slices, where
\begin{equation}
  E_{ij} := \sfrac{1}{2} \left(\dot{h}_{ij} - \nabla_{i} N_{j}
  -  \nabla_{j} N_{i}  \right) \quad \hbox{and} \quad E:= h^{ij} E_{ij} \ .
\end{equation}

Specializing to the gauge where the inflaton has no perturbation,
$\delta \varphi = 0$ and so $\varphi = \phi(t)$, the vanishing spatial
derivative $\partial_{j}\varphi =0$ allows the action to be simplified
to
\begin{align}
\label{actionApp2}
S & = \int \dd^4 x\; \sqrt{h} N
\left[ \frac{\Mp^2}{2} \left( \cR
  + \frac{E_{ij}E^{ij} - E^2}{N^2} \right) - V(\phi)
  + \frac{\dot{\phi}^2}{2 N^2} \right] \ .
\end{align}
The constraint equations obtained by varying $N$ and $N^{i}$ then become
\begin{equation}
\label{constraints}
\nabla_{i} \left( \frac{E^{i}_{\; j} - \delta^{i}_{\; j} E}{N} \right)
= 0, \quad \frac{\Mp^2}{2} \left( \cR + \frac{E_{ij}E^{ij} - E^2}{N^2}
\right) - V(\phi) + \frac{ \dot{\phi}^2}{2N^2} = 0 \,.
\end{equation}
These constraint equations are solved for $N$ and $N^{i}$ as functions
of the physical variables $\zeta$ and $\gamma_{ij}$, defined by
\begin{equation} \label{generalgauge}
  h_{ij} = a^2 e^{2 \zeta} \left( \delta_{ij} + \gamma_{ij}
  + \frac{1}{2} \gamma_{i\ell}\gamma_{\ell j} + \ldots \right) \,,
\end{equation}
with $\partial_{i} \gamma_{ij} = \gamma_{ii} = 0$. The goal is to
express the action~\eqref{actionApp2} as a function of these variables
after eliminating $N$ and $N^i$ using the constraints. Since our focus
is mainly on scalar fluctuations we drop $\gamma_{ij}$ in what
follows, simply quoting when needed the graviton-dependent terms found
elsewhere~\cite{Maldacena:2002vr}. For the metric (\ref{generalgauge})
the following relations prove useful:
\begin{equation}
\sqrt{h} = a^3 e^{3\zeta}, \quad   \cR   = a^{-2} e^{ - 2 \zeta}
\left[ - 4 ( \partial^2 \zeta ) - 2 ( \partial \zeta )^2 \right] \,,
\end{equation}
where ``$\partial$'' here denotes spatial differentiation.

\subsection{Quadratic scalar action}

We first verify the standard quadratic action for $\zeta$. At leading
order in $\zeta$ the lapse and shift are
\begin{equation}
\label{lapseshiftanswer}
N \simeq 1 + \frac{\dot{\zeta}}{H}, \quad
N_{i} \simeq -  \frac{\partial_{i} \zeta}{a^{2}H} + \partial_i \chi \,,
\end{equation}
where the field $\chi$ is defined as a solution to the equation
$\partial^2 \chi = \dot{\phi}^2 \dot{\zeta}/(2H^2\Mp^2)$, and so
\begin{equation}
\label{chidef}
\chi := \frac{\dot{\phi}^2}{2H^2\Mp^2} \partial^{-2}
\dot{\zeta} = \slrl  \partial^{-2} \dot{\zeta}  \ .
\end{equation}
Using the background equations of motion (\ref{background_App}) and
integrating by parts gives the quadratic action
\begin{equation}
\label{freescalaraction_App}
{}^{(2)}S = \int \dd t \; \dd^3 \bm{x}\; \Mp^2 \slrl
\left[ a^3 \dot{\zeta}^2 - a (\partial \zeta )^2 \right]
\end{equation}
as given as \Eq{freescalaraction} in the main text, with $\slrl = -
\dot H/H^2 = {\dot{\phi}^2}/({2H^2 \Mp^2})$. This becomes canonical
when expressed in terms of the Mukhanov-Sasaki field
\begin{equation}
\label{MS_field_App}
v  = a \Mp \sqrt{2 \slrl } \zeta
\end{equation}
and when specialized to near-de Sitter
geometries. \Eq{freescalaraction_App} has the same form as two of
the three divergent terms found in \Eq{renorm_operators} that require
renormalizing in the Lindblad equation:
\begin{equation}
{}^{(2)}S \simeq \int \dd \eta \; \dd^3 \bm{x}\; \left[ \frac{1}{2} (v')
+ \frac{1}{\eta^2} v^2 - \frac12 (\partial v)^2 \right]  \ .
\end{equation}

\subsection{Cubic scalar interactions}

We next record the cubic self-interactions contained in the cubic part
of the expansion $S ={}^{(2)}S +{}^{(3)}S + \ldots$ of the
action in powers of the fluctuations found in
\RRef{Maldacena:2002vr}. The part cubic in the scalar perturbation can
be written
\begin{align}
\label{allscalarcubics}
{}^{(3)}S & = \int \dd t\, \dd^3 \bm{x}\, \Biggl\lbrace
\slrl ^2 \Mp^2 \left[ a \left(\partial \zeta\right)^2  \zeta
+ a^3 \dot{\zeta}^2 \zeta \right] - 2 \slrl ^2 \Mp^2 a^3 \,
\dot{\zeta} \left(\partial_i \partial^{-2}\dot{\zeta}\right)
\left(\partial_i \zeta\right)
- \frac{1}{2} \epsilon_{1}^3 \Mp^2 a^3 \dot{\zeta}^2 \zeta
\nonumber \\ &
+2 \slrl  \Mp^4 a^3 \dot{\zeta} \zeta^2 \frac{\dd}{\dd t}
\left( \frac{\ddot{\phi}}{2\dot{\phi}H}
+ \frac{\dot{\phi}^2}{4 H^2 \Mp^2} \right)
+ \frac{1}{2} \slrl ^3 \Mp^2 a^3 \left(\partial_{i}\partial_{j}
\partial^{-2} \dot{\zeta} \right)
\left(\partial_{i}\partial_{j}  \partial^{-2} \dot{\zeta} \right)
\zeta  \Biggr\rbrace \, .
\end{align}
This way of writing the cubic action is organized in increasing powers
of the slow-roll parameter, with the first line containing the
dominant terms and the rest being subdominant. Although the first line
is naively $\mathcal{O}(\slrl ^2)$ the quadratic
action~\eqref{freescalaraction_App} shows that correlations of $\zeta$
are themselves enhanced by slow-roll parameters. For this reason
slow-roll behaviour is easier to read when the action is expressed in
terms of $v$, and shows that the leading term of
\Eq{allscalarcubics} is actually $\cO(\sqrt{\slrl })$. Notice also
that for super-Hubble modes each time derivative of $\zeta$ counts as
two spatial derivatives, so only the very first term dominates when
both slow-roll parameters and $k/(aH)$ are small.

\subsection{Tensor fluctuations}

Tensor fluctuations are also straightforward to include by expanding
the action (\ref{actionAppstart}), leading to the quadratic piece
\begin{equation}
{}^{(2)}S \supset \int \dd t \, \dd^3 \bm{x}\,
\frac{\Mp^2}{8} \bigg[ a^3 \dot{\gamma}_{ij} \dot{\gamma}_{ij}
- a (\partial_{k} \gamma_{ij} )(\partial_{k} \gamma_{ij} ) \bigg] \,,
\end{equation}
and revealing the canonically normalized field to be 
\begin{equation}
\label{canonicaltensor}
v_{ij} = \frac{1}{2} a \Mp \gamma_{ij} \,.
\end{equation}
The momentum-space mode expansion for this field then becomes
\begin{equation}
\label{tensormodeexp}
v_{ij}(\eta,\bm{x}) = \int \frac{\dd^3 \bm{k}}{(2\pi)^{3}}
\sum_{\ssP = +,\times} \epsilon_{ij}^{\ssP}(\bm{k}) v_{\bm{k}}^{\ssP}(\eta)
e^{ i \bm{k} \cdot \bm{x}}
\end{equation}
where $\epsilon_{ij}^\ssP({\bm k})$ is the polarization tensor with
properties $k^{i}\epsilon^{\ssP}_{ij}(\bm{k})=0$ and
$\epsilon_{ij}^{\ssP}\epsilon_{ij}^{\ssP'} = \delta^{\ssP \ssP'}$.

The cubic interactions that involve both scalar and tensor
fluctuations are obtained along the same lines as above, leading to
the following expressions~\cite{Maldacena:2002vr} for the
tensor-scalar-scalar interaction:
\begin{align}
\label{allgravitonscalarcubics}
{}^{(3)}S & \supset \int \dd t\ \dd^3 \bm{x}\, \Mp^2
\biggl[ \slrl  a \gamma_{ij} (\partial_i \zeta)( \partial_j \zeta)
  + \frac{1}{4} \, a^3 \partial^{2}\gamma_{ij}
  ( \partial_i \chi )( \partial_j \chi) 
  + \frac{1}{2} \slrl  a^3 \dot{\gamma}_{ij} (\partial_i \zeta)
  (\partial_{j} \chi)
\nonumber \\ &
+ \frac{1}{2}  H a^5 \dot{\gamma}_{ij} \dot{\gamma}_{ij} \chi \biggr],
\end{align}
which ignores redundant terms such as total derivatives and those that
can be removed with field re-definitions. The $\chi$-dependent terms
become non-local expressions once $\chi$ is eliminated using
\Eq{chidef}. Again only the very first term dominates in the
slow-roll limit, but this time arises unsuppressed by $\slrl $ as is
most easily seen when expressed using $v$ rather than $\zeta$.

The tensor-tensor-scalar interaction is found in an identical way and
is given by
\begin{equation}
\label{othergravitonscalarcubics}
{}^{(3)}S  \supset  \int \dd t\, \dd^3 \bm{x}\ \Mp^2
\left(  \frac{ \slrl }{8} \, a \, \zeta \, \partial_l\gamma_{ij} \partial_l
\gamma_{ij}+\frac{ \slrl }{8} a^3 \zeta \dot\gamma_{ij} \dot \gamma_{ij}
- \frac14 a^3 \dot \gamma_{ij} \partial_l \gamma_{ij} \partial_l \chi \right)\,,
\end{equation}
where $\chi$ is again given by \Eq{chidef} and redundant terms are
dropped. For super-Hubble modes only the first term dominates in the
slow-roll approximation and once powers of $k/(aH)$ are neglected,
showing that the leading result is in this case $\cO(\sqrt{\slrl })$.

\subsection{Curvature-squared counterterms}

We finally show that the final divergent contribution depends on
$\bm{k}$ and $\eta$ in a way consistent with it being absorbed into a
curvature-squared counter-term, which can have the general form
\begin{equation}
S \ \supset \ \int \dd^4 x\, \sqrt{ - g}
\left(c_0 R^2 + d_0 R_{\mu\nu} R^{\mu\nu} \right) ,
\end{equation}
where $c_0$ and $d_0$ are two constants. There is no
$R_{\mu\nu\sigma\lambda}R^{\mu\nu\sigma\lambda}$ term here because we
work in 4D where this term can be regarded as part of a topological
invariant.

In particular, the operator proportional to $k^4 \eta^2 \big(
{v}^{(\alpha)}_{\mathrm{\ssS}\bm{k}} \big)^2$ in \Eq{renorm_operators}
is found within the $R^2$ term, which when expanded in powers of
$\zeta$ contains the contribution
\begin{equation}
\int \dd^4 x\, \sqrt{ - g}\, R^2  =   \int \dd t\, \dd^3 \bm{x}\,
a^{3} (1 + 3 \zeta) \left( 1 + \frac{\dot{\zeta}}{H} \right)
\left\lbrace \frac{\Mp^2 a^{-2} e^{- 2 \zeta}
\left[- 4 ( \partial^2 \zeta )
- 2 ( \partial \zeta )^2 \right] }{2}
+ \cdots \right\rbrace^2 \,,
\end{equation}
which when expanded out involves the term
\begin{align}
\int \dd^4 x\, \sqrt{ - g}\, R^2 & \supset
\int \dd t\, \dd^3 \bm{x}\, a^{3} \left(1 + 3 \zeta\right)
\left( 1 + \frac{\dot{\zeta}}{H} \right) \, \Mp^4 a^{-4}
e^{- 4 \zeta} 4 \left( \partial^2 \zeta \right)^2 
\nonumber \\ &
\supset \int \dd t\, \dd^3 \bm{x}\,  4 \Mp^4 a^{-1} ( \partial^2 \zeta )^2 
=  \int \dd \eta\; \dd^3 \bm{x}\, \frac{2 \Mp^2}{\slrl  a^2}
( \partial^2 v )^2 \,.
\end{align}
Once translated to Fourier space (and using $a \propto \eta^{-1}$)
this has the same time- and momentum-dependence --- $k^4 \eta^2 \big(
{v}^{(\alpha)}_{\mathrm{\ssS}\bm{k}} \big)^2$ --- that appears in the
last operator of \Eq{renorm_operators}.

\section{Environmental Correlations}
\label{App:envcorr_eta}

\subsection{Correlations in real space}

In this Appendix, we explicitly compute the correlation function
$\mathscr{C}_{\bm{k}}(\eta,\eta')$ defined in \Eq{CR_FT} for each mode
$\bm{k}$ in the Lindblad equation associated with the environment
operator
\begin{equation}
{B}(\eta,\bm{x}) = \delta^{ij} \partial_{i} {v}_{\ssB}(\eta,\bm{x})
\partial_{j} {v}_{\ssB}(\eta,\bm{x}) \ . 
\end{equation}
Using the canonical commutation relations (\ref{equaltimeCCR}) it is
easy to see that the one-point function defined in \Eq{R_1pt_def} has
the integral representation
\begin{align}
\label{1ptfunctionvarying}
\mathscr{B}(\eta) \ : = \ \langle 0_{\ssB} | {B}(\eta,\bm{x}) | 0_{\ssB}
\rangle & = - \iint_{k,q>k_{\UV}} \frac{\dd^3\bm{k}\;
  \dd^{3}\bm{q}}{(2\pi)^3}\, (\bm{k} \cdot \bm{q})\,
u_{\bm{k}}(\eta)\, u^{\ast}_{\bm{q}}(\eta) \, \delta(\bm{k}
+ \bm{q}) \, e^{i ( \bm{k} + \bm{q} ) \cdot \bm{x}}  \\
& = \int_{k>k_{\UV}} \frac{\dd^3\bm{k}}{(2\pi)^3}\;
k^2 \left \vert u_{\bm{k}}(\eta) \right \vert ^2 \ .
\end{align}
This function is independent of the position $\bm{x}$ and is also
formally divergent, although we avoid regulating it since it does not
enter into any physical predictions in this work.

From here we similarly compute the two-point function:
\begin{align}
  \langle 0_{\ssB} | {B}(\eta,\bm{x}) {B}(\eta',\bm{x}') | 0_{\ssB} \rangle
 & = \iint_{k,q>k_{\UV} } \frac{\dd^3\bm{k}\; \dd^{3}\bm{q}}{(2\pi)^3}
  \iint_{p,\ell>k_{\UV} } \frac{ \dd^{3}\bm{p}\; \dd^{3}\bm{\ell}}{(2\pi)^3}\;
  \nonumber \\
  & \times \;  (\bm{k} \cdot \bm{q}) (\bm{p}
  \cdot \bm{\ell}) \langle 0_{\ssB} |
{v}_{\bm{k}}(\eta) {v}_{\bm{q}}(\eta)
{v}_{\bm{p}}(\eta') {v}_{\bm{\ell}}(\eta') | 0_{\ssB} \rangle \;
e^{ i ( \bm{k} + \bm{q} ) \cdot \bm{x} + i ( \bm{p} + \bm{\ell}  ) \cdot \bm{x}'} \, .
\end{align}
In the expectation value above there are many combinations of creation
and annihilation operators which occur, but only the following two
survive:
\begin{align}
\langle 0_{\ssB} | {c}_{\bm{k}} {c}^{\dagger}_{-\bm{q}} {c}_{\bm{p}}
{c}^{\dagger}_{-\bm{\ell}} | 0_{\ssB} \rangle & =
\delta(\bm{k} + \bm{q}) \delta(\bm{p} + \bm{\ell}) \\
\langle 0_{\ssB} | {c}_{\bm{k}} {c}_{\bm{q}}
{c}^{\dagger}_{-\bm{p}} {c}^{\dagger}_{-\bm{\ell}} | 0_{\ssB} \rangle & =
\delta(\bm{k} + \bm{\ell}) \delta(\bm{p}
+ \bm{q}) + \delta(\bm{k} + \bm{p}) \delta(\bm{\ell} + \bm{q}).
\end{align}
This gives
\begin{align}
\label{twopointvarying1}
\langle 0_{\ssB} | {B}(\eta,\bm{x}) {B}(\eta',\bm{x}') | 0_{\ssB} \rangle
&=  \iint_{k,q>k_{\UV}} \frac{\dd^3\bm{k} \,\dd^{3}\bm{q}}{(2\pi)^3}
\iint_{p,\ell>k_{\UV}} \frac{ \dd^{3}\bm{p} \,\dd^{3}\bm{\ell}}{(2\pi)^3} \,
(\bm{k} \cdot \bm{q}) (\bm{p} \cdot \bm{\ell}) \,
e^{ i ( \bm{k} + \bm{q} ) \cdot \bm{x} + i ( \bm{p} + \bm{\ell}  ) \cdot \bm{x}'}\notag  \\
& \times \big\lbrace u_{\bm{k}}(\eta) u^{\ast}_{\bm{q}}(\eta)
u_{\bm{p}}(\eta') u^{\ast}_{\bm{\ell}}(\eta') \delta(\bm{k} + \bm{q})
\delta(\bm{p} + \bm{\ell}) 
+ u_{\bm{k}}(\eta) u_{\bm{q}}(\eta) u^{\ast}_{\bm{p}}(\eta')
u^{\ast}_{\bm{\ell}}(\eta')
\nonumber \\ & \times
\left[ \delta(\bm{k} + \bm{\ell})
  \delta(\bm{p} + \bm{q}) + \delta(\bm{k} + \bm{p})
  \delta(\bm{\ell} + \bm{q})\right] \big\rbrace .
\end{align}
Using \Eq{1ptfunctionvarying} the first pair of $\delta$-functions can
be seen to give rise to a pair of 1-point functions. In addition to
this, the term involving the second pair of $\delta$-functions is
easily seen to be equal to the term involving the third pair after a
re-labeling of momenta giving
\begin{align}
  \langle 0_{\ssB} | {B}(\eta,\bm{x}) {B}(\eta',\bm{x}') | 0_{\ssB} \rangle
  & = \mathscr{B}(\eta) \mathscr{B}(\eta') 
  + \ 2 \iint_{k,q>k_{\UV}} \frac{\dd^3\bm{k}\, \dd^{3}\bm{q}}{(2\pi)^3}
  (\bm{k} \cdot \bm{q}) (\bm{k} \cdot \bm{q}) \,
  e^{ i ( \bm{k} + \bm{q} ) \cdot (\bm{x}  -  \bm{x}')}
  \nonumber \\ & \times
  u_{\bm{k}}(\eta) u_{\bm{q}}(\eta) u^{\ast}_{\bm{k}}(\eta')
  u^{\ast}_{\bm{q}}(\eta') ,
\end{align}
and so
\begin{align}
\label{2ptfunctionfluctuatingvarying}
C_\ssB(\eta,\eta' ; \bm{x} - \bm{x}') & =
\langle 0_{\ssB} | B(\eta,\bm{x}) B(\eta',\bm{x}')
| 0_{\ssB} \rangle - \mathscr{B}(\eta) \mathscr{B}(\eta') \nonumber \\
& = \iint_{q,p>k_{\UV}} \frac{\dd^3\bm{q}\, \dd^{3}\bm{p}}{(2\pi)^6}\,
2 (\bm{q} \cdot \bm{p})^2  u_{\bm{q}}(\eta) u_{\bm{p}}(\eta)
u^{\ast}_{\bm{q}}(\eta') u^{\ast}_{\bm{p}}(\eta')
e^{ i ( \bm{q} + \bm{p} ) \cdot ( \bm{x} - \bm{x}' )} .  
\end{align}

\subsection{Correlations for each mode $k$}
\label{App:Cketaetap}

\subsubsection{Integral expression}

Finally we here compute \Eq{CR_FT} defined in the main text, which is
the Fourier transform of the fluctuating part of the environment
correlation function, where
\begin{equation}
  \mathscr{C}_{\bm{k}}(\eta,\eta')
  = \int \frac{\dd^{3} \bm{y}}{(2\pi)^{3/2}} \, C_\ssB(\eta, \eta' ; \bm{y})
  e^{- i \bm{k} \cdot \bm{y}} \ .
\end{equation}
We note in particular that we are interested in
$\mathscr{C}_{\bm{k}}(\eta,\eta')$ for modes $\bm{k} \in
\mathbb{R}^{3+}$ in the open system with $ 0<k<k_{\UV} $ since the
Nakajima-Zwanzig equation (\ref{NZmodes}) is written in terms of these
modes. We begin with the position space representation
(\ref{2ptfunctionfluctuatingvarying})
\begin{align}
C_\ssB(\eta,\eta' ; \bm{y}) = \iint_{q,p>k_{\UV}}
\frac{\dd^3\bm{q}\, \dd^{3}\bm{p}}{(2\pi)^6}\,
2 (\bm{q} \cdot \bm{p})^2 \,  u_{\bm{q}}(\eta) u_{\bm{p}}(\eta)
u^{\ast}_{\bm{q}}(\eta') u^{\ast}_{\bm{p}}(\eta')\,
e^{ i ( \bm{q} + \bm{p} ) \cdot \bm{y} } .
\end{align}
We notice that the (double) momentum integration is restricted to be
in the environment only. Next notice that $C_\ssB(\eta,\eta';\bm{y})$
is a function of $y = |\bm{y}|$ only --- to see why, note that any
rotation $\bm{y} \to \bm{y}' = \mathbf{R} \bm{y}$ can be undone by a
rotation of coordinates in the integrals, such that $\bm{q}' =
\mathbf{R}^{-1} \bm{q}$ and $\bm{p}' = \mathbf{R}^{-1} \bm{p}$ (each
with Jacobian one) so that clearly $C_\ssB(\eta,\eta';\bm{y}) =
C_\ssB(\eta,\eta;\mathbf{R}\bm{y})$ for any rotation matrix
$\mathbf{R}$. This then of course implies the Fourier transform
$\mathscr{C}_{\bm{k}}(\eta,\eta')$ is a function of $k = |\bm{k}|$
only.

Inserting the above expression for $C_\ssB(\eta,\eta';\bm{y})$ into
our definition for $\mathscr{C}_{\bm{k}}(\eta,\eta')$ we get
\begin{equation}
\label{Ck1}
\mathscr{C}_{\bm{k}}(\eta,\eta') = \frac{2}{(2\pi)^{9/2}}
\iint_{q,p>k_{\UV}} \dd^3 \bm{q} \, \dd^3 \bm{p}\, \left(\bm{q}
\cdot\bm{p} \right)^2 u_{\bm{q} }\left(\eta\right)
u_{\bm{p}}\left(\eta\right) u_{\bm{p}}^*\left(\eta'\right)
u_{\bm{q}}^*\left(\eta'\right) \delta(\bm{q} + \bm{p} - \bm{k}).
\end{equation}
From here we convert to spherical coordinates $\bm{q} = (q,\theta_{q},
\varphi_{q})$ and $\bm{p} = (p,\theta_{p}, \varphi_{p})$, where we
have
\begin{equation}
  \bm{q} \cdot \bm{p} = q\, p \, \left[ \sin \theta_{q}  \sin \theta_{p}
    \cos(\varphi_{q}  - \varphi_{p}) + \cos \theta_{q}  \cos \theta_{p}
    \right],
\end{equation}
and, in addition, we have the identity (for any arbitrary vector
$\bm{\ell}$)
\begin{equation}
  \delta (\bm{q} - \bm{\ell}) = \frac{\delta(q - \ell)
    \delta(\theta_q - \theta_\ell) \delta(\varphi_q - \varphi_\ell)}
         {q^2 | \sin \theta_{q}| } \ .
\end{equation}
In \Eq{Ck1} we must express the vector $\bm{\ell} := \bm{k} - \bm{p}$
in spherical coordinates --- to this end, we exploit the rotational
symmetry of the function $\mathscr{C}_{\bm{k}}(\eta,\eta')$ in
$\bm{k}$ and pick $\bm{k} = (0,0,k)$ to be pointing along the
$3$-axis, so that in Cartesian coordinates $\bm{\ell} := \bm{k} -
\bm{p} = (-p_x, -p_y, k-p_z)$, which implies that
\begin{align}
\ell & = \sqrt{ p^2 + k^2  - 2 k p \cos\theta_p }\, , \quad 
\theta_{\ell} = \cos^{-1}\left( \frac{k-p \cos \theta_p}
  {\sqrt{p^2 + k^2  - 2 k p \cos \theta_p}} \right) \, , \quad 
\varphi_{\ell} =  \varphi_p + \pi \, .
\end{align}
With these identities, \Eq{Ck1} becomes
\begin{align}
\mathscr{C}_{\bm{k}}(\eta,\eta') & = \frac{2}{(2\pi)^{9/2}}
\int_{k_{\UV}}^{\infty} \dd q \, q^2 \int_{0}^{\pi} \dd \theta_{q} \sin \theta _q
\int_0^{2\pi} \dd \varphi_{q} \int_{k_{\UV}}^{\infty} \dd p \, p^2
\int_{0}^{\pi} \dd \theta_{p} \sin \theta_p \int_0^{2\pi} \dd \varphi_{p}
\nonumber \\ & \times 
(qp)^2 \left[ \sin \theta_{q} \sin \theta_{p}
  \cos(\varphi_{q}  - \varphi_{p}) + \cos \theta_{q} 
  \cos\theta_{p} \right]^2
u_{\bm{q} }\left(\eta\right)
u_{\bm{p}}\left(\eta\right) u_{\bm{p}}^*\left(\eta'\right)
u_{\bm{q}}^*\left(\eta'\right)
\nonumber \\ & \times
\frac{1}{q^2 \sin \theta_q}
\delta\left[q - \sqrt{ p^2 + k^2  - 2 k p \cos\theta_p } \right]
\delta\left[\theta_{q} -
\cos^{-1}\left(\frac{k-p \cos\theta_p}
    {\sqrt{p^2 + k^2  - 2 k p \cos\theta_p }} \right)  \right]
\nonumber \\ & \times
\delta\left[ \varphi_{q} - (\varphi_p + \pi ) \right] , 
\end{align}
Integrating over $\varphi_q$ and then $\varphi_{p}$ yields
\begin{align}
\mathscr{C}_{\bm{k}}(\eta,\eta') & =
\frac{2}{(2\pi)^{7/2}} \int_{k_{\UV}}^{\infty} \dd q \int_{0}^{\pi}
\dd \theta_{q} \int_{k_{\UV}}^{\infty} \dd p \int_{0}^{\pi} \dd \theta_{p} \, q^2\,
p^4 \sin \theta_p \left(- \sin \theta_{q}\sin \theta_{p}
  +\cos \theta_{q}  \cos\theta_{p}\right)^2
\nonumber \\ & \times
u_{\bm{q} }\left(\eta\right)
u_{\bm{p}}\left(\eta\right) u_{\bm{p}}^*\left(\eta'\right)
u_{\bm{q}}^*\left(\eta'\right)
\delta\left[q - \sqrt{ p^2 + k^2  - 2 k p \cos \theta_p } \right] \,
\nonumber \\ & \times
\delta\left[\theta_{q} - \cos^{-1}\left(\frac{k-p \cos \theta_p}
  {\sqrt{ p^2 + k^2  - 2 k p \cos\theta_p}} \right)  \right]. 
\end{align}
The next step consists in integrating over $\theta_q$. Using the
identities
\begin{align}
\cos\left[\cos^{-1}\left(\frac{k-p \cos\theta_p}
{\sqrt{ p^2 + k^2  - 2 k p \cos\theta_p}}  \right) \right]
=  & \frac{k-p \cos\theta_p}{\sqrt{ p^2 + k^2  - 2 k p \cos\theta_p}}, \\
\sin\left[\cos^{-1}\left(\frac{k-p \cos\theta_p}
{\sqrt{ p^2 + k^2  - 2 k p \cos\theta_p}}  \right) \right]
=& \frac{p \sin\theta_p}{\sqrt{ p^2 + k^2  - 2 k p \cos\theta_p}} ,
\end{align}
this leads to
\begin{align}
\mathscr{C}_{\bm{k}}(\eta,\eta') & = \sfrac{2}{(2\pi)^{7/2}}
\int_{k_{\UV}}^{\infty} \dd q\int_{k_{\UV}}^{\infty} \dd p
\int_{0}^{\pi} \dd \theta_{p} \,q^2 \,p^4 \sin \theta_p
\frac{ \left[ \left(k - p \cos \theta_{p} \right) \cos \theta_{p}
    - p \sin^2 \theta_{p}\right]^2}{ p^2 + k^2  - 2 k p \cos \theta_p }
\nonumber \\ &
u_{\bm{q} }\left(\eta\right)
u_{\bm{p}}\left(\eta\right) u_{\bm{p}}^*\left(\eta'\right)
u_{\bm{q}}^*\left(\eta'\right)
\delta\left(q - \sqrt{ p^2 + k^2  - 2 k p \cos \theta_p } \right).
\end{align}
Now switching coordinates to $\mu := \cos\ \theta_p$ turns the above into 
\begin{align}
\mathscr{C}_{\bm{k}}(\eta,\eta') &=
\frac{2}{(2\pi)^{7/2}} \int_{k_{\UV}}^{\infty} \dd q\int_{k_{\UV}}^{\infty}
\dd p \int_{-1}^{1} \dd \mu \,q^2 \,p^4
\frac{ ( p - k \mu )^2}{ p^2 + k^2  - 2 k p \mu } 
u_{\bm{q} }\left(\eta\right)
u_{\bm{p}}\left(\eta\right) u_{\bm{p}}^*\left(\eta'\right)
u_{\bm{q}}^*\left(\eta'\right)
\nonumber \\ & \times
\delta\left(q - \sqrt{ p^2 + k^2  - 2 k p \mu } \right) \ . 
\end{align}
Next, we notice that the $\delta$-function is actually easiest to
integrate over the $\mu$ variable, and so to this end we take note of
the rule $\delta\big[ f(\mu) \big] = \delta(\mu - \mu_0) / | f'(\mu_0)
|$ where $\mu_0 = (p^2 + k^2 - q^2)/(2pk)$ is the (only) zero of the
function $f(\mu) := q - \sqrt{ p^2 + k^2 - 2 p k \mu }$. This implies
\begin{equation}
  \delta\left( q - \sqrt{ p^2 + k^2 - 2 p k \mu } \right)
  = \frac{q}{kp} \, \delta\left( \mu - \frac{p^2 + k^2 - q^2}{2 p k} \right) 
\end{equation}
giving
\begin{align}
\label{Ck_beforemu2}
\mathscr{C}_{\bm{k}}(\eta,\eta') & = \frac{2}{(2\pi)^{7/2}k }
\int_{k_{\UV}}^{\infty} \dd q\int_{k_{\UV}}^{\infty} \dd p \, q^3 p^3
u_{\bm{q} }\left(\eta\right)
u_{\bm{p}}\left(\eta\right) u_{\bm{p}}^*\left(\eta'\right)
u_{\bm{q}}^*\left(\eta'\right)
\nonumber \\ & \times 
\int_{-1}^{1} \dd \mu \, \frac{ ( p - k \mu )^2}{ p^2 + k^2  - 2 k p \mu }
\, \delta\left( \mu - \frac{p^2 + k^2 - q^2}{2 p k} \right) \ .
\end{align}
The $\mu$-integration can now be performed in the sense that
\begin{equation}
  \int_{-1}^{1} \dd \mu \,
  \frac{ ( p - k \mu )^2}{ p^2 + k^2  - 2 k p \mu } \,
  \delta\left( \mu - \frac{p^2 + k^2 - q^2}{2 p k} \right)
  = \frac{( q^2 + p^2 - k^2 )^2}{4 q^2 p^2}
\end{equation}
but only in the region of momentum space where
\begin{equation}
- 1\ < \ \frac{p^2 + k^2 - q^2}{2 p k} \ < \ 1 \ ,
\end{equation}
which affects the region of integration in the $(p,q)$-plane. Note
that this means the region which the above inequality bounds is
actually rectangular --- to see why note that $- 1 < \frac{p^2 + k^2 -
  q^2}{2 p k}$ implies $(p+k)^2 > q^2$ while $ \frac{p^2 + k^2 -
  q^2}{2 p k} < 1$ implies $(p-k)^2 < q^2$, which means that this
region corresponds to
\begin{equation}
| p - k | < q < p + k \ . 
\end{equation}
\begin{figure}
\centering
\includegraphics[width=0.55\linewidth]{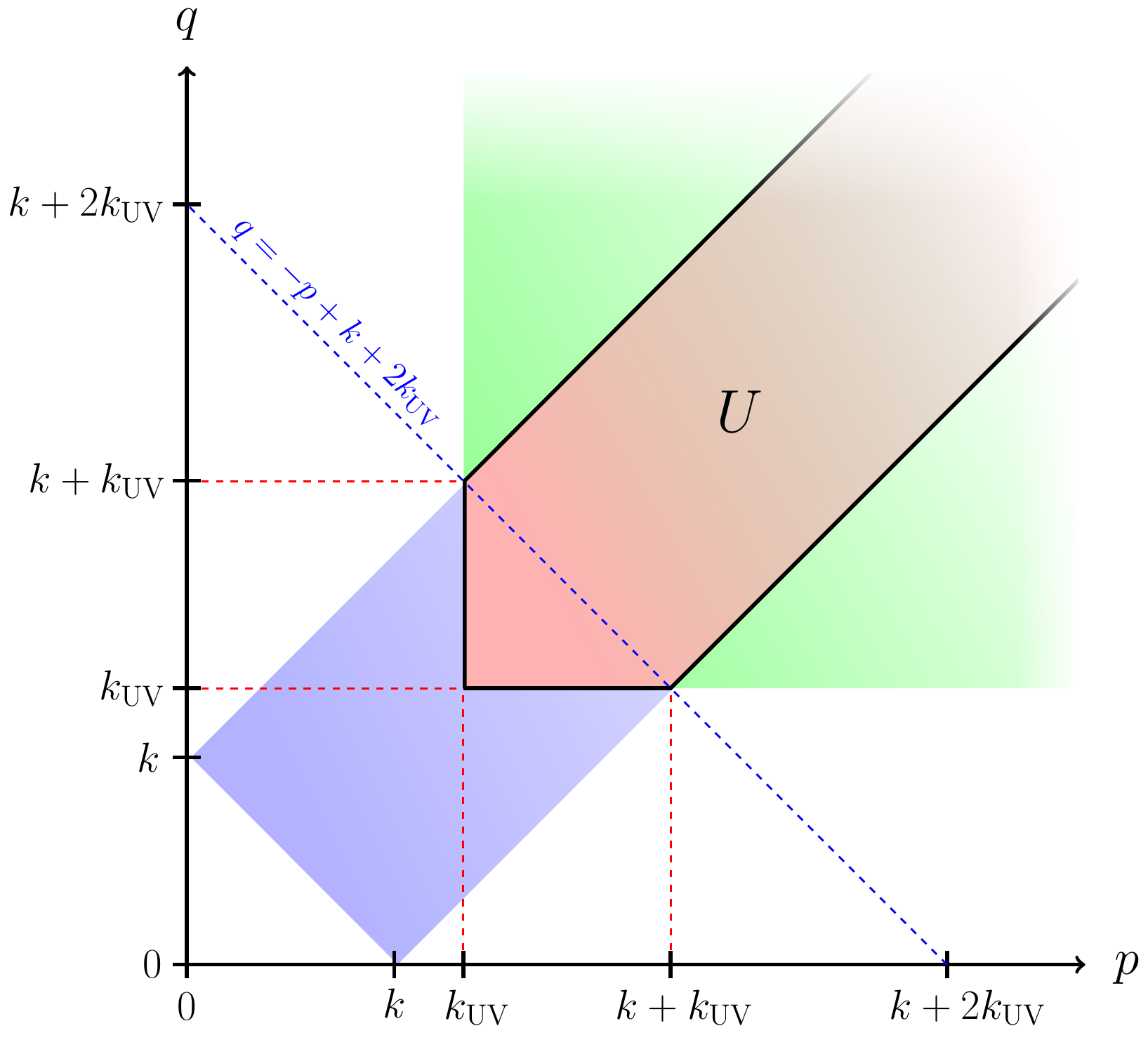}
\caption{Defining the integration region in $(p,q)$-space for
  computing $\mathscr{C}_{\bm{k}}(\eta,\eta')$ in \Eq{Ck_aftermu2} \ie
  after the $\mu$-integration is completed in \Eq{Ck_beforemu2}. The
  green region corresponds to $p,q>k_{\UV}$ and the blue region
  corresponds to $- 1 < (p^2 + k^2 - q^2)/(2 p k) < 1$ (assuming $k <
  k_{\UV}$) --- the intersection of these regions (in red) is the
  resulting integration region $U$ for computing
  $\mathscr{C}_{\bm{k}}(\eta,\eta')$ in $(p,q)$-space in
  \Eq{Ck_aftermu2}. }
\label{figure:Int_RegionU}
\end{figure}
Note that whenever $k < k_{\UV}$ (as we use in the main text), the
above simplifies to the region in which $p - k < q < p + k$ (since $p
> k_{\UV}>k$). Note however that we must have $p > k_{\UV}$ and $q >
k_{\UV}$ in addition to the earlier inequality being satisfied ---
this means that the actual region being integrated is the
quadrilateral region $U$ depicted in \Fig{figure:Int_RegionU}, giving
\begin{align}
\label{Ck_aftermu2}
\mathscr{C}_{\bm{k}}(\eta,\eta') & = \frac{1}{2(2\pi)^{7/2}k} \iint_{U}
\dd p\, \dd q\, p\, q \, ( q^2 + p^2 - k^2 )^2\, 
u_{\bm{q} }\left(\eta\right)
u_{\bm{p}}\left(\eta\right) u_{\bm{p}}^*\left(\eta'\right)
u_{\bm{q}}^*\left(\eta'\right).
\end{align}
In the current form it is complicated to integrate the above integrand
--- for this reason we transform integration variables to
\begin{equation} \label{PQtransformation}
p = \frac{P+Q}{2} \qquad \mathrm{and} \qquad q = \frac{P-Q}{2} 
\end{equation}
with the Jacobian
\begin{equation}
\left| \frac{\partial (p,q)}{\partial (P,Q)} \right| = \frac{1}{2} \ .
\end{equation}
The transformation $(p,q) \to (P,Q)$ rotates the region $U$ by $\pi/4$
(and rescales it as well) giving rise to the set $U'$ depicted in
\Fig{figure:Int_RegionUprime} below. Then, using the explicit form
of the mode functions, this leads to the following expression
\begin{align}
\label{Ck_Uprime}
\mathscr{C}_{\bm{k}}(\eta,\eta') & =  \frac{1}{64(2\pi)^{7/2}k}
\iint_{U'} \dd P\, \dd Q\,( P^2 + Q^2 - 2 k^2 )^2  
\left[1-\frac{2i}{(P-Q) \eta}\right]\left[1+\frac{2i}{(P-Q) \eta'}\right]
\nonumber \\ & \times
\left[1-\frac{2i}{(P+Q) \eta}\right] \left[1+\frac{2i}{(P+Q) \eta'}\right]
\; e^{- i P ( \eta - \eta' ) } .
\end{align}
\begin{figure}
\centering
  \includegraphics[width=0.60\linewidth]{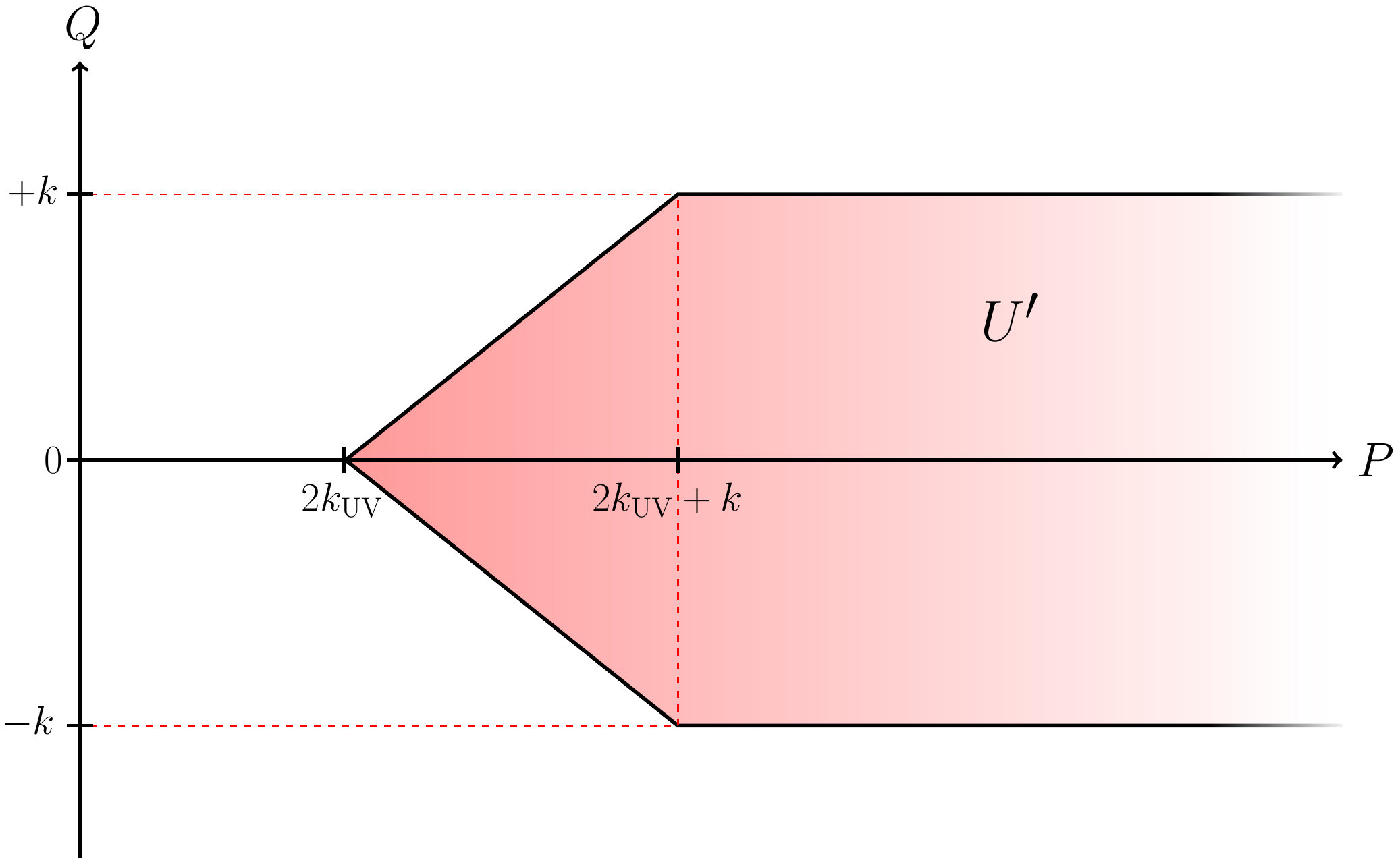}
\caption{Defining the integration region $U'$ in $(P,Q)$-space (after
  transforming $(p,q) \to (P,Q)$ in \Eq{PQtransformation}) for
  computing $\mathscr{C}_{\bm{k}}(\eta,\eta')$ in \Eq{Ck_Uprime}.  }
\label{figure:Int_RegionUprime}
\end{figure}
By noting that the integrand in \Eq{Ck_Uprime} is symmetric under $Q
\to -Q$, one can explicitly integrate over the region $U'$ as
\begin{align}
\label{Ck_PQ}
\mathscr{C}_{\bm{k}}(\eta,\eta') & = \frac{1}{32(2\pi)^{7/2}k}
\int_0^k \dd Q \int_{Q  + 2 k_{\UV}}^{\infty} \dd P\, ( P^2 + Q^2 - 2 k^2 )^2  
\left[1-\frac{2i}{(P-Q) \eta}\right]\left[1+\frac{2i}{(P-Q) \eta'}\right]
\nonumber \\ & \times
\left[1-\frac{2i}{(P+Q) \eta}\right] \left[1+\frac{2i}{(P+Q) \eta'}\right]
\, e^{- i P( \eta - \eta' )}. 
\end{align}
This representation of the correlator is most useful for computing the
Lindblad coefficient $\mathfrak{F}_{\bm{k}}$ appearing in the Lindblad equation.

\subsubsection{Integration of $\mathscr{C}_{\bm{k}}(\eta,\eta')$}

We begin with the double integral (\ref{Ck_PQ}), and note that it is
somewhat tricky to evaluate the $P$-integral because it is formally
divergent in the ultraviolet. All this means is that we are to
understand these integrals as distributions --- to deal with this we
organize the integrand in \Eq{Ck_PQ} in terms of decreasing powers of
$P$ such that
\begin{align}
\label{Ck_Uprime_DF}
\mathscr{C}_{\bm{k}}(\eta,\eta') & = \frac{1}{32(2\pi)^{7/2}k}
\int_0^k \dd Q \int_{Q  + 2 k_{\UV}}^{\infty} \dd P\,
\left[\mathcal{D}(P,Q) + \mathcal{F}(P,Q) \right]\, e^{-i P(\eta - \eta')}\ , 
\end{align}
where we define the function $\mathcal{D}(P,Q)$ (giving rise to a
formally divergent $P$-integral in the UV)
\begin{align}
\label{Ddefinition2}
\mathcal{D}(P,Q) & := \ P^4 + \frac{4i(\eta - \eta')}{\eta\eta'} P^3
+ 2 \left[ Q^2 - 2k^2 + \frac{8 \eta \eta'  - 2\eta^2 -  2 (\eta')^2}
  {\eta^2(\eta')^2} \right] P^2
\nonumber \\ &
+ \frac{16 i (\eta - \eta')
  \left[1 + \left( \frac{3}{4}Q^2 - k^2 \right) \eta \eta' \right]}
{ \eta^2 (\eta')^2 } P
\nonumber \\ &
+ \frac{ 16 \left\lbrace 1 - 4 (k^2-Q^2) \eta \eta'
  + \left( k^2 -\frac{3}{4}Q^2 \right) \left[ \eta^2 +  (\eta')^2 \right]
  + \frac{(Q^2 - 2 k^2)^2}{16} \eta^2 (\eta')^2 \right\rbrace}
{ \eta^2 (\eta')^2 },
\end{align}
as well as the function $\mathcal{F}(P,Q)$ (yielding a formally
convergent $P$-integral in the UV)
\begin{align}
\label{Fdefinition2}
\mathcal{F}(P,Q) & := \frac{16(k^2 - Q^2)}{\eta^2 (\eta')^2 (P^2 - Q^2)^2}
\bigg( i (\eta - \eta') \left[ (k^2 - Q^2) \eta \eta' - 4\right] P^3 
+ \bigl\lbrace \left[\eta^2 + (\eta')^2\right] (Q^2-k^2)
\nonumber \\ &
- 4 ( 2 Q^2 - k^2 ) \eta \eta'  - 4 \bigr\rbrace P^2 
+i (\eta - \eta') \left[ Q^4 \eta \eta'  + k^2 (4 - Q^2 \eta\eta')  \right] P
\nonumber \\ &
+\left\lbrace 4 + Q^2 [\eta^2 + (\eta')^2] \right\rbrace k^2
- Q^4
\left[ \eta^2 -4 \eta \eta' + (\eta')^2 \right]  \bigg). 
\end{align}
To further simplify the calculation, we compute the above under the
assumption that $\eta > \eta'$, which is the more useful case for the
calculation in the main text (and furthermore, we can easily extract
the opposing case $\eta < \eta'$ from the symmetry
$\mathscr{C}^{\ast}_{\bm{k}}(\eta,\eta') =
\mathscr{C}_{\bm{k}}(\eta',\eta)$ and so the calculation is performed
without any loss of generality).

First we compute the part of the double integral (\ref{Ck_Uprime_DF})
above involving $\mathcal{F}$ defined in \Eq{Fdefinition2}, namely the
quantity
\begin{align}
  \frac{1}{k}\int_0^k \dd Q \int_{Q  + 2 k_{\UV}}^{\infty} \dd P\,
  \mathcal{F}(P,Q) \, e^{-i P(\eta - \eta')} 
\end{align}
which we note yields a convergent $P$-integral in the ultraviolet
since for large $P$ the associated integrand behaves as:
\begin{equation}
  \mathcal{F}(P,Q)=\frac{16 i (k^2 - Q^2) (\eta - \eta')
    \left[ (k^2 - Q^2) \eta \eta' - 4 \right]}{\eta^2 (\eta')^2}
  \frac{1}{P} + \mathcal{O}\left(P^{-2}\right) . 
\end{equation}
It turns out that we can exactly write down the $P$-primitive of
$\mathcal{F}(P,Q)\, e^{- i P(\eta - \eta') }$, where
\begin{align}
\int \dd P \, \mathcal{F}(P,Q) & \, e^{- i P (\eta - \eta') }
= - \frac{ 32 (k^2 - Q^2)^2 \left[P( 1 + Q^2  \eta \eta' )
    + i Q^2 (\eta - \eta')\right]}{ \eta^2 (\eta')^2 Q^2
\left(P^2-Q^2\right)}\, e^{- i P(\eta - \eta')} \nonumber \\
& - \frac{8(k^2 -Q^2)(\eta-\eta')}{\eta^2 (\eta')^2 Q}
\biggl\lbrace \frac{k^2}{Q^2} \left[ \frac{2}{\eta-\eta'}
  + 2i Q - Q^2 (\eta - \eta') + i Q^3 \eta \eta' \right]
\nonumber \\ 
& + \frac{2}{\eta - \eta'} + 2i Q + Q^2 \frac{(\eta + \eta')^2}{\eta - \eta'}
- i Q^3 \eta \eta' \biggr\rbrace \mathrm{Ei}
\left[- i (P-Q) (\eta - \eta')  \right]
e^{ - i Q(\eta - \eta')} 
\nonumber \\
& +  \frac{8(k^2 -Q^2)(\eta-\eta')}{\eta^2 (\eta')^2Q}
\biggl\lbrace \frac{k^2}{Q^2} \left[ \frac{2}{\eta-\eta'}
  - 2i Q - Q^2 (\eta - \eta') - i Q^3 \eta \eta' \right]
\nonumber \\
& + 
\frac{2}{\eta - \eta'} - 2i Q + Q^2 \frac{(\eta + \eta')^2}{\eta - \eta'}
+ i Q^3 \eta \eta'    \biggr\rbrace \mathrm{Ei}
\left[- i (P+Q) (\eta - \eta')  \right]
e^{iQ(\eta - \eta')}, 
\end{align}
where $\mathrm{Ei}$ is the exponential integral function defined for
$z \in \mathbb{C} \backslash (-\infty, 0]$ (\ie there is a branch cut
  along the negative real axis) as
\begin{equation}
\label{eq:Ei:def}
\mathrm{Ei}(z) = - \int_{-z}^{\infty} \dd t\; \frac{e^{-t}}{t}  \ ,
\end{equation}
where the principal value of the integral is taken. 

To evaluate the above primitive at the endpoint $P \to \infty$,
we note the property
\begin{equation}
\mathrm{Ei}(-iy) \simeq - i \pi \qquad \mathrm{for} \ y \gg 1 \, ,
\end{equation}
which must be used since we are evaluating the correlator under the
assumption that $\eta - \eta'>0$. Evaluating the $P$-integral for the
required endpoints and simplifying yields
\begin{align}
\label{Pintegraldone2}
\int_{Q+2 k_{\UV}}^{\infty} & \dd P \, \mathcal{F}(P,Q)
\, e^{- i P (\eta - \eta')}= -\frac{16 \pi (k^2 - Q^2)}{\eta^2 (\eta')^2 Q^3}
\biggl( Q (\eta - \eta') \left[ 2 (k^2 + Q^2)
  + (k^2 - Q^2) Q^2 \eta \eta' \right]
 \nonumber \\ & \times
\cos[ Q(\eta - \eta') ]
+ \left\lbrace k^2 \left[Q^2 (\eta- \eta')^2 - 2 \right]
- Q^2\left[Q^2 (\eta + \eta')^2+2\right] \right\rbrace
\sin[ Q(\eta- \eta')] \biggr) \nonumber \\
& + \frac{ 32 (k^2 - Q^2)^2 }{ \eta^2 (\eta')^2 Q^2 }
\frac{  [Q+2k_{\UV}(\eta)]  ( 1+ Q^2  \eta \eta')
  + i Q^2 (\eta - \eta')}{[Q+2k_{\UV}(\eta)] ^2-Q^2}
e^{- i(Q+2k_{\UV})(\eta-\eta') } \nonumber \\
& + \frac{8(k^2 -Q^2)(\eta-\eta')}{\eta^2 (\eta')^2 Q^3}
\biggl\lbrace k^2 \left[ \frac{2}{\eta-\eta'} + 2 i Q - Q^2 (\eta - \eta')
  + i Q^3 \eta \eta' \right]
\nonumber \\ &
+ Q^2\left[\frac{2}{\eta - \eta'}
  + 2i Q + Q^2 \frac{(\eta + \eta')^2}{\eta - \eta'}
  - i Q^3 \eta \eta' \right] \biggr\rbrace \mathrm{Ei}
\left[- 2 i k_{\UV}(\eta - \eta')\right] e^{ - i Q(\eta - \eta') }
\nonumber  \\
& -\frac{8(k^2 -Q^2)(\eta-\eta')}{\eta^2 (\eta')^2 Q^3}
\biggl\lbrace k^2 \left[ \frac{2}{\eta-\eta'} - 2i Q
  - Q^2 (\eta - \eta') - i Q^3 \eta \eta' \right]
\nonumber \\ &
+ Q^2 \left[ \frac{2}{\eta - \eta'} - 2i Q + Q^2
  \frac{(\eta + \eta')^2}{\eta - \eta'} + i Q^3 \eta \eta' \right]
\biggr\rbrace \, \mathrm{Ei}
\left[  - 2 i (Q + k_{\UV}) (\eta - \eta')  \right]
e^{i Q (\eta - \eta') }.
\end{align}
Next we need to integrate the above function with respect to $Q$,
where we get exactly
\begin{align}
&\int  \dd Q \int_{Q+2 k_{\UV}}^{\infty} \dd P \, \mathcal{F}(P,Q) \,
e^{- i P (\eta - \eta')} = - \frac{16 \pi k^4\sin[ Q(\eta - \eta')]}
{\eta^2(\eta')^2} \frac{1}{Q^2}
\nonumber \\ &
+\frac{16k^4}{\eta^2 (\eta')^2} \left\lbrace
\pi (\eta - \eta') \cos[Q(\eta - \eta')]
- \frac{e^{- i (Q + 2k_{\UV} ) (\eta - \eta')}}{2k_{\UV}} \right\rbrace
\frac{1}{Q} + \frac{16 \pi k^2 (2 - k^2 \eta\eta')
\sin[ Q(\eta - \eta') ] }{\eta^2 (\eta')^2}
\nonumber \\ &
+ \frac{16 i e^{- i (Q + 2k_{\UV}) (\eta - \eta')} }{2 k_{\UV} \,
\eta^2 (\eta - \eta')^5 (\eta')^2 } \biggl\lbrace k^4
\eta (\eta - \eta')^4 \eta' - 4 k^2 (\eta - \eta')^2
\left[ \eta^2 - 3 \eta \eta' + (\eta')^2 \right]
\nonumber \\ &
- 8 \left[ \eta^2 - 5 \eta \eta' + (\eta')^2 \right] \biggr\rbrace
+ \frac{32k^2 (\eta - \eta')}{\eta^2 (\eta')^2}
\Biggl(-\pi \cos[Q(\eta - \eta')] +\frac{e^{- i (Q + 2k_{\UV}) (\eta - \eta')}}
{2k^2 k_{\UV}\, (\eta - \eta')^5 }
\nonumber \\ &
\times \left\lbrace k^2 (\eta - \eta')^2 \left[\eta^2 - 4 \eta \eta'
+ (\eta')^2\right]
+ 4 \left[\eta^2 - 5 \eta \eta' + (\eta')^2\right] \right\rbrace \Biggr) Q
\nonumber \\ &
+ \frac{16}{ \eta^2 (\eta')^2} \bigg( \pi (2 k^2 \eta\eta' - 1)
\sin[ Q(\eta - \eta') ]
- \frac{ i e^{- i (Q + 2k_{\UV}) (\eta - \eta') } }
{ k_{\UV} \, (\eta - \eta')^3 } \bigl\lbrace k^2 \eta (\eta - \eta')^2 \eta'
\nonumber \\ &
- 2 \left[\eta^2 - 5 \eta \eta' + (\eta')^2\right]
\bigr\rbrace \Biggr)Q^2 + \frac{16 (\eta - \eta')}
{\eta^2 (\eta')^2} \Biggl\lbrace  \pi \cos[Q(\eta - \eta')] -
\frac{ e^{- i (Q + 2k_{\UV}) (\eta - \eta')}}
{ 2 k_{\UV}\, (\eta - \eta')^3 } \bigl[ \eta^2 - 6 \eta \eta'
\nonumber  \\ &
+ (\eta')^2 \bigr] \Biggr\rbrace Q^3
+ \frac{16}{ \eta \eta'}
\left\lbrace - \pi \sin[Q(\eta - \eta')]
+\frac{ i  e^{- i (Q + 2k_{\UV}) (\eta - \eta')}}
{ 2 k_{\UV}\, (\eta - \eta') } \right\rbrace Q^4
\nonumber \\ &
+ \frac{8 (k^2 - Q^2)^2}{Q^2\eta^2(\eta')^2}
\biggl\lbrace (i - Q\eta)(i + Q\eta')
\mathrm{Ei}\left[ - 2 i k_{\UV}(\eta - \eta') \right]
e^{- i Q(\eta - \eta')} - (i + Q\eta)(i - Q\eta')
\nonumber \\ &
\times \mathrm{Ei}\left[ - 2 i (Q + k_{\UV})(\eta - \eta') \right]
e^{ i Q(\eta - \eta')} \biggr\rbrace .
\end{align}
The first few terms are organized by increasing powers of $Q$, and the
last term involves Ei functions. Finally, evaluating the above
primitive at the endpoints (from $Q=0$ to $Q = k$) gives
\begin{align}
\label{Fdouble_ans}
\int_0^k & \dd Q \int_{Q  + 2 k_{\UV}}^{\infty} \dd P\, \mathcal{F}(P,Q)
\, e^{-i P(\eta - \eta')}
= - \frac{192 i}{k_{\UV}\ \eta^2 (\eta - \eta')^5 }
\left[ e^{- 2 i k_{\UV} (\eta - \eta') }
- e^{- i ( k+ 2 k_{\UV}) (\eta - \eta') } \right] 
\nonumber \\ &
-\frac{192 i}{k_{\UV}\ \eta^3 (\eta - \eta')^4 } 
\left[ e^{- 2 i k_{\UV}(\eta - \eta') }
  - e^{- i ( k+ 2 k_{\UV} ) (\eta - \eta') } (1 + i k \eta) \right]
\nonumber \\ &
-\frac{32 i}{k_{\UV}\ \eta^4 (\eta - \eta')^3 }
\left[ e^{- 2 i k_{\UV} (\eta - \eta') }(4 + k^2 \eta^2)
- 2 \,e^{- i ( k+ 2 k_{\UV} ) (\eta - \eta') }
(1 + i k \eta)(2 + i k \eta) \right]
\nonumber \\ &
-\frac{32 i}{k_{\UV}\ \eta^5 (\eta - \eta')^2 }
\left[ e^{- 2 i k_{\UV} (\eta - \eta') }(2 + k^2 \eta^2)
- 2 \,e^{- i ( k+ 2 k_{\UV} ) (\eta - \eta') } (1 + i k \eta)^2 \right]
\nonumber  \\ &
- \frac{8 ik}{k_{\UV} \eta^5}
\left[ e^{- 2 i k_{\UV} (\eta - \eta') } k^3 \eta^3 - 8 i
\,e^{- i ( k+ 2 k_{\UV} ) (\eta - \eta') } (1 + i k \eta)\right] 
\left(\frac{1}{\eta - \eta'} + \frac{1}{\eta'} \right)
\nonumber  \\ &
+\frac{4 i}{k^2_{\UV} \eta^5 (\eta')^2 }
\left\lbrace e^{- 2 i k_{\UV} (\eta - \eta') }
\left[ 8 k_{\UV} ( 2 + k^2 \eta^2 ) - i k^4 \eta^3 \right]
- 16 \,e^{- i ( k+ 2 k_{\UV} ) (\eta - \eta') }
k_{\UV} (1 + i k \eta) \right\rbrace   \, . 
\end{align}

Our next move consists in computing the part of the double integral
(\ref{Ck_Uprime_DF}) above involving $\mathcal{D}$ defined in
\Eq{Ddefinition2}, namely
\begin{align}
\frac{1}{k}\int_0^k \dd Q \int_{Q  + 2 k_{\UV}}^{\infty} \dd P\,
\mathcal{D}(P,Q)
e^{-i P(\eta - \eta')}.
\end{align}
As mentioned previously, the above $P$-integral is formally divergent
(the integrand scaling as $\propto P^4$ in the ultraviolet) however it
is meaningful when understood as a distribution. Using the definition
(\ref{Ddefinition2}) we write the above integral over $P$ as
\begin{align}
\label{Dintegral_alpha2}
& \int_{Q  + 2 k_{\UV}}^{\infty} \dd P\,
\mathcal{D}(P,Q)\, e^{-i P(\eta - \eta')}
= \alpha_{4}\big( \eta - \eta', Q + 2k_{\UV} \big)
+ \frac{4i(\eta - \eta')}{\eta\eta'} \alpha_{3}
\big( \eta - \eta', Q + 2k_{\UV} \big)
\nonumber \\ &
+2 \left[ Q^2 - 2k^2 + \frac{8 \eta \eta'  - 2\eta^2 -  2 (\eta')^2}
  {\eta^2(\eta')^2} \right]\,
\alpha_{2}\big( \eta - \eta', Q + 2k_{\UV} \big)
+\frac{16 i (\eta - \eta')
  \left[1 + \left(\frac{3}{4}Q^2 - k^2 \right) \eta \eta' \right]}
{ \eta^2 (\eta')^2 }\,
\nonumber \\ & \times
\alpha_{1}\left( \eta - \eta', Q + 2k_{\UV}\right)
+ \frac{ 16 \left\lbrace 1 - 4 (k^2-Q^2) \eta \eta'
+ \big( k^2 -\frac{3}{4}Q^2 \big) \big[ \eta^2 +  (\eta')^2 \big]
+ \frac{(Q^2 - 2 k^2)^2}{16} \eta^2 (\eta')^2 \right\rbrace}
{\eta^2 (\eta')^2 }
\nonumber \\ & \times
\alpha_{0}\big( \eta - \eta', Q + 2k_{\UV} \big), 
\end{align}
where we define the distributions for $m \in \{ 0, 1, 2, 3, 4\}$
\begin{equation}
\alpha_{m}(x,y) : = \int_{y}^{\infty} \dd P\, P^{m} e^{ - i P x }\, ,
\end{equation}
which we must now compute. We first note the Fourier representation of
the Heaviside step function
\begin{equation}
  \Theta(P) = \lim_{\delta \to 0^{+}} \int_{-\infty}^{\infty} \dd x \,
  \frac{e^{ i P x }}{2\pi i( x - i \delta)} 
\end{equation}
where we recall that $\Theta(x) = 1$ for $x>0$ and $\Theta(x) = 0$ for
$x<0$. Inverting the above gives
\begin{equation}
\int_0^{\infty} \dd P \; e^{  - i P x }
= \frac{- i }{x - i \delta} = - \frac{i}{x} + \pi \delta(x)
\end{equation}
understood in the limit $\delta \to 0^{+}$ (in the last equality we
have used $(x\pm i \delta)^{-1} = x^{-1} \mp i \pi \delta(x)$ --- the
so called ``Sochocki-Plemelj'' theorem). From this we easily find that
\begin{equation}
\alpha_{0}(x,y) = \int_{y}^{\infty} \dd P \,e^{  - i P x}
= \frac{- i e^{- i x y} }{x - i \delta} ,
\end{equation}
which easily follows from a shift $P \to L$ in the integration
variable $L = P - y$. By taking the limit $\delta \to 0^{+}$ in the
above and using the property $f(x) \delta(x) = f(0) \delta(x)$ of
$\delta$-functions the above can be more simply written as
\begin{equation}
\label{alpha1}
\alpha_{0}(x,y) = \frac{- i e^{- i x y} }{x} + \pi \delta(x)\ .
\end{equation}
From here we can easily get the remaining set of required functions by
noticing that $i \partial_{x} \alpha_{m}(x,y) = \alpha_{m+1}(x,y)$,
giving:
\begin{align}
\label{alpham2}
\alpha_{1}(x,y)&=\int_{y}^{\infty} \dd P \, P e^{  - i P x}
=  e^{- i x y }\left( - \frac{1}{x^2}  - \frac{i y}{x}  \right)
+ i \pi \delta^{\prime}(x) \, , \\
\alpha_{2}(x,y) &= \int_{y}^{\infty} \dd P \, P^2 e^{  - i P x }
=  e^{- i x y }\left(  \frac{2i}{x^3} - \frac{2y}{x^2}  -
\frac{i y^2}{x}  \right) - \pi \delta^{\prime\prime}(x)\, , \\
\alpha_{3}(x,y) &= \int_{y}^{\infty} \dd P \, P^3 e^{  - i P x }
=  e^{- i x y }\left( \frac{6}{x^4} + \frac{6 i y}{x^3} - \frac{3y^2}{x^2}
- \frac{i y^3}{x}  \right) - i \pi \delta^{\prime\prime \prime}(x)\, , \\
\alpha_{4}(x,y) & = \int_{y}^{\infty} \dd P \, P^4 e^{  - i P x }
= e^{- i x y } \left( - \frac{24 i}{x^5} + \frac{24 y}{x^4}
+ \frac{12 i y^2}{x^3} - \frac{4 y^3}{x^2}  - \frac{i y^4}{x} \right)
+ \pi \delta^{\prime\prime \prime \prime}(x) \, .
\end{align} 
Inserting the distributions $\alpha_{m}(x,y)$ with $x=\eta-\eta'$ and
$y=Q+2k_\UV$ into \Eq{Dintegral_alpha2} yields the following expression
\begin{align}
& \int_{Q  + 2 k_{\UV} }^{\infty} \dd P\, \mathcal{D}(P,Q)
e^{-i P (\eta - \eta')}
= e^{- i (Q+2k_\UV)(\eta-\eta')}
\Bigg(
\biggl[-\frac{24 i}{(\eta-\eta')^5}
+ \frac{24 (Q+2k_{\UV})}{(\eta-\eta')^4}
\nonumber \\ &
+ \frac{12 i (Q+2k_\UV)^2}{(\eta-\eta')^3}
- \frac{4 (Q+2k_\UV)^3}{(\eta-\eta')^2}
- \frac{i (Q+2k_\UV)^4}{\eta-\eta'} \biggr]
+\frac{4i(\eta - \eta')}{\eta\eta'}
\biggl[\frac{6}{(\eta-\eta')^4}
+ \frac{6 i (Q+2k_\UV)}{(\eta-\eta')^3}
\nonumber \\ &
- \frac{3(Q+2k_\UV)^2}{(\eta-\eta')^2}
- \frac{i (Q+2k_\UV)^3}{\eta-\eta'}  \biggr]
+ 2\left[ Q^2 - 2k^2 + \frac{8 \eta \eta'  - 2\eta^2 - 2 (\eta')^2}
  {\eta^2(\eta')^2} \right]
\biggl[\frac{2i}{(\eta-\eta')^3}
\nonumber \\ &
- \frac{2(Q+2k_\UV)}{(\eta-\eta')^2}
- \frac{i (Q+2k_\UV)^2}{\eta-\eta'} \biggr]
-\frac{16 i (\eta - \eta') \left[1 + \left(\frac{3}{4}Q^2 - k^2 \right)
\eta \eta' \right]}{ \eta^2 (\eta')^2 }
\left[\frac{1}{(\eta-\eta')^2}
+\frac{i (Q+2k_\UV)}{\eta-\eta'} \right]
\nonumber  \\ &
+\frac{ 16 \left\lbrace 1 - 4 (k^2-Q^2) \eta \eta' +
  \big( k^2 -\tfrac{3}{4}Q^2 \big)
  \big[ \eta^2 +  (\eta')^2 \big]
  + \frac{(Q^2 - 2 k^2)^2}{16} \eta^2 (\eta')^2 \right\rbrace}
{ \eta^2 (\eta')^2 } \left( \frac{- i}{\eta-\eta'} \right) \, \Bigg) 
\nonumber \\ &
+ \pi \Bigg(
\delta^{\prime\prime\prime\prime}(\eta - \eta')
+ \frac{4(\eta - \eta')}{\eta\eta'} \delta^{\prime\prime\prime}(\eta - \eta')
- 2 \left[ Q^2 - 2k^2 + \frac{8 \eta \eta'  - 2\eta^2 - 2 (\eta')^2}
  {\eta^2(\eta')^2} \right]\delta^{\prime\prime}(\eta - \eta')
\nonumber \\ &
-\frac{16 (\eta - \eta') \left[1 + \left( \frac{3}{4}Q^2 - k^2 \right)
\eta \eta' \right]}{ \eta^2 (\eta')^2 } \delta^{\prime}(\eta - \eta')
\nonumber  \\ &
+\frac{ 16 \left\lbrace 1 - 4 (k^2-Q^2) \eta \eta'
+ \big( k^2 -\frac{3}{4}Q^2 \big) \big[ \eta^2 +  (\eta')^2 \big]
+ \frac{(Q^2 - 2 k^2)^2}{16} \eta^2 (\eta')^2 \right\rbrace}
{ \eta^2 (\eta')^2 }\delta(\eta - \eta') \Bigg)  . 
\end{align}
All that is left to do is to integrate over $Q$, which happens to be
straightforward in this case with use of the integrals:
\begin{align}
\int_0^k \dd Q \, Q^{n} & =
\frac{k^{n+1}}{n+1}, \\
\int_0^k \dd Q \, e^{- i ( Q + 2 k_{\UV} ) (\eta - \eta') }&=
\frac{i e^{- 2 i k_{\UV} (\eta-\eta')} \big[ - 1 + e^{- i k (\eta - \eta')}  \big] }
{(\eta - \eta')} \, ,
\\
\int_0^k \dd Q \, e^{- i ( Q + 2 k_{\UV} ) (\eta - \eta') } Q
&= \frac{e^{- 2 i k_{\UV} (\eta-\eta')} \big\{- 1 + [1+i k (\eta - \eta')]
\, e^{- i k (\eta - \eta')} \big\} }{(\eta - \eta')^2}\, ,
\\
\int_0^k \dd Q \, e^{- i ( Q + 2 k_{\UV} ) (\eta - \eta') } Q^2 & =
\frac{i \, e^{- 2 i k_{\UV} (\eta-\eta')} \big\{2+[ -2- 2 i k (\eta - \eta')
+ k^2 (\eta - \eta')^2]e^{- i k (\eta - \eta')} \big\} }
{(\eta - \eta')^3}\, ,
\\
\int_0^k \dd Q \, e^{- i ( Q + 2 k_{\UV} ) (\eta - \eta') } Q^3 & =
\frac{e^{- 2 i k_{\UV} (\eta-\eta')}}{(\eta - \eta' )^4}
\biggl\{6+\bigl[- 6- 6 i k (\eta - \eta')
  + 3 k^2 (\eta - \eta')^2
\nonumber \\ &
  + i k^3 (\eta - \eta')^3\bigr]
 e^{- i k (\eta - \eta')} \biggr\} \, ,
\\
\int_0^k \dd Q \, \frac{e^{- i ( Q + 2 k_{\UV} ) (\eta - \eta') }}{k} Q^4 & =
\frac{i \, e^{- 2 i k_{\UV} (\eta-\eta')}}{(\eta - \eta' )^5}
\biggl\{-24 +\bigl[24+24 i k (\eta - \eta')
  - 12 k^2 (\eta - \eta')^2
\nonumber \\ &
- 4 i k^3 (\eta - \eta')^3
+ k^4 (\eta - \eta')^4\bigr]e^{- i k (\eta - \eta')} \biggr\}\, .
\end{align}
The resulting expression contains many terms, but organizing the
expression as a partial fraction expansion in terms of $\eta'$ yields:
\begin{align}
\label{Ddouble_ans}
& \int_0^k \dd Q \int_{Q  + 2 k_{\UV}}^{\infty} \dd P\,
\mathcal{D}(P,Q)e^{-i P (\eta - \eta')P}
= - \frac{ 224 }{(\eta - \eta')^6}\left[ e^{- 2 i k_{\UV}(\eta - \eta') }
- e^{- i (k + 2 k_{\UV})(\eta - \eta') } \right]
\nonumber \\ &
-\frac{ 8 i}{(\eta - \eta')^5} \left[32 k_{\UV} e^{- 2 i k_{\UV}(\eta - \eta') }
- ( 25 k + 32 k_{\UV} ) e^{- i (k + 2 k_{\UV})(\eta - \eta') } \right]
\nonumber \\ &
+ \frac{1}{\eta^2 (\eta - \eta')^4}
\Biggl\{
32\left[ 10 + ( 5 k^2_{\mathrm{UV}} - k^2  ) \eta^2 \right]
e^{- 2 i k_{\UV} (\eta - \eta') }
- 8 \left[ 40 +  ( 7 k^2 + 26 k k_{\UV} + 20 k^2_{\mathrm{UV}} )
  \eta^2 \right]
\nonumber \\ & \times
e^{- i (k + 2 k_{\UV} )(\eta - \eta') } \Biggr\}
+\frac{1}{\eta^3 (\eta - \eta')^3}
\biggl\{ 32 \left[ 10 + 8 i  k_{\UV} \eta -i k_{\UV}( k^2 -2 k_{\UV}^2 )
  \eta^3 \right]  e^{- 2 i k_{\UV}(\eta - \eta') }
\nonumber \\ &
- 8 \left[ 40
    + i ( 35 k + 32 k_{\UV} ) \eta + 2 i  k_{\UV}
    ( 3 k^2 + 7 k k_{\UV} + 4 k_{\UV}^2 ) \eta^3 \right]
  e^{- i (k + 2 k_{\UV})(\eta - \eta') }\biggr\}
\nonumber \\ &
+\frac{4}{\eta^4 (\eta - \eta')^2}
\left[ 56 + 64 i k_{\UV} \eta + 16 (k^2 - 2 k_{\UV}^2) \eta^2
-  (k^2 - 2 k_{\UV}^2)^2 \eta^4  \right] e^{- 2 i k_{\UV}(\eta - \eta') } 
\nonumber \\ &
- \frac{8}{\eta^4 (\eta - \eta')^2}
\left[ 28 + i ( 35 k + 32 k_{\UV} ) \eta
- (k + 2 k_{\UV}) ( 7 k + 8 k_{\UV} ) \eta^2
- 2 k_{\UV}^2 (k + k_{\UV})^2 \eta^4 \right]
\nonumber \\ & \times
e^{- i (k + 2 k_{\UV})(\eta - \eta') } 
+ \frac{32}{\eta^5}
(2 + i k_{\UV}\eta) [ 2 + 2 i k_{\UV} \eta + (k^2 - k_{\UV}^2) \eta^2 ]
e^{- 2 i k_{\UV}(\eta - \eta') } \left( \frac{1}{\eta - \eta'} + \frac{1}{\eta'}
\right)
\nonumber \\ & - \frac{8}{\eta^5}
\left[ 16 + 4 i (7k + 6 k_{\UV}) \eta - (k+2 k_{\UV})(7k + 8 k_{\UV})
  \eta^2 - 2 i k_{\UV} (k + 2 k_{\UV}) (k + k_{\UV}) \eta^3 \right]
\nonumber \\ &
e^{- i (k + 2 k_{\UV})(\eta - \eta') } \left( \frac{1}{\eta - \eta'}
+ \frac{1}{\eta'} \right)
-\frac{1}{\eta^4 (\eta')^2}
\biggl\{
16 \left[ 6 + 4 i  k_{\UV} \eta + (k^2 - k_{\UV}^2 ) \eta^2 \right]
e^{- 2 i k_{\UV}(\eta - \eta') }
\nonumber \\ &
- 8 \left[ 12 + i ( 7 k + 8 k_{\UV} ) \eta
  - 2   k_{\UV}  (k + k_{\UV}  ) \eta^2 \right]
e^{- i (k + 2 k_{\UV})(\eta - \eta') } \biggr\}
+ \pi k \bigg[ \delta^{\prime\prime\prime\prime}(\eta - \eta') 
\nonumber \\ &
+\frac{4(\eta - \eta')}{\eta\eta'} \delta^{\prime\prime\prime}(\eta - \eta')
+ \frac{4\big[\eta^2 + (\eta')^2 + \big( \frac{5}{6} k^2 \eta \eta'
    - 4 \big) \eta \eta' \big] }{\eta^2 (\eta')^2}
\delta^{\prime\prime}(\eta - \eta')
\nonumber  \\ &
+\frac{4(\eta - \eta') \big( 3 k^2 \eta \eta' - 4 \big)}
{\eta^2 (\eta')^2} \delta^{\prime}(\eta - \eta')
+\frac{\tfrac{43}{15} k^4 \eta^2 (\eta')^2 +
  \frac{4}{3} k^2 \left[ 9 \eta^2 - 32 \eta \eta'
    + 9 (\eta')^2 \right] + 16}{\eta^2 (\eta')^2} \delta(\eta - \eta')
\bigg] . 
\end{align}
We are now in a position where the final result for
$\mathscr{C}_{\bm{k}}(\eta,\eta')$ can be obtained. The correlator we
seek is given by \Eq{Ck_Uprime_DF}. Using the derived results
(\ref{Fdouble_ans}) and (\ref{Ddouble_ans}), and organizing
in terms of a partial fraction expansion in $\eta'$ the result is at
last 
\begin{align}
\label{Ck_answer_Appendix}
\mathscr{C}_{\bm{k}}(\eta,\eta') &= \frac{1}{32(2\pi)^{7/2}} \Biggl(
-\frac{224}{k(\eta - \eta')^6}\left[ e^{- 2 i k_{\UV}(\eta - \eta' ) }
  - e^{- i ( k + 2 k_{\UV} ) ( \eta - \eta' ) } \right]
\nonumber \\ &
-\frac{1}{k\, \kUV \, \eta^2 \, (\eta - \eta')^5}
  \biggl\{64 i ( 3 + 4 k_{\UV}^2 \eta^2 ) e^{- 2 i \kUV ( \eta - \eta' )  }
  - 8 i [ 24 + \kUV (25k + 32 \kUV) \eta^2 ]
\nonumber \\ & \times
  e^{- i ( k + 2 k_{\UV} ) (\eta - \eta' ) }
  \biggr\}
+ \frac{1}{k\, \kUV \, \eta^3\, (\eta - \eta')^4}
\biggl\{
32 [ - 6 i + 10 \kUV \eta - \kUV (k^2 - 5 k_{\UV}^2) \eta^3 ]
\nonumber \\ & \times
e^{- 2 i \kUV ( \eta - \eta' )  } + 8 [ 24 i - 8 (3k + 5 \kUV) \eta
  - \kUV (7k^2 + 26 k \kUV + 20 k_{\UV}^2 ) \eta^3 ]
\nonumber \\ & \times
e^{- i ( k + 2 k_{\UV} ) (\eta - \eta' ) } \biggr\}
 +\frac{32}{k\, \kUV \, \eta^4\, (\eta - \eta')^3}
  (- i + \kUV \eta) [ 4 + 6 i \kUV \eta + (k^2 - 2 k_{\UV}^2) \eta^2
\nonumber \\ &
   - i \kUV(k^2 - 2 k_{\UV}^2) \eta^3 ] e^{- 2 i \kUV ( \eta - \eta' )  }  
 - \frac{8 i}{k\, \kUV \, \eta^4\, (\eta - \eta')^3}
 [ - 16 - 8 i (3k + 5 \kUV) \eta
\nonumber \\ &
   + ( 8 k^2 + 35 k \kUV + 32 k_{\UV}^2 )
   \eta^2 + 2 k_{\UV}^2 ( k + \kUV ) (3k + 4 \kUV) \eta^4  ]
 e^{- i ( k + 2 k_{\UV} ) (\eta - \eta' ) } 
 \nonumber \\ &
 - \frac{4}{k\, \kUV \, \eta^5\, (\eta - \eta')^2}
 [  16 i - 56 \kUV \eta + 8 i (k^2 - 8 k_{\UV}^2 ) \eta^2
   - 16 \kUV (k^2 - 2 k_{\UV}^2 ) \eta^3
   \nonumber \\ &
   + \kUV(k^2 - 2 k_{\UV}^2 )^2 \eta^5 ]  e^{- 2 i \kUV ( \eta - \eta' )  } 
+ \frac{8}{k\, \kUV \, \eta^5\, (\eta - \eta')^2}
[ 8 i  - 4 (4k + 7 \kUV ) \eta
\nonumber \\ &
  - i ( 8 k^2 + 35 k \kUV + 32 k_{\UV}^2 )
   \eta^2 + \kUV (k + 2\kUV) (7k + 8 \kUV ) \eta^3 + 2 k_{\UV}^3
   (k+ \kUV)^2 \eta^5  ]
\nonumber \\ & \times
e^{- i ( k + 2 k_{\UV} ) ( \eta - \eta' )  }
+ \frac{8}{k\,  \kUV \, \eta^5} [ 16 \kUV + 24 i k_{\UV}^2 \eta
  + 8 \kUV (k^2 - 2 k_{\UV}^2 ) \eta^2
  \nonumber \\ &
  - i ( k^2 - 2 k_{\UV}^2 )^2 \eta^3 ] e^{- 2 i \kUV ( \eta - \eta' )  }
\left( \frac{1}{\eta - \eta'} + \frac{1}{\eta'} \right)
+ \frac{8}{k\, \kUV \, \eta^5} ( k + 2 \kUV )
[ - 8
\nonumber \\ &
  - 4 i ( 2 k  + 3 \kUV ) \eta + \kUV ( 7 k + 8 \kUV )
  \eta^2 + 2 i k_{\UV}^2 (k + \kUV) \eta^3  ]
e^{- i ( k + 2 k_{\UV} ) ( \eta - \eta' )  }
\nonumber \\ & \times
\left( \frac{1}{\eta - \eta'}
+ \frac{1}{\eta'} \right)
+ \frac{4}{k\, k_{\UV}^2 \, \eta^5 \, (\eta')^2 }
[ 16 i \kUV - 24 k_{\UV}^2 \eta + 8 i \kUV (k^2 - 2 k_{\UV}^2 ) \eta^2
\nonumber \\ &
  + (k^2 - 2 k_{\UV}^2)^2 \eta^3  ]  e^{- 2 i \kUV ( \eta - \eta' )  }
- \frac{8}{k\, k_{\UV}^2 \, \eta^5\, (\eta')^2 }
\kUV [ 8 i - 4 ( 2 k + 3 \kUV ) \eta
\nonumber \\ &
  - i \kUV ( 7 k + 8 \kUV )
  \eta^2 + 2 k_{\UV}^2 (k + \kUV) \eta^3 ]
e^{- i ( k + 2 k_{\UV} ) ( \eta - \eta' )  }
+ \pi \bigg[ \delta^{\prime\prime\prime\prime}(\eta - \eta')
\nonumber \\ &
+\frac{4(\eta - \eta')}{\eta\eta'} \delta^{\prime\prime\prime}(\eta - \eta')
+ \frac{4\big[\eta^2 + (\eta')^2 + \big( \frac{5}{6} k^2 \eta \eta'
    - 4 \big) \eta \eta' \big] }{\eta^2 (\eta')^2}
\delta^{\prime\prime}(\eta - \eta')
\nonumber \\ &
+\frac{4(\eta - \eta')
  \big( 3 k^2 \eta \eta' - 4 \big)}{\eta^2 (\eta')^2}
\delta^{\prime}(\eta - \eta')
\nonumber \\ &
+ \frac{\frac{43}{15} k^4 \eta^2 (\eta')^2
  + \frac{4}{3} k^2 \left[ 9 \eta^2 - 32 \eta \eta'
    + 9 (\eta')^2 \right] + 16}{\eta^2 (\eta')^2}  \,
\delta(\eta - \eta') \bigg]\, \Biggr).
\end{align}
which assumes that $\eta > \eta'$ (the result for the opposing case of
$\eta < \eta'$ can be extracted from the symmetry $
\mathscr{C}_{\bm{k}}(\eta,\eta') =
\mathscr{C}^{\ast}_{\bm{k}}(\eta',\eta)$ easily). This is the result
used in the main text. In \Sec{ssec:EnvCorr}, we present the
coincidence and the early time $\eta'$ limits of the exact expression,
see \Eqs{Ck_coincident_etap} and~\eqref{Ck_early_etap}.

\section{Lindblad coefficients}
\label{App:Lindblad_F}

In this appendix we explicitly compute the Lindblad coefficient
$\mathfrak{F}_{\bm{k}}$ defined in \Eq{Ffrakdef} used to derive the
physical predictions of the Lindblad equation. We later also compute the
validity coefficient $\mathfrak{M}_{\bm{k}}$ defined in
\Eq{Mfrak_def}.

We begin with the integral $\mathfrak{F}_{\bm{k}}$ defined in
\Eq{Ffrakdef}, repeated here,
\begin{equation}
\mathfrak{F}_{\bm{k}}(\eta, \eta_{\mathrm{in}}) :=
(2\pi)^{3/2} \int_{ \eta_{\mathrm{in}} }^{\eta} \dd \eta' \,
G(\eta) G(\eta') \mathscr{C}_{\bm{k}}(\eta,\eta') \ ,
\end{equation}
where we here assume that $\eta_{\mathrm{in}}$ is fixed and arbitrary
(assuming only that $\eta_{\mathrm{in}} < \eta < 0$). To evaluate this
we use the representation (\ref{Ck_PQ}) of $\mathcal{C}_{\bm{k}}(\eta,
\eta')$, as well as $G(\eta) G(\eta') = \frac{\slrl H^2 }{8 \Mp^2}
\eta \eta'$, see \Eq{couplingdef}, which expresses
$\mathfrak{F}_{\bm{k}}$ as the triple integral
\begin{align}
\label{Fk_trip}
\mathfrak{F}_{\bm{k}}(\eta, \eta_{\mathrm{in}}) & =
\frac{\slrl  H^2}{1024 \pi^2 \Mp^2 k}
\int_{ \eta_{\mathrm{in}} }^{\eta} \dd \eta' \int_0^k \dd Q
\int_{Q  + 2 k_{\UV}}^{\infty} \dd P \, \eta\, \eta' \,
( P^2 + Q^2 - 2 k^2 )^2  
\left[1-\frac{2i}{(P-Q) \eta}\right]
\nonumber \\ & \times
\left[1+\frac{2i}{(P-Q) \eta'}\right]
\left[1-\frac{2i}{(P+Q) \eta}\right] \left[1+\frac{2i}{(P+Q) \eta'}\right]
e^{- i P (\eta - \eta')}.
\end{align}
The utility of this representation is that it allows for a more
standard/straightforward handling of the UV divergences that arise
(which can be understood as UV divergences arising in the $P \to
\infty$ limit). We proceed by computing the $\eta'$ integrals first,
where we note that
\begin{align}
\label{etap_integral_for_Fk}
\int_{ \eta_{\mathrm{in}} }^{\eta} \dd \eta' \, \eta' \, &
\left[1+\frac{2i}{(P-Q) \eta'}\right] \left[1+\frac{2i}{(P+Q) \eta'}\right]
e^{- i P( \eta - \eta' )} 
=\frac{1}{P^2} - \frac{i \eta}{P} + \frac{4}{P^2 - Q^2}
\nonumber \\ &
- \frac{4 e^{- i P \eta}  \, \mathrm{Ei}(i P \eta) }{P^2 - Q^2}
- e^{- i P (\eta - \eta_{\mathrm{in}})} \bigg( \frac{1}{P^2}
- \frac{i \eta_{\mathrm{in}}}{P} + \frac{4}{P^2 - Q^2} \bigg)
+ \frac{4 e^{- i P \eta}  \, \mathrm{Ei}(i P \eta_{\mathrm{in}})}{P^2 - Q^2}, 
\end{align}
where $\mathrm{Ei}$ is the exponential integral function defined
around \Eq{eq:Ei:def}.

To simplify the above expression we wish to express it in terms of
dimensionless and positive variables, so to this end we define
\begin{equation}
  z := - k \eta \ , \qquad \qquad z_{\mathrm{in}} := - k \eta_{\mathrm{in}} 
\end{equation}
and we also switch integration variables to
\begin{equation}
  a : = \frac{Q}{k} \qquad \qquad \mathrm{and} \qquad \qquad
  b : = \frac{P}{k} \ .
\end{equation}
After using the result (\ref{etap_integral_for_Fk}) as well as these
variable definitions, we find that \Eq{Fk_trip} becomes
\begin{align}
\label{Fk_ab}
\mathfrak{F}_{\bm{k}}(\eta, \eta_{\mathrm{in}}) &=
\frac{\slrl  H^2 k^2}{1024 \pi^2 \Mp^2 }  \int_0^1 \dd a
\int_{a  + 2 \kappa}^{\infty} \dd b \, z \, ( a^2 + b^2 - 2 )^2
\left[1+\frac{2i}{(b-a) z}\right]\left[1+\frac{2i}{(b+a) z}\right]
\nonumber \\ & \times
\biggl[ - \frac{1}{b^2} - \frac{i z}{b} - \frac{4}{b^2 - a^2}
  + \frac{4 e^{ i b z}  \mathrm{Ei}( - i b z )}{b^2 - a^2}
  + e^{ - i b ( \zin - z ) } \bigg( \frac{1}{b^2} + \frac{i \zin}{b}
  + \frac{4}{b^2 - a^2} \bigg)
\nonumber \\ &
  - \frac{4 e^{ i b z}
    \mathrm{Ei}( - i b \zin )}{b^2 - a^2}  \biggr] .
\end{align}
To emphasize the terms which depend on $\eta_{\mathrm{in}}$, we split
up the above function into two pieces,
\begin{align}
\label{eq:FintermsofIG}
  \mathfrak{F}_{\bm{k}}(\eta, \eta_{\mathrm{in}}) & =
  \frac{\slrl  H^2 k^2}{1024 \pi^2 \Mp^2 }
  \left[ \mathcal{I}_{\bm{k}}( \eta )
    - \mathcal{G}_{\bm{k}}(\eta,\eta_{\mathrm{in}})\right]
\end{align}
where
\begin{align}
  \label{I_forF}
  \mathcal{I}_{\bm{k}}(\eta) & := \int_0^1 \dd a \int_{a  + 2 \kappa}^{\infty}
  \dd b \, z \, ( a^2 + b^2 - 2 )^2
  \left[1+\frac{2i}{(b-a) z}\right]\left[1+\frac{2i}{(b+a) z}\right]
\nonumber \\ & \times
  \left[ - \frac{1}{b^2} - \frac{i z}{b}
    - \frac{4}{b^2 - a^2}
    + \frac{4 e^{ i b z}  \mathrm{Ei}( - i b z )}{b^2 - a^2}  \right]  
\\
 \label{G_forF}
 \mathcal{G}_{\bm{k}}(\eta,\eta_{\mathrm{in}}) & :=
 \int_0^1 \dd a \int_{a  + 2 \kappa}^{\infty} \dd b \, z \, ( a^2 + b^2 - 2 )^2  
 \left[1+\frac{2i}{(b-a) z}\right]\left[1+\frac{2i}{(b+a) z}\right]
 \nonumber \\ & \times
 \left[ - e^{ - i b ( \zin - z ) } \bigg( \frac{1}{b^2} + \frac{i \zin}{b}
   + \frac{4}{b^2 - a^2} \bigg) + \frac{4 e^{ i b z}
     \mathrm{Ei}( - i b \zin )}{b^2 - a^2}  \right].
\end{align}
We next evaluate each of these functions exactly, and also derive
asymptotic series for each. We start by evaluating
$\mathcal{G}_{\bm{k}}$ and then use this to get $\mathcal{I}_{\bm{k}}$
(which is straightforward to derive once we know what
$\mathcal{G}_{\bm{k}}$ is).

\subsection{The function $\mathcal{G}$}
\label{App:G_computation}

\begin{figure}
\centering
\includegraphics[width=0.60\linewidth]{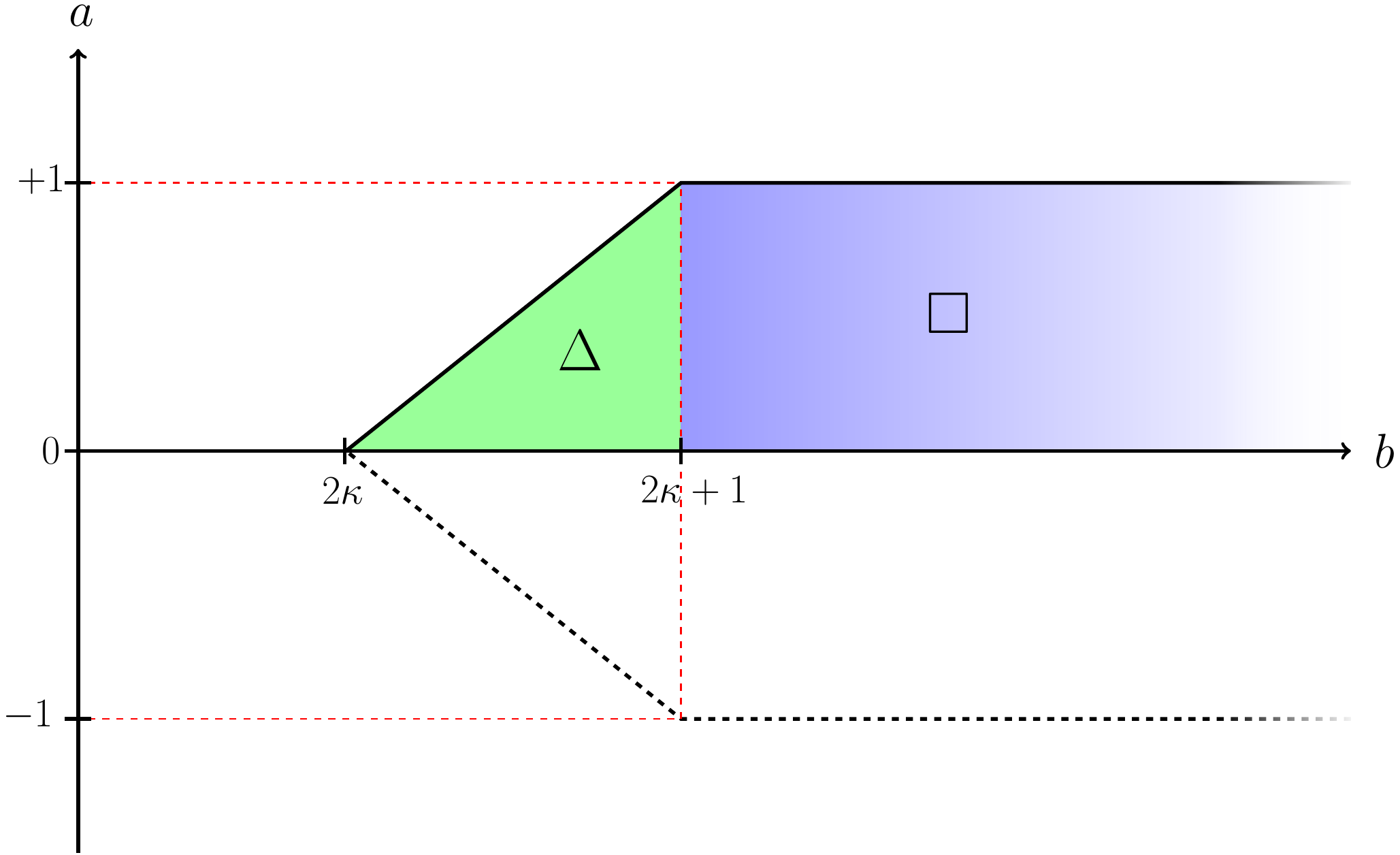}
\caption{The triangular region $\Delta$ integrated over in equation
  (\ref{GDelta_def}) is shown in green and the rectangular region
  $\square$ integrated over in equation (\ref{GSquare_def}) is shown
  in blue.}
\label{figure:Int_Tri_Square}
\end{figure}

Here we exactly compute
$\mathcal{G}_{\bm{k}}(\eta,\eta_{\mathrm{in}})$ defined in
\Eq{G_forF}. It turns out that it is easiest to do this by switching
the order of integration relative to the formula (\ref{G_forF}), and
so we compute it in two pieces
\begin{equation}
  \mathcal{G}_{\bm{k}}(\eta,\eta_{\mathrm{in}}) =
  \mathcal{G}^{\Delta}_{\bm{k}}(\eta,\eta_{\mathrm{in}})
  +  \mathcal{G}^{\square}_{\bm{k}}(\eta,\eta_{\mathrm{in}}) 
\end{equation}
where $\mathcal{G}^{\Delta}_{\bm{k}}$ integrates over a triangular
region
\begin{equation} \label{GDelta_def}
  \mathcal{G}^{\Delta}_{\bm{k}}(\eta, \eta_\mathrm{in}) : =
  \int_{2\kappa}^{2\kappa+1} \dd b \int_0^{b-2\kappa} \dd a\; f(a,b,z,\zin)
\end{equation}
and where $\mathcal{G}^{\square}_{\bm{k}}$ integrates over a
rectangular region
\begin{equation} \label{GSquare_def}
  \mathcal{G}^{\square}_{\bm{k}}(\eta,\eta_{\mathrm{in}}) : =
  \int_{2\kappa+1}^{\infty} \dd b \int_0^{1} \dd a\; f(a,b,z,\zin) \ ,
\end{equation}
using the shorthand 
\begin{align}
\label{fintegrand_def}
f(a,b,z,\zin)& : =  z \; ( a^2 + b^2 - 2 )^2
\left[1+\frac{2i}{(b-a) z}\right]\left[1+\frac{2i}{(b+a) z}\right]
\nonumber \\ & \times
\left[- e^{ - i b ( \zin - z ) } \left( \frac{1}{b^2}
  + \frac{i \zin}{b} + \frac{4}{b^2 - a^2} \right)
  + \frac{4 e^{ i b z} \, \mathrm{Ei}( - i b \zin )}{b^2 - a^2}  \right] 
\end{align}
for the integrand given in the definition (\ref{G_forF}). For a visual
representation of the domains of integration $\Delta$ and $\square$
see \Fig{figure:Int_Tri_Square}.

In order to proceed we note that $f$ has the following
$a$-primitive (assuming $b>a$)
\begin{align}
\label{f_antiderivative}
\int \dd a & \, f(a,b,z,\zin) =  e^{- i b (\zin - z)}
\biggl\lbrace - \frac{z(1+i b \zin)}{5b^2} a^5
+ \biggl[ \frac{4(i + b z)(i - b \zin)}{3 b^2 z}
\nonumber \\ &  
+\frac{2z(2+2 i b \zin + b^2  - i b^3 \zin)}{3b^2} \biggr] a^3
\nonumber \\ &
+ \frac{z(4 + 4 i b \zin + 12 b^2 - 4 i b^3
  \zin - 11 b^4 + i b^5 \zin)}{b^2} a
- \frac{8 (i + b z )(6 i - 6 b \zin + i b^2 + 5 b^3 \zin)}{3bz}
\nonumber \\ &
- \frac{4 z (15 + 15 i b \zin + 40 b^2 - 20 i b^3 \zin
  - 43 b^4 + 7 i b^5 \zin)}{15 b}
+ \frac{16 (b^2 - 1)^2 (-1 +i bz)}{b^2 z}
\nonumber \\ & 
\times \left( \frac{1}{a+b} + \frac{1}{a - b} \right)
+ \left[ - \frac{16 (b^2 - 1)^2 z}{b}
  + \frac{16 (b^2 - 1) (i + b z ) (3 i - b \zin
    + i b^2 + b^3 \zin)}{b^3 z } \right]
\nonumber \\ & \times
\mathrm{coth}^{-1}\left( \frac{b}{a} \right)  \biggr\rbrace
+  \biggl\lbrace - \frac{z}{3} a^3 + \left( - \frac{4}{z}  + 4 z
  + 4 i b - 3 b^2 z \right) a + \frac{2}{3} b \left( - \frac{6}{z}
  + 6 z + 6 i b - 5 b^2 z \right)
\nonumber \\ &
  + \frac{4(b^2 - 1)^2 (1 - i b z)}{b^2 z}
  \left( \frac{1}{a+b} + \frac{1}{a-b} \right)
+ 4 (b^2 - 1) \left[ \frac{2(b^2 + 1) (1 - i b z)}{b^3 z}
  + \frac{(b^2 - 1)z}{b} \right]
\nonumber \\ & \times
\mathrm{coth}^{-1}
  \left( \frac{b}{a} \right) \bigg\rbrace \, 4 \, e^{ i b z}
  \, \mathrm{Ei} ( - i b \zin ). 
\end{align}
Next we use this primitive to evaluate both
$\mathcal{G}^{\Delta}_{\bm{k}}(\eta, \eta_\mathrm{in})$ and
$\mathcal{G}^{\square}_{\bm{k}}(\eta, \eta_\mathrm{in})$.

Let us start with $\mathcal{G}^{\Delta}_{\bm{k}}$. Here we take the
double integral $\mathcal{G}^{\Delta}_{\bm{k}}$ defined in
\Eq{GDelta_def}, and perform the $a$-integration by evaluating the
$a$-primitive (\ref{f_antiderivative}) at the end points $a = 0 $ to
$a=b-2\kappa$:
\begin{align}
\label{GDelta_after_a}
\int_0^{b-2\kappa} \dd a & \, f(a,b,z,\zin) =
e^{- i b (\zin - z )} \Biggl( - \frac{28 i z \zin}{15} b^4
+ 4 \left[ - \frac{2i}{\kappa} - \frac{10 \zin}{3} + \left( \frac{43}{15}
  + 2 i \kappa \zin \right) z \right] b^3
\nonumber \\ &
+ 8 \left\lbrace \frac{3 - 5 i \kappa \zin}{3 \kappa z}
+ \frac{2(i+ 6 \kappa \zin)}{3} + \left[ - 3 \kappa
  + \frac{2 i (1 - 3 \kappa^2 ) \zin}{3} \right] z   \right\rbrace b^2
\nonumber \\ &
+ 8 \left\lbrace \frac{2( -1 + 6 i \kappa \zin)}{3 z}
+ \frac{2i}{\kappa} + i \kappa + 2 (1 - \kappa^2) \zin +
\frac{2\left[ - 2 + i \kappa (4 \kappa^2 - 3) \zin \right]}{3}
z \right\rbrace b
\nonumber \\ &
+ 4 \Biggl\lbrace  - \frac{2\left[2
    + \kappa^2 + 2 i \kappa (\kappa^2 - 1) \zin\right]}{\kappa z}
+ 2 i (3 \kappa^2 - 4) + \frac{8 \kappa (\kappa^2 - 3) \zin}{3}
\nonumber \\ &
+ \left[ \frac{4\kappa (2\kappa^2 + 3)}{3} - i (2\kappa^2 - 1)^2 \zin \right]
z \Biggr\rbrace
+ \frac{4}{3} \Biggl\lbrace
\frac{2\left[ 3 - 6 \kappa^2 (\kappa^2 - 1)+ 4 i \kappa^3 (\kappa^2 - 3)
    \zin \right]}{\kappa^2 z}
\nonumber \\ &
- \frac{4 i ( 2 \kappa^4 - 6 \kappa^2 + 3 )}{\kappa}
+ \frac{2 i \kappa (12 \kappa^4 - 20 \kappa^2 + 15) \zin - 15 (2\kappa^2 - 1)^2}{5} z
\Biggr\rbrace \frac{1}{b}
\nonumber \\ &
+ \frac{8 (\kappa^2 -1)^2 (-1 + i \kappa z )}{\kappa^2 z (b-\kappa)}
+ \frac{8}{3} \left[ \frac{4 \kappa^4 -12 \kappa^2 + 6}{\kappa z}
  + \frac{\kappa (12 \kappa^4 - 20 \kappa^2 + 15)}{5} z \right] \frac{1}{b^2}
\nonumber \\ &
+ 16\biggl[ (\zin -z) b^3 + \frac{i (\zin + z)}{z} b^2 -
  \left( \frac{1}{z} - 2 z + 2 \zin \right) b
  - \frac{2i (\zin - z)}{z} - \frac{2 + z^2 - z \zin}{b z}
\nonumber \\ &
  + \frac{ i(\zin - 3z)}{z b^2} + \frac{3}{z b^3} \biggr]
\mathrm{coth}^{-1}\left( \frac{b}{b-2\kappa} \right) \Biggr)
+ \Biggl\lbrace \frac{2(3i  - 5 \kappa z )}{3 \kappa} b^3
+ 2 \left( - \frac{1}{\kappa z} + i + 4 \kappa z \right) b^2
\nonumber \\ &
+ 2 \left[ - \frac{1}{z} - \frac{i (5 \kappa^2 +2 )}{\kappa}
  + 2 (1 - \kappa^2 )z \right] b + \frac{10 \kappa^2 +4}{\kappa z}
- 2 i (\kappa^2 - 2) + \frac{8 \kappa (\kappa^2 - 3)}{3} z
\nonumber \\ &
+ \frac{4 i \kappa z - 2}{\kappa^2 z b} + \frac{2(\kappa^2 - 1)^2
  (1 - i \kappa z)}{\kappa^2 z (b-\kappa)} - \frac{4}{\kappa z b^2}
+ 4 \biggl[ z b^3 - 2 i b^2 + \left( \frac{2}{z} - 2 z \right) b
  + \frac{z}{b}
\nonumber \\ &
  + \frac{2 i }{b^2} - \frac{2}{z b^3} \biggr]
\mathrm{coth}^{-1} \left(\frac{b}{b-2\kappa} \right)\Biggr\rbrace
\, 4 \, e^{+ i b z} \, \mathrm{Ei}( - i b \zin ) \, .
\end{align}
Next we find that we can explicitly write down the $b$-primitive
(up to a constant) of the above:
\begin{align}
\int  \dd b & \int_0^{b-2\kappa} \dd a \, f(a,b,z,\zin)
=e^{-i b (z_{\mathrm{in}}-z)} \Biggl(
\frac{224 z_{\mathrm{in}}^2}{5(\zin - z)^5}
+ \frac{1}{(\zin - z)^4}\biggl[\frac{48 i}{\kappa}-\frac{168 \zin}{5}
\nonumber \\ &
+ \frac{16 i (14 b-15 \kappa)z_\mathrm{in}^2}{5}\biggr]  
-\frac{1}{(\zin - z)^3}\biggl[ \frac{8 (90 b-19 \kappa)}{15 \kappa}
  + \frac{8 i (21 b-20 \kappa)\zin}{5} 
\nonumber \\ &
+\frac{16 (21 b^2-45 b \kappa+30 \kappa^2-10)z_\mathrm{in}^2}{15}\biggr]
+\frac{64 i}{\kappa z_{\mathrm{in}} z^3}
+\frac{1}{(\zin-z)^2}\biggl\{-\frac{32 i}{\kappa z_{\mathrm{in}}^2}
\nonumber \\ &
+ \frac{8 i [-45 b^2+19 b \kappa +30(\kappa^2+1)]}{15 \kappa}
+ \frac{4 ( 63 b^2-120 b \kappa+60 \kappa^2-20) \zin}{15}
\nonumber \\ &
-\frac{8 i \left[14 b^3-45 b^2 \kappa+20 b \left(3 \kappa^2-1\right)-40 \kappa^3
    +30 \kappa\right]z_\mathrm{in}^2}{15}\biggr\}
+\frac{32}{\kappa z^2}
\left(\frac{i}{z_{\mathrm{in}}^2} + \frac{b-\kappa}{\zin} \right)
\nonumber \\ &
+\frac{1}{\zin-z}\biggl\{
\frac{32 (b-\kappa)}{\kappa  z_{\mathrm{in}}^2}
+ \frac{8 (b^2-2 ) b}{\kappa}-\frac{4 ( 19 b^2+20 )}{15}
-16 \kappa (b-\kappa)
\nonumber \\ &
+ \frac{4 i b (21 b^2-60 b \kappa +60 \kappa^2-20)}{15} z_{\mathrm{in}}
+\frac{4}{15} \biggl[7 b^4+60 (b^2-1) \kappa^2-30 (b^2-2) b \kappa -20 b^2
  \nonumber \\ &
-80 b \kappa^3+60 \kappa^4+15\biggr] z_{\mathrm{in}}^2 \biggr\}
+\frac{8}{3z} \left[\frac{12 (b-\kappa)}{\kappa z_{\mathrm{in}}^2}
+ \frac{(b-2 \kappa) \left(5 b^2-2 b \kappa +2 \kappa^2-6\right)}{b} \right]
\nonumber \\ &
+\frac{1}{15}\biggl\{4i\left(-43 b^3+90 b^2 \kappa +40 b-40 \kappa^3-60 \kappa\right)
-4 z_{\mathrm{in}} \bigl[7 b^4+60 \left(b^2-1\right) \kappa^2
\nonumber \\ &  
  -30 \left(b^2-2\right) b \kappa
  -20 b^2-80 b \kappa^3+60 \kappa^4+15\bigr]
- \frac{8 \kappa \left(12 \kappa^4-20 \kappa^2+15\right)z}{b}  \biggr\}\Biggr)
\nonumber \\ &
+ \left[\frac{64 i}{\kappa z^4}+\frac{32 i}{\kappa z^2}-\frac{32}{z^3}
  + \frac{8 i \left(4 \kappa^4-12 \kappa^2-3\right)}{3 \kappa}
  -4 \left(2 \kappa^2-1\right)^2 z + \frac{8i \kappa
    \left(12 \kappa^4-20 \kappa^2+15\right)}{15} z^2 \right]
\nonumber \\ & \times
\mathrm{Ei}\left[-i b ( z_{\mathrm{in}}-z ) \right] 
  +\biggl[\frac{64 i}{\kappa z^4} -\frac{32}{z^3} +\frac{32 i}{\kappa z^2}
    +\frac{8 i \left(4 \kappa^4-12 \kappa^2-3\right)}{3 \kappa}
    -4 (2 \kappa^2-1)^2 z
\nonumber \\ &
+    \frac{8 i \kappa \left(12 \kappa^4-20 \kappa^2+15\right)}{15} z^2
    \biggr] \mathrm{Ei}\left[-i b (z_{\mathrm{in}}-z)\right] 
  +8 \bigg\lbrace \left(\frac{3}{\kappa z}+5 i\right)\frac{b^3}{3}
  + \left(\frac{4 i}{\kappa z^2}-\frac{3}{z}-4 i \kappa\right) b^2
\nonumber \\ &
  + 2 \left[ -\frac{4}{\kappa z^3}-\frac{2 i}{z^2}
    + \frac{2 \kappa^2-1}{\kappa z}+i (\kappa^2-1) \right] b - \frac{8 i}{\kappa z^4}
  +\frac{4}{z^3} -\frac{4 i}{\kappa z^2} -\frac{2 \left(\kappa^2-1\right)}{z}
  -\frac{4 i \kappa \left(\kappa^2-3\right) }{3}
\nonumber \\ &
  +\frac{1}{\kappa z b}
  \bigg\rbrace e^{i b z} \mathrm{Ei}(-i b z_{\mathrm{in}})
  + \frac{16 \left(b^2-1\right)^2 (1-i b z)}{b^2 z} \coth^{-1}
    \left(\frac{b}{b-2 \kappa}\right) 
    \biggl[e^{i b z} \mathrm{Ei}(-i b z_{\mathrm{in}})
\nonumber \\ &      
    -e^{-i b (z_{\mathrm{in}}-z)}\biggr]\, .
\end{align}
Finally we can evaluate this at the endpoints from $b=2\kappa$ to $b= 2\kappa+1$
and we are left with the answer
\begin{align}
\label{GDelta_answer}
\mathcal{G}^{\Delta}_{\bm{k}}(\eta, \eta_\mathrm{in}) &=
e^{-i (2 \kappa +1) (\zin-z)} \Biggl\lbrace \frac{224 z_{\mathrm{in}}^2}{5(\zin - z)^5}
+\frac{1}{(\zin-z)^4}\left[\frac{48 i}{\kappa}-\frac{168 \zin}{5}
+\frac{16 i (13 \kappa+14)}{5} z_{\mathrm{in}}^2\right]
\nonumber  \\ &
+ \frac{1}{(\zin-z)^3}\left[-\frac{1288}{15}-\frac{48}{\kappa}
-\frac{8 i (22 \kappa+21)\zin}{5} -\frac{16 (24 \kappa^2+39 \kappa+11)
z_\mathrm{in}^2}{15}\right]
+\frac{64 i}{\kappa \zin z^3}
\nonumber \\ &
+\frac{1}{(\zin-z)^2}\biggl[
-\frac{32 i}{\kappa z_{\mathrm{in}}^2}
-\frac{8 i (112 \kappa^2+161 \kappa+15 )}{15 \kappa}
+\frac{4 (72 \kappa^2+132 \kappa+43 )\zin}{15}
\nonumber \\ &
-\frac{8 i (12 \kappa^3+48 \kappa^2+29 \kappa-6)z_\mathrm{in}^2}{15}\biggr]
+\frac{32}{z^2} \left(\frac{i}{\kappa z_{\mathrm{in}}^2}
+ \frac{\kappa+1}{\kappa \zin}\right)
+\frac{1}{\zin-z}\biggl[
\frac{32 (\kappa+1)}{\kappa z_{\mathrm{in}}^2}
\nonumber \\ &
+\frac{4 (104 \kappa^3+224 \kappa^2+21 \kappa -30 )}{15 \kappa}
+\frac{4 i \left(48 \kappa^3+72 \kappa^2+26 \kappa +1\right)\zin}{15}
\nonumber \\ &
+\frac{8 (6 \kappa^4+12 \kappa^3+14 \kappa^2+3 \kappa+1)z_\mathrm{in}^2}{15}
\biggr]
+\frac{32 (\kappa+1)}{\kappa z_{\mathrm{in}}^2z}
+\frac{8 (18 \kappa^2+18 \kappa-1 )}{3 (2 \kappa+1)z}
\nonumber \\ &
-\frac{4 i (24 \kappa^3+156 \kappa^2+148 \kappa+3 ) }{15}
-\frac{ 8 \kappa ( 12 \kappa^4-20 \kappa^2+15 ) z}{15 (2 \kappa+1)}
\nonumber \\ &
-\frac{8 \left(6 \kappa^4+12 \kappa^3+14 \kappa^2+3 \kappa+1\right)\zin}{15}
+\frac{128 i (\kappa+1)^2 \kappa^2 [i + (2 \kappa+1) z]}{(2 \kappa+1)^2 z}
\log \left(\frac{\kappa+1}{\kappa}\right)\Biggr\rbrace
\nonumber \\ &
+ e^{-2 i \kappa (\zin-z)} \Biggl\{ - \frac{224 z_{\mathrm{in}}^2}{5(\zin - z)^5}
+\frac{1}{(\zin-z)^4}\left(
-\frac{48 i}{\kappa}+\frac{168 \zin}{5} -\frac{208 i \kappa z_\mathrm{in}^2}{5}
\right)
\nonumber \\ &
+\frac{1}{(\zin-z)^3}\left[\frac{1288}{15}
+\frac{176 i \kappa \zin}{5}
+\frac{32(12 \kappa^2-5)z_\mathrm{in}^2}{15}\right]
-\frac{64 i}{\kappa \zin z^3}
+\frac{1}{(\zin-z)^2}\biggl[
\frac{32 i}{\kappa z_{\mathrm{in}}^2}
\nonumber \\ &
+\frac{16 i (56 \kappa^2-15)}{15 \kappa}
-\frac{16(18 \kappa^2-5)\zin}{15} 
+\frac{16 i \kappa \left(6 \kappa^2-5\right)z_\mathrm{in}^2}{15}\biggr]
-\frac{32}{z^2}\left(\frac{i}{\kappa z_{\mathrm{in}}^2}
+\frac{1}{z_{\mathrm{in}}} \right)
\nonumber \\ &
-\frac{4}{\zin -z}\left[\frac{8}{z_{\mathrm{in}}^2}
+\frac{4(26 \kappa^2-35)}{15}
+\frac{8 i \kappa ( 6 \kappa^2-5 )\zin}{15} 
+\frac{(12 \kappa^4-20 \kappa^2+15)z_\mathrm{in}^2}{15}\right]
\nonumber \\ &
-\frac{32}{z z_{\mathrm{in}}^2}
+\frac{16i \kappa (6 \kappa^2-5) }{15}
+\frac{4 \left(12 \kappa^4-20 \kappa^2+15\right)}{15} (\zin+z)
\Biggr\}
+8\biggl[-\frac{8 i}{\kappa z^4} + \frac{4}{z^3} - \frac{4 i}{\kappa z^2}
\nonumber \\ &
-\frac{i(4 \kappa^4-12 \kappa^2-3)}{3 \kappa} + \frac{(2 \kappa^2-1)^2z}{2}  
-\frac{i \kappa ( 12 \kappa^4-20 \kappa^2+15 )z^2}{15}\biggr]
\bigl\lbrace \mathrm{Ei}\left[-2 i \kappa (\zin-z)\right]
\nonumber \\ &
-\mathrm{Ei}\left[-i (2 \kappa+1) (\zin-z)\right]
\bigr\rbrace 
+ 8 \biggl[\frac{8 i}{\kappa z^4}+\frac{12}{z^3}
  -\frac{4 i (2 \kappa^2-1)}{\kappa z^2}
  -\frac{4 \kappa^4-4 \kappa^2+1}{2 \kappa^2 z}  \biggr]
\nonumber \\ & \times
e^{2 i \kappa z}
\mathrm{Ei}(-2 i \kappa \zin)
+8  \biggl\lbrace -\frac{8 i}{\kappa z^4}-\frac{4 (3 \kappa+2)}{\kappa z^3}
+\frac{4 i (2 \kappa+3)}{z^2} + \frac{4 \kappa^3+10 \kappa^2
  +6 \kappa-1}{(2 \kappa+1) z}
\nonumber \\ &
+\frac{i (18 \kappa^2+18 \kappa-1 )}{3}
+ \frac{16 (\kappa+1)^2 \kappa^2 [1-i (2 \kappa+1) z]}
{(2 \kappa+1)^2 z}
\log \left(\frac{\kappa+1}{\kappa}\right)\biggr\rbrace
e^{i (2 \kappa+1) z}
\nonumber \\ & \times
\mathrm{Ei}\left[-i (2 \kappa+1) \zin\right] \, . 
\end{align}
We now turn to the calculation of $\mathcal{G}^{\square}_{\bm{k}}$. We
follow similar steps for evaluating $\mathcal{G}^{\square}_{\bm{k}}$
defined in \Eq{GSquare_def}. We first perform the $a$-integration by
evaluating the $a$-primitive in \Eq{f_antiderivative} at the endpoints
from $a =0$ to $a=1$ giving:
\begin{align}
\label{GSquare_after_a}
\mathcal{G}^{\square}_{\bm{k}}(\eta, & \eta_\mathrm{in}) =
\int _{2\kappa+1}^{+\infty}\, \dd b\, \Biggl(
e^{-i b (\zin-z)} \biggl\lbrace
\left(\frac{1}{z}-i b\right) \left[\frac{4 (27 b^2-13)}{3 b^2}
  -\frac{4 i (9 b^2-11)\zin}{3 b}\right]
\nonumber \\
& + \left[11 b^2  -  \frac{34}{3} - \frac{43}{15 b^2}
-\frac{i (15 b^4-50 b^2+43) \zin}{15 b}\right] z
+ \frac{16 \left(b^2-1\right) }{b}
\biggl[ \frac{i (b^2-1 ) b \zin-b^2-3}{b^2 z}
\nonumber \\ 
& +\frac{i (b^2+3 )}{b}
+ (b^2-1) (\zin-z) \biggr] \coth^{-1}(b) \biggr\rbrace 
+ 4 \bigg\{ \frac{4 (2-3 b^2) }{b} \left(\frac{1}{b z}-i\right)
- \frac{9 b^2-11}{3} z
\nonumber \\ 
& +\left[ \frac{ 8 (b^4-1) }{b^2} \left(\frac{1}{b z}
  -i\right) + \frac{4 ( b^2-1 )^2 }{b} z \right] \coth ^{-1}(b)
\biggr\rbrace \, e^{i b z} \, \mathrm{Ei}(-i b \zin )
\Biggr). 
\end{align}
At this stage we must deal with a subtlety occurring in the above
integration: the integral is formally divergent in the $b \to \infty$
limit (since $b \propto P$ this means these are UV divergences). To
deal with this we separate out the diverging $b \to \infty$ behaviour
from the rest of the integrand, so that
\begin{align}
\label{Gsquare_sum}
\mathcal{G}^{\square}_{\bm{k}}(\eta, \eta_\mathrm{in})
= \mathcal{G}^{\square,\mathrm{div}}_{\bm{k}}(\eta, \eta_\mathrm{in})
+ \mathcal{G}^{\square,\mathrm{reg}}_{\bm{k}}(\eta, \eta_\mathrm{in})
\end{align}
where we define 
\begin{align}
\label{Gsquare1_def}
\mathcal{G}^{\square,\mathrm{div}}_{\bm{k}}(\eta, \eta_\mathrm{in}) & : =
\int_{2\kappa+1}^{\infty} \dd b\,
\Biggl(4 \pi e^{i b z}\biggl[ -i  z b^2 +4 b+i
\left(3 z+\frac{4}{z}\right) \biggr]
+ e^{-i b (\zin-z)} \biggl\lbrace -i z \zin b^3
\nonumber \\ &
- (5 z-4 \zin) b^2
+ \frac{2 i b}{3}\left[ \frac{6 \zin}{z} -30
  + \left(\frac{6}{\zin} + 5 \zin\right) z \right]
+ 2 \biggl(\frac{10}{z} -\frac{8}{z_{\mathrm{in}}}
- 6 z_{\mathrm{in}}
\nonumber \\ &
+ \frac{23 z_{\mathrm{in}}^2-6}{3 z_{\mathrm{in}}^2} z \biggr)
+ \frac{4 i}{b} \left[ - \left( \frac{4}{z_{\mathrm{in}}}
    + 3 z_{\mathrm{in}} \right) \frac{1}{z}-\frac{4}{z_{\mathrm{in}}^2}
    +  \frac{41}{3} - \left( \frac{2}{z_{\mathrm{in}}^3}
    +\frac{3}{z_{\mathrm{in}}} + \frac{43 z_{\mathrm{in}}}{60} \right) z \right]
\bigg\rbrace  \, \Biggr)
\\
\label{Gsquare2_def}
\mathcal{G}^{\square,\mathrm{reg}}_{\bm{k}}(\eta, \eta_\mathrm{in}) & : =
\int_{2\kappa+1}^{\infty} \dd b\,
\bigg( - 4 \pi e^{i b z} \biggl[ -i  zb^2 +4 b
  +i \left(3 z+\sfrac{4}{z}\right) \biggr]
+ e^{-i b (\zin-z)} \biggl\lbrace -16  (z_{\mathrm{in}}-z) b^2
\nonumber \\ &
-\frac{4 i (z+2 z_{\mathrm{in}})^2 b}{z z_{\mathrm{in}}}
+ 4 \left[\frac{4}{z}+\frac{4}{z_{\mathrm{in}}}
  + \frac{z}{z_{\mathrm{in}}^2}+\frac{20}{3} (z_{\mathrm{in}}-z) \right]
+ \frac{4 i}{b}\biggl(\frac{2 z}{z_{\mathrm{in}}^3}+\frac{3 z}{z_{\mathrm{in}}}
+\frac{20 z_{\mathrm{in}}}{3 z}
\nonumber \\ &
+\frac{4}{z z_{\mathrm{in}}}
+\frac{4}{z_{\mathrm{in}}^2}-\frac{28}{3}\biggr)
- \frac{1}{b^2}\left(\frac{43 z}{15}+\frac{52}{3 z}\right)
+ \frac{16 \left(b^2-1\right) }{b}
\biggl[\frac{i (b^2-1 ) b \zin-b^2-3}{b^2 z}
\nonumber \\ &
  +\frac{i (b^2+3 )}{b} + (b^2-1) (\zin-z) \biggr] \coth ^{-1}(b)
\biggr\rbrace
+ 4 \biggl\lbrace \frac{4 (2-3 b^2) }{b}
\left(\frac{1}{b z}-i\right)
\nonumber \\ &
- \frac{9 b^2-11}{3} z
+\left[ \frac{ 8 (b^4-1) }{b^2} \left(\frac{1}{b z}-i\right)
  + \frac{4 ( b^2-1 )^2 }{b} z \right] \coth ^{-1}(b)
\biggr\rbrace e^{i b z} \mathrm{Ei}(-i b \zin ) \; \bigg)\, . 
\end{align}
One may check that summing
$\mathcal{G}^{\square,\mathrm{div}}_{\bm{k}} +
\mathcal{G}^{\square,\mathrm{reg}}_{\bm{k}}$ gives exactly
\Eq{GSquare_after_a}. The logic for this organization is that
$\mathcal{G}^{\square,\mathrm{div}}_{\bm{k}}$ contains all the
divergences, while $\mathcal{G}^{\square,\mathrm{reg}}_{\bm{k}}$ is a
formally convergent integral (one may check that it falls off fast
enough to converge, like $\propto b^{-1} e^{+ i b z } + \ldots$, in
the limit $b \to \infty$).

In order to make sense of the divergent integral
$\mathcal{G}^{\square,\mathrm{div}}_{\bm{k}}$, we must understand it
in the distributional sense --- using the results (\ref{alpha1}) and
(\ref{alpham2}) we have
\begin{align}
\label{b_distributions}
\int_{2\kappa+1}^{\infty} \dd b \,
e^{ - i b ( z_{\mathrm{in}} - z ) } b^3 & = e^{ - i (2\kappa+1) ( z_{\mathrm{in}} - z ) }
\biggl[ \frac{6}{(\zin - z)^4} + \frac{6 i (2\kappa+1)}{(\zin -z)^3} -
  \frac{3(2\kappa+1)^2}{(\zin -z)^2}  - \frac{i (2\kappa+1)^3}{\zin -z}  \biggr]
\nonumber \\ &
- i \pi \delta^{\prime\prime \prime}(\zin -z)\, ,
\\
\int_{2\kappa+1}^{\infty} \dd b \, e^{ - i b ( z_{\mathrm{in}} - z ) } b^2
& =  e^{ - i (2\kappa+1) ( z_{\mathrm{in}} - z ) } \biggl[\frac{2i}{(\zin-z)^3}
- \frac{2(2\kappa+1)}{(\zin - z)^2}  - \frac{i (2\kappa+1)^2}{\zin - z}  \biggr]
- \pi \delta^{\prime\prime}(z_{\mathrm{in}} - z)\, ,
\\
\int_{2\kappa+1}^{\infty} \dd b \, e^{ - i b ( z_{\mathrm{in}} - z ) } b
& =  e^{ - i (2\kappa+1) ( z_{\mathrm{in}} - z ) } \biggl[ - \frac{1}{(\zin - z)^2}
  - \frac{i (2\kappa+1)}{\zin - z}  \biggr] + i \pi \delta^{\prime}(\zin - z)\, ,
\\
\int_{2\kappa+1}^{\infty} \dd b \, e^{ - i b ( z_{\mathrm{in}} - z ) }
& =  \frac{- i e^{- i (2\kappa+1) (\zin - z)} }{\zin - z} + \pi \delta(\zin - z),
 \end{align}
and similarly 
\begin{align}
\label{eibz_dist}
\int_{2\kappa+1}^{\infty} \dd b \, e^{i b z } b^2
& = e^{ + i (2\kappa+1) z }  \bigg[ - \frac{2i}{z^3} - \frac{2(2\kappa+1)}{z^2}
  + \frac{i (2\kappa+1)^2}{z}  \bigg] - \pi \delta^{\prime\prime}(z)\, ,
\\
\int_{2\kappa+1}^{\infty} \dd b \, e^{ + i b  z  } b & =
e^{ + i (2\kappa+1) z } \biggl[ - \frac{1}{z^2} + \frac{i (2\kappa+1)}{z}
  \biggr] - i \pi \delta^{\prime}(z)\, ,
\\
\int_{2\kappa+1}^{\infty} \dd b \, e^{ + i b z } & =
\frac{i e^{+ i (2\kappa+1) z} }{z} + \pi \delta(z)\, .
\end{align}  
A comment on the $\delta$-functions is in order: we can ignore
$\delta(z)=0$ and its derivatives, since we always assume that $z >
0$. Since we will care about the coincident limit $\zin \to z \to 0$
later on, we need to keep $\delta(\zin - z)$ and its
derivatives. Finally we also need to use
\begin{equation}
\label{binverse_int_convergent}
\int_{2\kappa+1}^{\infty} \dd b \,
\frac{ e^{ - i b ( z_{\mathrm{in}} - z ) } }{b} =  -
\left\lbrace  \mathrm{Ei}\left[ -  i ( 2 \kappa + 1  )
(z_{\mathrm{in}} - z) \right] + i \pi \right\rbrace  \ .
\end{equation}
Putting the above pieces all together in \Eq{Gsquare1_def} leaves us with 
\begin{align}
\label{Gsquare1_answer}
\mathcal{G}^{\square,\mathrm{div}}_{\bm{k}}(\eta, \eta_\mathrm{in})
&= 8 \pi  e^{i (2 \kappa +1) z} \biggl[ -\frac{5}{z^2}
  +\frac{3 i (2 \kappa +1)}{z}+ 2 \kappa^2+2 \kappa-1 \biggr] +
e^{-i (2 \kappa+1) (\zin-z)} \bigg\lbrace -\frac{6 i z_{\mathrm{in}}^2}{(\zin-z)^4}
\nonumber \\ &
+ \frac{2}{(\zin-z)^3}[ 2 i + 3 (2 \kappa+1) \zin] \zin  +
\frac{1}{(\zin-z)^2}\biggl\{ 22 i - 4 (2 \kappa+1) \zin
\nonumber \\ &
+ \frac{i [36 \kappa (\kappa+1)-1]}{3} z_{\mathrm{in}}^2\biggr\}
- \frac{1}{\zin -z}(2\kappa+1)\left[ 22 + 2 i (2 \kappa+1) \zin
+ \frac{12 \kappa^2+12 \kappa-7}{3} z_{\mathrm{in}}^2 \right]
\nonumber \\ &
+ \frac{4}{z}\left( -\frac{6 i}{\zin} +  2 \kappa+1 \right)
-\frac{4 i}{z_{\mathrm{in}}^2} -\frac{4 (2 \kappa+1)}{z_{\mathrm{in}}}
-\frac{i \left(60 \kappa^2+60 \kappa-31\right)}{3}
\nonumber \\ &
+ \frac{(2 \kappa+1) \left(12 \kappa^2+12 \kappa-7\right)}{3}
z_{\mathrm{in}} \bigg\rbrace
+ 4 i \bigg[\frac{1}{z}\left(\frac{4}{z_{\mathrm{in}}} + 3 z_{\mathrm{in}}\right)
+\frac{4}{z_{\mathrm{in}}^2} - \frac{41}{3}
\nonumber \\ &
+ \left(\frac{2}{z_{\mathrm{in}}^3}+\frac{3}{z_{\mathrm{in}}}
+\frac{43 z_{\mathrm{in}}}{60} \right) z \bigg] \left\lbrace
\mathrm{Ei}\left[-i (2 \kappa+1) (\zin-z)\right] + i \pi \right\rbrace 
+ \pi \biggl\lbrace -z \zin \delta '''(\zin-z)
\nonumber \\ &
+ (5 z-4 \zin) \delta ''(\zin-z)
- \frac{2}{3}\left( 5 z z_{\mathrm{in}}+\frac{6 z}{z_{\mathrm{in}}}
+\frac{6 z_{\mathrm{in}}}{z}-30 \right)\delta '(\zin-z)
\nonumber \\ &
+ 2\left[  \frac{10}{z}-\frac{8}{\zin} -6 \zin + \left(\frac{23}{3}
  -\frac{2}{z_{\mathrm{in}}^2}\right) z \right] \delta (\zin-z) \bigg\rbrace .
\end{align}
We recall that, in the above expression, we have ignored the Dirac
functions $\delta(z)$ and its derivatives.

Next we compute $\mathcal{G}^{\square,\mathrm{reg}}_{\bm{k}}$. To this
end note that the integrand in the definition (\ref{Gsquare2_def}) has
an explicit $b$-primitive where
\begin{align}
  \mathcal{G}^{\square,\mathrm{reg}}_{\bm{k}}(\eta, \eta_\mathrm{in}) &=
\Biggl( 4 \pi  e^{i b z} \left(b^2+\frac{6 i b}{z}-\frac{10}{z^2}-3\right)
  + e^{-i b (\zin-z)} \biggl\lbrace -16 i b^2+
  \left(\frac{16}{z}-\frac{4}{\zin} \right) b
  \nonumber \\ &
  + \frac{4i}{3} \left[20-\frac{3 (z+6 \zin)}{z z_{\mathrm{in}}^2}\right]
  + \frac{43 z^2-220}{15 b z} + \frac{16 \left(b^2-1\right)^2
    (-1+i b z)}{b^2 z}  \coth ^{-1}(b)\biggr\rbrace
 \nonumber \\ &
+ \left[-\frac{40 i}{z^2 z_{\mathrm{in}}}
    +\frac{12 i \left(z_{\mathrm{in}}^2-2\right)}{z z_{\mathrm{in}}^2}
    +\frac{43 i }{15} z -\frac{4 i (3 z_{\mathrm{in}}^2+2 )}
    {z_{\mathrm{in}}^3}\right] (z_{\mathrm{in}}-z) \mathrm{Ei}
\left[-i b (z_{\mathrm{in}}-z)\right]
\nonumber \\ &
+ 4 \left[ \frac{10 i}{z^2} + \frac{2(b^2-2)}{b z}
    + \frac{ i (9 b^2-11)}{3} +\frac{4 (b^2-1 )^2 (1-i b z)}{b^2 z}
    \coth ^{-1}(b) \right]
\nonumber \\ & \times
e^{i b z} \mathrm{Ei}(-i b z_{\mathrm{in}})
\,   \Biggr)  \bigg|_{b=2\kappa+1}^{b \to \infty}\, . 
\end{align}
The integrand approaches $\pi (z-z_{\mathrm{in}})
\left[\frac{40}{z^2 z_{\mathrm{in}}}-\frac{12 (z_{\mathrm{in}}^2-2)}{z
    z_{\mathrm{in}}^2}+\frac{4 (3
    z_{\mathrm{in}}^2+2)}{z_{\mathrm{in}}^3} -\frac{43 z}{15} \right]$
at the upper limit $b \to \infty$. We find then that the result is
\begin{align}
\label{Gsquare2_answer}
\mathcal{G}^{\square,\mathrm{reg}}_{\bm{k}}(\eta, \eta_\mathrm{in})
&= \pi  (z-z_{\mathrm{in}}) \left[\frac{40}{z^2 z_{\mathrm{in}}}
-\frac{12 (z_{\mathrm{in}}^2-2)}{z z_{\mathrm{in}}^2}
+\frac{4 (3 z_{\mathrm{in}}^2+2)}{z_{\mathrm{in}}^3} -\frac{43 z}{15}
\right] -4 \pi  e^{i (2 \kappa+1) z} \biggl[-\frac{10}{z^2}
\nonumber \\ &
+\frac{6 i (2 \kappa+1)}{z} + 4 \kappa^2+4 \kappa-2 \biggr]
-e^{-i (2 \kappa+1) (\zin-z)} \biggl\lbrace -16 i (2 \kappa+1)^2 +
\frac{43 z^2-220}{15 (2 \kappa+1) z}
\nonumber \\ &
+(2 \kappa+1) \left(\frac{16}{z}-\frac{4}{\zin}\right)
+\frac{4 i}{3}\left[20-\frac{3 (z+6 \zin)}{z z_{\mathrm{in}}^2}\right]
+\frac{128 \kappa^2 (\kappa+1)^2 [-1+i (2 \kappa+1) z]}{(2 \kappa+1)^2 z}
\nonumber \\ & \times
\log \left(\frac{\kappa+1}{\kappa}\right) \biggr\rbrace 
-4 \biggl\lbrace \frac{10 i}{z^2}
+ \frac{2 \left(4 \kappa^2+4 \kappa-1\right)}{(2 \kappa+1) z}
+\frac{i (36 \kappa^2+36 \kappa-2)}{3}
\nonumber \\ &
+\frac{32 \kappa^2 (\kappa+1)^2 [1-i (2 \kappa+1) z] }{(2 \kappa+1)^2 z}
\log \left(\frac{\kappa+1}{\kappa}\right) \biggr\rbrace e^{i (2 \kappa+1) z}
\mathrm{Ei}\left[-i (2 \kappa+1) \zin\right]
\nonumber \\ &
- i \biggl[ -\frac{40}{z^2 z_{\mathrm{in}}}
  +\frac{12 \left(z_{\mathrm{in}}^2-2\right)}{z z_{\mathrm{in}}^2}
  - \frac{4 \left(3 z_{\mathrm{in}}^2+2\right)}{z_{\mathrm{in}}^3}
  + \frac{43}{15}z \bigg] (\zin-z)
\mathrm{Ei}\left[-i (2 \kappa+1) (\zin-z)\right].
\end{align}
We now have explicit expressions for
$\mathcal{G}^{\square,\mathrm{div}}_{\bm{k}}$ and
$\mathcal{G}^{\square,\mathrm{reg}}_{\bm{k}}$ and so we can finally
write down an expression for $\mathcal{G}_{\bm{k}}$ using the sum
\begin{equation}
\mathcal{G}_{\bm{k}}(\eta,\eta_\mathrm{in}) =
\mathcal{G}^{\Delta}_{\bm{k}}(\eta,\eta_\mathrm{in})
+\mathcal{G}^{\square,\mathrm{div}}_{\bm{k}}(\eta,\eta_\mathrm{in})
+\mathcal{G}^{\square,\mathrm{reg}}_{\bm{k}}(\eta,\eta_\mathrm{in}).
\end{equation}
Using the explicit formulae (\ref{GDelta_answer}),
(\ref{Gsquare1_answer}) and (\ref{Gsquare2_answer}) we arrive at last
to the formula:
\begin{align}
\mathcal{G}_{\bm{k}}(\eta, \eta_\mathrm{in}) & =
e^{- i (2 \kappa +1) (\zin-z)} \biggl\{
\frac{224 z_{\mathrm{in}}^2}{5(\zin - z)^5}
+\frac{1}{(\zin-z)^4}\biggl[\frac{48 i}{\kappa}
-\frac{168 \zin}{5}
+ \frac{2 i (104 \kappa + 97)}{5} z_{\mathrm{in}}^2\biggr]
\nonumber \\
& + \frac{64 i}{\kappa \zin z^3} 
+\frac{1}{(\zin-z)^3}\biggl[-\frac{8 (161 \kappa +90)}{15 \kappa}
-\frac{4 i (44 \kappa+37)}{5}  z_{\mathrm{in}}
-\frac{ 2 (192 \kappa^2+222 \kappa+43 ) }{15} z_{\mathrm{in}}^2\biggr]
\nonumber \\
& +\frac{32}{\kappa z^2}\left( \frac{i}{z_{\mathrm{in}}^2}
+ \frac{\kappa+1}{z_{\mathrm{in}}}\right)
+ \frac{1}{(\zin-z)^2}\biggl[
-\frac{32 i}{\kappa z_{\mathrm{in}}^2}
-\frac{2 i (448 \kappa^2+479 \kappa+60)}{15 \kappa}
\nonumber \\
& + \frac{8 (36 \kappa^2+51 \kappa+14 )}{15}z_{\mathrm{in}}
-\frac{i (96 \kappa^3+204 \kappa^2+52 \kappa-43 )}{15} z_{\mathrm{in}}^2 \biggr]
\nonumber  \\
& +\frac{32 (\kappa+1)}{\kappa z_{\mathrm{in}}^2 z}
+\frac{1}{\zin-z}\biggl[\frac{32 (\kappa+1)}{\kappa z_{\mathrm{in}}^2}
+\frac{2(208 \kappa^2+118 \kappa-\frac{60}{\kappa}-123)}{15}
\nonumber \\
& + \frac{2 i (2 \kappa+1) (48 \kappa^2+18 \kappa-13 )}{15} z_{\mathrm{in}}
+\frac{48 \kappa^4-24 \kappa^3-68 \kappa^2+34 \kappa+43}{15} z_{\mathrm{in}}^2 \biggr]
\nonumber \\
& -\frac{i}{15} \left( 96 \kappa^3-36 \kappa^2-68 \kappa+17\right)
-\frac{1}{15}\left(48 \kappa^4-24 \kappa^3-68 \kappa^2+34 \kappa+43\right)
( z_{\mathrm{in}} + z ) \biggr\} \nonumber \\
& +e^{- 2 i \kappa (\zin-z)} \biggl\lbrace
-\frac{224 z_{\mathrm{in}}^2 }{5 (\zin - z)^5}
+\frac{1}{(\zin-z)^4}\biggl[
-\frac{48 i}{\kappa}+\frac{168 }{5} \zin
-\frac{208 i \kappa}{5} z_{\mathrm{in}}^2\biggr]
-\frac{64 i}{\kappa \zin z^3}
\nonumber \\
& +\frac{1}{(\zin-z)^3}\biggl[\frac{1288}{15}
+\frac{176 i \kappa}{5}  z_{\mathrm{in}} + \frac{32(12 \kappa^2-5 )}{15}
z_{\mathrm{in}}^2\biggr]
-\frac{32}{z^2} \biggl(\frac{i}{\kappa z_{\mathrm{in}}^2}+ \frac{1}{\zin}
\biggr)
\nonumber \\
& +\frac{1}{(\zin-z)^2}\biggl[
\frac{32 i}{\kappa z_{\mathrm{in}}^2}
-\frac{16(18 \kappa^2-5)}{15} z_{\mathrm{in}}
+\frac{16 i \left(56 \kappa^2-15\right)}{15 \kappa}
+\frac{16 i \kappa (6 \kappa^2-5 )}{15}  z_{\mathrm{in}}^2\biggr]
\nonumber \\
& -\frac{4}{\zin-z}\biggl[
\frac{8}{z_{\mathrm{in}}^2} +\frac{4(26 \kappa^2-35)}{15}
+ \frac{8 i \kappa (6 \kappa^2-5 )}{15}z_{\mathrm{in}}
+\frac{12 \kappa^4-20 \kappa^2+15}{15} z_{\mathrm{in}}^2\biggr]
-\frac{32}{z_{\mathrm{in}}^2 z }
\nonumber \\ &
+ \frac{16 i (6 \kappa^2-5) \kappa }{15}
+\frac{4 (12 \kappa^4-20 \kappa^2+15)}{15}  (z+z_{\mathrm{in}}) \biggr\rbrace
+\left(-\frac{40}{z^2}+\frac{92}{3} -\frac{43}{15}
z^2 \right)
\nonumber \\ & \times
\bigl\lbrace \pi
-i \mathrm{Ei}
\left[-i (2 \kappa+1) (\zin-z)\right] \bigr\rbrace  
+ 4\biggl[-\frac{16 i}{\kappa z^4}+\frac{8}{z^3}-\frac{8 i}{\kappa z^2}
  -\frac{2 i (4 \kappa^4-12 \kappa^2-3)}{3 \kappa}
\nonumber \\ & 
+ (2 \kappa^2-1)^2 z
-\frac{2 i \kappa (12 \kappa^4-20 \kappa^2+15 )}{15} z^2\biggr]
\biggl\{\mathrm{Ei}\bigl[- 2 i \kappa (\zin - z)\bigr]
\nonumber \\ & 
-\mathrm{Ei}\bigl[-i (2 \kappa+1) (\zin-z)\bigr] \biggr\}
+ 4\biggl[ \frac{16 i}{\kappa z^4}+\frac{24}{z^3}
-\frac{8 i (2 \kappa^2-1)}{\kappa z^2}
+\frac{-4 \kappa^4+4 \kappa^2-1}{\kappa^2 z}\biggr]
\nonumber \\ & \times
e^{2 i \kappa z} \mathrm{Ei}(-2 i \kappa \zin)
+ 4\biggl[-\frac{16 i}{\kappa z^4}-\frac{8 (3 \kappa+2)}{\kappa z^3}
+\frac{2 i (8 \kappa+7)}{z^2}+\frac{4 \kappa (\kappa+1)}{z} \biggr]
\nonumber \\ & \times
e^{i (2 \kappa+1) z} \mathrm{Ei}\left[-i (2 \kappa+1) \zin \right]
+ \pi \biggl\lbrace -z \zin \delta'''(\zin-z)
+ (5 z-4 \zin) \delta ''(\zin-z)
\nonumber \\ &
- \frac{2}{3}\biggl(5 z z_{\mathrm{in}}
  +\frac{6 z}{z_{\mathrm{in}}}+\frac{6 z_{\mathrm{in}}}{z}-30\biggr)
\delta '(\zin-z) + 2\biggl[\frac{10}{z}-\frac{8}{\zin}-6 \zin
\nonumber \\ &
+ \biggl(\frac{23}{3}-\frac{2}{z_{\mathrm{in}}^2}\biggr)
z \biggr] \delta (\zin-z) \biggr\rbrace \ . 
 \end{align}
Next we explore the various asymptotic limits of
$\mathcal{G}_{\bm{k}}(\eta, \eta_\mathrm{in})$. First we note that in
the limit $\zin \gg 1$ (so that $\eta_{\mathrm{in}} \to -\infty$ in
the distant past) we have
\begin{align}
\mathcal{G}_{\bm{k}}(\eta, \eta_\mathrm{in}) & \simeq
4 \pi  e^{2 i \kappa z} \left[ \frac{16}{\kappa z^4}-\frac{24 i}{z^3}
-\frac{8 \left(2 \kappa^2-1\right)}{\kappa z^2}
+\frac{i \left(2 \kappa^2-1\right)^2}{\kappa^2 z}\right]
\nonumber \\ &
+8 \pi  e^{i (2 \kappa+1) z} \left[-\frac{8}{\kappa z^4}
+\frac{4 i (3 \kappa+2)}{\kappa z^3}+\frac{8 \kappa+7}{z^2}
-\frac{2 i \kappa (\kappa+1)}{z}\right]
+ \mathcal{O}\left( \frac{1}{\zin} \right), 
\end{align}
and note that if we furthermore take the super-Hubble limit of this
then we get
\begin{align}
\mathcal{G}_{\bm{k}}(\eta, \eta_\mathrm{in}) & \simeq -\frac{40 \pi }{z^2}
- \frac{4 i \pi  \left(24 \kappa^3+6 \kappa^2+16 \kappa-3\right)}{3 \kappa^2 z}
+ \frac{92 \pi }{3} + \mathcal{O}(z) \ .
\end{align}
Next let us fix arbitrary $\zin$ and take the super-Hubble limit $z
\ll 1$ so that
\begin{align}
\label{eq:limGsmallz}
\mathcal{G}_{\bm{k}}(\eta, \eta_\mathrm{in}) &\simeq
- \frac{40 \pi}{z^2} + \frac{16}{3z} \biggl\lbrace
e^{-2 i \kappa z_{\mathrm{in}}} \biggl[-\frac{2 i (\kappa^2-3)}{\kappa z_{\mathrm{in}}}
+\frac{4 i}{\kappa z_{\mathrm{in}}^3}-\frac{5}{z_{\mathrm{in}}^2}\biggr]
+e^{-i (2 \kappa+1) z_{\mathrm{in}}} \biggl[-\frac{4 i}{\kappa z_{\mathrm{in}}^3}
+\frac{5 \kappa+4}{\kappa z_{\mathrm{in}}^2}
\nonumber \\ &
+\frac{i (4 \kappa^2-5 \kappa-8)}{2 \kappa z_{\mathrm{in}}}\biggr] 
+\left(\kappa^2-9-\frac{3}{4 \kappa^2}\right)
\mathrm{Ei}(-2 i \kappa z_{\mathrm{in}})
+ \left(-\kappa^2+6 \kappa+\frac{4}{\kappa}+\frac{21}{2}\right)
\nonumber \\ & \times
\mathrm{Ei}
\left[-i (2 \kappa+1) z_{\mathrm{in}}\right]  \biggr\rbrace+ \frac{92\pi}{3}
  + \mathcal{O}(z) \ .
\end{align}
Let us now take the above expression and take $\zin \ll 1$ (assuming
the hierarchy $z \ll \zin \ll 1$)
\begin{align}
\mathcal{G}_{\bm{k}}(\eta, \eta_\mathrm{in}) &\simeq
-\frac{40 \pi}{z^2}+\frac{1}{z} \biggl[-\frac{40 i}{\zin}
+2\left(-\frac{1}{\kappa^2}+8 \kappa+\frac{16}{3 \kappa}+2\right) \left\lbrace
2 \log \left[e^{\gamma } (2 \kappa+1) \zin \right]
-i \pi \right\rbrace
\nonumber \\ &
+4 \left(-\frac{4 \kappa^2}{3}+\frac{1}{\kappa^2}+12\right)
\log \left(\frac{2\kappa+1}{2 \kappa}\right)
-\frac{16 (39 \kappa+15+{16}/{\kappa}) }{9} + \mathcal{O}(\zin)\biggr]
\nonumber \\ &
+ \frac{92\pi}{3} + \mathcal{O}(z) 
\end{align}
where $\gamma$ is Euler-Mascheroni constant. 

Finally let us take the coincident limit for $z \to \zin$ where we have  
\begin{align}
\mathcal{G}_{\bm{k}}(\eta, \eta_\mathrm{in}) & \simeq 
- \frac{6 i z_{\mathrm{in}}^2}{(z_{\mathrm{in}}-z)^4}
+ \frac{4 i z_{\mathrm{in}}}{(z_{\mathrm{in}}-z)^3}
+ \frac{2 i \left(33-5 z_{\mathrm{in}}^2\right)}{3 (z_{\mathrm{in}}-z)^2}
+ i \left(\frac{40}{z_{\mathrm{in}}^2}-\frac{92}{3}
+ \frac{43 }{15}z_{\mathrm{in}}^2 \right)
\nonumber \\ & \times 
\log \left[e^{\gamma } (2 \kappa +1) (z_{\mathrm{in}}-z) \right]
+\pi  \bigg\lbrace -z \zin \delta'''(\zin-z) + (5 z-4 \zin)
\delta ''(\zin-z)
\nonumber \\ 
& - \frac{2}{3}\left( 5 z z_{\mathrm{in}}
+\frac{6 z}{z_{\mathrm{in}}}
+\frac{6 z_{\mathrm{in}}}{z}-30 \right) \delta '(\zin-z)
+ 2\biggl[\frac{10}{z}-\frac{8}{\zin} -6 \zin
\nonumber \\ &
+ \left(\frac{23}{3}-\frac{2}{z_{\mathrm{in}}^2}\right) z \biggr]
\delta (\zin-z) \bigg\rbrace + \mathcal{O}\big[ (\zin - z)^0 \big] \ . 
\end{align}
Most important for us is the real part of the above, where we have
\begin{align}
\label{ReG_coincident}
\mathrm{Re}\left[ \mathcal{G}_{\bm{k}}(\eta, \eta_\mathrm{in}) \right]
& \simeq \pi \bigg\lbrace -z \zin \delta '''(\zin-z) + (5 z-4 \zin)
\delta ''(\zin-z)
- \frac{2}{3}\biggl( 5 z z_{\mathrm{in}}
+\frac{6 z}{z_{\mathrm{in}}}
+\frac{6 z_{\mathrm{in}}}{z}-30 \biggr)
\nonumber \\ & \times
\delta '(\zin-z)
+ 2\biggl[\frac{10}{z}-\frac{8}{\zin} -6 \zin
+ \left(\frac{23}{3}-\frac{2}{z_{\mathrm{in}}^2}\right) z \biggr]
\delta (\zin-z) \bigg\rbrace + \mathcal{O}\big[ (\zin - z)^0 \big] \, .
\end{align}
Notice that the singular part of the above is only coming from the
$\delta$-functions.

\subsection{The function $\mathcal{I}$}
\label{App:I_computation}

Here we compute the integral (\ref{I_forF}), repeated here for convenience
\begin{align} 
\mathcal{I}_{\bm{k}}(\eta) & := \int_0^1 \dd a \int_{a  + 2 \kappa}^{\infty}
\dd b \, z \, ( a^2 + b^2 - 2 )^2 \left[1+\frac{2i}{(b-a) z}\right]
\left[1+\frac{2i}{(b+a) z}\right]
\nonumber \\ & \times
\biggl[ - \frac{1}{b^2}
-\frac{i z}{b} - \frac{4}{b^2 - a^2}  + \frac{4 e^{ i b z}
\mathrm{Ei}( - i b z )}{b^2 - a^2}  \biggr].
\end{align}
Luckily the computation of this is straightforward since we already
know what the function $\mathcal{G}_{\bm{k}}(\eta,\eta_{\mathrm{in}})$
is --- the reason for this is that the integrand of
$\mathcal{I}_{\bm{k}}(\eta)$ and
$\mathcal{G}_{\bm{k}}(\eta,\eta_{\mathrm{in}})$ are identical in the
limit $\eta_{\mathrm{in}} \to \eta$ (one must only be careful about
the divergences in making this replacement).

As before, we split apart the integral $\mathcal{I}_{\bm{k}}(\eta)$
into a triangular and rectangular region,
\begin{align}
\mathcal{I}_{\bm{k}}(\eta) & = \mathcal{I}^{\Delta}_{\bm{k}}(\eta)
+ \mathcal{I}^{\square}_{\bm{k}}(\eta)
\end{align}
where we define
\begin{align}
\mathcal{I}^{\Delta}_{\bm{k}}(\eta) & : = \int_{2\kappa}^{2\kappa+1} \dd b
\int_0^{b-2r} \dd a\, f(a,b,z,z)
\\ 
\label{Isquare_def}
\mathcal{I}^{\square}_{\bm{k}}(\eta) & : =
\int_{2\kappa+1}^{\infty} \dd b \int_0^{1} \dd a\, f(a,b,z,z) \, ,
\end{align}
\cf \Eqs{GDelta_def} and (\ref{GSquare_def}) where we use
\begin{align}
f(a,b,z,z) &= z \, ( a^2 + b^2 - 2 )^2
\left[1+\frac{2i}{(b-a) z}\right]\left[1+\frac{2i}{(b+a) z}\right]
\nonumber \\ & \times
\biggl[ - \frac{1}{b^2} - \frac{i z}{b} - \frac{4}{b^2 - a^2}
    + \frac{4 e^{ i b z}  \mathrm{Ei}( - i b z )}{b^2 - a^2}  \biggr], 
\end{align}
where $f(a,b, z,\zin)$ is the integrand of
$\mathcal{G}_{\bm{k}}(\eta,\eta_{\mathrm{in}})$, defined in
\Eq{fintegrand_def}.

Note that 
\begin{equation}
\mathcal{I}^{\Delta}_{\bm{k}}(\eta) = \lim_{\zin \to z}
\left[\; \mathcal{G}^{\Delta}_{\bm{k}}(\eta, \eta_{\mathrm{in}}) \; \right]
\end{equation}
where we have the explicit formula for
$\mathcal{G}^{\Delta}_{\bm{k}}(\eta, \eta_{\mathrm{in}})$ given in
\Eq{GDelta_answer} and so we straightforwardly have an answer for
$\mathcal{I}^{\Delta}_{\bm{k}}(\eta)$ (for brevity we avoid writing it
down here).

For computing $\mathcal{I}^{\square}_{\bm{k}}(\eta)$ in
\Eq{Isquare_def} we use the formula (\ref{GSquare_after_a}) (after $a$
integration has been performed) and take the limit $\zin \to z$ giving
\begin{equation}
  \mathcal{I}^{\square}_{\bm{k}}(\eta) =
  \mathcal{I}^{\square,\mathrm{div}}_{\bm{k}}(\eta)
  + \mathcal{I}^{\square,\mathrm{reg}}_{\bm{k}}(\eta)\ ,
\end{equation}
where
\begin{align}
\label{Isquare1_def}
\mathcal{I}^{\square,\mathrm{div}}_{\bm{k}}(\eta) &: =
\int_{2\kappa+1}^{\infty} \dd b\, \bigg\{4 \pi e^{+ i b z}
\bigg[ -i  z b^2 +4 b+i \left(3 z+\frac{4}{z}\right) \bigg] 
+\bigg[ -i z^2 b^3  -  z b^2
\nonumber \\ &
+ \frac{ 2 i (5 z^2-18)}{3} b
+\frac{10 z}{3} +  \frac{i}{b}\left(-\frac{40}{z^2}+\frac{92}{3}
-\frac{43}{15} z^2 \right)\bigg]\bigg\} 
\\ 
\mathcal{I}^{\square,\mathrm{reg}}_{\bm{k}}(\eta) &: = \int_{2\kappa+1}^{\infty} \dd b\,
\Biggl(- 4 \pi e^{+ i b z} \bigg[ -i  z b^2 +4 b+i
  \left(3 z+\frac{4}{z}\right) \bigg] -36 i b
+\frac{36}{z} +\frac{4 i \left(z^2+30\right)}{3 b z^2}
\nonumber \\ &
+ \frac{-43 z^2-260}{15 b^2 z}
+ \frac{16 \left(2 i b^5 z-b^4-2 b^2-2 i b z+3\right) }{b^3 z}
\coth ^{-1}(b)+ 4  \biggl\lbrace-3 b^2 z+12 i b 
\nonumber \\ &
+\frac{11 z^2-36}{3 z} -\frac{8 i}{b}+\frac{8}{b^2 z}
+ \left[\frac{4 (b^2-1)^2 z}{b}+\frac{8}{b^2}(b^4-1)
    \left(\frac{1}{b z}-i\right)\right] \coth ^{-1}(b)
\biggr\rbrace
\nonumber \\ & \times
e^{i b z} \mathrm{Ei}(-i b z)\Biggr) 
\end{align}
\cf the definitions (\ref{Gsquare1_def}) and
(\ref{Gsquare2_def}). Note in particular that
\begin{equation}
\mathcal{I}^{\square,\mathrm{reg}}_{\bm{k}}(\eta) = \lim_{\zin \to z}
\left[ \, \mathcal{G}^{\square,\mathrm{reg}}_{\bm{k}}(\eta, \eta_{\mathrm{in}})
\, \right],
\end{equation}
where we can straightforwardly write down an expression for
$\mathcal{I}^{\square,\mathrm{reg}}_{\bm{k}}(\eta)$ given we know what
$\mathcal{G}^{\square,\mathrm{reg}}_{\bm{k}}(\eta, \eta_{\mathrm{in}})$ is in
\Eq{Gsquare2_answer} (although once again we avoid writing this down
here for brevity).

One can almost make a similar statement for
$\mathcal{I}^{\square,1}_{\bm{k}}(\eta)$, but not exactly because this
integral is formally divergent --- we need to use the distributions
(\ref{eibz_dist}) to evaluate the first terms in \Eq{Isquare1_def}, as
well as the following dimensional regularizations of various powers of
$b$:
\begin{align}
\int_{2\kappa+1}^{\infty} \dd b\, b^{j} \left( \frac{b}{M} \right)^\epsilon
& =  - \frac{(2\kappa+1)^{ \epsilon+ j +1}}{M^{\epsilon}(\epsilon + j + 1)} \simeq
- \frac{(2\kappa+1)^{j+1}}{j+1}  + \mathcal{O}(\epsilon),
\\
\int_{2\kappa+1}^{\infty} \dd b\, \frac{1}{b}\left( \frac{b}{M} \right)^\epsilon
& =  - \frac{(2\kappa+1)^{\epsilon}}{M^\epsilon \epsilon} \simeq
- \frac{1}{\epsilon}
- \log\left(\frac{2\kappa+1}{M} \right) + \mathcal{O}(\epsilon)
\end{align}
where the first relation applies for $j\in\{0,1,2,3\}$ and $M = \mu /
k > 0$ ($\epsilon \in \mathbb{C}$) for some mass scale $\mu > 0$ and where we
have expanded the above integrals for $0 < |\epsilon| \ll 1$ (although
the integrals are technically only convergent for
$\mathrm{Re}(\epsilon)< - j - 1$ and $\mathrm{Re}(\epsilon)<0$
respectively). With this, we find that
\begin{align}
\mathcal{I}^{\square,\mathrm{div}}_{\bm{k}}(\eta) & = i
\left(\frac{40}{z^2}-\frac{92}{3} + \frac{43}{15} z^2 \right)
\left[ \log \left(\frac{2 \kappa+1}{M}\right)  +\frac{1}{\epsilon}\right]
+ 8 \pi  e^{i (2 \kappa+1) z} \biggl[ -\frac{5}{z^2}
+\frac{3 i (2 \kappa+1)}{z}
\nonumber \\ &
+ 2 \kappa^2+2 \kappa-1 \biggr] 
+6 i (2 \kappa+1)^2+\frac{(2 \kappa+1)(4 \kappa^2+4 \kappa-9)}{3} z
\nonumber \\ &
+ \frac{i (2 \kappa+1)^2 (12 \kappa^2+12 \kappa-17)}{12} z^2\, ,
\end{align}
where we have ignored the Dirac functions. With the above, we can
at last sum together all the pieces to get our result
\begin{align}
\label{Ik_Answer}
\mathcal{I}_{\bm{k}}(\eta)&=  \mathcal{I}^{\Delta}_{\bm{k}}(\eta)
+ \mathcal{I}^{\square,\mathrm{div}}_{\bm{k}}(\eta)
+ \mathcal{I}^{\square,\mathrm{reg}}_{\bm{k}}(\eta)
\nonumber \\
&= i \left( \frac{40}{z^2} - \frac{92}{3} + \frac{43}{15} z^2 \right)
\left[ \frac{1}{\epsilon} + \log\left( \frac{2\kUV+k}{\mu} \right) \right]  
+ \frac{64}{\kappa z^3} - \frac{4i (\kappa+4)}{\kappa z^2}
+ \frac{2i}{9\kappa}\biggl(84 \kappa^3 + 60 \kappa^2
\nonumber \\ &
+ 79 \kappa + 27\biggr)
+ \frac{(32 \kappa^3 - 40 \kappa + 1)z}{3}
+ \frac{i}{900}\biggl(-3611-3360\kappa + 360\kappa^2+4320 \kappa^3
+ 720 \kappa^4\biggr)
\nonumber \\ &
+ \left[ \frac{64i}{\kappa z^4} - \frac{32}{z^3} + \frac{32i}{\kappa z^2}
+ \frac{8 i ( 4 \kappa^4 -12 \kappa^2 - 3 )}{3 \kappa} - 4( 2\kappa^2 -1 )^2 z
+ \frac{8i\kappa(12\kappa^4-20\kappa^2 +15)}{15} z^2 \right]
\nonumber \\ & \times
\log\left( \sfrac{2\kappa+1}{2\kappa} \right)
+ 8 e^{i ( 2 \kappa + 1 ) z } \mathrm{Ei}\left[-  i (2 \kappa+1 )z\right]
\biggl[ - \frac{8 i}{\kappa z^4} - \frac{4 (3\kappa+2)}{\kappa z^3}
 + \frac{i (8\kappa +7)}{z^2}
\nonumber \\ &
+ \frac{2\kappa(\kappa+1)}{z} \biggr]
+ 8 e^{2 i \kappa z } \mathrm{Ei}\left(-  2 i \kappa z \right)
\left[ \frac{8 i}{\kappa z^4} + \frac{12}{z^3}
 - \frac{4 i ( 2\kappa^2 - 1 )}{\kappa z^2}
 - \frac{(2\kappa^2-1)^2}{2\kappa^2 z}  \right] 
\end{align}

Writing out the real and imaginary parts of the above function gives
\begin{align}
\label{ReIk_Answer}
\mathrm{Re}\left[ \mathcal{I}_{\bm{k}}(\eta) \right] &=
\frac{64}{\kappa z^3} +  \frac{(32 \kappa^3 - 40 \kappa + 1)z}{3}
+ \left[ - \frac{32}{z^3} - 4( 2 \kappa^2 -1 )^2 z \right]
\log\left( \frac{2\kappa+1}{2\kappa} \right)
\nonumber \\ &
+  \mathrm{Re}\left\lbrace
8 e^{i ( 2 \kappa + 1 ) z }\mathrm{Ei}\left[-  i (2 \kappa+1 ) z\right]
\left[ - \frac{8 i}{\kappa z^4} - \frac{4 (3\kappa+2)}{\kappa z^3}
  + \frac{i (8\kappa+7)}{z^2}
+ \frac{2\kappa(\kappa+1)}{z} \right] \right\rbrace
\nonumber \\ &
+ \mathrm{Re}\left\lbrace 8 e^{2 i \kappa z } \mathrm{Ei}
\big(-  2 i \kappa z \big)  \left[ \frac{8 i}{\kappa z^4} + \frac{12}{z^3}
-\frac{4 i ( 2 \kappa^2 - 1 )}{\kappa z^2} - \frac{(2\kappa^2-1)^2}{2\kappa^2 z}
\right] \right\rbrace 
\end{align}
while the imaginary part is
\begin{align}
\label{ImIk_Answer}
\mathrm{Im}\left[ \mathcal{I}_{\bm{k}}(\eta) \right] & =
\left( \frac{40}{z^2} - \frac{92}{3} + \frac{43}{15} z^2 \right)
\left[ \frac{1}{\epsilon} + \log\left( \frac{2\kUV+k}{\mu} \right) \right]  
-\frac{4(\kappa+4)}{\kappa z^2} + \frac{2}{9\kappa}\biggl(84 \kappa^3 
+ 60 \kappa^2
\nonumber \\ &
 +79 \kappa + 27\biggr)
 + \frac{-3611-3360 \kappa + 360 \kappa^2+4320  \kappa^3
   + 720  \kappa^4}{900} z^2 
+ \biggl[ \frac{64}{ \kappa z^4} + \frac{32}{ \kappa z^2}
\nonumber \\ &
+ \frac{8 ( 4  \kappa^4 -12  \kappa^2 - 3 )}{3  \kappa}
+ \frac{8 \kappa (12  \kappa^4-20 \kappa^2 +15)}{15} z^2 \biggr]
\log\left( \frac{2 \kappa+1}{2 \kappa} \right)
\nonumber \\ &
+\mathrm{Im}\biggl\{8 e^{i ( 2  \kappa + 1 ) z }
\mathrm{Ei}\left[-  i (2  \kappa +1 ) z\right]
\left[ - \frac{8 i}{ \kappa z^4}
-\frac{4 (3  \kappa +2)}{ \kappa z^3} + \frac{i (8  \kappa+7)}{z^2}
+ \frac{2 \kappa( \kappa+1)}{z} \right] \biggr\}
\nonumber \\ &
+ \mathrm{Im}\left\lbrace 8 e^{2 i  \kappa z } \mathrm{Ei}
  \big(-  2 i  \kappa z \big)
\left[ \frac{8 i}{ \kappa z^4} + \frac{12}{z^3}
  - \frac{4 i ( 2 \kappa^2 - 1 )}{ \kappa z^2}
  - \frac{(2 \kappa^2-1)^2}{2 \kappa^2 z}  \right] \right\rbrace. 
\end{align}

Next we explore the various asymptotic limits of
$\mathcal{I}_{\bm{k}}(\eta)$. First we consider the super-Hubble limit
$z \ll 1$ (which also assumes $\kappa z = - \kUV \eta \ll 1$)
\begin{align}
\label{eq:limIsmallz}
\mathcal{I}_{\bm{k}}(\eta) & \simeq
i \left( \frac{40}{z^2} - \frac{92}{3}
\right)\left[ \frac{1}{\epsilon} + \log\left( \frac{2\kUV+k}{\mu}
\right) \right]
+\frac{- 20 \pi + 28i - 40i \log[ e^{\gamma}(2\kappa+1) z ]}
{z^2}
\nonumber \\ &
+ \frac{1}{z}\Biggl(\frac{4}{9}\biggl\{
-4 \left(39 \kappa+15 +\frac{16}{\kappa} \right)
+3 \left(-4 \kappa^2+ 36 +\frac{3}{\kappa^2}\right)
\log \left(\frac{2 \kappa+1}{2 \kappa}\right)
+\biggl(72 \kappa +18
\nonumber \\ &
+\frac{48}{\kappa}-\frac{9}{\kappa^2}\biggr)
\log \left[e^{\gamma } (2 \kappa+1) z\right] \biggr\}
-2 i \pi  \left(8 \kappa+2+\frac{16}{3 \kappa}
-\frac{1}{\kappa^2}\right)\Biggr)
+\frac{1}{3} \biggl\{46 \pi
\nonumber \\ &
+ 92  i \log[e^{\gamma} (2\kappa+1) z ]
-128i \biggr\rbrace \ \ + \mathcal{O}(z)
\end{align}
Next we consider the early-time expansion $z \gg 1$ (which implicitly
assumes that $\zin \gg z \gg 1$), where
\begin{align}
\mathcal{I}_{\bm{k}}(\eta) & \simeq
i \left(
- \frac{92}{3} + \frac{43}{15} z^2 \right)
\left[ \frac{1}{\epsilon} + \log\left( \frac{2\kUV+k}{\mu} \right) \right] 
+ i \biggl[ \frac{ 720 \kappa^4+4320 \kappa^3
    +360 \kappa^2-3360 \kappa -3611 }{900}
\nonumber \\ &
+ \frac{8 \kappa \left(12 \kappa^4-20 \kappa^2+15\right)}{15}
\log \left(\frac{2\kappa+1}{2 \kappa}\right)\biggr] z^2
+ \biggl[\frac{32 \kappa^3-40 \kappa+1}{3}-4 \left(1-2 \kappa^2\right)^2
\nonumber \\ & \times
\log \left(\frac{2\kappa+1}{2 \kappa} \right) \biggr] z  
+ \frac{2i}{9\kappa}\biggl[ 84 \kappa^3+60 \kappa^2+79 \kappa+27
+12 \left(4 \kappa^4-12 \kappa^2-3\right)
\log \left(\frac{2\kappa+1}{2 \kappa}\right) \biggr]
\nonumber \\ &
+ \mathcal{O}\left( \frac{1}{z} \right).
\end{align}

\subsection{Asymptotics of $\mathfrak{F}$}

We can now take the above asymptotic expressions for
$\mathcal{G}_{\bm{k}}(\eta,\eta_{\mathrm{in}})$ and
$\mathcal{I}_{\bm{k}}(\eta)$ and use formula (\ref{eq:FintermsofIG})
to obtain expressions for
$\mathfrak{F}_{\bm{k}}(\eta,\eta_{\mathrm{in}})$ in various limits
used in the main text. Most importantly we consider the limit $z \ll
\zin \ll 1$ in which, using \Eqs{eq:limGsmallz}
and~(\ref{eq:limIsmallz}), we have
\begin{align}
\label{F_superHubble_App}
\mathfrak{F}_{\bm{k}}(\eta, \eta_{\mathrm{in}}) & =
\frac{\slrl  H^2 k^2}{1024 \pi^2 \Mp^2 }
\Biggl(
\frac{40 i}{z^2}\left[\frac{1}{\epsilon}
+\log \left(\frac{2\kUV+k}{\mu}\right)\right]
-\frac{20\pi}{z^2}+\frac{i}{z^2}
\biggl\{28-40 \log \left[e^{\gamma}(2\kappa+1)z\right]\biggr\}
\nonumber \\ &
+{\cal O}\left(\frac{1}{z}\right)
-\left[-\frac{40\pi}{z^2}+{\cal O}\left(\frac{1}{z}\right)\right]
\Biggr)
\end{align}
This expression (in fact an even more accurate version of this
equation with higher order terms written explicitly) is given in
\Eq{F_superHubble_body}. \Eq{ReFSHresult}, which is also
used in the main text, also corresponds to the limit $z\ll 1$ but for
arbitrary $\zin$.

We now turn to address the question raised in Footnote
\ref{techfoot}. This footnote asks whether the singularity of Re
$[\mathfrak{F}_\bmk(\eta,\eta_{\rm in})]$ in the coincident limit
might mean that integrating through the coincident limit might compete
in the expression for $\Xi_\bmk$ even if $k\eta$ is not small. We show
here that these corrections are in fact subdominant in
$k\eta_{\mathrm{in}}$. In the coincidence limit, the integral in the
denominator of \Eq{eq:purity:exact} can be written as
\begin{align}
\label{divergingcoin}
& - \int_{z_0}^\zin \dd z' \,  \mathrm{Re}\left[ \mathcal{G}_{\bm{k}}
\left(-\frac{z'}{k},-\frac{\zin}{k} \right) \right]
\left[ 1 + \frac{1}{(z')^2}\right]
\simeq - \pi \int_{z_0}^\zin \dd z' \, \bigg\lbrace -z' \zin \delta '''(\zin-z')
\nonumber \\ & 
+ (5 z'-4 \zin) \delta ''(\zin-z')- \frac{2}{3}
\left( 5 z' z_{\mathrm{in}}+\frac{6 z'}{z_{\mathrm{in}}}
+\frac{6 z_{\mathrm{in}}}{z'}-30 \right)\delta '(\zin-z')
+ 2\biggl[\frac{10}{z'}-\frac{8}{\zin} -6 \zin
\nonumber \\ &
+ \left(\frac{23}{3}-\frac{2}{z_{\mathrm{in}}^2}\right) z' \biggr]
\delta (\zin-z') \bigg\rbrace \left[ 1 + \frac{1}{(z')^2}\right] 
\end{align}
where we have used $\mathfrak{F}_{\bm{k}} \propto \mathcal{I}_{\bm{k}}
-\mathcal{G}_{\bm{k}}$ and where $z \simeq z_0$ is close to the
coincident limit so that $|z_0 - \zin|\ll 1$ and where we keep the
$\delta$-functions because all other terms are regular in the
coincident limit. The above integral can be calculated exactly by
using the regularization $\delta_\epsilon(x)=\pi^{-1}\lim_{\epsilon
  \rightarrow 0} \epsilon/(x^2+\epsilon^2)$. Then, one obtains
\begin{align}
- \int_{z_0}^\zin \dd z' \,  \mathrm{Re}\left[ \mathcal{G}_{\bm{k}}
\left(-\frac{z'}{k},-\frac{\zin}{k} \right) \right]
\left[ 1 + \frac{1}{(z')^2}\right]
& \simeq 
\frac{2 \left(z_{\mathrm{in}}^2+1\right)}{\epsilon ^3}
-\frac{10 z_{\mathrm{in}}^4-41 z_{\mathrm{in}}^2-39}{3 z_{\mathrm{in}}^2 \epsilon }
\nonumber \\ &
-\frac{2 \pi  \left(5 z_{\mathrm{in}}^2-24\right)}{3 z_{\mathrm{in}}^3}
+ \mathcal{O}(\epsilon).
\end{align}
We see that the $\epsilon$-dependence is power-law, and any
$\zin$-dependence is subdominant to the result
quoted in the main text. 

\subsection{The validity coefficient}
\label{App:ValidityM}

Here we compute the integral $\mathfrak{M}_{\bm{k}}$ defined in
\Eq{Mfrak_def}
\begin{equation}
\mathfrak{M}_{\bm{k}}(\eta,\eta_{\mathrm{in}}): =
(2\pi)^{3/2} \int_{ \eta_{\mathrm{in}} }^{\eta} \dd \eta' \,
G(\eta) G(\eta') \mathscr{C}_{\bm{k}}(\eta,\eta') \, (\eta' - \eta) \ .
\end{equation}
Noting the earlier definition (\ref{Ffrakdef}) of
$\mathfrak{F}_{\bm{k}}(\eta,\eta_{\mathrm{in}})$, it turns out that we
can decompose $\mathfrak{M}_{\bm{k}}$ as follows
\begin{align}
\mathfrak{M}_{\bm{k}}(\eta,\eta_\mathrm{in}) & =
\mathfrak{N}_{\bm{k}}(\eta,\eta_\mathrm{in})
- \eta \mathfrak{F}_{\bm{k}}(\eta,\eta_\mathrm{in})
\end{align}
with $\mathfrak{N}_{\bm{k}}$ is defined by
\begin{align}
\mathfrak{N}_{\bm{k}}(\eta,\eta_{\mathrm{in}}) & : =
(2\pi)^{3/2} \int_{\eta_\mathrm{in}}^{\eta}
\dd \eta' \, G(\eta) \, G(\eta') \, \mathscr{C}_{\bm{k}}(\eta,\eta')
\, \eta' \ .
\end{align}
Following similar steps for the computation as for
$\mathfrak{F}_{\bm{k}}(\eta,\eta_{\mathrm{in}})$ as in Appendix
\ref{App:Lindblad_F}, we find that
$\mathfrak{N}_{\bm{k}}(\eta,\eta_{\mathrm{in}})$ is given by the
following expression 
\begin{align}
\mathfrak{N}_{\bm{k}} (\eta, &\eta_{\mathrm{in}}) =
\frac{\slrl  H^2 k}{1024 \pi ^2 \Mp^2} \Biggl(
-32 \kappa -8-\frac{64}{3 \kappa}+\frac{4}{\kappa^2}
-\frac{2z^2}{3}\biggl[32 \kappa^3-40 \kappa
  +1 -12 \left(2 \kappa^2-1\right)^2
\nonumber \\ & \times
\log \left(\frac{2 \kappa+1}{2 r}\right)\biggr]
- i z\left(\frac{40}{z^2}-24 +\frac{43 z^2}{15}\right)
\left[ \frac{1}{\epsilon}+\log \left(\frac{2 \kappa+1}{M}\right)\right]
+ \frac{i}{z}\biggl[2 \left(\frac{1}{\kappa^3}+\frac{2}{2 \kappa+1}
+\frac{4}{\kappa}\right)
\nonumber \\ &
-\frac{32}{\kappa}\log
\left(\frac{2 \kappa+1}{2 \kappa}\right)\biggr]
+\frac{iz}{\kappa}\biggl[8 \left(4 \kappa^4+1\right) \log
\left(\frac{2 \kappa+1}{2 \kappa}\right)
-\frac{2 \left(36 \kappa^4+24 \kappa^3+6 \kappa^2+9 \kappa+4\right)}{2 \kappa+1}
\biggr]
\nonumber \\ & 
+ \frac{iz^3}{15} \biggl[-12 \kappa^4-72 \kappa^3-6 \kappa^2+56 \kappa
+\frac{3611}{60}-8 \left(12 \kappa^4-20 \kappa^2+15\right) \kappa
\log \left(\frac{2 \kappa +1}{2 r}\right)\biggr]
\nonumber \\ &
- 8 \left[ \frac{4 i}{\kappa z} -i \left(4 \kappa^3+\frac{1}{\kappa}\right) z
-\left(1-2 \kappa^2\right)^2 z^2
+\frac{i \kappa (12 \kappa^4-20 \kappa^2+15 )}{15} z^3\right]
\mathrm{Ei}\left[-2 i \kappa (\zin-z)\right]
\nonumber \\ &
- 8 \biggl[-\frac{i}{z}\left(\frac{4}{\kappa}+5\right)
+ \frac{iz}{\kappa} \left( 4 \kappa^4+3 \kappa+1\right)
+ \left(1-2 \kappa^2\right)^2 z^2
-\frac{iz^3}{120} \bigl(96 \kappa^5-160 \kappa^3+120 \kappa
\nonumber \\ &
+43 \bigr)\biggr]
\mathrm{Ei}\left[i (2 \kappa +1) (z-\zin)\right]
- \pi  \left(\frac{43 z^3}{15}-24 z+\frac{40}{z}\right)
\nonumber \\ &
+ e^{-2 i \kappa (\zin-z)} \bigg\{
-\frac{224 z_{\mathrm{in}}^3}{5 (z_{\mathrm{in}}-z)^5}
+\frac{16 z_{\mathrm{in}}}{5(\zin-z)^4}
\left(-13 i \kappa z_{\mathrm{in}}^2-\frac{15 i}{\kappa}+7 z_{\mathrm{in}}\right)
+ \frac{16}{15(\zin-z)^3}
\nonumber \\ & \times
\biggl[2 \left(12 \kappa^2-5\right)
z_{\mathrm{in}}^3+27 i \kappa z_{\mathrm{in}}^2-\frac{15 i}{\kappa}
+84 z_{\mathrm{in}}\biggr]
+ \frac{16}{15(\zin-z)^2} \biggl[i \kappa \left(6 \kappa^2-5\right)
z_{\mathrm{in}}^3-12 \kappa^2 z_{\mathrm{in}}^2
\nonumber \\ &
+\frac{3 i \left(21 \kappa^2-5\right) z_{\mathrm{in}}}{\kappa}
+\frac{30 i}{\kappa z_{\mathrm{in}}}+27\biggr]
+ \frac{4}{15(\zin-z)} \biggl[-12 i \kappa \left(6 \kappa^2-5\right)
z_{\mathrm{in}}^2-24 \left(3 \kappa^2-5\right) z_{\mathrm{in}}
\nonumber \\ &
+\frac{12 i \left(3 \kappa^2-5\right)}{\kappa}
-\left(12 \kappa^4-20 \kappa^2+15\right) z_{\mathrm{in}}^3
+\frac{120 i}{\kappa z_{\mathrm{in}}^2}
-\frac{120}{z_{\mathrm{in}}}\biggr]
+\frac{2}{z} \biggl[-\frac{i \left(2 \kappa^2-1\right)^2}{\kappa^3}
+\frac{16 i}{\kappa z_{\mathrm{in}}^2}
\nonumber \\ &
-\frac{16}{z_{\mathrm{in}}}\biggr]
+ \frac{4}{15}\biggl[8 i \kappa \left(6 \kappa^2-5\right) z_{\mathrm{in}}
+\left(12 \kappa^4-20 \kappa^2+15\right) z_{\mathrm{in}}^2
-\frac{36 \kappa^4+20 \kappa^2+15}{\kappa^2}\biggr]
\nonumber \\ &
+ \frac{4z^2}{15} \left(12 \kappa^4-20 \kappa^2+15 \right)
+\frac{2z}{15} \biggl[2 \left(12 \kappa^4-20 \kappa^2+15\right) \zin
+\frac{i \left(108 \kappa^4-100 \kappa^2+15\right)}{\kappa}\biggr] \bigg\}
\nonumber \\ &
+ e^{ - i (2 \kappa+1) (\zin - z)} \bigg\{
\frac{224 z_{\mathrm{in}}^3}{5 (z_{\mathrm{in}}-z)^5}
+\frac{2 z_{\mathrm{in}}}{5(\zin-z)^4}
\left[i (104 \kappa+97) z_{\mathrm{in}}^2+\frac{120 i}{\kappa}
-56 z_{\mathrm{in}}\right]
\nonumber \\ &
+  \frac{1}{15(\zin-z)^3}\biggl[-2 \left(192 \kappa^2+222 \kappa+43\right)
  z_{\mathrm{in}}^3-18 i (24 \kappa+17) z_{\mathrm{in}}^2
  -\frac{48 (28 \kappa+15) z_{\mathrm{in}}}{\kappa}
\nonumber \\ &
  +\frac{240 i}{\kappa}\biggr]
+ \frac{1}{15(\zin-z)^2}\biggl[-\frac{6 i
    \left(168 \kappa^2+169 \kappa+20\right) z_{\mathrm{in}}}{\kappa}
  -i \left(96 \kappa^3+204 \kappa^2+52 \kappa-43\right) z_{\mathrm{in}}^3
\nonumber \\ &
  +6 (2 \kappa +1) (16 \kappa +23) z_{\mathrm{in}}^2
  -\frac{480 i}{\kappa z_{\mathrm{in}}}
  -\frac{48 (9 \kappa+5)}{\kappa}\biggr]
+\frac{1}{15(\zin-z)}\biggl[3 i \bigl(96 \kappa^3
  +44 \kappa^2
\nonumber \\ &
  -28 \kappa-3\bigr) z_{\mathrm{in}}^2
  +\left(48 \kappa^4-24 \kappa^3-68 \kappa^2+34 \kappa +43\right) z_{\mathrm{in}}^3
  -\frac{480 i}{\kappa z_{\mathrm{in}}^2}
\nonumber \\ &
  +\frac{6 (2 \kappa +1) (3 \kappa +4) (8 \kappa -5) z_{\mathrm{in}}}{\kappa}
  +\frac{480 (\kappa+1)}{\kappa z_{\mathrm{in}}}
  -\frac{6 i (3 \kappa+4) (8 \kappa-5)}{\kappa}\biggr]
\nonumber \\ &
+\frac{16}{z}\biggl[-\frac{2 i}{\kappa z_{\mathrm{in}}^2}
  +\frac{2 (\kappa +1)}{\kappa z_{\mathrm{in}}}
  +\frac{i \kappa(\kappa+1)}{2 \kappa+1}\biggr]
+ \frac{1}{15}\biggl[-2 i \left(96 \kappa^3-36 \kappa^2-68 \kappa+17\right)
  z_{\mathrm{in}}
\nonumber \\ &
  +\frac{2 \left(72 \kappa^3-18 \kappa^2+43 \kappa+60\right)}{\kappa}
  -\left(48 \kappa^4-24 \kappa^3-68 \kappa^2+34 \kappa +43\right)
  z_{\mathrm{in}}^2\biggr]
-\frac{z}{15}\biggl[\bigl(48 \kappa^4
\nonumber \\ &
-24 \kappa^3-68 \kappa^2+34 \kappa +43\bigr)
z_{\mathrm{in}}+\frac{i \left(432 \kappa^4+24 \kappa^3-412 \kappa^2-34 \kappa
  +77\right)}{2 \kappa +1}\biggr]
\nonumber \\ &
-\frac{z^2}{15}\bigl(48 \kappa^4-24 \kappa^3-68 \kappa^2+34 \kappa+43\bigr)
\bigg\}
+ \pi \biggl\lbrace-z z_{\mathrm{in}}^2 \delta '''(z_{\mathrm{in}}-z)
\nonumber \\ &
+  2 z_{\mathrm{in}} (3 z-2 z_{\mathrm{in}}) \delta ''(z_{\mathrm{in}}-z)
+ \frac{2}{3}\biggl[-\frac{6 z_{\mathrm{in}}^2}{z}
  -5 z \left(z_{\mathrm{in}}^2+3\right)
  +36 z_{\mathrm{in}}\biggr] \delta '(z_{\mathrm{in}}-z)
\nonumber \\ &
+ \frac{4}{3} \biggl(14 z z_{\mathrm{in}}
+\frac{18 z_{\mathrm{in}}}{z}-9 z_{\mathrm{in}}^2-30\biggr)
\delta (z_{\mathrm{in}}-z) \biggr\rbrace \Biggr). 
\end{align}
The limit $z \ll \zin \ll 1$ of the above is
\begin{align}
\mathfrak{N}_{\bm{k}}(\eta,\eta_{\mathrm{in}}) &\simeq
\frac{\slrl  H^2 k}{1024 \pi ^2 \Mp^2}
\Biggl(-\frac{20 \pi}{z}  + \frac{4 i}{z} \left\lbrace10
\log \left[e^{\gamma } (2 \kappa +1) z_{\mathrm{in}}\right]-7\right\rbrace
+\left(-32 \kappa - 8 -\frac{64}{3 \kappa} +  \frac{4}{\kappa^2} \right)
\nonumber \\ &
- i\frac{40}{z}
\left[ \frac{1}{\epsilon}+\log
\left(\frac{2 \kappa+1}{M}\right) \right] +{\cal O}(z)\Biggr) .
\end{align}
With this formula, along with the earlier formula
(\ref{F_superHubble_App}) in the same limit, we get
\begin{align}
\label{M_superHubble_App}
\mathfrak{M}_{\bm{k}}(\eta,\eta_\mathrm{in})  & =
\mathfrak{N}_{\bm{k}}(\eta,\eta_\mathrm{in}) -
\eta \, \mathfrak{F}_{\bm{k}}(\eta,\eta_\mathrm{in})
=\mathfrak{N}_{\bm{k}}(\eta,\eta_\mathrm{in}) +\frac{z}{k}
\mathfrak{F}_{\bm{k}}(\eta,\eta_\mathrm{in})
\nonumber \\
& =   \frac{\slrl  H^2 k}{1024 \pi ^2 \Mp^2} \bigg\{
- \frac{40 i}{z}\log \left(\frac{z}{\zin}\right) +\frac{40 i}{\zin}
-\frac{20i}{3} z \left[\frac{1}{\epsilon}
+\log \left(\frac{2 \kappa+1}{M}\right)\right] 
\nonumber \\ &
+ 4 \left(-\frac{1}{\kappa^2}+8 \kappa+\frac{16}{3 \kappa}+2\right)
\log \left(\frac{z}{e \zin}\right) + \mathcal{O}(z,\zin)
\bigg\} \,,
\end{align}
where (as before) $M = \mu/k$.

\section{Infrared volume factors}
\label{App:Dimensions}

After splitting apart the Nakajima-Zwanzig equation into different
(continuous) momenta $\bm{k} \in \mathbb{R}^{3+}$ in \Eq{NZmodes}
there appear volume factors $\mathcal{V}$ on the LHS so that the
equations make sense dimensionally. This Appendix derives the
necessity of these volume factors, in the simpler setting where there
are no interactions at all where
\begin{equation}
\label{free_modebymode_App}
\frac{\mathcal{V}}{(2\pi)^3} \frac{\partial {\varrho}_{\mathrm{\ssS}\bm{k}}}
{\partial \eta} = - i \left[ {\mathcal{H}}_{\mathrm{\ssS}\bm{k}}(\eta),
{\varrho}_{\mathrm{\ssS}\bm{k}}(\eta) \right] 
\end{equation}
which is derived from the free Liouville equation. For clarity of
notation, we omit the label $\alpha = \mathrm{R},\mathrm{I}$ from
\Eq{alphaRI} in the main text in this Appendix. This is done by
assuming an ansatz of the form (\ref{rhofactor}) for the reduced
density matrix, repeated here,
\begin{equation}
\label{rhofactorApp}
{\varrho}(\eta) = \bigotimes_{\substack{q < k_{\UV} }} {\varrho}_{\bm{q}}(\eta) \ ,
\end{equation}
which assumes the momentum label is continuous. We justify the
presence of the volume factors in \Eq{free_modebymode_App} here by
passing to the limit where discrete momenta are considered, so that
the system is placed inside a box of volume $\mathcal{V}$. In this
case the conversion between discrete and continuum normalization is
given by
\begin{equation}  
\sum_{\bm{k}} = \frac{\mathcal{V}}{(2\pi)^3} \int \dd^3\bm{k} \, ,
\quad {C}_{\bm{k}}(\eta_{\mathrm{in}}) =
\left[ \frac{(2\pi)^3}{\mathcal{V}} \right]^{1/2} {c}_{\bm{k}}(\eta_{\mathrm{in}})\, ,
\quad {V}_{\bm{k}}(\eta) = \left[ \frac{(2\pi)^3}{\mathcal{V}} \right]^{1/2}
{v}_{\bm{k}}(\eta)\, ,
\end{equation}
where ${C}_{\bm{k}}(\eta_{\mathrm{in}})$ are the discretely normalized
annihilation operators, and ${V}_{\bm{k}}(\eta)$ are the discretely
normalized Fourier transforms of the Mukhanov-Sasaki field, and so
on. With this, continuum momentum field expansions like
\begin{equation} 
{v}(\eta,\bm{x}) = \int \frac{\dd^3 \bm{k}}{(2\pi)^{3/2}}
{v}_{\bm{k}}(\eta) e^{i \bm{k} \cdot \bm{x} } = \int
\frac{\dd^3\bm{k}}{(2\pi)^{3/2}} \left[ u_{\bm{k}}(\eta)
  {c}_{\bm{k}}(\eta_{\mathrm{in}}) + u^{\ast}_{\bm{k}}(\eta)
  {c}^{\dagger}_{-\bm{k}}(\eta_{\mathrm{in}}) \right] e^{i \bm{k}
  \cdot \bm{x} }
\end{equation}
become instead
\begin{equation} 
{v}(\eta,\bm{x}) = \frac{1}{\sqrt{\mathcal{V}}} \sum_{\bm{k}}
{V}_{\bm{k}}(\eta) e^{i \bm{k} \cdot \bm{x} } = \frac{1}{\sqrt{\cV}}
\sum_{\bm{k}} \left[ u_{\bm{k}}(\eta) {C}_{\bm{k}}(\eta_{\mathrm{in}})
  + u^{\ast}_{\bm{k}}(\eta)
  {C}^{\dagger}_{-\bm{k}}(\eta_{\mathrm{in}}) \right] e^{i \bm{k}
  \cdot \bm{x} } \ .
\end{equation} 
The tensor product structure of the ansatz (\ref{rhofactorApp}) is now
the same but over a discrete label --- this allows us to compute the
time-derivative of ${\varrho}_{\mathrm{\ssS}}$ as
\begin{align}
\frac{\partial {\varrho}_{\mathrm{\ssS}}}{\partial \eta} & =
\frac{\partial}{\partial \eta} \left[ \bigotimes_{\substack{ k < k_{\UV} }}
{\varrho}_{\mathrm{\ssS} \bm{k}}(\eta) \right]
=\sum_{k < \kUV}  \frac{\partial {\varrho}_{\ssA\mathrm{\ssS}\bm{k}}}{\partial \eta}
\bigotimes_{\substack{ q < k_{\UV}, \bm{q} \neq \bm{k} } }
{\varrho}_{\mathrm{\ssS} \bm{q}}(\eta)
\nonumber \\ &
=\frac{\mathcal{V}}{(2\pi)^3} \int_{k < k_{\UV} } \dd^3 \bm{k} \,
\frac{\partial {\varrho}_{\ssA\mathrm{\ssS}\bm{k}}}{\partial \eta}
\bigotimes_{\substack{ q < k_{\UV}, \bm{q} \neq \bm{k} } }
{\varrho}_{\mathrm{\ssS} \bm{q}}(\eta)\, .
\end{align}
On the other hand, the RHS of the free Liouville equation {\it does
  not} end up with a volume factor by the above logic. To see why,
recall the form of the free Hamiltonian~(\ref{freeHclassical})
\begin{align}
{\cal H}_{\ssA\mathrm{\ssS}}(\eta) &= \frac12 \int_{k < k_{\UV}} \dd^3\bm{k} \,
\left[ {p}_{\mathrm{\ssS}\bm{k}}(\eta){p}_{\mathrm{\ssS}\bm{k}}^\dagger (\eta)
+ \omega^2(\bm{k},\eta) {v}_{\mathrm{\ssS}\bm{k}}(\eta)
{v}_{\mathrm{\ssS}\bm{k}}^\dagger(\eta)  \right] =\frac12 \int_{k < k_{\UV}} \dd^3\bm{k}
\, {\mathcal{H}}_{\mathrm{\ssS}\bm{k}}(\eta)
\nonumber \\ &
=  \frac12 \sum_{k < k_{\UV} }  \left[ {P}_{\mathrm{\ssS}\bm{k}}(\eta)
  {P}^{\dagger}_{\mathrm{\ssS}\bm{k}}(\eta)
  + \omega^2(\bm{k},\eta)  {V}_{\mathrm{\ssS}\bm{k}}(\eta)
  {V}^{\dagger}_{\mathrm{\ssS}\bm{k}}(\eta)   \right]
= \sum_{k < k_{\UV} } {\mathfrak{h}}_{\mathrm{\ssS}\bm{k}}(\eta) 
\end{align}
where we define the shorthand ${\mathfrak{h}}_{\mathrm{\ssS}\bm{k}}(\eta):=\frac{1}{2} \big[ {P}_{\mathrm{\ssS}\bm{k}}(\eta)
{P}^{\dagger}_{\mathrm{\ssS}\bm{k}}(\eta) + \omega^2(\bm{k},\eta)
{V}_{\mathrm{\ssS}\bm{k}}(\eta)
{V}^{\dagger}_{\mathrm{\ssS}\bm{k}}(\eta) \big]$ in the last line, which is the discrete normalized version of the (free) Hamiltonian density
${\mathcal{H}}_{\mathrm{\ssS}\bm{k}}(\eta)$ --- note that there is no
volume factor since this operator is built from two fields. The above
implies that:
\begin{align}
- i \left[ {\cal H}_{\ssA\mathrm{\ssS}}(\eta),
\varrho_{\ssA\mathrm{\ssS}}(\eta) \right]  &=
- i \left[ \sum_{\substack{ k < k_{\UV} }}
\mathfrak{h}_{\mathrm{\ssS}\bm{k}}(\eta) , \,
\bigotimes_{\substack{ q < k_{\UV} }} \varrho_{\mathrm{\ssS} \bm{q}}(\eta)
\right]
\\
&=- i \sum\limits_{k< \kUV} \left[ 
\mathfrak{h}_{\mathrm{\ssS}\bm{k}}(\eta),
\varrho_{\mathrm{\ssS}\bm{k}}(\eta) \right] 
\bigotimes_{\substack{ q < k_{\UV}, \bm{q} \neq \bm{k} }}
\varrho_{\mathrm{\ssS} \bm{q}}(\eta)
\\
&= - i  \int_{k < k_{\UV} }  \dd^3 \bm{k} \,
\left[ \mathcal{H}_{\mathrm{\ssS}\bm{k}}(\eta),
  \varrho_{\mathrm{\ssS}\bm{k}}(\eta) \right]
\bigotimes_{\substack{ q < k_{\UV}, \bm{q} \neq \bm{k} }}
\varrho_{\mathrm{\ssS} \bm{q}}(\eta)\, .
\end{align}
Using the above in the free Liouville equation yields
\Eq{free_modebymode_App} with the desired volume factors.

\section{Scalar decoherence from a tensor environment}
\label{sec:decotensor}

\subsection{Correlation function in real space}

Here we consider the $\zeta\gamma\gamma$ interaction, 
\begin{equation}
\label{action:tss}
S_{\mathrm{int}} \ = \frac{\Mp^2}{8} \int \dd t \, \dd^3 \bm{x} \, a \,\slrl
\, \zeta \partial_\ell \gamma_{ij} \partial_{\ell} \gamma_{ij}
\end{equation}
which, after using $ a \, \dd \eta = \dd t$ corresponds to the interaction
Hamiltonian
\begin{equation}
\cH_{\mathrm{int}}(\eta) \ = \ - \frac{\Mp^2 \slrl }{8} a^2 \int \dd^3 \bm{x} \,
\zeta (\eta,\bm{x} ) \otimes \partial_\ell
\gamma_{  ij}(\eta,\bm{x}) \partial_{\ell} \gamma_{ ij}(\eta,\bm{x}) \ .
\end{equation} 
Our interest is in how the environment of short-wavelength tensors decohering long-wavelength scalar fluctuations. 

In terms of the canonical fields $v = a \Mp \sqrt{2\slrl } \zeta$ and
$v_{ij} = \frac{1}{2} a \Mp \gamma_{ij}$ -- see 
\Eq{canonicaltensor} and using $a^{-1}
= - H \eta$ -- we have
\bea
\cH_{\mathrm{int}}(\eta) &=& - \frac{\sqrt{\slrl }}
{2 \sqrt{2} a(\eta) \Mp} \int \dd^3 \bm{x} \, v (\eta,\bm{x} )
\otimes \partial_\ell v_{  ij}(\eta,\bm{x}) \partial_{\ell}
v_{ ij}(\eta,\bm{x}) \nn\\
&=&  G(\eta) \int \dd^3 \bm{x} \,
v (\eta,\bm{x} ) \otimes B_{\ssT}(\eta,\bm{x}) 
\eea
where $G(\eta)$ is the same as that defined in \Eq{couplingdef} and  
\begin{equation}
B_{\ssT}(\eta,\bm{x}) \ := \ \partial_\ell v_{  ij}(\eta,\bm{x})
\partial_{\ell} v_{ ij}(\eta,\bm{x}) \ .
\end{equation} 

Next we concern ourselves with the mode expansion for the graviton,
noting that the free part of the graviton action is
\begin{align}
\label{freegrav}
^{(2)}S[\gamma] & = \frac{\Mp^2}{8}
\int \dd \eta \, \dd^3\bm{x} \, a^2
\left( \gamma'_{ij} \gamma'_{ij}
- \partial_\ell \gamma_{ij} \partial_\ell \gamma_{ij} \right) =
\frac{1}{2} \int \dd \eta \, \dd^3\bm{x} \, \left( v_{ij}' v_{ij}'
- \partial_\ell v_{ij} \partial_\ell v_{ij} + \frac{2 }{\eta^2} v_{ij}v_{ij}\right) 
\nonumber \\ &
=\frac{1}{2} \sum_{\ssP=+,\times} \int \dd \eta\, \dd^3\bm{k} \,
\left[ \left \vert ( v_{\bm{k}}^{\ssP})^\prime\right \vert ^2
  - \left( k^2 - \frac{2}{\eta^2} \right) \left| v_{\bm{k}}^{\ssP} \right|^2
  \right]  
\end{align}
which uses $\dot{\gamma}_{ij} = a^{-1} \gamma_{ij}'$ as well as
$a\gamma_{ij}' = \frac{2}{\Mp} ( v_{ij}' + v_{ij} / \eta )$, and the
expansion in terms of graviton modes, already introduced in
\Eq{tensormodeexp}, and repeated here for convenience
\begin{equation}
\label{tensormodeexp2}
v_{ij}(\eta,\bm{x}) = \int \frac{\dd^3 \bm{k}}{(2\pi)^{3/2}} \sum_{\ssP = +,\times}
\epsilon_{ij}^{\ssP}(\bm{k}) v_{\bm{k}}^{\ssP}(\eta) e^{i \bm{k} \cdot \bm{x}}
\end{equation}
where we recall that $\epsilon^{\ssP}_{ij}(\bm{k})$ is the
polarization tensor. Let us explain how they are defined
concretely. We here follow the conventions of \RRef{Fujita:2018zbr},
up to numerical factors. For a given momentum vector $\hat{\bm{k}} =
\bm{k} / k$ pointing along the direction $(\theta,\varphi)$ in polar
coordinates we define the vectors
\begin{equation}
\label{polarizationconventions}
\bm{e}^{x}(\hat{\bm{k}}) =
\begin{pmatrix}
\cos\theta \cos\varphi
\\ \cos\theta\sin\varphi\\
-\sin\theta \end{pmatrix},
\qquad
\bm{e}^{y}(\hat{\bm{k}}) \ = \
\begin{pmatrix}
  - \sin\varphi \\ \cos\varphi \\ 0
\end{pmatrix}
\ .
\end{equation}
These vectors are perpendicular to $\bm{k}$ as well as to each other
where
\begin{equation}
\bm{e}^{L}(\hat{\bm{k}}) \cdot \bm{e}^{L'}(\hat{\bm{k}})  \
= \  \delta_{L L'}, \qquad
\bm{k} \cdot \bm{e}^{L}(\hat{\bm{k}}) = 0, \qquad L,L' \in (x,y) \ ,
\end{equation}
and under reflection $\bm{k} \to - \bm{k}$ these satisfy (this means
taking $\theta \to \pi - \theta$ and $\varphi \to \varphi + \pi$)
\begin{equation}
\bm{e}^{x}(-\hat{\bm{k}}) = \bm{e}^{x}(\hat{\bm{k}}), \qquad
\bm{e}^{y}(-\hat{\bm{k}}) = - \bm{e}^{y}(\hat{\bm{k}}) \ .
\end{equation}
Note that when $\hat{\bm{k}} = \hat{\bm{z}}$ with $\varphi = \theta =
0 $ then the above reduce to $\bm{e}^{x}(\hat{\bm{z}}) = (1,0,0)$ and
$\bm{e}^{y}(\hat{\bm{z}}) = (0,1,0)$. Next construct
$\bm{\epsilon}^{+} := \frac{1}{\sqrt{2}} \left( \bm{e}^x \otimes
\bm{e}^x - \bm{e}^y \otimes \bm{e}^y \right)$ and
$\bm{\epsilon}^{\times} := \frac{1}{\sqrt{2}} \left( \bm{e}^x
\otimes \bm{e}^y + \bm{e}^y \otimes \bm{e}^x \right)$, or more simply
in terms of components as:
\begin{align}
\epsilon_{ij}^{+}(\bm{k}) & :=
\frac{1}{\sqrt{2}} \left[ e_i^{x}(\hat{\bm{k}})e_j^{x}(\hat{\bm{k}})
- e_i^{y}(\hat{\bm{k}}) e_j^{y}(\hat{\bm{k}}) \right],
\\
\epsilon_{ij}^{\times}(\bm{k}) & := \frac{1}{\sqrt{2}}
\left[  e_i^{x}(\hat{\bm{k}}) e_j^{y}(\hat{\bm{k}})
+ e_i^{y}(\hat{\bm{k}}) e_j^{x}(\hat{\bm{k}}) \right].
\end{align}
These are the standard linear polarizations, which for the familiar
case of $\bm{k} \propto \hat{\bm{z}}$ simplify to
\begin{equation}
\bm{\epsilon}^{+}(\hat{\bm{z}}) = \frac{1}{\sqrt{2}}
\begin{pmatrix}
1 & 0 & 0 \\
0 & - 1 & 0 \\
0 & 0 & 0
\end{pmatrix},
\qquad
\bm{\epsilon}^{\times}(\hat{\bm{z}})
= \frac{1}{\sqrt{2}}
\begin{pmatrix}
0 & 1 & 0 \\
1 & 0 & 0 \\
0 & 0 & 0
\end{pmatrix} \ .
\end{equation}
Note that these are normalized as
\begin{equation}
\epsilon_{ij}^{\ssP}(\bm{k})  \epsilon_{ij}^{\ssP'}(\bm{k}) = \delta_{\ssP\ssP'} \ . 
\end{equation}
Importantly the symmetry under $\bm{k} \to - \bm{k}$ means that
$\epsilon_{ij}^{+}(-\bm{k}) = \epsilon_{ij}^{+}(\bm{k})$ as well
as $\epsilon_{ij}^{\times}(- \bm{k}) = -
\epsilon_{ij}^{\times}(\bm{k})$. This implies that
\begin{equation}
\epsilon_{ij}^{+}(-\bm{k}) \epsilon_{ij}^{+}(\bm{k}) = 1,
\qquad \epsilon_{ij}^{\times}(-\bm{k}) \epsilon_{ij}^{\times}(\bm{k}) = - 1 \ . 
\end{equation}
Furthermore we have the identify, see Eq.~(2.21) of \RRef{Kanno:2020usf}
\begin{equation}
\label{projectionidentity}
\sum_{\ssP} \epsilon_{ij}^{\ssP}(\bm{k}) \epsilon_{nm}^{\ssP}(\bm{k}) 
= \frac{1}{2} \left[\perp_{in}(\bm{k})  \perp_{j m}(\bm{k})
+ \perp_{i m}(\bm{k})  \perp_{j n}(\bm{k})  - \perp_{ij}(\bm{k})
\perp_{nm}(\bm{k})\right]
\end{equation}
where $\perp_{ij}(\bm{k}) = \delta_{ij} - k_i k_j / k^2$ is a (symmetric)
projection tensor such that $k_i \perp_{ij} = 0$.

As for scalars, the reality of $v_{ij}(\eta,\bm{x})$ implies that
$v_{\bm{k}}^{+\ast}(\eta) = v_{-\bm{k}}^{+\ast}(\eta)$ and
$v_{\bm{k}}^{\times\ast}(\eta) = -
v_{-\bm{k}}^{\times\ast}(\eta)$. This has the same mode expansion as
the Mukhanov-Sasaki field with an extra polarization label summing
over $P = +, \times$. We then expand
\begin{equation}
v_{\bm{k}}^{\ssP}(\eta) \ = \ u_{\bm{k}}(\eta) \, c_{\bm{k}}^{\ssP}  \,
+ \, s_{P}\, u_{\bm{k}}^{\ast}(\eta) \, c_{-\bm{k}}^{\ssP\dagger}
\end{equation}
with $s_\ssP=1$ for $P=+$ and $s_\ssP=-1$ for $P=\times$, where
$u_{\bm{k}}$ are the Bunch-Davies mode functions (same as for the
scalar) and the ladder operators satisfy $[ c_{\bm{k}}^{\ssP},
  c_{\bm{k}}^{\ssP'\dagger} ] = \delta(\bm{k} - \bm{q})
\delta_{\ssP\ssP'}$ and so on. The above then implies that the
operator $B_{\ssT}(\eta,\bm{x})$ has the expansion
\begin{equation}
B_{\ssT}(\eta,\bm{x}) \ := \ - \sum_{\ssP, \ssP' = +,\times}
\int_{k,q>\kUV} \frac{\dd^3 \bm{k} \, \dd^3 \bm{q}}{(2\pi)^{3}} \,
(\bm{k} \cdot \bm{q}) \,
\epsilon_{ij}^{\ssP}(\bm{k}) \epsilon_{ij}^{\ssP'}(\bm{q})
v_{\bm{k}}^{\ssP}(\eta) v_{\bm{q}}^{\ssP'}(\eta) e^{i ( \bm{k} + \bm{q} ) \cdot \bm{x}} \ ,
\end{equation} 
which can be used to simplify the one-point function,
\begin{align}
\mathscr{B}_{\ssT}(\eta) & := \left \langle 0_\ssB \left \vert
B_{\ssT}(\eta,\bm{x}) \right \vert  0_\ssB \right \rangle 
\\ 
&
= - \sum_{\ssP, \ssP' = +,\times}
\int_{k,q>\kUV} \frac{\dd^3 \bm{k} \, \dd^3 \bm{q}}{(2\pi)^{3}}
\, (\bm{k} \cdot \bm{q}) \,  \epsilon_{ij}^{\ssP}(\bm{k})
\epsilon_{ij}^{\ssP'}(\bm{q})
s_{P'}  u_{\bm{k}}(\eta) u_{\bm{q}}^{\ast}(\eta)
\delta(\bm{k} + \bm{q}) \delta_{\ssP\ssP'} e^{i ( \bm{k} + \bm{q} ) \cdot \bm{x}} 
\\
& = 2 \int_{k>\kUV} \frac{\dd^3 \bm{k}}{(2\pi)^{3}} \, k^2 \left \vert
u_{\bm{k}}(\eta) \right \vert ^2 \ , 
\end{align}
where we have used $ \epsilon_{ij}^{\ssP}(- \bm{k}) s_P =
\epsilon_{ij}^{\ssP}( \bm{k}) $ and then
$\epsilon_{ij}^{\ssP}(\bm{k})\epsilon_{ij}^{\ssP}(\bm{k}) = 1$
for each $P = +,\times$. Using the one-point function
$\mathscr{B}(\eta)$ defined in \Eq{1ptfunctionvarying}, we find that
\begin{equation}
\mathscr{B}_{\ssT}(\eta) = 2 \; \mathscr{B}(\eta) \ .
\end{equation}
Let us now turn to the calculation of the two-point correlation
function. Using the expectation values
\begin{align}
\left \langle 0_{\ssB} \left \vert
{c}_{\bm{k}}^{\ssP} {c}^{\ssP' \dagger}_{-\bm{q}} {c}_{\bm{p}}^{\ssP''}
{c}^{\ssP''' \dagger}_{-\bm{\ell}} \right \vert 0_{\ssB} \right \rangle & =
\delta(\bm{k} + \bm{q}) \delta(\bm{p} + \bm{\ell}) \delta_{PP'} \delta_{P''P'''},
\\
\left \langle 0_{\ssB} \left \vert {c}_{\bm{k}}^{\ssP} {c}_{\bm{q}}^{\ssP'}
{c}^{\ssP'' \dagger}_{-\bm{p}} {c}^{\ssP''' \dagger}_{-\bm{\ell}} \right \vert 0_{\ssB}
\right \rangle & = \delta(\bm{k} + \bm{\ell})
\delta(\bm{p} + \bm{q}) \delta_{PP'''}
\delta_{P'P''}
+ \delta(\bm{k} + \bm{p}) \delta(\bm{\ell} + \bm{q}) \delta_{PP''} \delta_{P'P'''}
\end{align}
this quantity simplifies to
\begin{align}
\label{2pttensors}
\left \langle 0_{\ssB} \left \vert  {B}_{\ssT}(\eta,\bm{x})
{B}_{\ssT}(\eta',\bm{x}') \right \vert 0_{\ssB} \right \rangle 
&= \mathscr{B}_{\ssT}(\eta) \mathscr{B}_{\ssT}(\eta')
+ 2 \sum_{\ssP, \ssP',\ssP'',\ssP'''}  \int_{k,q,p,\ell>\kUV}
\frac{\dd^3 \bm{k} \, \dd^3 \bm{q}\, \dd^3 \bm{p} \,
\dd^3 \bm{\ell}}{(2\pi)^{6}}
\,
\nonumber \\ & \times 
\epsilon_{ij}^{\ssP}(\bm{k}) \epsilon_{ij}^{\ssP'}(\bm{q})
\epsilon_{nm}^{\ssP''}(\bm{p})
\epsilon_{nm}^{\ssP'''}(\bm{\ell}) \,
(\bm{k} \cdot \bm{q}) (\bm{p} \cdot \bm{\ell}) 
e^{i (\bm{k}+\bm{q}) \cdot \bm{x} + i (\bm{p}+\bm{\ell}) \cdot \bm{x}' }  \,
\nonumber \\ & \times
u_{\bm{k}}(\eta) u_{\bm{q}}(\eta) u^{\ast}_{\bm{p}}(\eta')
u^{\ast}_{\bm{\ell}}(\eta') s_{P''} s_{P'''}
\delta(\bm{k} + \bm{\ell}) \delta(\bm{p} + \bm{q})
\delta_{PP'''} \delta_{P'P''}\, .
\end{align}
Our interest is in the centred two-point function, defined by
\begin{align}
C_{\ssT}(\eta,\eta'; \bm{x} - \bm{x}' )  & :=
\left \langle 0_{\ssB} \left \vert
\left[ {B}_{\ssT}(\eta,\bm{x}) -  \mathscr{B}_{\ssT}(\eta) \right]
\left[ {B}_{\ssT}(\eta',\bm{x}') -  \mathscr{B}_{\ssT}(\eta') \right]
\right \vert  0_{\ssB} \right \rangle
\nonumber \\ &
= \left \langle 0_{\ssB} \left \vert  {B}_{\ssT}(\eta,\bm{x})
{B}_{\ssT}(\eta',\bm{x}') \right \vert 0_{\ssB} \right \rangle
- \mathscr{B}_{\ssT}(\eta)\mathscr{B}_{\ssT}(\eta')  \ ,
\end{align}
which after using \Eq{2pttensors}, integrating over the
$\delta$-functions and using $s_{P} \epsilon_{nm}^{\ssP}(-\bm{k}) =
\epsilon_{nm}^{\ssP}(\bm{k})$ gives rise to
\begin{align}
C_{\ssT}(\eta,\eta'; \bm{y} )  & = \sum_{\ssP, \ssP'}
\int_{k,q>\kUV} \frac{\dd^3 \bm{k} \, \dd^3 \bm{q}}{(2\pi)^{6}} \,
\epsilon_{ij}^{\ssP}(\bm{k}) \epsilon_{ij}^{\ssP'}(\bm{q})
\epsilon_{nm}^{\ssP'}(\bm{q}) \epsilon_{nm}^{\ssP}(\bm{k})
\, 2 \, (\bm{k} \cdot \bm{q})^2 \,
\nonumber \\ & \times
u_{\bm{k}}(\eta) u_{\bm{q}}(\eta)
u^{\ast}_{\bm{q}}(\eta') u^{\ast}_{\bm{k}}(\eta')
e^{i (\bm{k}+\bm{q}) \cdot \bm{y} } \, . 
\end{align}
Using the identity (\ref{projectionidentity}) involving the projection
$\perp_{ij}(\bm{k}) := \delta_{ij} - k_i k_j / k^2$ the summations
over polarizations can be simplified to
\begin{align}
C_{\ssT}(\eta,\eta'; \bm{y} )&= \int_{k,q>\kUV} \frac{\dd^3 \bm{k} \,
  \dd^3 \bm{q}}{(2\pi)^{6}} \, \frac{1}{2}\left[\perp_{in}(\bm{k})
  \perp_{j m}(\bm{k})
+ \perp_{i m}(\bm{k})  \perp_{j n}(\bm{k})  - \perp_{ij}(\bm{k})
\perp_{nm}(\bm{k})\right]
\nonumber \\ & \times
\frac{1}{2}\left[\perp_{in}(\bm{q})  \perp_{j m}(\bm{q})  + \perp_{i m}(\bm{q})
    \perp_{j n}(\bm{q})  - \perp_{ij}(\bm{q})  \perp_{nm}(\bm{q})\right]
2 \, (\bm{k} \cdot \bm{q})^2 \,
\nonumber \\ & \times 
u_{\bm{k}}(\eta) u_{\bm{q}}(\eta)
  u^{\ast}_{\bm{q}}(\eta') u^{\ast}_{\bm{k}}(\eta') e^{i (\bm{k}+\bm{q}) \cdot \bm{y} } \, . 
\end{align}
There is a sum over the indices $i,j,n,m$, and so to this end we note that
\begin{align}
\perp_{in}(\bm{k}) \perp_{in}(\bm{q})&=1 +
\frac{ (\bm{k} \cdot \bm{q})^2 }{k^2 q^2},
\\ 
\perp_{in}(\bm{k}) \perp_{in}(\bm{q}) \perp_{jm}(\bm{k}) \perp_{jm}(\bm{q})
&= \left[ 1 + \frac{ (\bm{k} \cdot \bm{q})^2 }{k^2 q^2} \right]^2,
\\ 
\perp_{in}(\bm{k}) \perp_{im}(\bm{q}) \perp_{jm}(\bm{k}) \perp_{jn}(\bm{q})
&= 1 + \frac{ (\bm{k} \cdot \bm{q})^4 }{k^4 q^4},
\end{align}
which are the only two types of contractions that occur in the above
(after re-labeling $i,j,n,m$ in various ways). After some
manipulation the above implies that
\begin{align}
\label{CTmomentum} 
C_{\ssT}(\eta,\eta'; \bm{y} ) &= \int_{k,q>\kUV} \frac{\dd^3 \bm{k} \,
  \dd^3 \bm{q}}{(2\pi)^{6}} \, \frac{1}{4}\left[ 1
  + 6 \frac{ (\bm{k} \cdot \bm{q})^2 }{k^2 q^2}
  + \frac{ (\bm{k} \cdot \bm{q})^4 }{k^4 q^4}\right] \,
2 (\bm{k} \cdot \bm{q})^2
\nonumber \\ & \times
u_{\bm{k}}(\eta) u_{\bm{q}}(\eta)
u^{\ast}_{\bm{q}}(\eta') u^{\ast}_{\bm{k}}(\eta')\, e^{i (\bm{k}+\bm{q}) \cdot \bm{y} }  \ .
\end{align}

\subsection{Fourier Transform of $C_{\ssT}$}

What appears in the Lindblad equation for this interaction is of
course the Fourier transform of the above correlator which we define
as
\begin{equation}
\mathscr{T}_{\bm{k}}(\eta,\eta') : =
\int \frac{\dd^{3} \bm{y}}{(2\pi)^{3/2}} \ C_\ssT(\eta, \eta' ; \bm{y})
\, e^{- i \bm{k} \cdot \bm{y}} \ .
\end{equation}
We again are restricted to modes with $0 < k < \kUV$ and using
\Eq{CTmomentum} we find
\begin{align}
\label{C4k1}
\mathscr{T}_{\bm{k}}(\eta,\eta') & = \frac{2}{(2\pi)^{9/2}}
\int_{q,p>k_{\UV}} \dd^3 \bm{q} \, \dd^3 \bm{p} \, (\bm{q} \cdot \bm{p})^2
\left[ \frac{1}{4} + \frac{3}{2}  \frac{ (\bm{q} \cdot \bm{p})^2 }{q^2 p^2}
+ \frac{1}{4} \frac{ (\bm{q} \cdot \bm{p})^4 }{q^4 p^4}  \right]
\nonumber \\ & \times 
u_{\bm{q} }\left(\eta\right) u_{\bm{p}}\left(\eta\right)
u_{\bm{p}}^*\left(\eta'\right) u_{\bm{q}}^*\left(\eta'\right)
\delta(\bm{q} + \bm{p} - \bm{k}) \, . 
\end{align}
From here the integration over the angles goes over exactly in the
same manner as in Appendix \ref{App:Cketaetap}, eventually giving rise
to the expression
\begin{align}
\label{CEk_PQ}
\mathscr{T}_{\bm{k}}(\eta,\eta') & = \frac{1}{32(2\pi)^{7/2}k}
\int_0^k \dd Q \int_{Q  + 2 k_{\UV}}^{\infty} \dd P\, \left[ \frac{1}{4}
  + \frac{3}{2}  \frac{( P^2 + Q^2 - 2 k^2 )^2}{(P^2-Q^2)^2}
  + \frac{1}{4}  \frac{( P^2 + Q^2 - 2 k^2 )^4}{(P^2-Q^2)^4} \right]
\nonumber \\ & \times
\left( P^2 + Q^2 - 2 k^2 \right)^2\left[1-\frac{2i}{(P-Q) \eta}\right]
\left[1+\frac{2i}{(P-Q) \eta'}\right] \left[1-\frac{2i}{(P+Q) \eta}\right]
\nonumber \\ & \times
\left[1+\frac{2i}{(P+Q) \eta'}\right] e^{- i ( \eta - \eta' ) P } . 
\end{align}

\subsection{Super-Hubble limit of Lindblad coefficient}

Next we must compute the super-Hubble limit of the Lindblad
coefficient
\begin{equation}
\mathfrak{F}_{\bm{k}}(\eta,\eta_{\mathrm{in}})  : =
(2\pi)^{3/2} \int_{ \eta_{\rm in} }^{\eta} \dd \eta' \;
G(\eta) G(\eta') \left[ \mathscr{C}_{\bm{k}}(\eta,\eta')
+\mathscr{T}_{\bm{k}}(\eta,\eta') \right] \ .
\end{equation}
We use the earlier representations (\ref{Ck_PQ}) and (\ref{CEk_PQ}) of
$\mathcal{C}_{\bm{k}}$ and $\mathscr{T}_{\bm{k}}$ above, as well as
$G(\eta) G(\eta') = \slrl H^2 \eta \eta'/(8\Mp^2)$, which expresses
$\mathfrak{F}_{\bm{k}}$ as the triple integral
\begin{align}
\label{Fk_trip_tensors}
\mathfrak{F}_{\bm{k}}(\eta, \eta_{\mathrm{in}}) &=
\frac{\slrl  H^2}{1024 \pi^2 \Mp^2 k}
\int_{ \eta_{\mathrm{in}} }^{\eta} \dd \eta' \int_0^k \dd Q \int_{Q  + 2 k_{\UV}}^{\infty}
\dd P \, \eta\, \eta' \, \bigg[ \frac{5}{4} + \frac{3}{2}
\frac{( P^2 + Q^2 - 2 k^2 )^2}{(P^2-Q^2)^2}
\nonumber \\ &   
  + \frac{1}{4} \frac{( P^2 + Q^2 - 2 k^2 )^4}{(P^2-Q^2)^4} \bigg]
\left( P^2 + Q^2 - 2 k^2 \right)^2\,
\left[1-\frac{2i}{(P-Q) \eta}\right]\left[1+\frac{2i}{(P-Q) \eta'}\right]
\nonumber \\ & \times
\left[1-\frac{2i}{(P+Q) \eta}\right] \left[1+\frac{2i}{(P+Q) \eta'}\right]
e^{- i P (\eta - \eta')} \, .
\end{align}
We first evaluate the $\eta'$-integral using the formula
(\ref{etap_integral_for_Fk}),
\begin{align}
\int_{ \eta_{\mathrm{in}} }^{\eta} \dd \eta' \, & \eta' \,
\left[1+\frac{2i}{(P-Q) \eta'}\right] \left[1+\frac{2i}{(P+Q) \eta'}\right]
e^{- i ( \eta - \eta' ) P}
\nonumber \\ &
=\bigg[e^{- i P (\eta - \eta')} \bigg( \frac{1}{P^2} - \frac{i \eta'}{P}
+ \frac{4}{P^2 - Q^2} \bigg) - \frac{4 e^{- i P \eta}
  \mathrm{Ei}(i P \eta')  }{P^2 - Q^2} \bigg]\;
\bigg|_{\eta' \to \eta_{\mathrm{in}}}^{\eta' \to \eta}
\\
&=\frac{1}{P^2} - \frac{i \eta}{P} + \frac{4}{P^2 - Q^2}
- \frac{4 e^{- i P \eta} \left[ \mathrm{Ei}(i P \eta)
+ i \pi \right] }{P^2 - Q^2}
- e^{- i P (\eta - \eta_{\mathrm{in}})} \bigg( \frac{1}{P^2}
- \frac{i \eta_{\mathrm{in}}}{P}
+ \frac{4}{P^2 - Q^2} \bigg)
\nonumber \\ &
+ \frac{4 e^{- i P \eta}
  \left[\mathrm{Ei}(i P \eta_{\mathrm{in}}) + i \pi \right]}{P^2 - Q^2}\, ,  
\end{align}
where we have added and subtracted an extra factor of $4 i \pi e^{- i
  P \eta}/(P^2-Q^2)$ for convenience later on. With this we find that
\begin{equation}
\label{Ftensorhg}
\mathfrak{F}_{\bm{k}}(\eta,\eta_{\mathrm{in}}) =
\frac{\slrl  H^2}{1024 \pi^2 \Mp^2 }
\left[h( \eta ) + g(\eta,\eta_{\mathrm{in} } ) \right]
\end{equation}
with the definitions
\begin{align}
\label{littleh_def}
h( \eta ) & : = \int_0^k \dd Q \int_{Q  + 2 k_{\UV}}^{\infty}
\dd P \, \eta \, \left[\frac{5}{4} + \frac{3}{2}
  \frac{( P^2 + Q^2 - 2 k^2 )^2}{(P^2-Q^2)^2} + \frac{1}{4}
  \frac{( P^2 + Q^2 - 2 k^2 )^4}{(P^2-Q^2)^4} \right] 
\nonumber \\ & \times 
\frac{( P^2 + Q^2 - 2 k^2 )^2}{k} 
\left[1-\frac{2i}{(P-Q) \eta}\right]
\left[1-\frac{2i}{(P+Q) \eta}\right]
\bigg\{ \frac{1}{P^2} - \frac{i \eta}{P} + \frac{4}{P^2 - Q^2}
\nonumber \\ &  
- \frac{4 e^{- i P \eta} \left[ \mathrm{Ei}(i P \eta) + i \pi \right] }
  {P^2 - Q^2} \bigg\} \ , 
\\
\label{littleg_def}
g( \eta, \eta_{\mathrm{in}} ) & : =
\int_0^k \dd Q \int_{Q  + 2 k_{\UV}}^{\infty} \dd P \, \eta \,
\bigg[ \frac{5}{4} + \frac{3}{2} \frac{( P^2 + Q^2 - 2 k^2 )^2}{(P^2-Q^2)^2}
+ \frac{1}{4} \frac{( P^2 + Q^2 - 2 k^2 )^4}{(P^2-Q^2)^4} \bigg]
\nonumber \\ & \times
\frac{( P^2 + Q^2 - 2 k^2 )^2}{k}
\left[1-\frac{2i}{(P-Q) \eta}\right] \left[1-\frac{2i}{(P+Q) \eta}\right]
\Bigg\{- e^{- i P (\eta - \eta_{\mathrm{in}})} \bigg( \frac{1}{P^2}
\nonumber \\ &
-\frac{i \eta_{\mathrm{in}}}{P}
+ \frac{4}{P^2 - Q^2} \bigg)
+ \frac{4 e^{- i P \eta} \big[ \mathrm{Ei}(i P \eta_{\mathrm{in}})+ i \pi \big] }
{P^2 - Q^2} \Bigg\} \ .
\end{align}
Our goal will be write down the $0 \ll - k \eta \ll - k
\eta_{\mathrm{in}} \ll 1$ limit of
$\mathfrak{F}_{\bm{k}}(\eta,\eta_{\mathrm{in}})$.

Let us now study the super-Hubble limit of $h$. As before, we make use
of the variables $z=-k\eta$, $z_{\mathrm{in}}=- k\eta_{\mathrm{in}}$
and $\kappa := \kUV/k$ and change the (positive) integration variables
to $x = - Q \eta $ and $y=- P \eta$  which turns \Eq{littleh_def} into 
\begin{equation}
\label{h_intermsof_f}
h(\eta) := \int_0^z \exd x\; \frac{f(x,z)}{z^3}
\end{equation}
with $f$ defined by
\begin{align}
\label{tensorf_def}
f(x,z) & : = \int_{x + 2 \kappa z }^{\infty} \dd y \,
\left[ \frac{5}{4} + \frac{3}{2} \frac{( x^2 + y^2 - 2 z^2 )^2}{(x^2-y^2)^2}
+\frac{1}{4} \frac{( x^2 + y^2 - 2 z^2 )^4}{(x^2-y^2)^4} \right]
( x^2 + y^2 - 2 z^2  )^2
\nonumber \\ & \times
\left( 1 - \frac{2i}{x-y} \right)
\left( 1 + \frac{2i}{x + y} \right) \left\{  - \frac{1}{y^2 }
-\frac{i}{y} + \frac{4}{x^2 - y^2} - \frac{4 e^{i y }
\left[  \mathrm{Ei}( - i y ) + i \pi \right] }{x^2 - y^2}  \right\} \, .
\end{align}
We next notice the Taylor series about $z=0$, 
\begin{equation}
\int_0^z \dd x\, f(x,z) \simeq \ f(0,0) z \, + \, \mathcal{O}(z^{-1})\, ,
\end{equation}
for $0<z\ll 1$, which implies that $f$ contributes to $h$ in
\Eq{h_intermsof_f} at order $z^{-2}$ for small $z$, such that
\begin{equation}
\label{h_Taylor}
h(\eta) \simeq \frac{f(0,0)}{z^2} + \mathcal{O}(z^{-1})
\end{equation}
with coefficient through \Eq{tensorf_def} given as
\begin{equation}
\label{f00eq}
f(0,0) := 3 \int_{0}^{\infty} \dd y \; y^4 \left( 1 + \frac{2i}{y} \right)^2
\left\lbrace - \frac{5 + i y}{y^2 } + \frac{4 e^{i y } }{y^2}
\left[ \mathrm{Ei}( - i y ) + i \pi \right] \right\rbrace \ .
\end{equation}
In order to compare to the earlier section, let us focus on the real
part of the above where
\begin{align}
\mathrm{Re}\left[ f(0,0) \right] &= 3 \int_{0}^{\infty} \dd y
\biggl( 20 - y^2 + 4 (y^2 - 4) \, \mathrm{Re}\left\lbrace e^{i y }
\left[ \mathrm{Ei}( - i y ) + i \pi \right] \right\rbrace
\nonumber \\ &
- 16 y \, \mathrm{Im}\left\lbrace e^{i y } \left[\mathrm{Ei}( - i y )
+ i \pi \right] \right\rbrace \biggr)\, .
\end{align}
We next use (for $y>0$)
\begin{align}
e^{i y }  \left[\mathrm{Ei}( - i y ) + i \pi \right]=
\mathrm{Ci}(y) \cos y + \left[ \mathrm{Si}(y) - \frac{\pi}{2} \right]
\sin(y) + i \left\lbrace \mathrm{Ci}(y) \sin y - \left[ \mathrm{Si}(y)
- \frac{\pi}{2} \right] \cos y \right\rbrace \, ,
\end{align}
where the functions $\mathrm{Ci}(y)$ and $\mathrm{Si}(y)$ are defined by
\begin{equation}
\mathrm{Ci}(y) \ = \ - \int_y^\infty \dd t \ \frac{\cos t }{t},
\qquad  \mathrm{Si}(y) \ = \ \int_0^y \dd t\; \frac{\sin t}{t} \,,
\end{equation}
to write
\begin{align}
\mathrm{Re}\left[ f(0,0) \right] &= 3 \int_{0}^{\infty} \dd y \,
\biggl(  - y^2 + 20 + 4 (y^2 - 4) \left\lbrace \mathrm{Ci}(y) \cos y
+ \left[ \mathrm{Si}(y) - \frac{\pi}{2} \right] \sin y  \right\rbrace
\nonumber \\ & 
- 16 y \left\lbrace  \mathrm{Ci}(y) \sin y
- \left[ \mathrm{Si}(y) - \frac{\pi}{2} \right] \cos y \right\rbrace
\biggr) \ .
\end{align}
For $y \gg 1$ the integrand in this expression behaves as
\begin{align}
&- y^2 + 20 + 4 (y^2 - 4) \left\lbrace \mathrm{Ci}(y) \cos y
+ \left[ \mathrm{Si}(y) - \frac{\pi}{2} \right] \sin y  \right\rbrace
- 16 y \left\lbrace \mathrm{Ci}(y) \sin y
- \left[ \mathrm{Si}(y) - \frac{\pi}{2} \right] \cos y \right\rbrace
\nonumber \\ &
\simeq  - y^2 + \frac{72}{y^2} + \mathcal{O}\left(y^{-4}\right)  
\end{align}
and so the integral diverges in the UV.  A similar divergence also
arose when summing over scalar fluctuations, and corresponded to a
distributional singularity in the correlation function near $\eta =
\eta'$, see the discussion surrounding \Eq{b_distributions}. The
distributional singularities do not contribute to the integration over
$\eta'$ performed here, and for the present purposes it is convenient
to have an alternative approach that provides a short-cut to the
small-$z$ behaviour without fully evaluating the position-space
correlation function for all values of its arguments. We adopt here a
regularization of this divergence that properly reproduces the
small-$z$ behaviour found in the more complete treatment of Appendix
\ref{App:Lindblad_F}.

To this end we regulate the divergence by first isolating the
UV-divergent part $\propto y^2 $ of the function, leading to
\begin{align}
\mathrm{Re}\left[ f(0,0) \right] &= - 3  \int_{0}^{\infty} \dd y \, y^2 
+ 3 \int_{0}^{\infty} \dd y \,  \biggl( 20 - 4 (y^2 - 4)
\left\lbrace \mathrm{Ci}(y) \cos y + \left[ \mathrm{Si}(y)
- \frac{\pi}{2} \right] \sin y  \right\rbrace
\nonumber \\ &
- 16 y \left\lbrace
\mathrm{Ci}(y) \sin(y) - \left[ \mathrm{Si}(y)
- \frac{\pi}{2} \right] \cos y
\right\rbrace \biggr) \ .
\end{align}
We regulate the divergent integral $\int_{0}^{\infty} \dd y \; y^2$ in
the spirit of dimensional regularization by writing
\begin{equation}
\label{dimregfdef}
\int_{0}^{\infty} \dd y \, y^2  \to  \lim_{q \to 0^{+}}
\lim_{n\to 0}\int_{0}^{\infty} \dd y \,
y^n \left( \frac{ y^4 }{y^2 + q^2 }\right)
=\lim_{q \to 0^{+}} \lim_{n\to 0}\left[ \frac{\pi q^{n+3}}{2 \cos( n \pi/2 )} \right]
= 0  ,
\end{equation}
where the initial integral only converges when $\mathrm{Re}(n) < -5$
and we introduce a parameter $q> 0$ to regulate the associated
divergence in the IR (whose presence ultimately cancels the UV
divergence). This leaves the convergent integral
\begin{align}
\label{f00eval}
\mathrm{Re}\left[f^{\rm reg}(0,0)\right] & = 3\int_{0}^{\infty} \dd y \,
\biggl( 20 + 4 (y^2 - 4) \left\lbrace \mathrm{Ci}(y) \cos y
+ \left[\mathrm{Si}(y) - \frac{\pi}{2} \right] \sin y  \right\rbrace 
\nonumber \\ &
- 16 y \left\lbrace  \mathrm{Ci}(y) \sin y - \left[\mathrm{Si}(y)
- \frac{\pi}{2} \right] \cos y \right\rbrace  \biggr)
\\ 
& = \biggl\{- 4 y + 4 \mathrm{Ci}(y) \left[ 6 y \cos y
  + (y^2 - 10) \sin y \right] - 4 \left[ \left(y^2-10\right) \cos y
  - 6y \sin y \right]
\nonumber \\ & \times
\left[\mathrm{Si}(y)-\frac{\pi}{2} \right] \bigg\} \; \biggr
\vert_{y \to 0}^{y \to \infty} = 60 \pi
\end{align}
which is the result used in \Eq{h_Taylor}.

Let us finally compute the super-Hubble behaviour of $g$ defined in
\Eq{littleg_def}. We again use the variables $z$ and $z_\mathrm{in}$
and we here argue that
\begin{equation}
\label{gsubdom}
g(\eta,\eta_{\mathrm{in}})  \ \sim \ \mathcal{O}(z^{-1})
\end{equation}
in the $ z\ll 1$ limit and so is subdominant to $h$. To see why it is
easier to use the integration $a := Q/k$ and $b:= P/k$ which turns
\Eq{littleg_def} into
\begin{align}
\label{tensorg_def}
g(\eta,\eta_{\mathrm{in}}) & : =
\int_0^{1} \dd a \int_{a + 2 \kappa }^{\infty} \dd b \,
\bigg[ \frac{5}{4} + \frac{3}{2} \frac{( a^2 + b^2 - 2 )^2}{(a^2-b^2)^2}
  + \frac{1}{4} \frac{( a^2 + b^2 - 2 )^4}{(a^2-b^2)^4} \bigg]
( a^2 + b^2 - 2  )^2 z
\nonumber \\ & \times
\left[ 1 - \frac{2i}{(a-b)z} \right]
\left[ 1 + \frac{2i}{(a + b)z} \right]
\biggl\{e^{i b \left(z - \zin \right)} \left(\frac{1}{b^2}
+ \frac{i z_{\mathrm{in}}}{b} - \frac{4}{a^2 - b^2} \right)
\nonumber \\ &  
+ \frac{4 e^{i z b} \left[\mathrm{Ei}(- i z_{\mathrm{in}} b )
+ i \pi \right] }{a^2 - b^2} \biggr\} .  
\end{align}
We then write the above as
\begin{align}
\label{tensorg_def}
g(z,\zin) & : =  \int_0^{1} \exd a\, \mathscr{G}(a,z,\zin) 
\end{align}
where we have
\begin{equation}
\label{gG_FT}
\mathscr{G}(a,z,\zin) := \int_{a+2\kappa}^\infty \dd b\, \left[
\frac{ \chi_{-1}(a,b,\zin) }{z} + \chi_{0}(a,b,\zin)
+ \chi_{+1}(a,b,\zin)  z  \right] e^{i b z}
\end{equation}
with the definitions:
\begin{align}
\chi_{-1}(a,b,\zin) &:= \frac{4}{a^2-b^2} \left[ \frac{5}{4}
+ \frac{3}{2} \frac{( a^2 + b^2 - 2 )^2}{(a^2-b^2)^2}
+ \frac{1}{4} \frac{( a^2 + b^2 - 2 )^4}{(a^2-b^2)^4} \right]
( a^2 + b^2 - 2  )^2
\nonumber \\ & \times
\left\{e^{- i b \zin } \bigg( \frac{1}{b^2} + \frac{i z_{\mathrm{in}}}{b}
- \frac{4}{a^2 - b^2} \bigg) + \frac{4\left[ \mathrm{Ei}(- i z_{\mathrm{in}} b )
+ i \pi \right] }{a^2 - b^2} \right\}
\\
\chi_{0}(a,b,\zin) &:=  -\frac{4i b}{a^2-b^2} \left[ \frac{5}{4}
  + \frac{3}{2} \frac{( a^2 + b^2 - 2 )^2}{(a^2-b^2)^2} + \frac{1}{4}
  \frac{( a^2 + b^2 - 2 )^4}{(a^2-b^2)^4} \right] ( a^2 + b^2 - 2  )^2
\nonumber \\ & \times
\left\{e^{- i b \zin } \bigg( \frac{1}{b^2} + \frac{i z_{\mathrm{in}}}{b}
- \frac{4}{a^2 - b^2} \bigg) + \frac{4\left[ \mathrm{Ei}(- i z_{\mathrm{in}} b )
+ i \pi \right] }{a^2 - b^2} \right\}
\\ 
\chi_{+1}(a,b,\zin) &:= \left[ \frac{5}{4} + \frac{3}{2}
\frac{( a^2 + b^2 - 2 )^2}{(a^2-b^2)^2} + \frac{1}{4}
\frac{( a^2 + b^2 - 2 )^4}{(a^2-b^2)^4} \right] ( a^2 + b^2 - 2  )^2
\nonumber \\ & \times
\left\{e^{- i b \zin } \bigg( \frac{1}{b^2} + \frac{i z_{\mathrm{in}}}{b}
- \frac{4}{a^2 - b^2} \bigg) + \frac{4
  \left[ \mathrm{Ei}(- i z_{\mathrm{in}} b )+ i \pi \right] }
{a^2 - b^2} \right\} 
\end{align}
Since \Eq{gG_FT} reveals that $\mathscr{G}$ is a Fourier transform,
we note that its $0 < z \ll 1$ limit (for fixed $\zin$ and $\kappa$)
is governed by the $b\gg 1$ behaviour of the functions $\chi_{j}$, in
which limit we have:
\begin{align}
\chi_{-1}(a,b,\zin) &\simeq e^{-i b \zin} \left\{ -12 i b - 60 \zin
+ \mathcal{O}\left(b^{-1}\right)  \right\} \, ,
\label{chi_asym}
\\
\chi_{0}(a,b,\zin) &\simeq e^{-i b z_{\mathrm{in}}}
\left\{-12 z_{\mathrm{in}} b^2 + 60 i b + \left[\left(80-68 a^2\right)
z_{\mathrm{in}}+\frac{48}{z_{\mathrm{in}}} \right]
+ \mathcal{O}\left(b^{-1}\right) \right\}\, ,
\\
\chi_{+1}(a,b,\zin) & \simeq  e^{-i b z_{\mathrm{in}}} \left\{
3 i z_{\mathrm{in}} b^3  +15 b^2 + \sfrac{2 i \left[\left(7 a^2-10\right)
    z_{\mathrm{in}}^2-6\right]}{z_{\mathrm{in}}} b + \frac{12}{z_{\mathrm{in}}^2}
+ 82 a^2-100 +  \mathcal{O}\left(b^{-1}\right) \right\}\, .
\end{align}
To derive the required asymptotics of each of the above Fourier
transforms, we write
\begin{align}
\int_{a+2\kappa}^{\infty} \dd b\,  \chi_{-1}(a,b,\zin) e^{i b z}
&=\int_{a+2\kappa}^{\infty} \dd b\, e^{-i b \zin}
\left( -12 i b - 60 \zin  \right)  e^{i b z} \, + \, \Xi_{-1}(z,a,\zin)
\end{align}
with the definition
\begin{equation}
\Xi_{-1}(z,a,\zin) := \int_{a+2\kappa}^{\infty} \dd b\,
\left[ \chi_{-1}(a,b,\zin) - e^{-i b \zin}
\left( -12 i b - 60 \zin  \right)  \right] e^{i b z} \, .
\end{equation}
First note that $\Xi_{-1}(z,a,\zin)\sim \mathcal{O}(z^0)$ in the $z\ll
1$ limit, which follows from the fact that its integrand converges to
a $z$-independent constant when $z \to 0^{+}$. What remains then is to
evaluate the distribution
\begin{align}
\int_{a+2\kappa}^{\infty} \dd b\, e^{-i b \zin}
\left( -12 i b - 60 \zin  \right)  e^{ i b z} & =
e^{ - i (\zin - z) (a + 2\kappa)} \left[\frac{12 i}{(z_{\mathrm{in}} - z)^2}
-\frac{12 \left(a+2 \kappa -5 i\right)}{z_{\mathrm{in}} - z} \right]
\nonumber \\ &
+ 12 \pi \delta'(\zin - z)  - 60 \pi \zin \delta(\zin - z)\, ,
 \end{align}
which follows from \Eqs{alpha1} and (\ref{alpham2}). Fixing $a$ and
$\zin$ in the above, and taking the $0< z\ll 1$ limits then implies
that (also assuming that $\kappa z \ll 1$)
\begin{align}
\int_{a+2\kappa}^{\infty} \dd b\, e^{-i b \zin} \left( -12 i b - 60 \zin  \right)
e^{+ i b z} & \simeq e^{ - i \zin (a + 2\kappa)}
\left[\frac{12 i}{z_{\mathrm{in}}^2}-\frac{12 \left(a+2 \kappa -5 i\right)}
  {z_{\mathrm{in}}} \right] + \mathcal{O}\left(z\right)  \, , 
 \end{align}
which implies that
\begin{align}
\int_{a+2\kappa}^{\infty} \dd b\,  \chi_{-1}(a,b,\zin) e^{i b z} & \sim
\mathcal{O}\left(z^0\right) \ .
\end{align}
Similar computations show that in the super-Hubble limit $0<z \ll 1$
with $\kappa z \ll 1$, $\int_{a+2\kappa}^{\infty} \dd b\;
\chi_{0,+1}(a,b,\zin) e^{i b z} \sim \mathcal{O}(z^0) $, which when
combined with \Eq{gG_FT} show that $ \mathscr{G}(a,z,\zin) \sim
\mathcal{O}(z^{-1})$ in the same limit. Integrating from $a=0$ to
$a=1$ in \Eq{tensorg_def} leaves the same $z$-dependence and so we
conclude that $ g(z,\zin) \sim \mathcal{O}(z^{-1})$.

Combining the above dependences for $h$ and $g$ in \Eq{Ftensorhg} implies
\begin{equation}
\label{eq:FappendixE}
\mathfrak{F}_{\bm{k}}(\eta,\eta_{\mathrm{in}}) \ \simeq \ \frac{\slrl
  H^2 k^2} {1024 \pi^2 \Mp^2 }\frac{60 \pi}{(-k\eta)^2} +
\cO\left[\left(-k\eta\right)^{-1}\right]
\end{equation}
in the super-Hubble limit, as claimed in the main text, see
\Eq{ReFSHresultz}.

\section{Tensor decoherence from a scalar environment}
\label{App:Tensordecohere}

\subsection{General considerations}

Here we consider the interaction Hamiltonian (\ref{HintRdefy_body}) 
\begin{equation}
\label{HintRdefy}
{\cH}_{\mathrm{int}}(\eta) = \tilde{G}(\eta) \int\dd^3 \bm{x}\;
{v}_{ij} (\eta, \bm{x}) \otimes {B^{ij}}(\eta,\bm{x}) \,,
\end{equation}
with $ \tilde G = - (2\Mp a)^{-1}$ defined in \Eq{tildeGvsG}, $v_{ij
}$ the canonical tensor mode for the system modes (related to
$\gamma_{ij }$ by \Eq{canonicaltensor}) and the environmental operator
${B}^{ij}(\bm{x}) = \delta^{ik} \delta^{jl} \partial_{k} {v}
(\eta,\bm{x}) \partial_{l} {v} (\eta,\bm{x}) $ defined in
\Eq{couplingdefy}.

Using an analogous setup as that used for the scalar, the
Nakajima-Zwanzig equation for the reduced density matrix (now for the
graviton) at second-order in $\tilde{G}$ is given by
\begin{align}
\label{NZ_grav}
\frac{\partial {\varrho}}{\partial \eta} &\simeq
- i \tilde{G}(\eta) \mathscr{B}^{ia}(\eta) \int \dd^3\bm{x} \;
\Bigl[  {v}_{ia }(\eta,\bm{x}),  {\varrho}(\eta) \Bigr]
\nonumber \\ &
-  \int \dd^3\bm{x} \int \dd^3\bm{x}'
\int_{\eta_{\mathrm{in} } }^{\eta} \dd \eta' \; \tilde{G}(\eta)
\tilde{G}(\eta')
\bigg\lbrace \Bigl[  {v}_{ia }(\eta,\bm{x}) ,
  {v}_{jb }(\eta',\bm{x}')  {\varrho}(\eta') \Bigr]
C^{iajb}(\eta, \eta'; \bm{x} - \bm{x}')  
\nonumber \\ &
+ \Bigl[  {\varrho}(\eta') {v}_{jb }(\eta',\bm{x}')  ,
  {v}_{ia}(\eta,\bm{x}) \Bigr] C^{iajb\ast}(\eta, \eta';\bm{x} - \bm{x}')
\bigg\rbrace 
\end{align}
\cf \Eq{NZ}, with the definitions
\begin{equation}
\mathscr{B}^{ia}(\eta) := \langle 0_{\ssB} \vert
{B}^{ia}(\eta,\bm{x}) \vert 0_{\ssB} \rangle 
\end{equation}
and 
\begin{equation}
C^{iajb}(\eta,\eta' ; \bm{x} - \bm{x}') =  \langle 0_{\ssB} \left \vert
\left[ {B}^{ia}(\eta,\bm{x}) - \mathscr{B}^{ia}(\eta) \right]
\left[ {B}^{jb}(\eta',\bm{x}') - \mathscr{B}^{jb}(\eta') \right]
\right \vert  0_{\ssB} \rangle \ .
\end{equation}
Using the mode expansion of the scalar as usual one finds that 
\begin{equation}
\mathscr{B}^{ia}(\eta) = \int_{k>k_{\UV}} \frac{\dd^3 \bm{k}}{(2\pi)^3}
k^i k^a |u_{\bm{k}}(\eta)|^2 \ = \ \frac{1}{3} \delta^{ia} \mathscr{B}(\eta)
\end{equation}
with $\mathscr{B}$ given in \Eq{R_1pt_def}\footnote{Notice that since
  $\mathscr{B}^{ia} \propto \delta^{ia}$ then the contraction in the
  first term of \Eq{NZ_grav} vanishes since we assume the graviton is
  traceless, where $\delta^{ia} \gamma_{ia} = 0$} and
\begin{align}
\label{Ctensor_app}
C^{iajb}(\eta,\eta' ; \bm{x} - \bm{x}') & =
2\int_{q,p>k_{\UV}} \frac{\dd^3\bm{q}\; \dd^{3}\bm{p}}{(2\pi)^6}
\; p^i q^a p^j q^b \,  u_{{q}}(\eta) u_{{p}}(\eta) u^{\ast}_{{q}}(\eta')
u^{\ast}_{{p}}(\eta') \, e^{ i ( \bm{q} + \bm{p} ) \cdot ( \bm{x} - \bm{x}' )}.
\end{align}
Using the mode expansion for $v_{ij}$ given in \Eq{tensormodeexp},
repeated here for convenience
\begin{equation}
v_{ij}(\eta,\bm{x}) = \int \frac{\dd^3 \bm{k}}{(2\pi)^{3/2}}
\sum_{\ssP = +,\times} \epsilon_{ij}^{\ssP}(\bm{k}) v_{\bm{k}}^{\ssP}(\eta)
\, e^{i \bm{k} \cdot \bm{x}} \ ,
\end{equation}
one can re-write the above equation as
\begin{align}
\label{NZmom_tensor}
\frac{\partial {\varrho}}{\partial \eta} = &
- (2\pi)^{3/2} \sum_{\ssP,\ssP'} \int_{k < k_{\mathrm{UV}} }
\dd^3\bm{k} \int_{\eta_{\mathrm{in}}}^{\eta} \dd \eta' \ \tilde{G}(\eta)
\tilde{G}(\eta') \epsilon_{ia}^{\ssP}(\bm{k})
\epsilon_{jb}^{\ssP'}(-\bm{k}) \nonumber 
\\ & \times
\left\lbrace \left[ {v}^{\ssP}_{\bm{k}}(\eta) ,
  {v}^{\ssP'}_{-\bm{k}}(\eta') {\varrho}(\eta') \right]
\mathscr{C}^{iajb}_{-\bm{k}}(\eta,\eta')
+ \left[ {\varrho}(\eta') {v}^{\ssP'}_{-\bm{k}}(\eta'),
{v}^{\ssP}_{\bm{k}}(\eta) \right] \mathscr{C}^{iajb\ast}_{-\bm{k}}(\eta,\eta')
\right\rbrace , 
\end{align}
where we define the Fourier transform $\mathscr{C}^{iajb}_{\bm{k}}$ of
the correlator $C^{iajb}(\eta,\eta' ; \bm{y})$ in the variable
$\bm{y}$ where
\begin{align}
\mathscr{C}^{iajb}_{\bm{k}}(\eta,\eta') = & \int
\frac{\dd^3 \bm{y}}{(2\pi)^{3/2}} \; C^{iajb}(\eta,\eta' ; \bm{y})\,
e^{- i \bm{k} \cdot \bm{y}} \nonumber
\\ = & \frac{2}{(2\pi)^{9/2}} \int_{q,p>k_{\UV}} \dd^3\bm{q}\;
\dd^{3}\bm{p} \; p^i q^a p^j q^b \,  u_{{q}}(\eta) u_{{p}}(\eta)
u^{\ast}_{{q}}(\eta') u^{\ast}_{{p}}(\eta') \,
\delta\left(\bm{p} + \bm{q} - \bm{k}\right) \ .\label{App_Ciajb}
\end{align}
From here we note that \Eq{App_Ciajb} implies
$\mathscr{C}^{iajb}_{-\bm{k}}(\eta,\eta')=
\mathscr{C}^{iajb}_{\bm{k}}(\eta,\eta')$, and we also use the symmetry
$\epsilon_{jb}^{\ssP'}(-\bm{k}) {v}^{\ssP'}_{-\bm{k}}(\eta') =
\epsilon_{jb}^{\ssP'}(\bm{k}) {v}^{\ssP'\ast}_{\bm{k}}(\eta')$ for
each polarization $P'$ giving
\begin{align}
\frac{\partial {\varrho}}{\partial \eta} = & - (2\pi)^{3/2}
\sum_{\ssP,\ssP'} \int_{k < k_{\mathrm{UV}} } \dd^3\bm{k}
\int_{\eta_{\mathrm{in}}}^{\eta} \dd \eta' \ \tilde{G}(\eta)
\tilde{G}(\eta') \, \epsilon_{ia}^{\ssP}(\bm{k})
\epsilon_{jb}^{\ssP'}(\bm{k}) \nonumber 
\\ & \times
\left\lbrace \left[ {v}^{\ssP}_{\bm{k}}(\eta) ,
  {v}^{\ssP' \dagger}_{\bm{k}}(\eta') {\varrho}(\eta') \right]
\mathscr{C}^{iajb}_{\bm{k}}(\eta,\eta')
+ \left[ {\varrho}(\eta') {v}^{\ssP'\dagger}_{\bm{k}}(\eta'),
{v}^{\ssP}_{\bm{k}}(\eta) \right] \mathscr{C}^{iajb\ast}_{\bm{k}}(\eta,\eta')
\right\rbrace . 
\end{align}
We next split the density matrix into a product over modes as before, 
with mode labels $\bm{k}$ and polarizations $P = +,\times$  
\begin{equation}
\rho \ = \ \bigotimes_{\bm{k}, \ssP } \rho_{\bm{k}}^{\ssP }
\end{equation}
and repeating the arguments used in the scalar case leads to the
separate evolution equation for each label:
\begin{equation}
\label{NZ_tensor_final}
\frac{\mathcal{V}}{(2\pi)^{3/2}}
\frac{\partial {\varrho}_{\bm{k}}^{\ssP}}{\partial \eta}  = 
- (2\pi)^{3/2} \int_{\eta_{\mathrm{in}}}^{\eta} \dd \eta' 
\tilde{G}(\eta) \tilde{G}(\eta') \left\lbrace
\mathscr{S}_{\bm{k}}(\eta,\eta')\left[ \tilde{v}^{\ssP}_{\bm{k}}(\eta) ,
\tilde{v}^{\ssP}_{\bm{k}}(\eta') {\varrho}_{\bm{k}}^{\ssP}(\eta') \right]
+ \mathrm{h.c.} \right\rbrace .
\end{equation}
with $\tilde{v}$ a proxy for the real and imaginary parts of the field (as in the text below (\ref{alphaRI})) and where we define
\begin{equation}
\label{correlator_contracted}
\epsilon_{ia}^{\ssP}(\bm{k}) \epsilon_{jb}^{\ssP'}(\bm{k}) \;
\mathscr{C}^{iajb}_{\bm{k}}(\eta,\eta') = \delta^{\ssP\ssP'}
\mathscr{S}_{\bm{k}}(\eta,\eta') \,.
\end{equation}

\subsection{Computing the correlation function}

We first note that \Eq{App_Ciajb} implies
$\mathscr{C}^{iajb}_{\bm{k}}(\eta,\eta')$ has the symmetries
$\mathscr{C}_{\bm{k}}^{iajb} = \mathscr{C}_{\bm{k}}^{jaib} =
\mathscr{C}_{\bm{k}}^{ibja} = \mathscr{C}_{\bm{k}}^{aibj}$. The most
general form possible consistent with these symmetries is
\begin{align}
\mathscr{C}_{\bm{k}}^{iajb} &=
\frac{A}{2}\left( \delta^{ia} \delta^{jb} + \delta^{ib} \delta^{ja} \right)
+ B \, \delta^{ij} \delta^{ab} + \frac{C}{4} \,\left( \delta^{ia} k^j k^b
+ \delta^{ja} k^i k^b + \delta^{ib} k^j k^a + \delta^{jb} k^i k^a \right)
\nonumber \\
& + \frac{D}2 \,\left(\delta^{ij} k^a k^b + \delta^{ab} k^i k^j \right)
+ E \, k^i k^j k^a k^b\,,
\end{align}
where the coefficients $A$ through $E$ are functions of $\eta$,
$\eta'$ and $k^2$. Contracting the above form with two polarization
tensors as in~\Eq{correlator_contracted} yields
\begin{equation}
\epsilon_{ia}^{\ssP}(\bm{k}) \epsilon_{jb}^{\ssP'}(\bm{k}) \;
\mathscr{C}^{iajb}_{\bm{k}}(\eta,\eta')
= \delta^{\ssP\ssP'} \left( \frac{A}{2}  + B \right)  \ ,
\end{equation}
which uses
$\epsilon^{\ssP}_{ij}(\bm{k})=\epsilon^{\ssP}_{ji}(\bm{k})$,
$\epsilon^{\ssP}_{ij}(\bm{k}) \epsilon_{ij}^{\ssP'}(\bm{k}) =
\delta^{\ssP\ssP'}$ and $\epsilon^{\ssP}_{ii}(\bm{k}) = k^i
\epsilon_{ij}(\bm{k}) = 0$, showing that
\begin{equation}
\label{curlyS_AB}
\mathscr{S}_{\bm{k}}(\eta,\eta') =  \frac{A}{2}  + B 
\end{equation}
is the function appearing in \Eq{correlator_contracted}. The required
coefficients $A$ and $B$ can be computed by inverting the following
five expressions for rotation-invariant integrals:
\begin{align}
\delta_{ij} \delta_{ab} \mathscr{C}_{\bm{k}}^{iajb}
&= 3A + 9B + k^2 C + 3 k^2 D + k^4 E \,,
\\
\delta_{ia} \delta_{jb} \mathscr{C}_{\bm{k}}^{iajb} &=\mathscr{C}_{\bm{k}}=
6A + 3B + 2 k^2 C + k^2 D + k^4 E  \,,
\\
k_i k_j \delta_{ab} \mathscr{C}_{\bm{k}}^{iajb}
&= k^2 A + 3 k^2 B + k^4 C + 2 k^4 D + k^6 E  \,,
\\
\delta_{ia} k_j k_b \mathscr{C}_{\bm{k}}^{iajb} & =
2 k^2 A + k^2 B + \frac32 \,k^4 C + k^4 D + k^6 E \,,
\end{align}
and
\begin{equation}
k_i k_a k_j k_b \mathscr{C}_{\bm{k}}^{iajb} =
k^4 A + k^4 B + k^6 C + k^6 D + k^8 E \,.
\end{equation}
Inverting these formulas gives the coefficients $A$ through $E$ in
terms of integrals, and using the result in \Eq{curlyS_AB} yields
\begin{align}
\label{Sintform}
\mathscr{S}_{\bm{k}}(\eta,\eta') =& \frac{\delta_{ij} \delta_{ab}
\mathscr{C}_{\bm{k}}^{iajb}}{4}  - \frac{k_i k_j \delta_{ab}
\mathscr{C}_{\bm{k}}^{iajb}}{2 k^2} + \frac{k_i k_a k_j k_b
\mathscr{C}_{\bm{k}}^{iajb}}{4k^4}
\\
=& \frac{2}{(2\pi)^{9/2}} \int_{q,p>k_{\UV}} \dd^3\bm{q}\;
\dd^{3}\bm{p}\, u_{{q}}(\eta) u_{{p}}(\eta) u^{\ast}_{{q}}(\eta')
u^{\ast}_{{p}}(\eta')
\bigg[ \frac{q^2 p^2}{4} -
\frac{ (\bm{k} \cdot \bm{p} )^2 q^2 }{2k^2}
\nonumber \\  &
+ \frac{(\bm{k} \cdot \bm{q})^2 (\bm{k} \cdot \bm{p})^2}{4k^4} \bigg]
\delta( \bm{q} + \bm{p} -\bm{k})
\\
=&\frac{2}{(2\pi)^{9/2}} \int_{q,p>k_{\UV}} \dd q\; \dd p\;
p^4 q^4 u_{{q}}(\eta) u_{{p}}(\eta) u^{\ast}_{{q}}(\eta') u^{\ast}_{{p}}(\eta')
\int_0^{4\pi} \dd^2\Omega_q \, \dd^2\Omega_p \;
\bigg[ \frac{1}{4}
\nonumber \\ &
- \frac{ ({\bm{k}} \cdot {\bm{p}} )^2   }{2 k^2p^2}
+ \frac{({\bm{k}} \cdot {\bm{q}})^2 ({\bm{k}}
  \cdot {\bm{p}})^2}{4k^4 p^2 q^2} \bigg]
\delta( \bm{q} + \bm{p} -\bm{k})  \,, 
\end{align}
which writes the integrals in polar coordinates. The three terms
mainly differ in their angular integrations. Rotation invariance
allows us to choose $\bm{k}$ (as in Appendix \ref{App:envcorr_eta}) to
point along the $z$ axis: $\bm{k} = (0,0,k)$. Using this choice, the
angular integrals become
\begin{align}
\label{angles}
\cJ(p,q,k) :=& \int_0^{4\pi} \dd^2\Omega_q \, \dd^2\Omega_p \;
\bigg[ \frac{1}{4} - \frac{ ({\bm{k}} \cdot {\bm{p}} )^2   }{2k^2 p^2 }
+ \frac{({\bm{k}} \cdot {\bm{q}})^2 ({\bm{k}} \cdot {\bm{p}})^2}
{4k^4 p^2 q^2} \bigg]\delta( \bm{q} + \bm{p} -\bm{k})
\\
= & \frac{1}{q^2}\int_{-1}^{1} \dd \cos\theta_{q} \int_0^{2\pi}
\dd \varphi_{q}\,   \int_{-1}^{1} \dd \cos \theta_{p}
\int_0^{2\pi} \dd \varphi_{p}\,
\bigg[ \frac{1}{4} - \frac{  \cos^2 \theta_p  }{2}
+ \frac{ \cos^2 \theta_q   \cos^2 \theta_p }{4} \bigg] \nonumber
\\
&\times   \delta\left(q - \sqrt{ p^2 + k^2  - 2 k p \cos\theta_p} \right)
\delta\left( \cos\theta_{q} -  \frac{k-p \cos\theta_p}
{\sqrt{p^2 + k^2  - 2 k p \cos\theta_p}}  \right)
\nonumber \\ & \times
\delta\left[\varphi_{q} - (\varphi_p + \pi ) \right]  . 
\end{align}
Integrating over $\varphi_q$, $\varphi_p$ and then $\theta_q$, and
writing $\mu = \cos \theta_p$ yields
\begin{equation} 
\cJ(p,q,k)  = \frac{2\pi}{pqk} \int_{-1}^{1} \dd \mu\;
\bigg[\frac{1}{4} - \frac{\mu^2}{2} + \frac{ ( k-p \mu )^2   \mu^2}
  { 4 ( p^2 + k^2  - 2 k p \mu)} \bigg]
\delta\left( \mu - \frac{p^2 + k^2 - q^2}{2 p k} \right)  \,,
\end{equation}
which uses the identity $\delta\big(q - \sqrt{ p^2 + k^2 - 2 k p \mu }
\; \big) = \frac{q}{kp} \; \delta\left( \mu - \frac{p^2 + k^2 - q^2}{2
  p k} \right)$. The result vanishes unless $-2pk \leq p^2 + k^2 - q^2
\leq 2pk$, which requires the $p$ and $q$ integrals to run over the
region $U$ depicted in \Fig{figure:Int_RegionU}. For $p,q$ in
this region the final angular integral gives
\begin{equation} 
\cJ(p,q,k)  = \frac{\pi}{2pqk} \left\lbrace 1
- \frac{(p^2+k^2-q^2)^2}{2p^2k^2}
+ \frac{[ k^4-(p^2-q^2)^2]^2 }{16  p^2 q^2 k^4} \right\rbrace    \,,
\end{equation}
which when used in \Eq{Sintform} together with the Bunch-Davies mode
functions~\eqref{eq:deSitter:BD:vk} gives
\begin{align} 
\mathscr{S}_{\bm{k}}(\eta,\eta') = & \frac{1}{8(2\pi)^{7/2}k}
\iint_{U}  \dd q  \dd p \left\lbrace q^2 p^2
- \frac{(k^2 + p^2 - q^2)^2 q^2}{2k^2}
+ \frac{\big[k^4 - (p^2 - q^2)^2\big]^2 }{16 k^4} \right\rbrace
\nonumber \\
& \times \left(1-\frac{i}{q \eta}\right)\left(1+\frac{i}{q \eta'}\right)
\left(1-\frac{i}{p \eta}\right) \left(1+\frac{i}{p \eta'}\right)
\, e^{-i (q+p) ( \eta - \eta' ) }  \,. 
\end{align}
The integral over the region $U$ is performed using the coordinate
change $p = \frac12(P+Q)$ and $q = \frac12(P-Q)$, leading to
\begin{align}
\label{Sk_PQ_App}
\mathscr{S}_{\bm{k}}(\eta,\eta') = & \frac{1}{32(2\pi)^{7/2}k }
\int_{-k}^k \dd Q \int_{|Q|  + 2 k_{\UV}}^{\infty} \dd P\;
\bigg[ \frac{(P^2-Q^2)^2}{8} - \frac{ (k^2 + P Q)^2 (P-Q)^2 }{4 k^2}
\nonumber \\ &  
  + \frac{(k^4 - P^2 Q^2)^2}{8 k^4} \bigg] 
\left[1-\frac{2i}{(P-Q) \eta}\right]\left[1+\frac{2i}{(P-Q) \eta'}\right]
\nonumber \\ & \times
\left[1-\frac{2i}{(P+Q) \eta}\right] \left[1+\frac{2i}{(P+Q) \eta'}\right]
\, e^{- i ( \eta - \eta' ) P }.
\end{align}
Note that this integrand is {\it not} symmetric under $Q \to - Q$. By
splitting up the $Q$-integral to be over $[-k,0]$ and $[0,k]$ and then
taking $Q \to -Q$ in the first piece the above simplifies to
\begin{align}
\label{Sk_PQ_App}
\mathscr{S}_{\bm{k}}(\eta,\eta') = &
\frac{1}{128(2\pi)^{7/2} k^5 } \int_{0}^k \dd Q \,
(k^2-Q^2)^2\int_{Q  + 2 k_{\UV}}^{\infty} \dd P\; (k^2-P^2)^2
\left[1-\frac{2i}{(P-Q) \eta}\right]
\nonumber \\ & \times
\left[1+\frac{2i}{(P-Q) \eta'}\right]
\left[1-\frac{2i}{(P+Q) \eta}\right]\left[1+\frac{2i}{(P+Q) \eta'}\right]
\, e^{- i ( \eta - \eta' ) P } .
\end{align}

Our interest is in the leading behaviour as for small $-k\eta$ and so we focus on the $k \to 0$ (and so also $Q \to 0$) limit of \Eq{Sk_PQ_App}, which is
\begin{align}
\label{Sk_PQ_App0}
\mathscr{S}_{\bm{k}}(\eta,\eta')  \simeq &  \frac{1}{128(2\pi)^{7/2} k^5 }
\int_{0}^k \dd Q \; (k^2-Q^2)^2\int_{ 2 k_{\UV}}^{\infty} \dd P\, P^4
\left(1-\frac{2i}{P \eta}\right)^2\left(1+\frac{2i}{P \eta'}\right)^2
e^{- i ( \eta - \eta' ) P }
\\ \simeq &
\frac{1}{240 (2\pi)^{7/2}  } \int_{ 2 k_{\UV}}^{\infty} \dd P\, P^4
\left(1-\frac{2i}{P \eta}\right)^2\left(1+\frac{2i}{P \eta'}\right)^2
e^{- i ( \eta - \eta' ) P } \,,
\end{align}
where the last expression is valid in the limit $k\rightarrow 0$. This
is to be compared with the same limit for the scalar
contribution~\eqref{Ck_PQ}, which is
\begin{align}
  \label{Ck_PQRRR}
  \mathscr{C}_{\bm{k}}(\eta,\eta') = & \frac{1}{32(2\pi)^{7/2}}
  \int_0^k \dd Q \int_{Q  + 2 k_{\UV}}^{\infty} \dd P\;
  \frac{( P^2 + Q^2 - 2 k^2 )^2}{k}  
\left[1-\frac{2i}{(P-Q) \eta}\right]
\left[1+\frac{2i}{(P-Q) \eta'}\right]
\nonumber \\ & \times 
\left[1-\frac{2i}{(P+Q) \eta}\right]
 \left[1+\frac{2i}{(P+Q) \eta'}\right] \; e^{- i ( \eta - \eta' ) P }
 \\ & \simeq \frac{1}{32(2\pi)^{7/2}}   \int_{ 2 k_{\UV}}^{\infty} \dd P\;
 P^4   \left(1-\frac{2i}{P \eta}\right)^2
 \left(1+\frac{2i}{P \eta'}\right)^2 \; e^{- i ( \eta - \eta' ) P },
\end{align}
where, again, the last expression is valid for $k\rightarrow
0$. Therefore, we see that for small $k$
\begin{equation}
\label{415result}
\mathscr{S}_{\bm{k}}(\eta,\eta') \simeq \frac{2}{15} \,
\mathscr{C}_{\bm{k}}(\eta,\eta') \,.
\end{equation}
This can also be derived directly by taking the $\bm{k} \to 0$ limit
of \Eq{angles}, which becomes
\begin{align}
\label{angles0}
\cJ(p,q,k \to 0) := &\frac{1}{q^2} \delta(p-q) \int_0^{4\pi}
\dd^2\Omega_q \, \dd^2\Omega_p \; \bigg[ \frac{1}{4}
  - \frac{ ({\bm{k}} \cdot {\bm{p}} )^2   }{2k^2p^2 }
  + \frac{({\bm{k}} \cdot {\bm{q}})^2
    ({\bm{k}} \cdot {\bm{p}})^2}{4k^4 p^2 q^2} \bigg]
\delta\left( \frac{\bm{q}}{q}+ \frac{\bm{p}}{p}\right)
\\
= & \frac{2\pi}{q^2}\, \delta(p-q) \int_{-1}^1
\dd \mu \; \left( \frac{1}{4} - \frac{ \mu^2   }{2 }
+ \frac{\mu^4}{4} \right) = \frac{4}{15}
\left( \frac{2\pi}{q^2} \right) \, \delta(p-q) \,,
\end{align}
which is $\frac{2}{15}$ times the result obtained when the square
bracket is $1$.

\subsection{Lindblad Coefficient}

Beginning with the Nakajima-Zwanzig equation~\eqref{NZ_tensor_final},
and applying the Markovian approximation in the same spirit as earlier
gives
\begin{equation}
\label{NZ_tensor_final}
\frac{\mathcal{V}}{(2\pi)^{3/2}}
\frac{\partial {\varrho}_{\bm{k}}^{\ssP}}{\partial \eta}  =
- \mathfrak{T}_{\bm{k}}(\eta,\eta_{\mathrm{in}})
\left[ \tilde{v}^{\ssP}_{\bm{k}}(\eta) , \tilde{v}^{\ssP}_{\bm{k}}(\eta')
  {\varrho}_{\bm{k}}^{\ssP}(\eta') \right] -
\mathfrak{T}^{\ast}_{\bm{k}}(\eta,\eta_{\mathrm{in}})
\left[ {\varrho}_{\bm{k}}^{\ssP}(\eta') \tilde{v}^{\ssP}_{\bm{k}}(\eta') ,
  \tilde{v}^{\ssP}_{\bm{k}}(\eta)  \right] ,
\end{equation}
where we define 
\begin{equation}
\label{LindbladT_def}
\mathfrak{T}_{\bm{k}}(\eta,\eta_{\mathrm{in}}) \ = \ (2\pi)^{3/2}
\int_{\eta_{\mathrm{in}}}^{\eta} \dd \eta'\; \tilde{G}(\eta) \,
\tilde{G}(\eta') \, \mathscr{S}_{\bm{k}}(\eta,\eta')
\end{equation}
and we note that again $\mathrm{Re}\left[
  \mathfrak{T}_{\bm{k}}(\eta,\eta_{\mathrm{in}}) \right]$ drives the
decoherence. Using $\tilde{G}(\eta) = H \eta / \Mp$ as well as
\Eq{Sk_PQ_App} means that \Eq{LindbladT_def} can be written as the
triple integral
\begin{align}
  \mathfrak{T}_{\bm{k}}(\eta,\eta_{\mathrm{in}}) &=
  \frac{H^2}{512 \pi^2 \Mp^2 k^5 } \int_{\eta_{\mathrm{in}}}^{\eta} \dd \eta'
  \int_{0}^k \dd Q \int_{Q  + 2 k_{\UV}}^{\infty} \dd P\; \eta \eta' \,
  (k^2-P^2)^2 (k^2-Q^2)^2 \left[1-\frac{2i}{(P-Q) \eta}\right]
  \nonumber \\ & \times
  \left[1+\frac{2i}{(P-Q) \eta'}\right]
  \left[1-\frac{2i}{(P+Q) \eta}\right]
  \left[1+\frac{2i}{(P+Q) \eta'}\right] \; e^{- i ( \eta - \eta' ) P } \; . 
\end{align}
We first evaluate the $\eta'$-integral using the formula
(\ref{etap_integral_for_Fk}), giving
\begin{align}
\mathfrak{T}_{\bm{k}}(\eta,\eta_{\mathrm{in}}) &=
\frac{H^2}{512 \pi^2 \Mp^2 k^5 } \int_{0}^k \dd Q
\int_{Q  + 2 k_{\UV}}^{\infty} \dd P\; \eta \; (k^2-P^2)^2 (k^2-Q^2)^2
\left[1-\frac{2i}{(P-Q) \eta}\right]
\nonumber \\ & \times  
  \left[1-\frac{2i}{(P+Q) \eta}\right] 
  \biggl\lbrace \frac{1}{P^2} - \frac{i \eta}{P} + \frac{4}{P^2 - Q^2}
  - \frac{4 e^{- i P \eta}  \big[ \mathrm{Ei}(i P \eta) + i \pi \big]}
  {P^2 - Q^2}
\nonumber \\ &
  - e^{- i P (\eta - \eta_{\mathrm{in}})}
  \Bigg( \frac{1}{P^2} - \frac{i \eta_{\mathrm{in}}}{P}
  + \frac{4}{P^2 - Q^2} \bigg) + \frac{4 e^{- i P \eta}
    \big[ \mathrm{Ei}(i P \eta_{\rm in}) + i \pi \big]}{P^2 - Q^2}
  \Biggr\rbrace .
\end{align}
The terms involving $\eta_{\rm in}$ can be handled in precisely the
same way as in the discussion around \Eq{gsubdom}, showing that they
contribute only subdominantly for small $(-k\eta)$, and so for our
purposes can be neglected. This leaves
\begin{align}
  \mathfrak{T}_{\bm{k}}(\eta,\eta_{\mathrm{in}}) &\simeq
  \frac{H^2}{512 \pi^2 \Mp^2 k^5 } \int_{0}^k
  \dd Q \int_{Q  + 2 k_{\UV}}^{\infty}
  \dd P\; \eta \; (k^2-P^2)^2 (k^2-Q^2)^2
  \left[1-\frac{2i}{(P-Q) \eta}\right]
\nonumber \\ & \times 
  \left[1-\frac{2i}{(P+Q) \eta}\right] 
  \left\lbrace \frac{1}{P^2} - \frac{i \eta}{P} + \frac{4}{P^2 - Q^2}
  - \frac{4 e^{- i P \eta} \big[ \mathrm{Ei}(i P \eta) + i \pi \big] }
  {P^2 - Q^2} \right\rbrace \,,
\end{align}
which can be rewritten using the change of variables already used
before, reproduced here for convenience, $z = - k \eta$, $\kappa =
{\kUV}/{k}$, $x = - Q \eta$ and $y = - P \eta$ as
\begin{equation}
\mathfrak{T}_{\bm{k}}   \simeq
\frac{H^2 k^2}{512 \pi^2 \Mp^2 z^3} \int_0^{z}
\dd x \;\left( \frac{x^2}{z^2} - 1 \right)^2 h(x,z)
\end{equation}
with 
\begin{align}
h(x,z) := \int_{x + 2 \kappa z }^{\infty} \dd y \;
( y^2 - z^2 )^2 \left( 1 - \frac{2i}{x-y} \right)
\left( 1 + \frac{2i}{x + y} \right) \left\lbrace - \frac{1}{y^2 }
- \frac{i}{y} + \sfrac{4 - 4 e^{i y } \left[ \mathrm{Ei}( - i y )
+ i \pi \right]  }{x^2 - y^2} \right\rbrace \ .  
\end{align}
This integral diverges (as did the previous examples), since the
integrand approaches $- y^2$ for large $y$. We handle this divergence
the same way as before by subtracting the divergent large-$y$ form,
leaving a convergent result (see the discussion around
\Eq{dimregfdef}).

Taking the $z \to 0$ limit (which also implies $x \to 0$) then gives
the leading contribution
\begin{align}
\mathfrak{T}_{\bm{k}} &\simeq
\frac{H^2 k^2}{512 \pi^2 \Mp^2 z^3} h^{\rm reg}(0,0) \int_0^{z}
\dd x \;\left( \frac{x^2}{z^2} - 1 \right)^2 =
\frac{H^2 k^2}{512 \pi^2 \Mp^2 z^2}
\left[ \frac{8h^{\rm reg}(0,0)}{15} \right]
\end{align}
where (compare to \Eqs{f00eq} and \pref{f00eval})
\begin{align}
h^{\rm reg}(0,0) &= \int_{0}^{\infty} \dd y
\left( y^4  \left( 1 + \frac{2i}{ y} \right)^2
\left\lbrace- \frac{1}{y^2 } - \frac{i}{y}
- \frac{4 - 4 e^{i y } \left[ \mathrm{Ei}( - i y )
+ i \pi \right]}{  y^2} \right\rbrace + y^2 \right) \\
&= \frac{f^{\rm reg}(0,0)}{3} = 20 \pi  \,.  
\end{align}
This leads to the following $\eta_{\rm in}$-independent leading behaviour
\begin{align} 
\mathrm{Re}\left[ \mathfrak{T}_{\bm{k}}(\eta,\eta_{\mathrm{in}}) \right] & \simeq
\frac{H^2 k^2}{48 \pi \Mp^2 z^2}
= \frac{H^2 k^2}{1024 \pi^2 \Mp^2} \bigg[ \frac{64\pi}{3z^2}
+ \mathcal{O}(z^{-1}) \bigg] 
\end{align}
and so comparing with \Eq{ReFSHresult} we see that 
\begin{align} 
\mathrm{Re}\left[ \mathfrak{T}_{\bm{k}}(\eta,\eta_{\rm in}) \right]
& \simeq \frac{16}{15\, \slrl } \mathrm{Re}
\left[\mathfrak{F}_{\bm{k}}(\eta,\eta_{\mathrm{in}})\right] 
\end{align}
in the super-Hubble limit, as claimed in the main text. The factor
$16/(15\slrl )$ has two sources: the factor of $\frac{2}{15}$ seen in
\Eq{415result} together with the factor of $8/\slrl $ coming from
the change in effective coupling noted in \Eq{Ffrakdefy}.

\bibliographystyle{JHEP}
\bibliography{dSDecoherence.bib}

\providecommand{\href}[2]{#2}\begingroup\raggedright\begin{thebibliography}{10}

\bibitem{eBOSS:2020yzd}
{\scshape eBOSS} collaboration, S.~Alam et~al., \emph{{Completed SDSS-IV
  extended Baryon Oscillation Spectroscopic Survey: Cosmological implications
  from two decades of spectroscopic surveys at the Apache Point Observatory}},
  \href{http://dx.doi.org/10.1103/PhysRevD.103.083533}{\emph{Phys. Rev. D} {\bf
  103} (2021) 083533}, [\href{http://arxiv.org/abs/2007.08991}{{\tt
  2007.08991}}].

\bibitem{Planck:2018vyg}
{\scshape Planck} collaboration, N.~Aghanim et~al., \emph{{Planck 2018 results.
  VI. Cosmological parameters}},
  \href{http://dx.doi.org/10.1051/0004-6361/201833910}{\emph{Astron.
  Astrophys.} {\bf 641} (2020) A6},
  [\href{http://arxiv.org/abs/1807.06209}{{\tt 1807.06209}}].

\bibitem{Mukhanov:1981xt}
V.~F. Mukhanov and G.~Chibisov, \emph{{Quantum Fluctuation and Nonsingular
  Universe.}}, {\emph{JETP Lett.} {\bf 33} (1981) 532--535}.

\bibitem{Guth:1982ec}
A.~H. Guth and S.~Pi, \emph{{Fluctuations in the New Inflationary Universe}},
  \href{http://dx.doi.org/10.1103/PhysRevLett.49.1110}{\emph{Phys. Rev. Lett.}
  {\bf 49} (1982) 1110--1113}.

\bibitem{Hawking:1982cz}
S.~Hawking, \emph{{The Development of Irregularities in a Single Bubble
  Inflationary Universe}},
  \href{http://dx.doi.org/10.1016/0370-2693(82)90373-2}{\emph{Phys.Lett.} {\bf
  B115} (1982) 295}.

\bibitem{Starobinsky:1982ee}
A.~A. Starobinsky, \emph{{Dynamics of Phase Transition in the New Inflationary
  Universe Scenario and Generation of Perturbations}},
  \href{http://dx.doi.org/10.1016/0370-2693(82)90541-X}{\emph{Phys. Lett.} {\bf
  117B} (1982) 175--178}.

\bibitem{Bardeen:1983qw}
J.~M. Bardeen, P.~J. Steinhardt and M.~S. Turner, \emph{{Spontaneous Creation
  of Almost Scale - Free Density Perturbations in an Inflationary Universe}},
  \href{http://dx.doi.org/10.1103/PhysRevD.28.679}{\emph{Phys. Rev.} {\bf D28}
  (1983) 679}.

\bibitem{Mukhanov:1988jd}
V.~F. Mukhanov, \emph{{Quantum Theory of Gauge Invariant Cosmological
  Perturbations}}, {\emph{Sov. Phys. JETP} {\bf 67} (1988) 1297--1302}.

\bibitem{Brandenberger:1990bx}
R.~H. Brandenberger, R.~Laflamme and M.~Mijic, \emph{{Classical Perturbations
  From Decoherence of Quantum Fluctuations in the Inflationary Universe}},
  \href{http://dx.doi.org/10.1142/S0217732390002651}{\emph{Mod. Phys. Lett. A}
  {\bf 5} (1990) 2311--2318}.

\bibitem{Kiefer:2008ku}
C.~Kiefer and D.~Polarski, \emph{{Why do cosmological perturbations look
  classical to us?}}, \href{http://dx.doi.org/10.1166/asl.2009.1023}{\emph{Adv.
  Sci. Lett.} {\bf 2} (2009) 164--173},
  [\href{http://arxiv.org/abs/0810.0087}{{\tt 0810.0087}}].

\bibitem{Polarski:1995jg}
D.~Polarski and A.~A. Starobinsky, \emph{{Semiclassicality and decoherence of
  cosmological perturbations}},
  \href{http://dx.doi.org/10.1088/0264-9381/13/3/006}{\emph{Class. Quant.
  Grav.} {\bf 13} (1996) 377--392},
  [\href{http://arxiv.org/abs/gr-qc/9504030}{{\tt gr-qc/9504030}}].

\bibitem{Kiefer:1998qe}
C.~Kiefer, D.~Polarski and A.~A. Starobinsky, \emph{{Quantum to classical
  transition for fluctuations in the early universe}},
  \href{http://dx.doi.org/10.1142/S0218271898000292}{\emph{Int. J. Mod. Phys.
  D} {\bf 07} (1998) 455--462}, [\href{http://arxiv.org/abs/gr-qc/9802003}{{\tt
  gr-qc/9802003}}].

\bibitem{Lombardo:2005iz}
F.~C. Lombardo and D.~Lopez~Nacir, \emph{{Decoherence during inflation: The
  Generation of classical inhomogeneities}},
  \href{http://dx.doi.org/10.1103/PhysRevD.72.063506}{\emph{Phys. Rev.} {\bf
  D72} (2005) 063506}, [\href{http://arxiv.org/abs/gr-qc/0506051}{{\tt
  gr-qc/0506051}}].

\bibitem{Burgess:2006jn}
C.~P. Burgess, R.~Holman and D.~Hoover, \emph{{Decoherence of inflationary
  primordial fluctuations}},
  \href{http://dx.doi.org/10.1103/PhysRevD.77.063534}{\emph{Phys.Rev.} {\bf
  D77} (2008) 063534}, [\href{http://arxiv.org/abs/astro-ph/0601646}{{\tt
  astro-ph/0601646}}].

\bibitem{Martineau:2006ki}
P.~Martineau, \emph{{On the decoherence of primordial fluctuations during
  inflation}},
  \href{http://dx.doi.org/10.1088/0264-9381/24/23/006}{\emph{Class. Quant.
  Grav.} {\bf 24} (2007) 5817--5834},
  [\href{http://arxiv.org/abs/astro-ph/0601134}{{\tt astro-ph/0601134}}].

\bibitem{Sharman:2007gi}
J.~W. Sharman and G.~D. Moore, \emph{{Decoherence due to the Horizon after
  Inflation}},
  \href{http://dx.doi.org/10.1088/1475-7516/2007/11/020}{\emph{JCAP} {\bf 0711}
  (2007) 020}, [\href{http://arxiv.org/abs/0708.3353}{{\tt 0708.3353}}].

\bibitem{Burgess:2014eoa}
C.~P. Burgess, R.~Holman, G.~Tasinato and M.~Williams, \emph{{EFT Beyond the
  Horizon: Stochastic Inflation and How Primordial Quantum Fluctuations Go
  Classical}}, \href{http://dx.doi.org/10.1007/JHEP03(2015)090}{\emph{JHEP}
  {\bf 03} (2015) 090}, [\href{http://arxiv.org/abs/1408.5002}{{\tt
  1408.5002}}].

\bibitem{Martin:2015qta}
J.~Martin and V.~Vennin, \emph{{Quantum Discord of Cosmic Inflation: Can we
  show that CMB Anisotropies are of Quantum-Mechanical Origin?}},
  \href{http://dx.doi.org/10.1103/PhysRevD.93.023505}{\emph{Phys. Rev.} {\bf
  D93} (2016) 023505}, [\href{http://arxiv.org/abs/1510.04038}{{\tt
  1510.04038}}].

\bibitem{Martin:2017zxs}
J.~Martin and V.~Vennin, \emph{{Obstructions to Bell CMB Experiments}},
  \href{http://dx.doi.org/10.1103/PhysRevD.96.063501}{\emph{Phys. Rev.} {\bf
  D96} (2017) 063501}, [\href{http://arxiv.org/abs/1706.05001}{{\tt
  1706.05001}}].

\bibitem{Campo:2005sv}
D.~Campo and R.~Parentani, \emph{{Inflationary spectra and violations of Bell
  inequalities}},
  \href{http://dx.doi.org/10.1103/PhysRevD.74.025001}{\emph{Phys. Rev.} {\bf
  D74} (2006) 025001}, [\href{http://arxiv.org/abs/astro-ph/0505376}{{\tt
  astro-ph/0505376}}].

\bibitem{Maldacena:2015bha}
J.~Maldacena, \emph{{A model with cosmological Bell inequalities}},
  \href{http://arxiv.org/abs/1508.01082}{{\tt 1508.01082}}.

\bibitem{Martin:2021znx}
J.~Martin, A.~Micheli and V.~Vennin, \emph{{Discord and decoherence}},
  \href{http://dx.doi.org/10.1088/1475-7516/2022/04/051}{\emph{JCAP} {\bf 04}
  (2022) 051}, [\href{http://arxiv.org/abs/2112.05037}{{\tt 2112.05037}}].

\bibitem{Tsamis:2005hd}
N.~C. Tsamis and R.~P. Woodard, \emph{{Stochastic quantum gravitational
  inflation}},
  \href{http://dx.doi.org/10.1016/j.nuclphysb.2005.06.031}{\emph{Nucl. Phys. B}
  {\bf 724} (2005) 295--328}, [\href{http://arxiv.org/abs/gr-qc/0505115}{{\tt
  gr-qc/0505115}}].

\bibitem{Burgess:2009bs}
C.~P. Burgess, L.~Leblond, R.~Holman and S.~Shandera, \emph{{Super-Hubble de
  Sitter Fluctuations and the Dynamical RG}},
  \href{http://dx.doi.org/10.1088/1475-7516/2010/03/033}{\emph{JCAP} {\bf 03}
  (2010) 033}, [\href{http://arxiv.org/abs/0912.1608}{{\tt 0912.1608}}].

\bibitem{Giddings:2011zd}
S.~B. Giddings and M.~S. Sloth, \emph{{Cosmological observables, IR growth of
  fluctuations, and scale-dependent anisotropies}},
  \href{http://dx.doi.org/10.1103/PhysRevD.84.063528}{\emph{Phys. Rev. D} {\bf
  84} (2011) 063528}, [\href{http://arxiv.org/abs/1104.0002}{{\tt 1104.0002}}].

\bibitem{Burgess:2020tbq}
C.~P. Burgess, \emph{{Introduction to Effective Field Theory}}.
\newblock Cambridge University Press, 12, 2020,
  \href{http://dx.doi.org/10.1017/9781139048040}{10.1017/9781139048040}.

\bibitem{Colas:2022hlq}
T.~Colas, J.~Grain and V.~Vennin, \emph{{Benchmarking the cosmological master
  equations}},  \href{http://arxiv.org/abs/2209.01929}{{\tt 2209.01929}}.

\bibitem{Kaplanek:2019dqu}
G.~Kaplanek and C.~P. Burgess, \emph{{Hot Accelerated Qubits: Decoherence,
  Thermalization, Secular Growth and Reliable Late-time Predictions}},
  \href{http://dx.doi.org/10.1007/JHEP03(2020)008}{\emph{JHEP} {\bf 03} (2020)
  008}, [\href{http://arxiv.org/abs/1912.12951}{{\tt 1912.12951}}].

\bibitem{Kaplanek:2019vzj}
G.~Kaplanek and C.~P. Burgess, \emph{{Hot Cosmic Qubits: Late-Time de Sitter
  Evolution and Critical Slowing Down}},
  \href{http://dx.doi.org/10.1007/JHEP02(2020)053}{\emph{JHEP} {\bf 02} (2020)
  053}, [\href{http://arxiv.org/abs/1912.12955}{{\tt 1912.12955}}].

\bibitem{Kaplanek:2020iay}
G.~Kaplanek and C.~P. Burgess, \emph{{Qubits on the Horizon: Decoherence and
  Thermalization near Black Holes}},
  \href{http://dx.doi.org/10.1007/JHEP01(2021)098}{\emph{JHEP} {\bf 01} (2021)
  098}, [\href{http://arxiv.org/abs/2007.05984}{{\tt 2007.05984}}].

\bibitem{Kaplanek:2021fnl}
G.~Kaplanek, C.~P. Burgess and R.~Holman, \emph{{Qubit heating near a
  hotspot}}, \href{http://dx.doi.org/10.1007/JHEP08(2021)132}{\emph{JHEP} {\bf
  08} (2021) 132}, [\href{http://arxiv.org/abs/2106.10803}{{\tt 2106.10803}}].

\bibitem{Unruh:1976db}
W.~G. Unruh, \emph{{Notes on black hole evaporation}},
  \href{http://dx.doi.org/10.1103/PhysRevD.14.870}{\emph{Phys. Rev. D} {\bf 14}
  (1976) 870}.

\bibitem{DeWitt:1980hx}
B.~S. DeWitt, \emph{{Quantum gravity: the new synthesis}}, pp.~680--745.
\newblock 1980.

\bibitem{Agon:2014uxa}
C.~Agon, V.~Balasubramanian, S.~Kasko and A.~Lawrence, \emph{{Coarse Grained
  Quantum Dynamics}},
  \href{http://dx.doi.org/10.1103/PhysRevD.98.025019}{\emph{Phys. Rev. D} {\bf
  98} (2018) 025019}, [\href{http://arxiv.org/abs/1412.3148}{{\tt 1412.3148}}].

\bibitem{Burgess:2015ajz}
C.~P. Burgess, R.~Holman and G.~Tasinato, \emph{{Open EFTs, IR effects \&
  late-time resummations: systematic corrections in stochastic inflation}},
  \href{http://dx.doi.org/10.1007/JHEP01(2016)153}{\emph{JHEP} {\bf 01} (2016)
  153}, [\href{http://arxiv.org/abs/1512.00169}{{\tt 1512.00169}}].

\bibitem{Braaten:2016sja}
E.~Braaten, H.~W. Hammer and G.~P. Lepage, \emph{{Open Effective Field Theories
  from Deeply Inelastic Reactions}},
  \href{http://dx.doi.org/10.1103/PhysRevD.94.056006}{\emph{Phys. Rev. D} {\bf
  94} (2016) 056006}, [\href{http://arxiv.org/abs/1607.02939}{{\tt
  1607.02939}}].

\bibitem{Martin:2018zbe}
J.~Martin and V.~Vennin, \emph{{Observational constraints on quantum
  decoherence during inflation}},  \href{http://arxiv.org/abs/1801.09949}{{\tt
  1801.09949}}.

\bibitem{Martin:2018lin}
J.~Martin and V.~Vennin, \emph{{Non Gaussianities from Quantum Decoherence
  during Inflation}},
  \href{http://dx.doi.org/10.1088/1475-7516/2018/06/037}{\emph{JCAP} {\bf 06}
  (2018) 037}, [\href{http://arxiv.org/abs/1805.05609}{{\tt 1805.05609}}].

\bibitem{Burgess:2021luo}
C.~P. Burgess, R.~Holman and G.~Kaplanek, \emph{{Quantum Hotspots: Mean Fields,
  Open EFTs, Nonlocality and Decoherence Near Black Holes}},
  \href{http://dx.doi.org/10.1002/prop.202200019}{\emph{Fortsch. Phys.} {\bf
  70} (2022) 2200019}, [\href{http://arxiv.org/abs/2106.10804}{{\tt
  2106.10804}}].

\bibitem{brahma2022universal}
S.~Brahma, A.~Berera and J.~C. Figueroa, \emph{Universal signature of quantum
  entanglement across cosmological distances}, {\emph{Classical and Quantum
  Gravity} (2022) }.

\bibitem{Hammou:2022sol}
A.~D. Hammou and N.~Bartolo, \emph{{Cosmic decoherence: primordial power
  spectra and non-Gaussianities}},  \href{http://arxiv.org/abs/2211.07598}{{\tt
  2211.07598}}.

\bibitem{Lindblad:1975ef}
G.~Lindblad, \emph{{On the Generators of Quantum Dynamical Semigroups}},
  \href{http://dx.doi.org/10.1007/BF01608499}{\emph{Commun. Math. Phys.} {\bf
  48} (1976) 119}.

\bibitem{Gorini:1976cm}
V.~Gorini, A.~Frigerio, M.~Verri, A.~Kossakowski and E.~C.~G. Sudarshan,
  \emph{{Properties of Quantum Markovian Master Equations}},
  \href{http://dx.doi.org/10.1016/0034-4877(78)90050-2}{\emph{Rept. Math.
  Phys.} {\bf 13} (1978) 149}.

\bibitem{Baumann:2009ds}
D.~Baumann, \emph{{Inflation}},  in \emph{{Theoretical Advanced Study Institute
  in Elementary Particle Physics}: {Physics of the Large and the Small}},
  pp.~523--686, 2011.
\newblock \href{http://arxiv.org/abs/0907.5424}{{\tt 0907.5424}}.
\newblock \href{http://dx.doi.org/10.1142/9789814327183_0010}{DOI}.

\bibitem{Adshead:2017srh}
P.~Adshead, C.~P. Burgess, R.~Holman and S.~Shandera, \emph{{Power-counting
  during single-field slow-roll inflation}},
  \href{http://dx.doi.org/10.1088/1475-7516/2018/02/016}{\emph{JCAP} {\bf 02}
  (2018) 016}, [\href{http://arxiv.org/abs/1708.07443}{{\tt 1708.07443}}].

\bibitem{Babic:2019ify}
I.~Babic, C.~P. Burgess and G.~Geshnizjani, \emph{{Keeping an eye on DBI:
  power-counting for small-c$_s$ cosmology}},
  \href{http://dx.doi.org/10.1088/1475-7516/2020/05/023}{\emph{JCAP} {\bf 05}
  (2020) 023}, [\href{http://arxiv.org/abs/1910.05277}{{\tt 1910.05277}}].

\bibitem{Maldacena:2002vr}
J.~M. Maldacena, \emph{{Non-Gaussian features of primordial fluctuations in
  single field inflationary models}},
  \href{http://dx.doi.org/10.1088/1126-6708/2003/05/013}{\emph{JHEP} {\bf 05}
  (2003) 013}, [\href{http://arxiv.org/abs/astro-ph/0210603}{{\tt
  astro-ph/0210603}}].

\bibitem{Kiefer:2006je}
C.~Kiefer, I.~Lohmar, D.~Polarski and A.~A. Starobinsky, \emph{{Pointer states
  for primordial fluctuations in inflationary cosmology}},
  \href{http://dx.doi.org/10.1088/0264-9381/24/7/002}{\emph{Class. Quant.
  Grav.} {\bf 24} (2007) 1699--1718},
  [\href{http://arxiv.org/abs/astro-ph/0610700}{{\tt astro-ph/0610700}}].

\bibitem{Albrecht:1992kf}
A.~Albrecht, P.~Ferreira, M.~Joyce and T.~Prokopec, \emph{{Inflation and
  squeezed quantum states}},
  \href{http://dx.doi.org/10.1103/PhysRevD.50.4807}{\emph{Phys. Rev. D} {\bf
  50} (1994) 4807--4820}, [\href{http://arxiv.org/abs/astro-ph/9303001}{{\tt
  astro-ph/9303001}}].

\bibitem{Starobinsky:1986fx}
A.~A. Starobinsky, \emph{{Stochastic de Sitter (inflationary) stage in the
  early Universe}},
  \href{http://dx.doi.org/10.1007/3-540-16452-9_6}{\emph{Lect. Notes Phys.}
  {\bf 246} (1986) 107--126}.

\bibitem{Starobinsky:1994bd}
A.~A. Starobinsky and J.~Yokoyama, \emph{{Equilibrium state of a
  selfinteracting scalar field in the De Sitter background}},
  \href{http://dx.doi.org/10.1103/PhysRevD.50.6357}{\emph{Phys. Rev. D} {\bf
  50} (1994) 6357--6368}, [\href{http://arxiv.org/abs/astro-ph/9407016}{{\tt
  astro-ph/9407016}}].

\bibitem{Mijic:1994vv}
M.~Mijic, \emph{{Stochastic dynamics of coarse grained quantum fields in the
  inflationary universe}},
  \href{http://dx.doi.org/10.1103/PhysRevD.49.6434}{\emph{Phys. Rev. D} {\bf
  49} (1994) 6434--6441}, [\href{http://arxiv.org/abs/gr-qc/9401030}{{\tt
  gr-qc/9401030}}].

\bibitem{Seery:2010kh}
D.~Seery, \emph{{Infrared effects in inflationary correlation functions}},
  \href{http://dx.doi.org/10.1088/0264-9381/27/12/124005}{\emph{Class. Quant.
  Grav.} {\bf 27} (2010) 124005}, [\href{http://arxiv.org/abs/1005.1649}{{\tt
  1005.1649}}].

\bibitem{Prokopec:2008gw}
T.~Prokopec, N.~C. Tsamis and R.~P. Woodard, \emph{{Two loop stress-energy
  tensor for inflationary scalar electrodynamics}},
  \href{http://dx.doi.org/10.1103/PhysRevD.78.043523}{\emph{Phys. Rev. D} {\bf
  78} (2008) 043523}, [\href{http://arxiv.org/abs/0802.3673}{{\tt 0802.3673}}].

\bibitem{Cohen:2020php}
T.~Cohen and D.~Green, \emph{{Soft de Sitter Effective Theory}},
  \href{http://dx.doi.org/10.1007/JHEP12(2020)041}{\emph{JHEP} {\bf 12} (2020)
  041}, [\href{http://arxiv.org/abs/2007.03693}{{\tt 2007.03693}}].

\bibitem{Cohen:2021fzf}
T.~Cohen, D.~Green, A.~Premkumar and A.~Ridgway, \emph{{Stochastic Inflation at
  NNLO}}, \href{http://dx.doi.org/10.1007/JHEP09(2021)159}{\emph{JHEP} {\bf 09}
  (2021) 159}, [\href{http://arxiv.org/abs/2106.09728}{{\tt 2106.09728}}].

\bibitem{Baumgart:2019clc}
M.~Baumgart and R.~Sundrum, \emph{{De Sitter Diagrammar and the Resummation of
  Time}}, \href{http://dx.doi.org/10.1007/JHEP07(2020)119}{\emph{JHEP} {\bf 07}
  (2020) 119}, [\href{http://arxiv.org/abs/1912.09502}{{\tt 1912.09502}}].

\bibitem{Liddle:1994dx}
A.~R. Liddle, P.~Parsons and J.~D. Barrow, \emph{{Formalizing the slow roll
  approximation in inflation}},
  \href{http://dx.doi.org/10.1103/PhysRevD.50.7222}{\emph{Phys. Rev. D} {\bf
  50} (1994) 7222--7232}, [\href{http://arxiv.org/abs/astro-ph/9408015}{{\tt
  astro-ph/9408015}}].

\bibitem{Schwarz:2001vv}
D.~J. Schwarz, C.~A. Terrero-Escalante and A.~A. Garcia, \emph{{Higher order
  corrections to primordial spectra from cosmological inflation}},
  \href{http://dx.doi.org/10.1016/S0370-2693(01)01036-X}{\emph{Phys. Lett.}
  {\bf B517} (2001) 243--249},
  [\href{http://arxiv.org/abs/astro-ph/0106020}{{\tt astro-ph/0106020}}].

\bibitem{Leach:2002ar}
S.~M. Leach, A.~R. Liddle, J.~Martin and D.~J. Schwarz, \emph{{Cosmological
  parameter estimation and the inflationary cosmology}},
  \href{http://dx.doi.org/10.1103/PhysRevD.66.023515}{\emph{Phys. Rev.} {\bf
  D66} (2002) 023515}, [\href{http://arxiv.org/abs/astro-ph/0202094}{{\tt
  astro-ph/0202094}}].

\bibitem{Planck:2018jri}
{\scshape Planck} collaboration, Y.~Akrami et~al., \emph{{Planck 2018 results.
  X. Constraints on inflation}},
  \href{http://dx.doi.org/10.1051/0004-6361/201833887}{\emph{Astron.
  Astrophys.} {\bf 641} (2020) A10},
  [\href{http://arxiv.org/abs/1807.06211}{{\tt 1807.06211}}].

\bibitem{Kodama:1985bj}
H.~Kodama and M.~Sasaki, \emph{{Cosmological Perturbation Theory}},
  \href{http://dx.doi.org/10.1143/PTPS.78.1}{\emph{Prog. Theor. Phys. Suppl.}
  {\bf 78} (1984) 1--166}.

\bibitem{Mukhanov:1990me}
V.~F. Mukhanov, H.~Feldman and R.~H. Brandenberger, \emph{{Theory of
  cosmological perturbations. Part 1. Classical perturbations. Part 2. Quantum
  theory of perturbations. Part 3. Extensions}},
  \href{http://dx.doi.org/10.1016/0370-1573(92)90044-Z}{\emph{Phys. Rept.} {\bf
  215} (1992) 203--333}.

\bibitem{Shandera:2017qkg}
S.~Shandera, N.~Agarwal and A.~Kamal, \emph{{Open quantum cosmological
  system}}, \href{http://dx.doi.org/10.1103/PhysRevD.98.083535}{\emph{Phys.
  Rev. D} {\bf 98} (2018) 083535}, [\href{http://arxiv.org/abs/1708.00493}{{\tt
  1708.00493}}].

\bibitem{Bunch:1978yq}
T.~Bunch and P.~Davies, \emph{{Quantum Field Theory in de Sitter Space:
  Renormalization by Point Splitting}},
  \href{http://dx.doi.org/10.1098/rspa.1978.0060}{\emph{Proc.Roy.Soc.Lond.}
  {\bf A360} (1978) 117--134}.

\bibitem{Burgess:2003jk}
C.~P. Burgess, \emph{{Quantum gravity in everyday life: General relativity as
  an effective field theory}},
  \href{http://dx.doi.org/10.12942/lrr-2004-5}{\emph{Living Rev. Rel.} {\bf 7}
  (2004) 5--56}, [\href{http://arxiv.org/abs/gr-qc/0311082}{{\tt
  gr-qc/0311082}}].

\bibitem{Serafini:2003ke}
A.~Serafini, F.~Illuminati and S.~De~Siena, \emph{{Von Neumann entropy, mutual
  information and total correlations of Gaussian states}},
  \href{http://dx.doi.org/10.1088/0953-4075/37/2/L02}{\emph{J. Phys. B} {\bf
  37} (2004) L21}, [\href{http://arxiv.org/abs/quant-ph/0307073}{{\tt
  quant-ph/0307073}}].

\bibitem{Grain:2019vnq}
J.~Grain and V.~Vennin, \emph{{Canonical transformations and squeezing
  formalism in cosmology}},
  \href{http://dx.doi.org/10.1088/1475-7516/2020/02/022}{\emph{JCAP} {\bf 02}
  (2020) 022}, [\href{http://arxiv.org/abs/1910.01916}{{\tt 1910.01916}}].

\bibitem{Colas:2021llj}
T.~Colas, J.~Grain and V.~Vennin, \emph{{Four-mode squeezed states: two-field
  quantum systems and the symplectic group $\mathrm {Sp}(4,{\mathbb {R}})$}},
  \href{http://dx.doi.org/10.1140/epjc/s10052-021-09922-y}{\emph{Eur. Phys. J.
  C} {\bf 82} (2022) 6}, [\href{http://arxiv.org/abs/2104.14942}{{\tt
  2104.14942}}].

\bibitem{Nelson:2016kjm}
E.~Nelson, \emph{{Quantum Decoherence During Inflation from Gravitational
  Nonlinearities}},
  \href{http://dx.doi.org/10.1088/1475-7516/2016/03/022}{\emph{JCAP} {\bf 1603}
  (2016) 022}, [\href{http://arxiv.org/abs/1601.03734}{{\tt 1601.03734}}].

\bibitem{Hazumi:2019lys}
M.~Hazumi et~al., \emph{{LiteBIRD: A Satellite for the Studies of B-Mode
  Polarization and Inflation from Cosmic Background Radiation Detection}},
  \href{http://dx.doi.org/10.1007/s10909-019-02150-5}{\emph{J. Low Temp. Phys.}
  {\bf 194} (2019) 443--452}.

\bibitem{gong2019quantum}
J.-O. Gong and M.-S. Seo, \emph{Quantum non-linear evolution of inflationary
  tensor perturbations}, {\emph{Journal of High Energy Physics} {\bf 2019}
  (2019) 1--39}.

\bibitem{ye2018quantum}
G.~Ye and Y.-S. Piao, \emph{Quantum decoherence of primordial perturbations
  through nonlinear scaler-tensor interaction}, {\emph{arXiv:1806.07672} (2018)
  }.

\bibitem{Cheung:2007st}
C.~Cheung, P.~Creminelli, A.~L. Fitzpatrick, J.~Kaplan and L.~Senatore,
  \emph{{The Effective Field Theory of Inflation}},
  \href{http://dx.doi.org/10.1088/1126-6708/2008/03/014}{\emph{JHEP} {\bf 03}
  (2008) 014}, [\href{http://arxiv.org/abs/0709.0293}{{\tt 0709.0293}}].

\bibitem{Fujita:2018zbr}
T.~Fujita, I.~Obata, T.~Tanaka and S.~Yokoyama, \emph{{Statistically
  Anisotropic Tensor Modes from Inflation}},
  \href{http://dx.doi.org/10.1088/1475-7516/2018/07/023}{\emph{JCAP} {\bf 07}
  (2018) 023}, [\href{http://arxiv.org/abs/1801.02778}{{\tt 1801.02778}}].

\bibitem{Kanno:2020usf}
S.~Kanno, J.~Soda and J.~Tokuda, \emph{{Noise and decoherence induced by
  gravitons}}, \href{http://dx.doi.org/10.1103/PhysRevD.103.044017}{\emph{Phys.
  Rev. D} {\bf 103} (2021) 044017},
  [\href{http://arxiv.org/abs/2007.09838}{{\tt 2007.09838}}].

\end{thebibliography}\endgroup

\end{document}